\makeatletter\def\input@path{
{graphics/}
}\makeatother%

\documentclass[final]{article}
\usepackage{ragged2e}
\usepackage{caption}
\usepackage[margin=0.5in]{geometry}
\usepackage{fancyhdr}
\usepackage{gensymb}
\makeatletter%

\RaggedRight

\renewcommand{\maketitle}{
\huge
\@title
\medskip
\global\@topnum\z@
   }
   
 \renewenvironment{author} {
 \large
 \emph
  }

\newcommand{\keywords}[1]{
\medskip
Keywords: \textit{#1}
}

\newcommand{\dedication}[1]{
\medskip
\textit{#1}
}

\newenvironment{affiliations}{
\medskip
\large
}

\renewenvironment{abstract}{
\small
\medskip\medskip
}

\makeatother%

\usepackage{iftex}
\ifPDFTeX
   \usepackage[pdftex]{graphicx}
   \usepackage[utf8]{inputenc}
  \usepackage{lmodern}
\else 
   \ifXeTeX
     \usepackage[xetex]{graphicx}
     \usepackage{fontspec}
   \else 
     \usepackage[luatex]{graphicx}
     \usepackage{luatextra}
   \fi
   \defaultfontfeatures{Ligatures=TeX}
\fi

\usepackage{amsmath,amsfonts,amssymb,amsthm,latexsym,float,lscape, mathtools,revsymb, booktabs}
\usepackage[labelformat=simple]{subcaption}

\usepackage{tabularx,multirow}
\usepackage{microtype}

\usepackage[detect-all=true,group-minimum-digits=4]{siunitx}
\usepackage{chemfig}
\setchemfig{atom style={scale=0.4}, double bond sep = 4pt}
\usepackage[version=4,arrows=pgf-filled,textfontname=sffamily,
mathfontname=mathsf]{mhchem}

\ifPDFTeX
   \usepackage[pdftex]{hyperref}
   \usepackage[paper=A4,pagesize=pdftex,DIV=14]{typearea}
\else 
   \ifXeTeX
     \usepackage{hyperref}
     \usepackage[paper=A4,pagesize=pdftex,DIV=14]{typearea}
   \else 
     \usepackage[luatex]{hyperref}
     \usepackage[paper=A4,pagesize=luatex,DIV=14]{typearea}
   \fi
\fi

\usepackage{cleveref}

\usepackage[PGF]{adjustbox}
\usepackage{colortbl}

\newcommand{\newsecnumstyle}{%
  \renewcommand{\thesection}{S} 
  \renewcommand{\thesubsection}{\thesection.\arabic{subsection}}%
  \renewcommand{\thetable}{\thesection. \arabic{table}~}
  \renewcommand{\thefigure}{\thesection. \arabic{figure}~}
}

\newcommand{\xtb}{\textit{x}\texttt{TB}\,}
\newcommand{\crest}{\texttt{CREST}\,}
\newcommand{\gfn}{GFNi-\xtb}
\newcommand{\gf}{GFN2-\xtb}
\newcommand{\stda}{\textit{s}\texttt{TDA}\,}
\newcommand{\stddft}{\textit{s}\texttt{TD-DFT}\,}
\newcommand{\tddft}{\texttt{TD-DFT}\,}
\newcommand{\tda}{\texttt{TDA}\,}
\newcommand{\mmf}{\texttt{MMFF94s}\,}
\newcommand{\CcelCo}[2]{\cellcolor[RGB]{#2}{\color{#1}(\textbf{#2})}}
\DeclareSIUnit\calorie{cal}

\newcolumntype{Y}{>{\small\raggedright\arraybackslash}X}
\newcolumntype{Z}{>{\footnotesize\centering\arraybackslash}X}

\begin{document}

\pagestyle{fancy}
\rhead{\includegraphics[width=2.5cm]{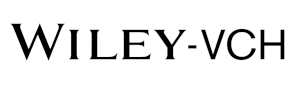}}

\title{Advancing excited-state simulations for TADF emitters: An eXtended Tight-Binding framework for high-throughput screening and design}

\maketitle


\author{Jean-Pierre Tchapet Njafa*, Aissatou Maghame, Elvira Vanelle Kameni Tcheuffa, Serge Guy Nana Engo*}


\dedication{}


\begin{affiliations}
Dr. J-P. Tchapet Njafa\\
Department of Physics, Faculty of Science, University of Yaounde 1,
Po. Box 812, Yaounde, Cameroon\\
Email Address: jean-pierre.tchapet@facsciences-uy1.cm

A. Maghame\\
Department of Physics, Faculty of Science, University of Yaounde 1,
Po. Box 812, Yaounde, Cameroon

E. V. Kameni Tcheuffa\\
Department of Physics, Faculty of Science, University of Yaounde 1,
Po. Box 812, Yaounde, Cameroon

Prof. S. G. Nana Engo\\
Department of Physics, Faculty of Science, University of Yaounde 1,
Po. Box 812, Yaounde, Cameroon\\
Email Address: serge.nana-engo@facsciences-uy1.cm

\end{affiliations}



 \begin{abstract}
 We present a computationally efficient framework for predicting the excited-state properties of thermally activated delayed fluorescence (TADF) emitters, 
integrating extended tight-binding (\xtb), simplified Tamm-Dancoff approximation (\stda), and simplified time-dependent density functional theory (\stddft) 
methods. Benchmarking against Tamm-Dancoff approximation (noted full \tda) demonstrates that this approach accurately captures key photophysical properties, 
including singlet-triplet energy gaps, excitation energies, and fluorescence spectra, in both vacuum and solvent environments, while achieving over 99\% 
reduction in computational cost. We analyze a series of representative TADF emitters, revealing a strong correlation between the torsional angle between donor 
and acceptor units and the solvent-induced redshift in the emission spectrum. This work highlights the potential of semi-empirical methods for high-throughput 
screening of TADF materials and provides valuable insights for designing next-generation optoelectronic devices. The multi-objective function is one of a kind, 
and it further enhances our results with an original solution. While acknowledging the limitations of semi-empirical methods for highly complex systems, we 
outline promising future directions, including hybrid computational approaches and integration with machine learning techniques, to further improve predictive 
accuracy and accelerate the discovery of advanced functional materials.
  \end{abstract}

\section{Introduction}

The quest for sustainable, high-performance lighting has ignited intense research in thermally activated delayed fluorescence (TADF) materials, promising a 
route to next-generation organic light-emitting diodes (OLEDs) that circumvent the need for scarce and costly heavy-metal dopants 
\cite{Endo2009,Dias2017,Li2023,Nakanotani2014,Wong2017}. Realizing the full potential of TADF-OLEDs hinges on precise computational modeling of their complex 
photophysics, particularly the accurate prediction of key excited-state properties like singlet-triplet energy gaps ($\Delta E_{ST}$) and excited-state 
lifetimes \cite{Dias2016,Grimme2022}. However, the computational cost of high-accuracy methods, such as time-dependent density functional theory (TD-DFT) 
\cite{Casida1995,Dreuw2004}, often becomes prohibitive for large-scale material screening and optimization, a bottleneck hindering the discovery of novel TADF 
emitters with tailored properties. This challenge is amplified by the well-known limitations of TD-DFT in accurately describing systems with significant 
charge-transfer character \cite{Grimme2022}, further motivating the need for computationally efficient yet reliable alternatives. Addressing this challenge has 
implications for both improved technology as well as reducing computational power consumption, thus making this relevant for the Sustainable Development Goal 
(SDG) 13 on climate action.

This work introduces a robust and rigorously validated semi-empirical computational framework designed to strike an optimal balance between accuracy and 
efficiency in predicting the key photophysical properties of TADF molecules. Our approach integrates extended tight-binding 
(xTB)~\cite{Grimme2013,Grimme2017,Bannwarth2019}, simplified Tamm-Dancoff approximation (sTDA)~\cite{Grimme2013,Grimme2016,Grimme2022}, and simplified 
time-dependent density functional theory (sTD-DFT)~\cite{Grimme2021} methods and incorporates an implicit solvent model to account for environmental effects. 
Uniquely, we combine these techniques to enable large-scale screening and optimization of TADF materials, often precluded by computationally expensive methods. 
The multi-objective approach allows us to see which are the best combinations of qualities and aspects and get data in an easier manner. The performance and 
reliability of this framework are evaluated via a comprehensive benchmarking study (\Cref{sec:Ben_Compt_Meth}) against full \tda calculations and a detailed 
comparison to experimental trends for a representative set of TADF emitters. Furthermore, we explore the correlations between calculated molecular properties, 
such as donor-acceptor separation and torsional angles (illustrated in 
\Cref{tab:DAnunits}), and observed photophysical behavior (\Cref{sec:MolProp}). A detailed analysis of singlet-triplet energy gaps (\Cref{sec:STGap}), 
excitation energies, and oscillator strengths (\Cref{sec:EST_Osc}), along with an investigation of the critical impact of solvent effects on the excited-state 
properties (\Cref{sec:SolvEff}), is also provided. We believe this approach provides a valuable tool for investigating TADF emitters and, more generally, for 
studying the interplay of electronic structure, molecular geometry, and environmental effects in photoactive molecular systems, with the ultimate goal of 
accelerating the design and discovery of high-performance, sustainable optoelectronic materials.

\section{Computational methods}\label{sec:Methods}

\begin{figure}[!htbp]
\centering
 \includegraphics{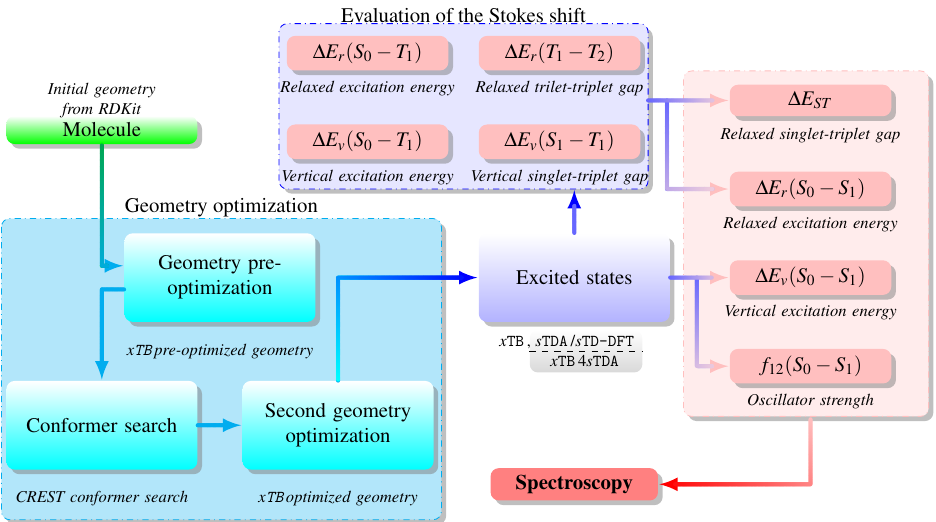}
\caption{Overview of the simulation workflow. Starting with a SMILES string, the code performs conformer search and geometry optimisation via \xtb for the 
singlet ground state $S_0$ and the triplet state $T_1$. It allows the extraction of the relaxed triplet excitation energy. Simplified time-dependent DFT 
calculation with \stda/\stddft extracts the vertical singlet-triplet gap, the relaxed triplet-triplet gap, the oscillator strength, the vertical excitation 
energy and the fluorescence absorption and emission spectra, while incorporating solvent effects to enhance the realism of our simulations. The Stokes shift is 
evaluated and then allows the relaxed singlet-triplet gap to be estimated.}
\label{fig:workflow}
\end{figure}

Accurately modeling the excited-state properties of TADF emitters requires a careful balance between computational efficiency and predictive accuracy, 
particularly when considering the challenges of solvent effects and the need for high-throughput screening. This study introduces a comprehensive computational 
framework that integrates semi-empirical and density functional theory-based methods to efficiently predict key photophysical properties. Our framework 
incorporates extended tight-binding (xTB)~\cite{Grimme2013, Grimme2017, Bannwarth2019}, simplified Tamm-Dancoff approximation (sTDA)~\cite{Grimme2013, 
Grimme2016, Grimme2022}, and simplified time-dependent density functional theory (sTD-DFT)~\cite{Grimme2021, Sun2018}. These methods were carefully selected to 
provide a computationally tractable and reliable approach for excited-state simulations. The workflow is specifically designed to evaluate key photophysical 
properties, such as singlet-triplet energy gaps ($\Delta E_{ST}$) and fluorescence emission spectra, while incorporating solvent effects to enhance the realism 
of our simulations. This methodology is applied to a set of representative TADF emitters, enabling a comprehensive benchmarking of computational efficiency and 
predictive accuracy under both vacuum and solvated conditions. A schematic representation of the workflow is presented in \Cref{fig:workflow}.

\subsection{Molecular geometry optimization}\label{sec:MolGOpt}

Precise molecular geometries are paramount for reliable predictions of excited-state properties, as they directly influence key parameters like energy gaps and 
oscillator strengths. We employed a three-step geometry optimization process to obtain accurate geometries for both the ground state ($S_0$) and the first 
triplet state ($T_1$). This approach comprises:
\begin{enumerate}
    \item \textbf{Initial structure generation}. Initial molecular structures are generated from SMILES strings using the RDKit library (version 
\verb|2024.3.1|)~\cite{rdkit}. These 2D structures are converted to 3D geometries using the \mmf force field~\cite{Halgren99} for a preliminary optimization. 
Subsequently, geometric refinement is performed using the \gfn semi-empirical method (with $i = 0, 1, 2$), as implemented in the xTB package (version 
\verb|6.7.0|)~\cite{grimme2019exploration}.

    \item \textbf{Conformer search}. To account for molecular flexibility and identify the most stable conformers, we performed a conformer search using the 
\crest program (version \verb|3.0|) \cite{crest2}.  The \gfn method (version \verb|6.7.0| of \xtb)~\cite{grimme2019exploration, Bannwarth2019} is used for an 
efficient sampling of conformational space, generating an ensemble of possible geometries. The lowest-energy conformer from this ensemble is selected for 
subsequent calculations, reducing the risk of overlooking relevant stable conformations, that could significantly influence excited-state properties.

    \item \textbf{Final geometry optimization}. The selected conformer(s) undergo final geometry optimization using the same \gfn method (version \verb|6.7.0| 
of \xtb) under both vacuum and implicit solvent conditions. Toluene is chosen as the solvent, owing to its relevance in OLED applications. An implicit solvent 
model, implemented within \xtb, is employed to simulate the solvent environment. The root-mean-square deviation (RMSD) between vacuum- and solvent-optimized 
geometries is computed, using the module \verb|rdkit.Chem.rdMolAlign| of RDKit~\cite{rdkit}, to quantify the structural changes induced by solvation. A 
convergence threshold of \num{e-7} a.u. for the energy and \num{2e-4} a.u. for the gradient norm is applied for all optimizations. This process ensures that 
the obtained geometries are stable, physically meaningful, and reflective of realistic conditions.
\end{enumerate}
\Cref{tab:rmsd} in the Supporting Information summarizes the RMSD values obtained for each molecule. \Cref{fig:4CzIPN-Sem} shows an example (4CzIPN) of a 
pre-optimized geometry evolving through conformer search to the final optimization.

\subsection{Excited-state property calculations}

Accurate prediction of excited-state properties is critical for understanding the photophysical behavior of TADF emitters. To this end, the present study 
utilizes a combination of semi-empirical and density functional theory (DFT)-based methodologies, encompassing:
\begin{enumerate}
    \item \textbf{\xtb4\stda implementation} (version \verb|1.0|). The \xtb4\stda method performs ground-state calculations using the \xtb model, followed by a 
simplified Tamm-Dancoff approximation (\stda) and simplified Time-Dependent Density Functional Theory (\stddft) for excited-state calculations.

    \item \textbf{Simplified Tamm-Dancoff Approximation (\stda)} (version \verb|1.6.3| of the sTDA software)~\cite{Grimme2013, Grimme2016, Grimme2022}. \stda 
calculations are performed to efficiently determine the excitation energies and oscillator strengths of low-lying singlet ($S_1$) and triplet ($T_1$) states.

    \item \textbf{Simplified Time-Dependent Density Functional Theory (\stddft)} (version \verb|1.6.3| of the \stda software)~\cite{Grimme2021}. \stddft 
calculations are employed for a more refined calculation of excitation energies and oscillator strengths.

    \item \textbf{Full \tda (for benchmarking objectives)}. To benchmark the accuracy of our semi-empirical methods, full \tda (in opposition of \stda) 
calculations are performed using the PySCF framework (version \verb|2.7.0|)~\cite{Sun2018,Sun2020} with the B3LYP and CAM-B3LYP functionals and the def2-TZVP 
basis set, on a subset of representative molecules to validate the accuracy of the \stda and \stddft methods, 
\end{enumerate}

For each molecule, we assessed the following photophysical properties:
\begin{enumerate}
    \item Singlet-triplet energy gap ($\Delta E_{ST}$), delineated as the energy difference between the lowest singlet ($S_1$) and triplet ($T_1$) excited 
states, which is a key parameter for determining TADF performance. Smaller $\Delta E_{ST}$ values facilitate enhanced reverse intersystem crossing (rISC).

    \item The vertical excitation energies, calculated for singlet-singlet ($S_0 \rightarrow S_1$) and singlet-triplet ($S_0 \rightarrow T_1$) transitions; the 
relaxed excitation energies, including singlet-triplet ($S_0 \rightarrow T_1$) and triplet-triplet ($T_1 \rightarrow T_2$) transitions, are also evaluated. 
These calculations aid in elucidating the electronic properties that influence absorption and emission spectra.

    \item Oscillator strengths, which measure the probability of radiative transitions and provide valuable insights into fluorescence efficiency.

    \item Fluorescence spectra, predicted for $S_1 \rightarrow S_0$ transitions, are used to model the emission behavior. Peak wavelengths and intensities are 
determined to enable comparisons with experimental trends when available. UV-Vis spectra, which provide critical insights into electronic transitions, are also 
evaluated (see \Cref{subSec-UV}). Additionally, the fluorescence emission spectrum is estimated based on \textit{Kasha's rule}, which assumes that emission 
primarily occurs from the lowest excited state.
\end{enumerate}
All calculations are performed on the optimized molecular geometries obtained from \Cref{sec:MolGOpt}. While these methods were applied to all molecules in the 
study, the calculations performed on the 4CzIPN molecule play a particularly important benchmarking role, since the availability of extensive experimental data 
for this well-characterized TADF emitter enabled a direct comparison of calculated and measured values. These comparisons provide a robust validation of the 
accuracy and predictive power of the sTDA and sTD-DFT methods.

Further details of these calculations and the rationale behind the selection of these specific methods are discussed in the Supporting Information.

\subsection{Solvent effects}

Excited-state calculations are systematically conducted under both vacuum and solvated conditions. The vacuum environment serves as a baseline for determining 
the intrinsic excited-state properties of each molecule. In contrast, solvation effects facilitated by toluene are incorporated through an implicit solvation 
model based on the analytical linearized Poisson–Boltzmann (ALPB) model, as implemented in \xtb~\cite{ehlert2021robust}. Toluene is selected as the solvent due 
to its relevance in OLED applications. The influence of solvation on both molecular geometries and excited-state properties is thoroughly investigated and 
analyzed. To quantify these effects, the differences between vacuum and solvation conditions are characterized using the method: $\Delta E_{\rm solv} = 
\left|\Delta E_{\rm vacuum}-\Delta E_{\rm toluene}\right|$, which measures the absolute energy differences between the excited states in vacuum and solvated 
environments.

\subsection{Stokes shifts and relaxed singlet-triplet excitation energies}

Since \xtb does not permit geometry optimization of molecules in the $S_1$ state, a direct evaluation of the relaxed singlet-triplet excitation energy, $\Delta 
E_{ST}$, is not possible. However, vertical and relaxed excitation energies for the $S_0 \rightarrow T_1$ transition are computed. To estimate $\Delta E_{ST}$, 
we first obtain the geometric energy relaxation for the $T_1$ state as the difference: $\Delta E_v(S_0\to T_1) - \Delta E_r(S_0\to T_1)$. Based on this value, 
and considering a simplified Jablonski diagram with relaxed triplet-triplet transition energies (as detailed in \Cref{fig:jablonski} and 
\Cref{tab:T1T2,tab:T1T2geo}), we employed the following approximations to estimate the Stokes shifts ($S_{shift}$) and the relaxed singlet-triplet excitation 
energies ($\Delta E_{ST}$):
\begin{itemize}
 \item If $\Delta E_v(S_1\gets T_1)<\Delta E_r(T_2\gets T_1)$,
 \begin{align*}
&\Delta E_{ST}\approx \Delta E_v(S_1\gets T_1),
& S_{shift}\approx \Delta E_v(S_0\to T_1) - \Delta E_r(S_0\to T_1);
 \end{align*}
\item If $\Delta E_v(S_1\gets T_1)>\Delta E_r(T_2\gets T_1)$,
\begin{align*}
& \Delta E_{ST}\approx\dfrac{1}{2}\Delta E_r(T_2\gets T_1),
& S_{shift}\approx \dfrac{1}{2}\Delta E_r(T_2\gets T_1)+\left[\Delta E_v(S_0\to T_1) - \Delta E_r(S_0\to T_1)\right].
\end{align*}
\end{itemize}

The analysis of fluorescence spectra, which incorporates solvent effects, shows how the solvent modulates emission characteristics. Each molecule’s spectrum is 
presented through two distinct plots: a UV-Vis absorption spectrum and a fluorescence spectrum which incorporates the Stokes shift. In these plots, findings 
from \stda are shown using green and magenta lines for vacuum and toluene, respectively, while results from \stddft calculations are shown using blue and red 
lines for the corresponding environments. The spectral data are derived using Multiwfn (version \verb|3.8|(dev))~\cite{lu2024multiwfn}, a comprehensive tool 
popular for its ability to analyze electronic transitions. Multiwfn provides a wide range of state function analysis techniques and it is recognized for its 
operational efficiency, ease of use, and flexibility. To ensure precise spectral characterization, a Full Width at Half Maximum (FWHM) of 
$\qty{0.15}{\electronvolt}$, is utilized.

\subsection{Computational details}

All calculations were performed using specific computational resources, the details of which, along with information on parallelization and timings, are 
provided in the Supporting Information.

\section{Results and discussion}\label{sec:Results}

This section presents the results of our computational study on selected Thermally Activated Delayed Fluorescence (TADF) emitters. We begin by benchmarking the 
accuracy and efficiency of the employed semi-empirical methods against full \tda calculations using the B3LYP and CAM-B3LYP functionals and the def2-TZVP basis 
set. We focused on the well-characterized 4CzIPN molecule as documented in experimental studies \cite{Liu2018,Samanta2017}, and assessed the performance of the 
semi-empirical methods by comparing singlet and triplet excitation energies, as well as oscillator strengths. Having validated our computational approach using 
metrics such as absolute energy differences and root mean square error, we now turn our attention to the photophysical properties of the selected TADF 
molecules. We will analyze key properties including singlet-triplet splitting ($\Delta E_{ST}$), oscillator strengths, and charge transfer character of the 
excited states to establish correlations between the molecular structure, the nature of the donor moiety, and photophysical behavior. We define the donor type 
in terms of the electron-donating ability of the substituted groups. Finally, we discuss the broader implications of our findings for computational material 
design, particularly regarding its potential to reduce the time and cost associated with the discovery of novel TADF emitters, and to create more fundamental 
design rules for efficient TADF emitters.

\begin{table}[!htbp]
 \centering
\caption{Calculated photophysical properties of TADF emitters in vacuum. Energies are given in \unit{\electronvolt}, wavelengths in \unit{\nano\meter}, and 
radiative lifetimes in \unit{\nano\second}. Results are shown for both the semi-empirical \xtb method, and the Tamm-Dancoff Approximation (TDA) and Time 
Dependent Density Functional Theory (TD-DFT) within a simplified scheme (\stda and \stddft, respectively).}
{\scriptsize
\begin{tabular}{l*{8}{@{ }S@{ }}}
\toprule
 Molecule & {DMAC-TRZ} & {DMAC-DPS} & {PSPCz} & {4CzIPN} & {Px2BP} & {CzS2} & 
{2TCz-DPS} & {TDBA-DI} \\
\midrule
HOMO-LUMO gap & 1.539 & 1.960 & 2.840 & 1.950 & 1.447 & 2.671 & 2.619 & 2.070 \\
$\Delta E_r(S_0\to T_1)$ & 2.090 & 2.579 & 3.113 & 2.268 & 2.033 & 2.688 & 
2.459 & 2.558 \\
$\Delta E_v(S_0\to T_1)$ (\stda) & 3.386 & 3.471 & 3.399 & 3.086 & 2.573 & 3.435 & 
3.413 & 3.092 \\
$\Delta E_v(S_0\to T_1)$ (\stddft) & 3.373 & 3.441 & 3.376 & 3.077 & 2.563 & 3.412 & 
3.391 & 3.073 \\
$\Delta E_v(S_0\to S_1)$ (\stda) & 3.483 & 3.878 & 3.849 & 3.298 & 2.911 & 3.902 & 
3.874 & 3.573 \\
$\Delta E_v(S_0\to S_1)$ (\stddft) & 3.476 & 3.873 & 3.777 & 3.268 & 2.879 & 3.839 & 
3.809 & 3.502 \\
$\Delta E_v(S_1\gets T_1)$ (\stda) & 0.097 & 0.407 & 0.450 & 0.212 & 0.338 & 0.467 & 
0.461 & 0.481 \\
$\Delta E_v(S_1\gets T_1)$ (\stddft) & 0.103 & 0.432 & 0.401 & 0.191 & 0.316 & 0.427 & 
0.418 & 0.429 \\
$S_{shift}$ (\stda) & 1.296& 1.055& 1.012& 0.818& 0.540& 0.788& 
1.032&0.534\\
$S_{shift}$ (\stddft) & 1.283& 1.025& 0.934& 0.809& 0.530& 
0.751& 0.990&0.515\\
$\Delta E_r(S_0\to S_1)$ (\stda) & 2.187 & 2.742 & 3.233 & 2.480 & 2.371 & 2.728 & 2.537 & 3.039 \\
$\Delta E_r(S_0\to S_1)$ (\stddft) & 2.138 & 2.625 & 3.215 & 2.403 & 2.262 & 2.846 & 2.787 & 2.979 \\
$\Delta E_{ST}$ (\stda) & 0.097& 0.163& 0.120& 0.212& 0.338& 0.041& 0.078&0.481\\
$\Delta E_{ST}$ (\stddft) & 0.103& 0.163& 0.072& 0.191& 0.316&0.027 & 0.058&0.429\\
$\lambda_{abs}$ (\stda) & 356.010& 319.690& 322.083& 375.928& 425.926& 317.712& 320.070& 347.012\\
$\lambda_{abs}$ (\stddft) & 356.681& 320.143& 328.303& 379.335& 430.677& 322.929& 325.530& 354.057\\
$\lambda_{PL}$ (\stda) & 566.890 & 452.233 & 383.543 & 499.966 & 522.944 & 454.445 & 488.800 & 407.965\\
$\lambda_{PL}$ (\stddft) & 565.339 & 452.233 & 389.324 & 504.236 & 527.842 & 456.789 & 492.685 & 415.067\\

$f_{12}(S_0\to S_1)$ (\stda) & 0.146 & 0.021 & 0.189 & 0.152 & 0.103 & 0.351 & 0.151 
& 0.194 \\
$f_{12}(S_0\to S_1)$ (\stddft) & 0.140 & 0.019 & 0.159 & 0.137 & 0.086 & 0.309 & 
0.131 & 0.219 \\
$\tau$ (\stda) & 33.045 & 148.121 & 11.675 & 24.590 & 39.689 & 8.816 & 23.659 & 12.855 \\
$\tau$ (\stddft) & 36.016 & 174.136 & 14.012 & 29.115 & 52.684 & 9.191 & 22.720 & 11.853 \\
MOF (\stda) & -0.964 & -0.600 & 0.036 & -0.780 & -1.064 & -0.161 & -0.590 & -0.448 \\
MOF (\stddft) & -1.025 & -0.718 & 0.072 & -0.851 & -1.169 & -0.071 & -0.340 & -0.431 \\
\bottomrule
\end{tabular}}
\label{tab:resultGas}
\end{table}

\begin{table}[!htbp]
 \centering
\caption{Calculated photophysical properties of TADF emitters in toluene solvent. Energies are given in \unit{\electronvolt}, wavelengths in 
\unit{\nano\meter}, and radiative lifetimes in \unit{\nano\second}. Results are shown for both the semi-empirical \xtb method, and the Tamm-Dancoff 
Approximation (TDA) and Time Dependent Density Functional Theory (TD-DFT) within a simplified scheme (\stda and \stddft, respectively).}
{\scriptsize
\begin{tabular}{l*{8}{@{ }S@{ }}}
\toprule
 Molecule & {DMAC-TRZ} & {DMAC-DPS} & {PSPCz} & {4CzIPN} & {Px2BP} & {CzS2} & 
{2TCz-DPS} & {TDBA-DI} \\
\midrule
HOMO-LUMO gap & 1.512 & 1.822 & 2.719 & 1.871 & 1.373 & 2.615 & 2.333 & 2.090 \\
$\Delta E_r(S_0\to T_1)$ & 2.058 & 2.383 & 3.081 & 2.186 & 1.942 & 2.614 & 
2.322 & 2.552 \\
$\Delta E_v(S_0\to T_1)$ (\stda) & 3.377 & 3.471 & 3.413 & 3.026 & 2.661 & 3.450 & 
3.425 & 3.106 \\
$\Delta E_v(S_0\to T_1)$ (\stddft) & 3.345 & 3.441 & 3.390 & 3.016 & 2.650 & 3.428 & 
3.404 & 3.087 \\
$\Delta E_v(S_0\to S_1)$ (\stda) & 3.457 & 3.713 & 3.859 & 3.243 & 2.981 & 3.918 & 
3.890 & 3.564 \\
$\Delta E_v(S_0\to S_1)$ (\stddft) & 3.453 & 3.708 & 3.789 & 3.214 & 2.952 & 3.858 & 
3.831 & 3.508 \\
$\Delta E_v(S_1\gets T_1)$ (\stda) & 0.080 & 0.242 & 0.446 & 0.217 & 0.320 & 0.468 & 
0.465 & 0.458 \\
$\Delta E_v(S_1\gets T_1)$ (\stddft) & 0.108 & 0.267 & 0.399 & 0.198 & 0.302 & 0.430 & 
0.427 & 0.421 \\
$S_{shift}$ (\stda) & 1.319& 1.088& 0.466& 0.840& 0.719& 1.069& 1.103&0.554\\
$S_{shift}$ (\stddft) & 1.287& 1.058& 0.443& 0.830& 0.708& 0.948& 1.279&0.535\\
$\Delta E_r(S_0\to S_1)$ (\stda) & 2.193 & 2.742 & 3.185 & 2.459 & 2.349 & 2.714 & 2.517 & 2.987 \\
$\Delta E_r(S_0\to S_1)$ (\stddft) & 2.166 & 2.650 & 3.215 & 2.384 & 2.244 & 2.748 & 2.518 & 2.973 \\
$\Delta E_{ST}$ (\stda) & 0.080& 0.242& 0.134& 0.217& 0.320& 0.233& 0.465&0.458\\
$\Delta E_{ST}$ (\stddft) & 0.108& 0.267& 0.134& 0.198& 0.302& 0.134& 0.197&0.421\\
$\lambda_{abs}$ (\stda) & 358.669& 333.937& 321.259 & 382.365& 415.887& 316.409& 318.690& 347.929\\
$\lambda_{abs}$ (\stddft) & 359.072& 334.417& 327.264& 385.715& 419.976& 321.394& 323.676& 353.446\\
$\lambda_{PL}$ (\stda) & 579.934 & 472.241 & 385.613 & 515.994 & 548.141 & 435.592 & 444.877 & 416.201\\
$\lambda_{PL}$ (\stddft) & 572.436 & 467.787 & 385.613 & 520.106 & 552.538 & 451.207 & 492.308 & 417.041\\

$f_{12}(S_0\to S_1)$ (\stda) & 0.066 & 0.034 & 0.212 & 0.138 & 0.100 & 0.389 & 0.156 
& 0.011 \\
$f_{12}(S_0\to S_1)$ (\stddft) & 0.064 & 0.033 & 0.178 & 0.125 & 0.084 & 0.340 & 
0.149 & 0.221 \\
$\tau$ (\stda) & 73.043 & 91.526 & 10.714 & 27.722 & 41.771 & 8.042 & 23.268 & 243.665 \\
$\tau$ (\stddft) & 77.243 & 99.113 & 12.517 & 32.496 & 54.619 & 8.980 & 24.305 & 11.809 \\
MOF (\stda) & -1.021 & -0.667 & 0.063 & -0.821 & -1.071 & -0.329 & -0.992 & -0.660  \\
MOF (\stddft) & -1.078 & -0.783 & 0.029 & -0.889 & -1.174 & -0.246 & -0.729 & -0.427 \\
\bottomrule
\end{tabular}}
\label{tab:resultTol}
\end{table}

\begin{figure}[!htbp]
\centering
\leavevmode
  \subfloat[Root Mean Square Deviation (RMSD) of molecular geometries.]{\adjustbox{width=0.5\textwidth}{\includegraphics{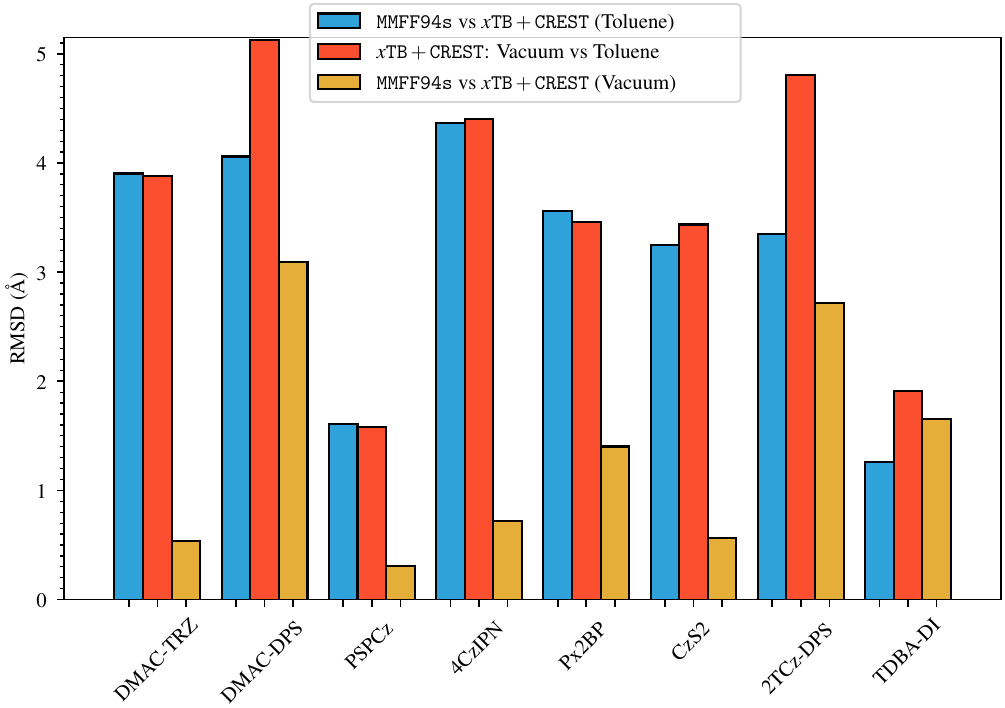}}\label{fig:rmsd}}%
\hfill%
  \subfloat[HOMO-LUMO gap.]{\adjustbox{width=0.475\textwidth}{\includegraphics{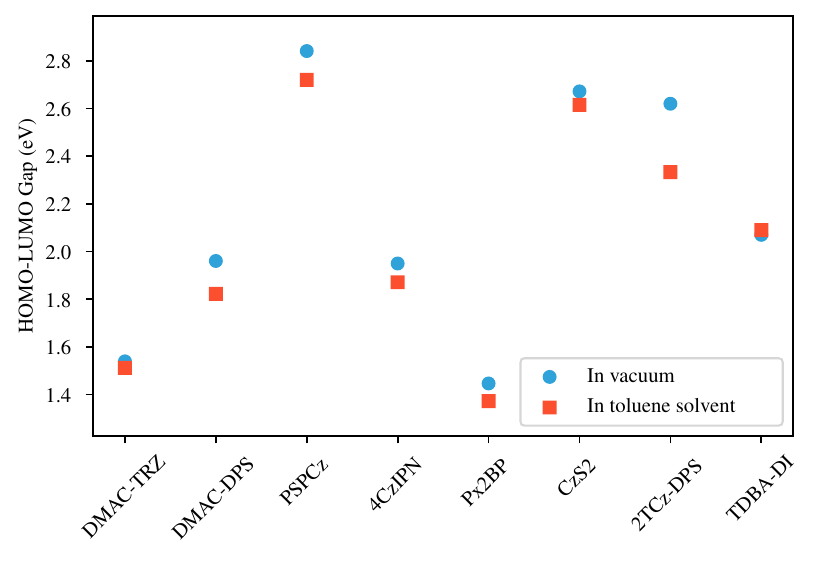}}\label{fig:HLG}}%
 \caption{Solvent effects on molecular geometry and electronic structure. Panel (a) shows the RMSD between geometries optimized in vacuum and in toluene, 
indicating the solvent-induced changes in molecular structure. Panel (b) illustrates the HOMO-LUMO gap of the molecules in both environments, showcasing the 
stabilizing effect of toluene on their electronic properties.}
\label{fig:solEffect}
\end{figure}

\begin{figure}[!htbp]
 \centering
 \leavevmode
 \subfloat[Singlet-Triplet transition energies ($\Delta E(S_0\to T_1)$) in 
vacuum.]{\adjustbox{width=0.4\textwidth}{\includegraphics{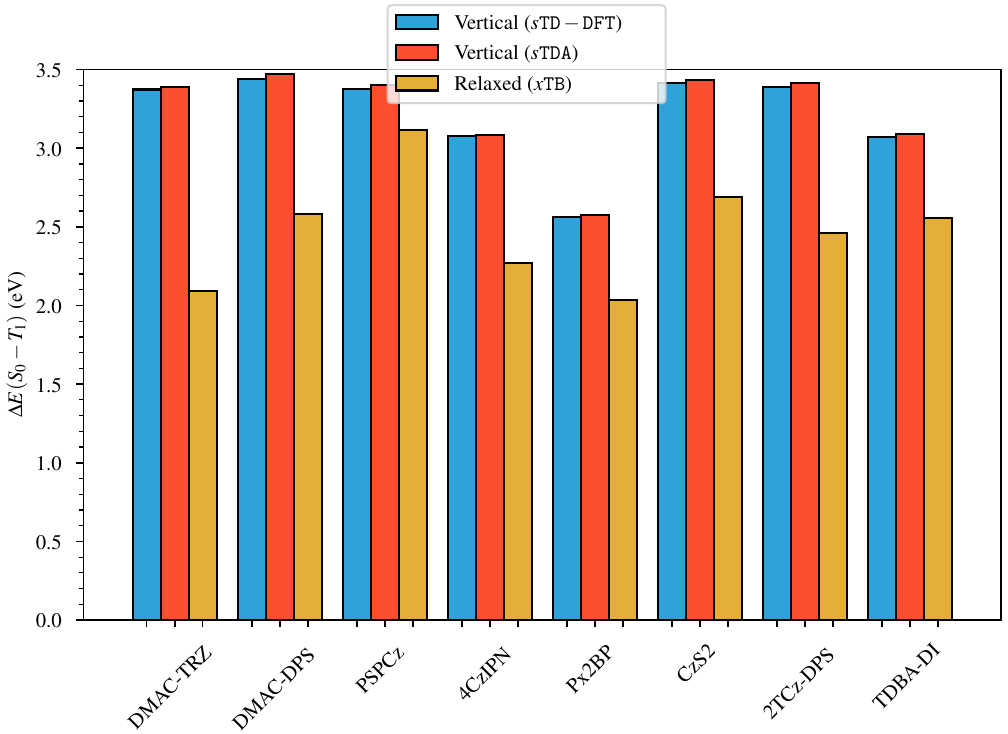}}\label{fig:ST}}\hfill
 \subfloat[Singlet-Triplet transition energies ($\Delta E(S_0\to T_1)$) in toluene 
solvent.]{\adjustbox{width=0.4\textwidth}{\includegraphics{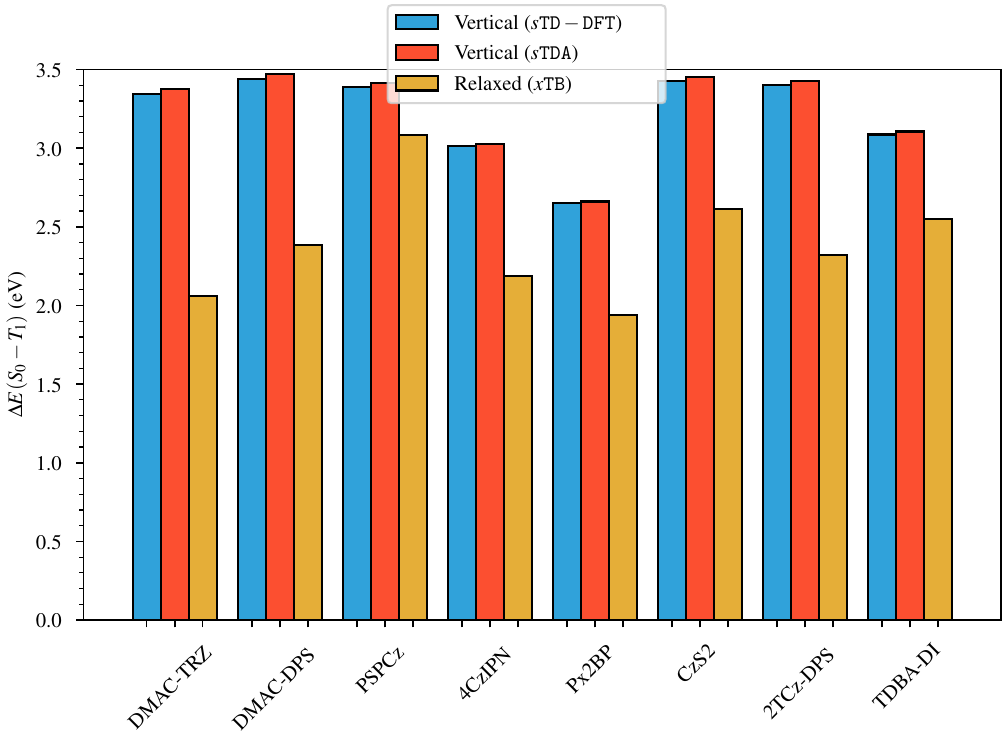}}\label{fig:STt}}\\
 \subfloat[Singlet-Singlet transition energies ($\Delta E(S_0\to S_1)$) in 
vacuum.]{\adjustbox{width=0.4\textwidth}{\includegraphics{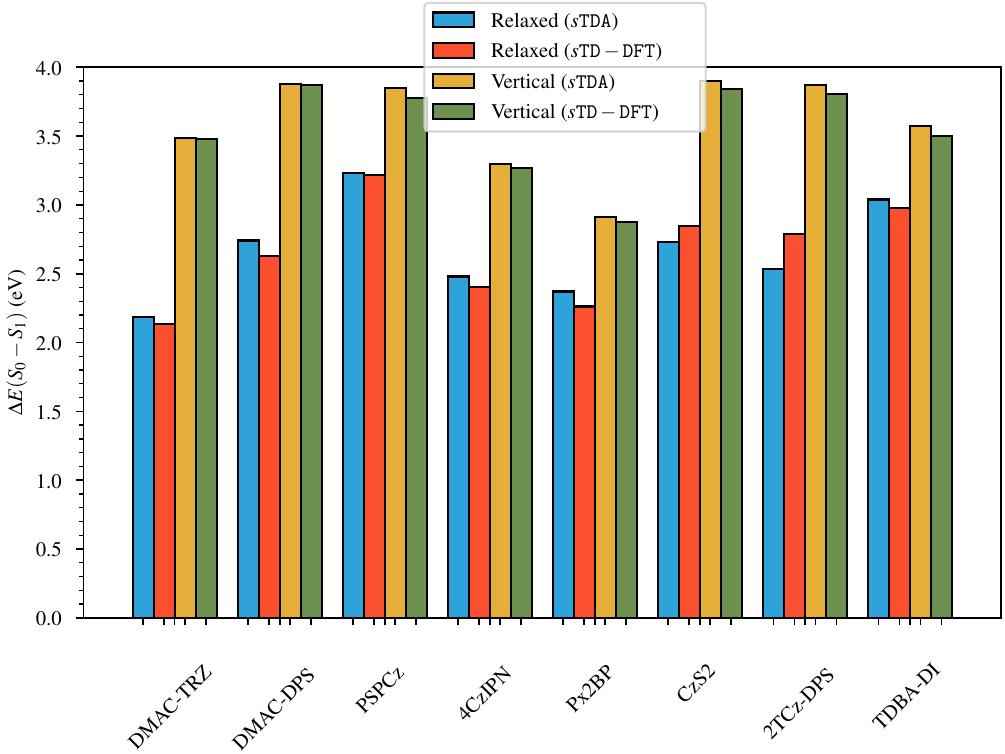}}\label{fig:SS}}\hfill
 \subfloat[Singlet-Singlet transition energies ($\Delta E(S_0\to S_1)$) in toluene 
solvent.]{\adjustbox{width=0.4\textwidth}{\includegraphics{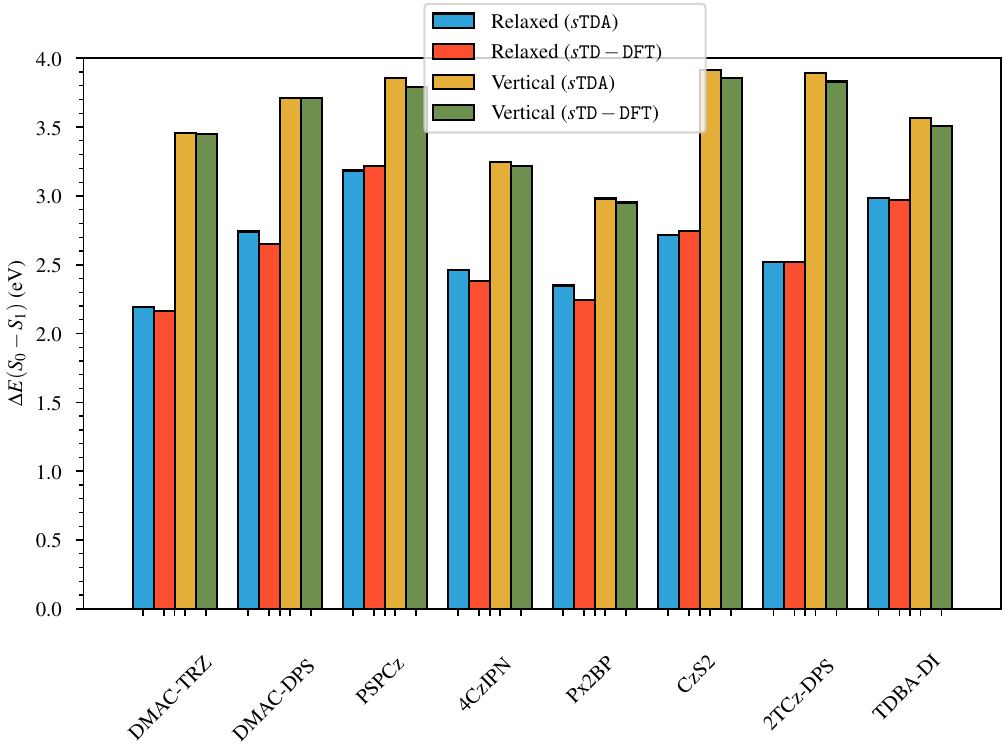}}\label{fig:SSt}}
 \caption{Comparison of calculated singlet-triplet ($(S_0\to T_1)$) and singlet-singlet ($(S_0\to S_1)$) transition energies using different methods (\xtb, 
simplified Tamm-Dancoff Approximation (\stda), and simplified Time-Dependent Density Functional Theory (\stddft)) in vacuum and toluene solvent. The figure 
highlights the effect of the computational method and solvation on the calculated transition energies.}
 \label{fig:transEnerg}
\end{figure}

\begin{figure}[!htbp]
 \centering
 \leavevmode
  \subfloat[Vertical excitation energies calculated with \stda.]{\adjustbox{width=0.4\textwidth}{\includegraphics{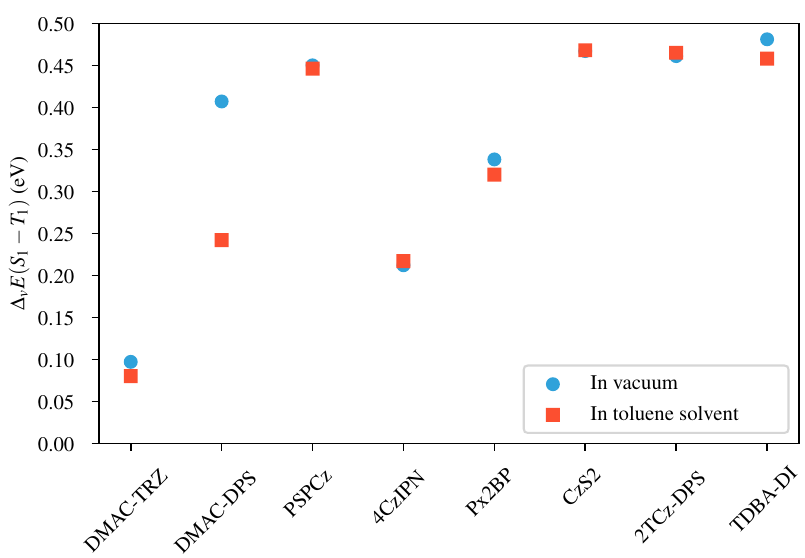}}\label{fig:S1T1-stda}}\hfill
 \subfloat[Vertical excitation energies calculated with \stddft.]{\adjustbox{width=0.4\textwidth}{\includegraphics{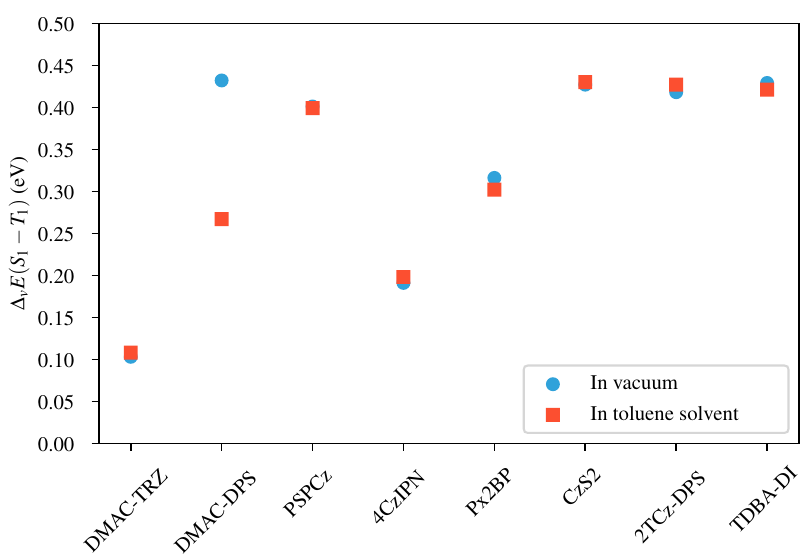}}\label{fig:S1T1-stddft}}\\
  \subfloat[Relaxed excitation energies calculated with 
\stda.]{\adjustbox{width=0.4\textwidth}{\includegraphics{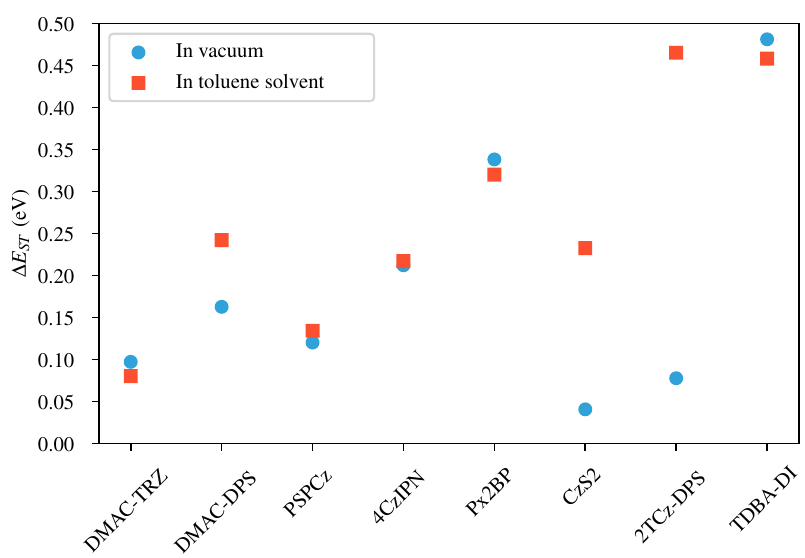}}\label{fig:deltaS1T1-stda}}\hfill
 \subfloat[Relaxed excitation energies calculated with 
\stddft.]{\adjustbox{width=0.4\textwidth}{\includegraphics{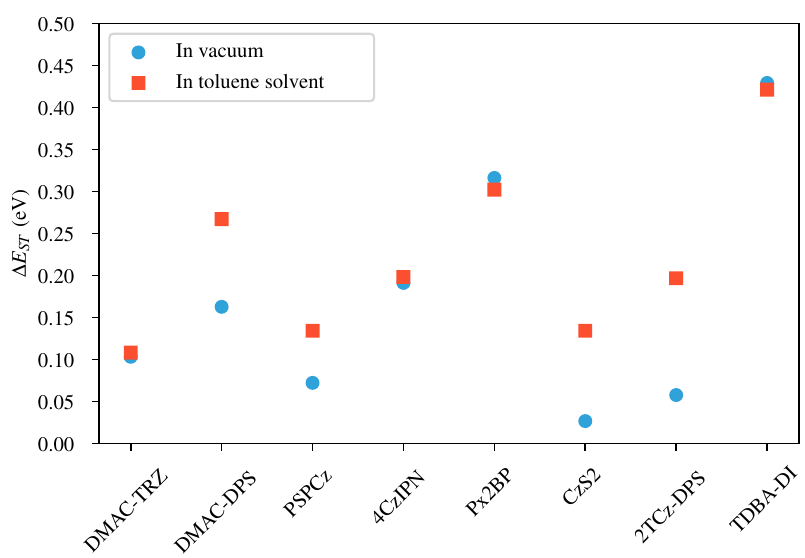}}\label{fig:deltaS1T1-stddft}}
 \caption{Comparison of vertical and relaxed excitation energies for the $S_1\gets T_1$ transition. The $\Delta E_v(S_1\gets T_1)$ and $\Delta E_{ST}$ values 
for the chosen TADF molecules are shown for both the simplified Tamm-Dancoff Approximation (\stda) and simplified Time-Dependent Density Functional Theory 
(\stddft) methods. Both vacuum and toluene solvent conditions are presented. Note that vertical energies refer to transitions from the ground state geometry, 
while relaxed energies refer to transition between the lowest-energy triplet and singlet excited states. The $\Delta E_{ST}$ is a key parameter influencing the 
efficiency of 
reverse intersystem crossing (rISC) in TADF materials.}
 \label{fig:S1T1}
\end{figure}

\subsection{Benchmarking computational methods}\label{sec:Ben_Compt_Meth}

\begin{figure}[htbp]
 \leavevmode
 \centering
 \subfloat[In vacuum.]{\includegraphics[width=0.45\textwidth]{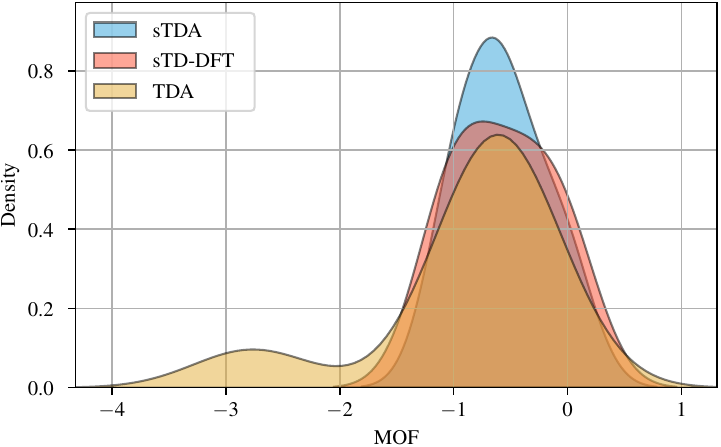}\label{fig:comparMOFVac}}\hfill
 \subfloat[In toluene solvent.]{\includegraphics[width=0.45\textwidth]{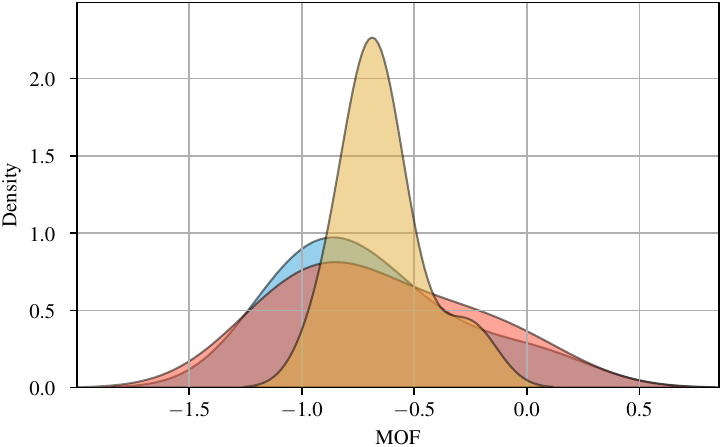}\label{fig:comparMOFTol}}
 \caption{Comparison of the multi-objective function (MOF) values calculated using simplified Tamm-Dancoff Approximation (\stda) and simplified Time-Dependent 
Density Functional Theory (\stddft) methods, against full Tamm-Dancoff Approximation (\tda) as a benchmark. Panels \subref{fig:comparMOFVac} and 
\subref{fig:comparMOFTol} show the results in vacuum and toluene solvent, respectively. The kernel density estimate plots show that \stda and \stddft exhibit 
similar distributions, but they differ from the benchmark full \tda results, indicating that these methods do not fully reproduce the range of MOF values, 
particularly in the toluene solvent. The deviations are less prominent in vacuum.}
 \label{fig:comparMOF}
\end{figure}

To evaluate the reliability of the semi-empirical methods used in this study, we benchmarked the results (see the 
\Cref{tab:resultGas,tab:resultTol,tab:benchGas,tab:benchTol}
 and the \Cref{fig:solEffect,fig:transEnerg,fig:S1T1,fig:comparDeST,fig:comparEssvstv,fig:CompEssrstr,fig:comparMOF}) of simplified Tamm-Dancoff approximation 
(\stda) and simplified time-dependent density functional theory (\stddft) against full Tamm-Dancoff Approximation (\tda) calculations, using the B3LYP and 
CAM-B3LYP functionals and the def2-TZVP basis set, for a subset of TADF emitters. The key excited-state properties compared include the singlet-triplet energy 
gap ($\Delta E_{ST}$), vertical and relaxed excitation energies, and oscillator strengths. These properties are critical for understanding and optimizing the 
performance of TADF materials in OLED applications. MOF=$+f_{12} - \Delta E_{ST} - |\Delta E_r(S_0\to S_1)-\qty{3.2}{\electronvolt}|$ stands for 
multi-objective function introduced by \cite{Nigam2023}. This function assigns a numerical score based on the oscillator strength ($f_{12}$), the relaxed 
singlet-triplet energy gap ($\Delta E_{ST}$), and the relaxed singlet-singlet transition energy  ($\Delta E_r(S_0\to S_1)$), seeking to balance the desirable 
properties of high $f_{12}$ and low $\Delta E_{ST}$ for efficient TADF emitters.

\begin{table}[!htbp]
 \centering
\caption{Comparison of calculated excited-state properties and computational times for TADF emitters in vacuum. Energies are reported in \unit{\electronvolt} 
and computation times in \unit{\second}. The table presents results from full \tda, simplified TDA (\stda), and simplified TD-DFT (\stddft) methods. Mean 
absolute error (MAE) and root mean square error (RMSE) are calculated by comparison with the full \tda results. 't' represents the computation time for each 
method.}
{\scriptsize
\begin{tabularx}{\textwidth}{m{2cm}*{10}{@{}Z@{}}}
\toprule
 Molecule & {DMAC-TRZ} & {DMAC-DPS} & {PSPCz} & {4CzIPN} & {Px2BP} & {CzS2} & 
{2TCz-DPS} & {TDBA-DI} & \textbf{MAE} & \textbf{RMSE}\\
\midrule
$\Delta E_v(S_1\gets T_1)\sim\Delta E_{ST}$ (\tda) & 0.081 & 0.069 & 0.257 & 0.187 & 0.147 & 0.317 & 0.238 & 1.185 & -  &  - \\
$\Delta E_{ST}$ (\stda) & 0.097 & 0.163 & 0.120 & 0.212 & 0.338 & 0.041 & 0.077 & 0.481 & 0.145 & 0.113 \\
$\Delta E_{ST}$ (\stddft) & 0.103 & 0.163 & 0.072 & 0.191 & 0.316 & 0.026 & 0.058 & 0.429 & 0.154 & 0.118 \\
$\Delta E_v(S_1\gets T_1)$ (\stda) & 0.097 & 0.407 & 0.450 & 0.212 & 0.338 & 0.467 & 0.461 & 0.481 & 0.135 & 0.142 \\
$\Delta E_v(S_1\gets T_1)$ (\stddft) & 0.103 & 0.432 & 0.401 & 0.191 & 0.316 & 0.427 & 0.418 & 0.429 & 0.119 & 0.124 \\
$\Delta E_v(S_0\to S_1)$ (\tda) & 2.479 & 2.546 & 3.661 & 2.678 & 2.401 & 3.489 & 2.934 & 1.597 & - & - \\
$\Delta E_v(S_0\to S_1)$ (\stda) & 3.483 & 3.878 & 3.849 & 3.298 & 2.911 & 3.902 & 3.874 & 3.573 & 0.751 & 0.626 \\
$\Delta E_v(S_0\to S_1)$ (\stddft) & 3.476 & 3.873 & 3.777 & 3.268 & 2.879 & 3.839 & 3.809 & 3.502 & 0.728 & 0.592 \\
$\Delta E_v(S_0\to T_1)$ (\tda) & 2.398 & 2.477 & 3.404 & 2.491 & 2.254 & 3.172 & 2.696 & 0.413 & - & -\\
$\Delta E_v(S_0\to T_1)$ (\stda) & 3.386 & 3.471 & 3.399 & 3.086 & 2.573 & 3.435 & 3.413 & 3.092 & 0.613 & 0.485 \\
$\Delta E_v(S_0\to T_1)$ (\stddft) & 3.373 & 3.441 & 3.376 & 3.077 & 2.563 & 3.412 & 3.391 & 3.073 & 0.598 & 0.475 \\
$\Delta E_r(S_0\to S_1)$ (\stda) & 2.187 & 2.742 & 3.233 & 2.480 & 2.371 & 2.728 & 2.537 & 3.039 & 0.368 & 0.288 \\
$\Delta E_r(S_0\to S_1)$ (\stddft) & 2.138 & 2.625 & 3.215 & 2.403 & 2.262 & 2.846 & 2.787 & 2.979 & 0.326 & 0.259 \\
$\Delta E_r(S_0\to T_1)$ & 2.090 & 2.579 & 3.113 & 2.268 & 2.033 & 2.688 & 2.459 & 2.558 & 0.269 & 0.233 \\
$f_{12}$ (\tda) & 0.074 & 0.021 & 0.026 & 0.064 & 0.215 & 0.366 & 0.061 & 0.023 & -  & -\\
$f_{12}$ (\stda) & 0.146 & 0.021 & 0.189 & 0.152 & 0.103 & 0.351 & 0.151 & 0.194 & 0.087 & 0.068 \\
$f_{12}$ (\stddft) & 0.140 & 0.019 & 0.159 & 0.137 & 0.086 & 0.309 & 0.131 & 0.219 & 0.081 & 0.066 \\
MOF (\tda) & -0.729 & -0.701 & -0.692 & -0.645 & -0.731 & -0.240 & -0.443 & -2.765 & - & - \\
MOF (\stda) & -0.964 & -0.600 & 0.036 & -0.780 & -1.064 & -0.161 & -0.590 & -0.448 & 0.307 & 0.220 \\
MOF (\stddft) & -1.025 & -0.718 & 0.072 & -0.851 & -1.169 & -0.071 & -0.340 & -0.431 & 0.344 & 0.249 \\
t (\tda) & 39153.007 & 71422.600 & 10799.822 & 107266.372 & 29068.778 & 23497.945 & 100249.382 & 124363.604 & - & - \\
t (\stda) & 7.238 & 25.811 & 8.149 & 40.449 & 8.226 & 10.301 & 20.095 & 28.707 & - & - \\
t (\stddft) & 42.847 & 87.329 & 19.707 & 208.962 & 40.692 & 40.074 & 59.932 & 98.676 & - & - \\
\bottomrule
\end{tabularx}}
\label{tab:benchGas}
\end{table}

\begin{table}[!htbp]
 \centering
\caption{Comparison of calculated excited-state properties and computational times for TADF emitters in toluene solvent. Energies are reported in 
\unit{\electronvolt} and computation times in \unit{\second}. The table presents results from full \tda, simplified TDA (\stda), and simplified TD-DFT 
(\stddft) methods. Mean absolute error (MAE) and root mean square error (RMSE) are calculated by comparison with the full \tda results. 't' represents the 
computation time for each method.}
{\scriptsize
\begin{tabularx}{\textwidth}{m{2cm}*{10}{@{}Z@{}}}
\toprule
 Molecule & {DMAC-TRZ} & {DMAC-DPS} & {PSPCz} & {4CzIPN} & {Px2BP} & {CzS2} & 
{2TCz-DPS} & {TDBA-DI} & \textbf{MAE} & \textbf{RMSE}\\
\midrule
$\Delta E_v(S_1\gets T_1)\sim\Delta E_{ST}$ (\tda) & 0.064 & 0.078 & 0.271 & 0.193 & 0.150 & 0.325 & 0.269 & 0.281 & - & - \\
$\Delta E_{ST}$ (\stda) & 0.080 & 0.242 & 0.134 & 0.217 & 0.320 & 0.233 & 0.465 & 0.458 & 0.124 & 0.100 \\
$\Delta E_{ST}$ (\stddft) & 0.108 & 0.267 & 0.134 & 0.198 & 0.302 & 0.134 & 0.197 & 0.421 & 0.123 & 0.099 \\
$\Delta E_v(S_1\gets T_1)$ (\stda) & 0.080 & 0.242 & 0.446 & 0.217 & 0.320 & 0.468 & 0.465 & 0.458 & 0.135 & 0.111 \\
$\Delta E_v(S_1\gets T_1)$ (\stddft) & 0.108 & 0.267 & 0.399 & 0.198 & 0.302 & 0.430 & 0.427 & 0.421 & 0.119 & 0.098 \\
$\Delta E_v(S_0\to S_1)$ (\tda) & 2.509 & 2.575 & 3.700 & 2.671 & 2.448 & 3.515 & 2.895 & 2.582 & - & - \\
$\Delta E_v(S_0\to S_1)$ (\stda) & 3.457 & 3.713 & 3.859 & 3.243 & 2.981 & 3.918 & 3.890 & 3.564 & 0.706 & 0.593 \\
$\Delta E_v(S_0\to S_1)$ (\stddft) & 3.453 & 3.708 & 3.789 & 3.214 & 2.952 & 3.858 & 3.831 & 3.508 & 0.682 & 0.561 \\
$\Delta E_v(S_0\to T_1)$ (\tda) & 2.445 & 2.497 & 3.429 & 2.478 & 2.299 & 3.189 & 2.626 & 2.301 & - & - \\
$\Delta E_v(S_0\to T_1)$ (\stda) & 3.377 & 3.471 & 3.413 & 3.026 & 2.661 & 3.450 & 3.425 & 3.106 & 0.608 & 0.487 \\
$\Delta E_v(S_0\to T_1)$ (\stddft) & 3.345 & 3.441 & 3.390 & 3.016 & 2.650 & 3.428 & 3.404 & 3.087 & 0.589 & 0.474 \\
$\Delta E_r(S_0\to S_1)$ (\stda) & 2.193 & 2.742 & 3.185 & 2.459 & 2.349 & 2.714 & 2.517 & 2.987 & 0.393 & 0.311 \\
$\Delta E_r(S_0\to S_1)$ (\stddft) & 2.166 & 2.650 & 3.215 & 2.384 & 2.244 & 2.748 & 2.518 & 2.973 & 0.389 & 0.317 \\
$\Delta E_r(S_0\to T_1)$ & 2.058 & 2.383 & 3.081 & 2.186 & 1.942 & 2.614 & 2.322 & 2.552 & 0.339 & 0.297 \\
$f_{12}(S_0\to S_1)$ (\tda) & 0.056 & 0.024 & 0.027 & 0.067 & 0.219 & 0.387 & 0.041 & 0.010 & - & - \\
$f_{12}(S_0\to S_1)$ (\stda) & 0.066 & 0.034 & 0.212 & 0.138 & 0.100 & 0.389 & 0.156 & 0.011 & 0.092 & 0.064 \\
$f_{12}(S_0\to S_1)$ (\stddft) & 0.064 & 0.033 & 0.178 & 0.125 & 0.084 & 0.340 & 0.149 & 0.221 & 0.086 & 0.065 \\
MOF (\tda) & -0.699 & -0.679 & -0.744 & -0.654 & -0.682 & -0.253 & -0.533 & -0.889 & - & - \\
MOF (\stda) & -1.021 & -0.667 & 0.063 & -0.821 & -1.071 & -0.329 & -0.992 & -0.660 & 0.379 & 0.279 \\
MOF (\stddft) & -1.078 & -0.783 & 0.029 & -0.889 & -1.174 & -0.246 & -0.729 & -0.427 & 0.369 & 0.273 \\
t (\tda) & 25808.543 & 43183.527 & 6735.599 & 69263.481 & 21527.554 & 20657.614 & 91689.638 & 75083.643 & - & - \\
t (\stda) & 9.632 & 15.789 & 6.532 & 32.071 & 10.949 & 7.408 & 54.402 & 45.517 & - & - \\
t (\stddft) & 37.151 & 90.607 & 24.684 & 170.211 & 39.369 & 29.184 & 87.140 & 101.857 & - & - \\
\bottomrule
\end{tabularx}}
\label{tab:benchTol}
\end{table}

\subsubsection{Singlet-triplet energy gap ($\Delta E_{ST}$)}

\begin{figure}[htbp]
 \leavevmode
 \centering
 \subfloat[In vacuum.]{\includegraphics[width=0.45\textwidth]{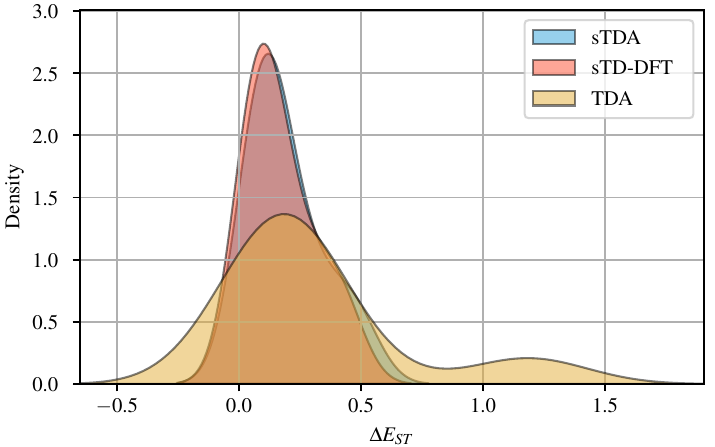}\label{fig:comparDeSTVac}}\hfill
 \subfloat[In toluene solvent.]{\includegraphics[width=0.45\textwidth]{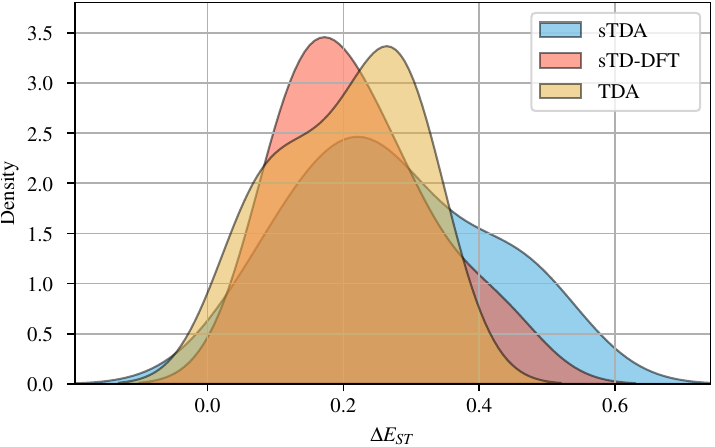}\label{fig:comparDeSTTol}}
 \caption{Comparison of the singlet-triplet energy gap ($\Delta E_{ST}$) calculated using simplified Tamm-Dancoff Approximation (\stda) and simplified 
Time-Dependent Density Functional Theory (\stddft) methods, against full Tamm-Dancoff Approximation (\tda) as a benchmark. Panels \subref{fig:comparDeSTVac} 
and \subref{fig:comparDeSTTol} show the results in vacuum and toluene solvent, respectively. The kernel density estimation plots show that \stda and \stddft 
exhibit similar distributions, but they differ from the benchmark full \tda results, indicating that these methods do not fully reproduce the range of $\Delta 
E_{ST}$ values. }
 \label{fig:comparDeST}
\end{figure}

The $\Delta E_{ST}$ values computed by \stda, \stddft, and full \tda (using vertical energies) exhibit consistent trends across the test set (see 
\Cref{tab:resultGas,tab:resultTol,fig:comparDeST}). On average, \stda shows a mean absolute error (MAE) of \qty{0.135}{\electronvolt} and a root-mean-square 
error (RMSE) of \qty{0.107}{\electronvolt} compared to full \tda, while \stddft yields an MAE of \qty{0.139}{\electronvolt} and an RMSE of 
\qty{0.109}{\electronvolt}. These deviations are within the acceptable range for TADF emitter design, where $\Delta E_{ST}$ values below 
\qty{0.3}{\electronvolt} are generally sufficient to facilitate reverse intersystem crossing (rISC).

\subsubsection{Vertical excitation energies}

\begin{figure}[htbp]
 \leavevmode
 \centering
 \subfloat[In vacuum.]{\includegraphics[width=.9\textwidth]{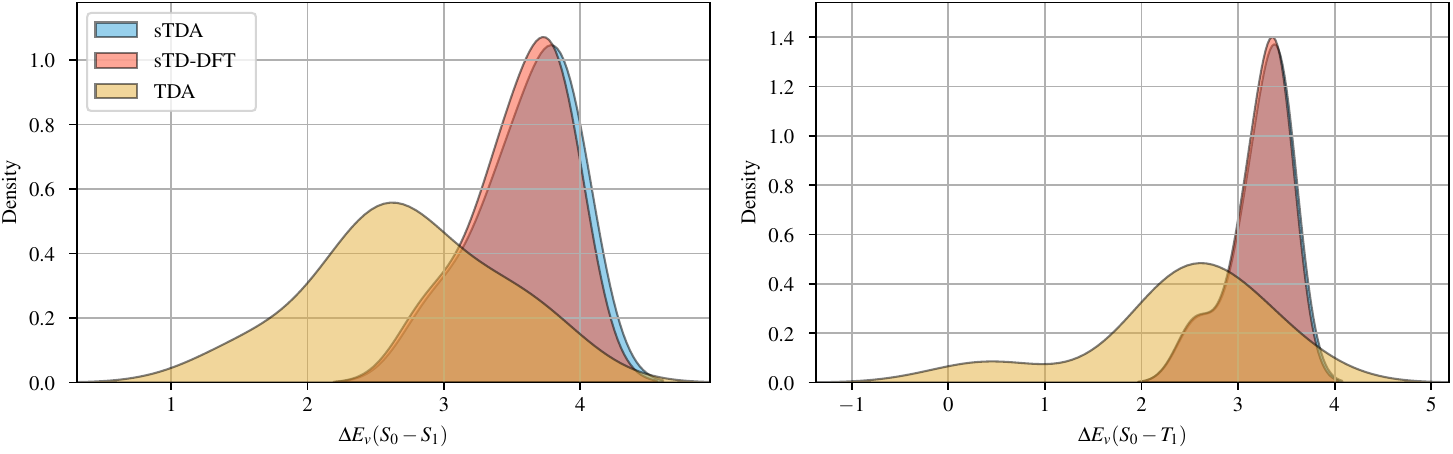}\label{fig:comparEssvstvVac}}\\
 \subfloat[In toluene solvent.]{\includegraphics[width=.9\textwidth]{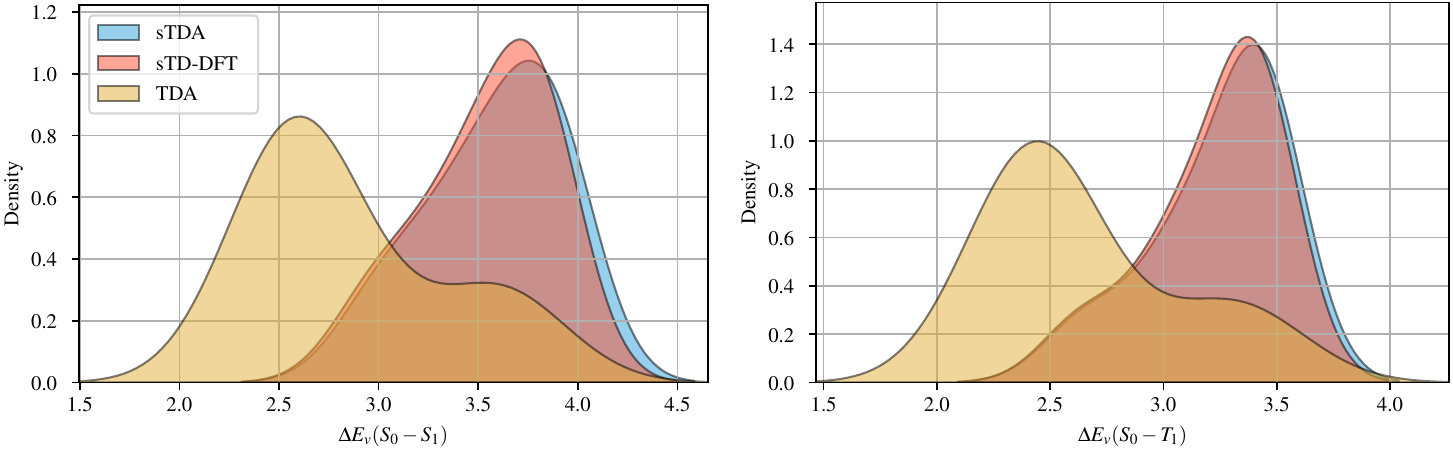}\label{fig:comparEssvstvTol}}
 \caption{Comparison of the vertical excitation energies for the singlet-singlet ($S_0\to S_1$) and singlet-triplet ($S_0\to T_1$) transitions calculated using 
simplified Tamm-Dancoff Approximation (\stda) and simplified Time-Dependent Density Functional Theory (\stddft) methods, against full Tamm-Dancoff 
Approximation (\tda) as a benchmark. Panels \subref{fig:comparEssvstvVac} and \subref{fig:comparEssvstvTol} show results in vacuum and toluene solvent, 
respectively. The kernel density estimation plots show that \stda and \stddft exhibit similar distributions, but they differ significantly from the benchmark 
full \tda results in both the overall shape of the distributions and the peak locations.}
 \label{fig:comparEssvstv}
\end{figure}

For singlet-singlet ($S_0 \to S_1$) and singlet-triplet ($S_0 \to T_1$) transitions, both \stda and \stddft do not accurately reproduce the trends from full 
\tda calculations (see \Cref{fig:comparEssvstv}), although \stddft shows slightly improved accuracy. Compared to full \tda results, the mean absolute error 
(MAE) for $S_0 \to S_1$ excitation energies is $\qty{0.751}{\electronvolt}$ for \stda and $\qty{0.728}{\electronvolt}$ for \stddft in vacuum. In toluene 
solvent, the MAE is respectively $\qty{0.706}{\electronvolt}$ and $\qty{0.682}{\electronvolt}$. For $S_0 \to T_1$, \stda and \stddft predictions deviate by 
more than $\qty{0.5}{\electronvolt}$. These differences likely arise from the approximations in the semi-empirical methods used to calculate the vertical 
excitation energies, specifically regarding the excited-state wave function. Nevertheless, things are different in the case of relaxed energies as mentioned in 
\Cref{subsec:relaxEe}.

\subsubsection{Relaxed excitation energies}\label{subsec:relaxEe}

\begin{figure}[htbp]
 \leavevmode
 \centering
 \subfloat[In vacuum.]{\includegraphics[width=.9\textwidth]{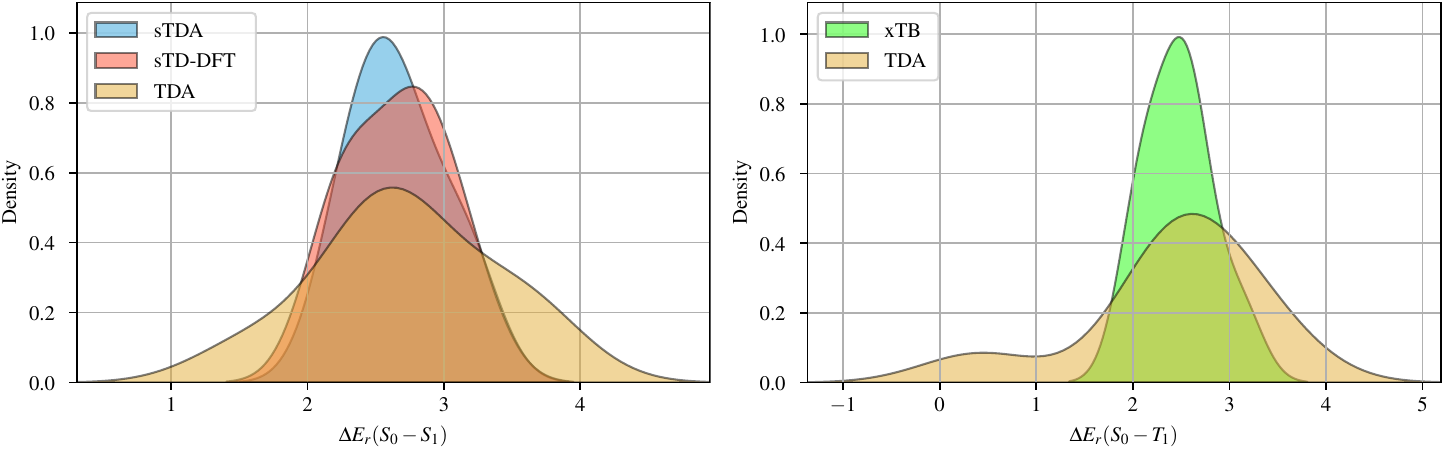}\label{fig:CompEssrstrVac}}\\
 \subfloat[In toluene solvent.]{\includegraphics[width=.9\textwidth]{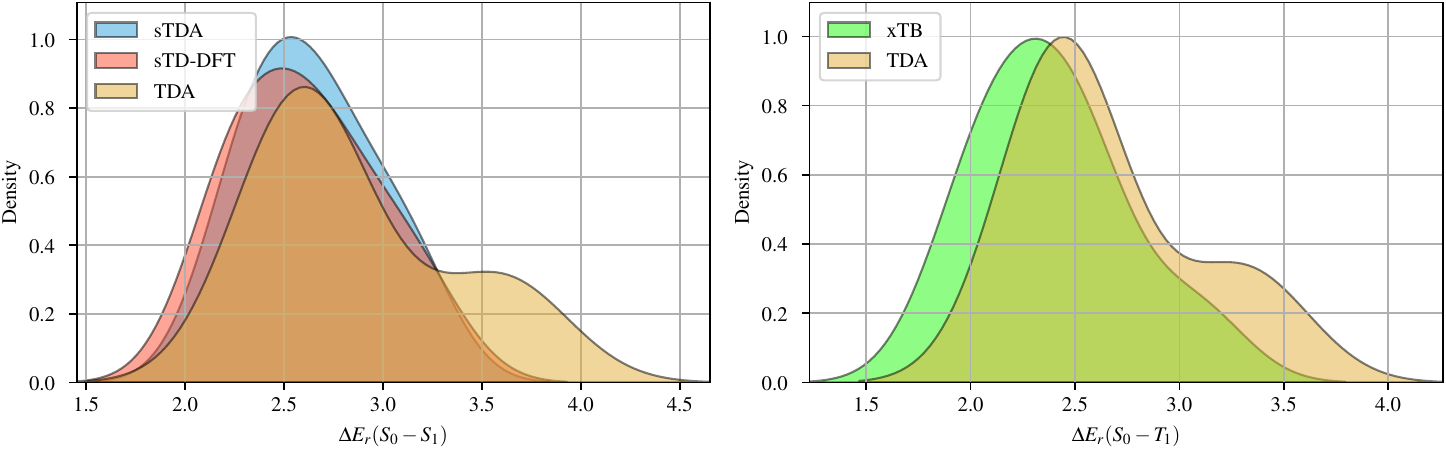}\label{fig:CompEssrstrTol}}
 \caption{Comparison of the relaxed singlet-singlet transition energy ($S_0\to S_1$) calculated using simplified Tamm-Dancoff Approximation (\stda) and 
simplified Time-Dependent Density Functional Theory (\stddft) methods, and the relaxed singlet-triplet transition energy ($S_0\to T_1$) calculated using \xtb 
method for geometry optimization, with the vertical energies obtained from full Tamm-Dancoff Approximation (\tda) calculations as a benchmark. Panels 
\subref{fig:CompEssrstrVac} and \subref{fig:CompEssrstrTol} show the results in vacuum and toluene solvent, respectively. The kernel density estimate plots 
show that \stda and \stddft exhibit similar distributions which align more closely with the full \tda benchmark compared to the vertical excitation energies. 
This suggests that the relaxed energies from \stda and \stddft are more similar to the vertical energies of full TDA than their respective vertical energies.}
 \label{fig:CompEssrstr}
\end{figure}

For singlet-singlet ($S_0 \to S_1$) and singlet-triplet ($S_0 \to T_1$) transitions, both \stda and \stddft reliably reproduce the full \tda trends (see 
\Cref{fig:CompEssrstr}), with \stddft showing slightly improved accuracy. The mean absolute error (MAE) for $S_0 \to S_1$ excitation energies is 
$\qty{0.381}{\electronvolt}$ for \stda and $\qty{0.356}{\electronvolt}$ for \stddft, compared to the full TDA results. For $S_0 \to T_1$, predictions using 
relaxed energies deviate from \tda by an average of $\qty{0.304}{\electronvolt}$. The geometry optimization was performed at the \xtb level of theory before 
the relaxed energies calculation. These results highlight the importance of using relaxed energies for capturing more appropriate photophysical transitions 
compared to those calculated from the ground state geometry.

\subsubsection{Oscillator strengths}

The computed oscillator strengths, which reflect the intensity of radiative transitions, show minor deviations between methods. The relative trends between 
molecules are preserved across all three approaches, with \stddft providing the closest match to full \tda values. The differences observed (MAE < \num{0.09}) 
are unlikely to impact practical interpretations of fluorescence efficiency.

\subsubsection{Multi-objective function}

The multi-objective function (MOF) \cite{Nigam2023} is employed to quantify how well a proposed molecule meets our design goals for TADF emitters. It 
simultaneously optimizes several key parameters: (1) a small singlet-triplet energy gap ($\Delta E_{ST}$) to promote efficient reverse intersystem crossing 
(rISC); (2) a high oscillator strength ($f_{12}$) for the $S_0 \to S_1$ transition to ensure strong radiative emission; (3) a targeted $S_0 \to S_1$ excitation 
energy ($E_{S1}$) to achieve the desired emission wavelength (color tunability); and (4) a low computational cost to enable high-throughput screening.

By assigning a numerical score based on these factors, the MOF allows for easy comparison between different molecules. In this section, the MOF serves as a 
benchmark to evaluate the accuracy and efficiency of the simplified \stda and \stddft methods by comparing their results to those obtained with the full \tda 
method. Specifically, this scoring system helps identify which method reproduces the same molecular properties as predicted by the full \tda method, which 
serves as our benchmark for accuracy. This section is not focused on the absolute MOF values, but rather on the relative performance of the computational 
methods in replicating the \tda results.

For the \stda method, the average mean absolute error (MAE) in the MOF value, compared to \tda, is \num{0.343}, and the average root-mean-square error (RMSE) 
is \num{0.250}. The average mean absolute error (MAE) for the \stddft method is \num{0.356}, and the average root-mean-square error (RMSE) is \num{0.261}. 
These results suggest that both \stda and \stddft provide reasonably accurate approximations of the full \tda method, with \stda exhibiting slightly lower 
errors overall. 

\subsubsection{Computational cost}

A key advantage of semi-empirical methods is their computational efficiency. \stda and \stddft required an average of \qty{20.705}{\second} and 
\qty{43.651}{\second} per molecule, respectively, compared to an average of over \qty{15}{hours} (or \qty{54000}{\second}) for full \tda, when using 
single-core CPU calculations. These times refer to the single point calculations with the optimized geometry. The geometry optimization with \xtb required much 
less time. This represents a runtime reduction of approximately 3-4 orders of magnitude, making \stda and \stddft considerably faster than full \tda 
calculations. This substantial reduction in runtime enables the large-scale screening of TADF emitters without sacrificing predictive reliability.

\vspace{.5cm}

These benchmarks demonstrate that \stda and \stddft are effective for predicting the photophysical properties of TADF emitters, providing a reliable compromise 
between computational efficiency and accuracy. The observed deviations from full \tda, which are generally below \qty{0.4}{\electronvolt} for transition 
energies and above \num{0.1} for oscillator strengths, are within acceptable limits for practical applications, making these methods suitable for large-scale 
TADF material design. These deviations have a minimal effect on the relative trends and general conclusions of the study.

\subsection{Molecular properties}\label{sec:MolProp}

\begin{table}[!htbp]
    \centering
    \caption{Summary of key molecular properties relevant to TADF emission, calculated in vacuum and toluene solvent. Energies are reported in 
\unit{\electronvolt}, and centroid distances are reported in \unit{\angstrom}. The HOMO-LUMO absolute spatial overlap ($S^\prime_{HL}\in[0,1]$) quantifies the 
overlap between the spatial densities of the HOMO and LUMO, defined as $S^\prime_{HL}=\int |\psi_{HOMO}(\mathbf{r})|\cdot |\psi_{LUMO}(\mathbf{r})| \, 
d\mathbf{r}$ where $\psi_{HOMO}(\mathbf{r})$ and $\psi_{LUMO}(\mathbf{r})$ are the HOMO and LUMO state functions respectively. The centroid distance represents 
the spatial separation of the electron and hole, calculated as the distance between the centers of mass of the HOMO and LUMO. The singlet-triplet energy gap 
($\Delta E_{ST}$) and oscillator strengths ($f_{12}(S_0\to S_1)$) are calculated using the simplified Tamm-Dancoff Approximation (\stda) and simplified 
Time-Dependent Density Functional Theory (\stddft) methods. This data highlights the trade-off between minimizing $\Delta E_{ST}$ and maximizing oscillator 
strength for optimal TADF performance.}
    \label{tab:SumProperties}
    {%
    \newcommand{\mc}[3]{\multicolumn{#1}{#2}{#3}}
    \scriptsize
    \begin{tabularx}{\linewidth}{Ym{3cm}*{8}{@{ }S@{ }}}
    \toprule
    &Molecule & {DMAC-TRZ} & {DMAC-DPS} & {PSPCz} & {4CzIPN} & {Px2BP} & {CzS2} & {2TCz-DPS} & {TDBA-DI}\\
    \midrule
    \multirow{6}{=}{In vacuum} & $S^\prime_{HL}$ & 0.199 & 0.267 & 0.319 & 0.325 & 0.293 & 0.452 & 0.437 & 0.133 \\
    &Centroid distance (\unit{\angstrom}) & 6.702 & 3.536 & 4.186 & 0.032 & 2.712 & 2.293 & 3.708 & 7.519 \\
    &$\Delta E_{ST}$ (\stda) & 0.097 & 0.163 & 0.120 & 0.212 & 0.338 & 0.041 & 0.078 & 0.481 \\
    &$\Delta E_{ST}$ (\stddft) & 0.103 & 0.163 & 0.072 & 0.191 & 0.316 & 0.027 & 0.058 & 0.429 \\
    &$f_{12}(S_0\to S_1)$ (\stda) & 0.146 & 0.021 & 0.189 & 0.152 & 0.103 & 0.351 & 0.151 & 0.194 \\
    &$f_{12}(S_0\to S_1)$ (\stddft) & 0.140 & 0.019 & 0.159 & 0.137 & 0.086 & 0.309 & 0.131 & 0.219 \\\midrule
    \multirow{6}{=}{In toluene\\ solvent} & $S^\prime_{HL}$ & 0.159 & 0.267 & 0.318 & 0.318 & 0.287 & 0.455 & 0.424 & 0.126 \\
    &Centroid distance (\unit{\angstrom}) & 6.772 & 3.542 & 4.183 & 0.168 & 2.725 & 2.406 & 3.939 & 7.535 \\
    &$\Delta E_{ST}$ (\stda) & 0.080 & 0.242 & 0.134 & 0.217 & 0.320 & 0.233 & 0.465 & 0.458 \\
    &$\Delta E_{ST}$ (\stddft) & 0.108 & 0.267 & 0.134 & 0.198 & 0.302 & 0.134 & 0.197 & 0.421 \\
    &$f_{12}(S_0\to S_1)$ (\stda) & 0.066 & 0.034 & 0.212 & 0.138 & 0.100 & 0.389 & 0.156 & 0.011 \\
    &$f_{12}(S_0\to S_1)$ (\stddft) & 0.064 & 0.033 & 0.178 & 0.125 & 0.084 & 0.340 & 0.149 & 0.221 \\
    \bottomrule
    \end{tabularx}
    }
\end{table}

\begin{table}[!htbp]
    \centering
    \caption{Chemical structures of the donor and acceptor units for the studied TADF emitters. These units determine the charge-transfer character of the 
excited states, influencing HOMO/LUMO separation and excited state energies.}
    \label{tab:DAunits}
    {\newcommand{\mc}[3]{\multicolumn{#1}{#2}{#3}}
    \footnotesize
    \begin{tabularx}{0.8\textwidth}{*{5}{Z}}
        \toprule
        Molecule & DMAC-TRZ & DMAC-DPS & PSPCz & 4CzIPN \\
        \midrule
        Donor & \mc{1}{c}{\includegraphics[width=0.1\textwidth]{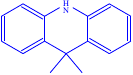}} & \mc{1}{c}{\includegraphics[width=0.1\textwidth]{DMAC.pdf}$\times2$} & 
        \mc{1}{c}{\includegraphics[width=0.1\textwidth]{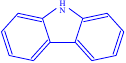}} & \mc{1}{c}{\includegraphics[width=0.1\textwidth]{Cz.pdf}$\times4$} \\
        Acceptor & \mc{1}{c}{\includegraphics[width=0.1\textwidth]{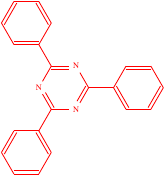}} & \mc{1}{c}{\includegraphics[width=0.1\textwidth]{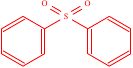}} & 
        \mc{1}{c}{\includegraphics[width=0.1\textwidth]{DPS.pdf}} & \mc{1}{c}{\includegraphics[width=0.1\textwidth]{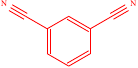}}
        \\\midrule
        Molecule & Px2BP & CzS2 & 2TCz-DPS & TDBA-DI\\
        \midrule
        Donor & 
        \mc{1}{c}{\includegraphics[width=0.1\textwidth]{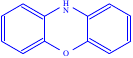}$\times2$} & \mc{1}{c}{\includegraphics[width=0.1\textwidth]{Cz.pdf}$\times2$} & 
        \mc{1}{c}{\includegraphics[width=0.1\textwidth]{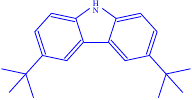}$\times2$} & \mc{1}{c}{\includegraphics[width=0.1\textwidth]{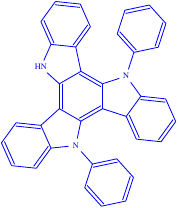}} \\
        Acceptor &
        \mc{1}{c}{\includegraphics[width=0.1\textwidth]{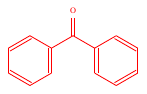}} & \mc{1}{c}{\includegraphics[width=0.1\textwidth]{DPS.pdf}} & 
        \mc{1}{c}{\includegraphics[width=0.1\textwidth]{DPS.pdf}} & \mc{1}{c}{\includegraphics[width=0.1\textwidth]{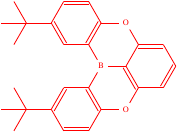}}\\
        \bottomrule
    \end{tabularx}
    }
\end{table}

The molecular properties of TADF emitters, particularly their donor-acceptor interactions, electronic distributions, and geometrical features, critically 
influence their excited-state behavior and, consequently, their TADF performance. This section explores how these intrinsic properties correlate with key 
photophysical characteristics such as singlet-triplet energy gaps ($\Delta E_{ST}$), excitation energies, and fluorescence efficiency. Relevant properties for 
the studied molecules are summarized in \Cref{tab:SumProperties} and \Cref{tab:DAunits}. The chemical structures of the donor and acceptor units in 
\Cref{tab:DAunits} dictate the charge-transfer character of the molecules. Stronger electron-donating groups and electron-withdrawing groups will enhance 
charge transfer, leading to more efficient TADF, as we will show in the following sections.

\subsubsection{Donor-acceptor interactions and HOMO-LUMO distributions}\label{sec:MolProp1}

\begin{figure}[!htbp]
\centering
\leavevmode
\subfloat[In vacuum.]{\includegraphics[width=0.47\textwidth]{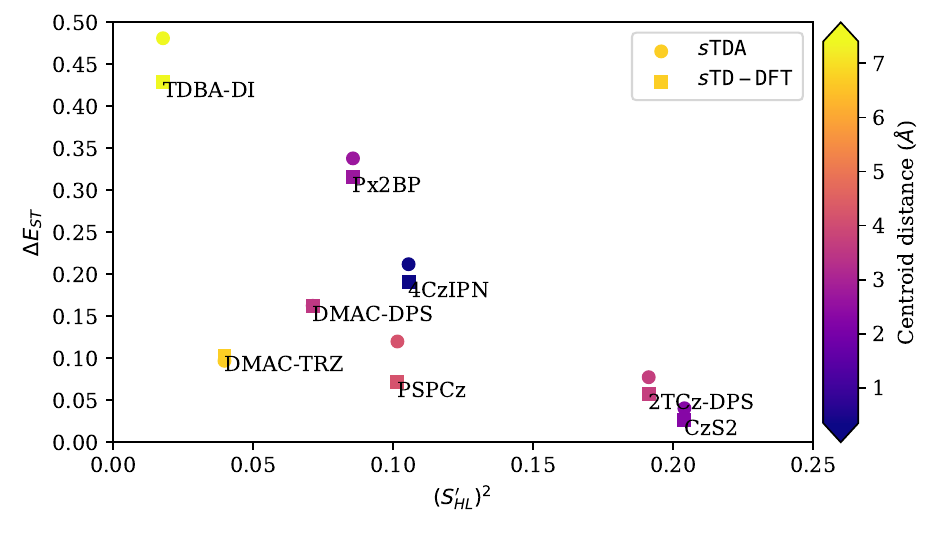}\label{fig:overlapVac}}
\subfloat[In toluene solvent.]{\includegraphics[width=0.47\textwidth]{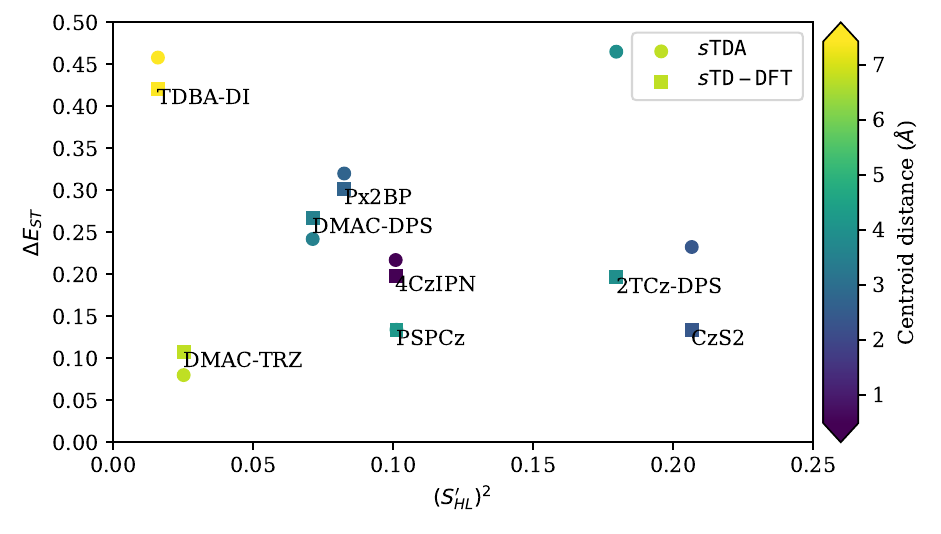}\label{fig:overlapTol}}
\caption{Relationship between singlet-triplet energy gap ($\Delta E_{ST}$) and HOMO-LUMO overlap ($S'_{HL}$). Scatter plots show $\Delta E_{ST}$ as a function 
of the square of the HOMO-LUMO absolute spatial overlap ($S'^2_{HL}$) in (a) vacuum and (b) toluene. Data points are colored according to their centroid 
distance, providing further insight into the spatial separation of the HOMO and LUMO. The general trend indicates that smaller HOMO-LUMO overlap correlates 
with a smaller singlet-triplet energy gap.}
\label{fig:HMoverST}
\end{figure}

A key feature of TADF emitters is the spatial separation of the highest occupied molecular orbital (HOMO) and lowest unoccupied molecular orbital (LUMO). This 
separation, quantified by the HOMO-LUMO overlap ($S'_{HL}$) and the centroid distance (\Cref{tab:SumProperties}), promotes a small singlet-triplet energy gap 
($\Delta E_{ST}$), which is essential for efficient reverse intersystem crossing (rISC). A small $\Delta E_{ST}$ is crucial because it allows for thermal 
activation from the triplet state ($T_1$) back to the singlet state ($S_1$), enabling the delayed fluorescence characteristic of TADF. Molecules with larger 
centroid distances and smaller $S'_{HL}$ values generally exhibit lower $\Delta E_{ST}$ values (see \Cref{fig:HMoverST}), supporting their suitability for TADF 
applications. The scatter plot in \Cref{fig:HMoverST} illustrates this inverse relationship, showing a general trend of decreasing $\Delta E_{ST}$ with 
decreasing $S'_{HL}$.

To characterize the spatial separation between HOMO and LUMO for each molecule, we evaluated the HOMO-LUMO absolute spatial overlap and the centroid distance 
using Multiwfn (\Cref{tab:SumProperties}). Across all studied molecules and both environments (vacuum and toluene), the $S'_{HL}$ values range from 0.1 to 0.5, 
indicating a low-to-moderate spatial overlap. This is consistent with the observed $\Delta E_{ST}$ values, which are generally below \qty{0.3}{\electronvolt} 
(except for TDBA-DI) for both \stda and \stddft methods. The calculated oscillator strengths ($f_{12}(S_0 \to S_1)$) are also sufficiently high (typically > 
0.01) to support radiative transitions from the $S_1$ state to the ground state ($S_0$). This means that once the molecule reaches the singlet excited state 
(either directly by absorption or via rISC from the triplet state), it has a reasonable probability of emitting a photon.

A closer examination of the data reveals some interesting trends. DMAC-TRZ exhibits a centroid distance of \qty{6.702}{\angstrom} in vacuum, coupled with an 
$S'_{HL}$ of \num{0.199} and a $\Delta E_{ST}$ of \qty{0.097}{\electronvolt} (\stda). This large spatial separation is a direct consequence of the strong donor 
(DMAC) and acceptor (TRZ) units, which force the HOMO and LUMO to localize on different parts of the molecule. The $S'_{HL}$ value suggests weak electronic 
coupling, leading to a smaller $\Delta E_{ST}$. This small $\Delta E_{ST}$ makes DMAC-TRZ a promising candidate for efficient rISC.

In contrast, 4CzIPN has a much smaller centroid distance of \qty{0.032}{\angstrom}, a higher $S'_{HL}$ of \num{0.325}, and a larger $\Delta E_{ST}$ of 
\qty{0.212}{\electronvolt} (\stda). As shown in \Cref{fig:4CzIPN-Sem}, the HOMO and LUMO are significantly delocalized throughout the molecule, leading to a 
greater spatial overlap and a larger $\Delta E_{ST}$. This delocalization arises from the four carbazole (Cz) donor units directly connected to the IPN 
acceptor, promoting strong electronic communication between the donor and acceptor regions. While the larger $\Delta E_{ST}$ might suggest less efficient RISC 
compared to DMAC-TRZ, 4CzIPN benefits from a higher oscillator strength (\num{0.152} in vacuum using \stda), indicating a potentially faster radiative decay 
rate and a higher fluorescence quantum yield if RISC is still reasonably efficient.

The centroid distance provides further insights into the nature of the excited state and allows us to categorize the studied molecules:
\begin{itemize}
    \item Small centroid distance (<\qty{1}{\angstrom}). 4CzIPN falls into this category, exhibiting a small centroid distance and, as seen in 
\Cref{fig:4CzIPN-Sem}, significant spatial overlap between the HOMO and LUMO. The HOMO is distributed across the four carbazole donors, while the LUMO is 
primarily located on the IPN acceptor, but with some delocalization onto the carbazole units. This indicates a more localized excited state with some degree of 
charge transfer character.

    \item Intermediate centroid distance (\qtyrange{2}{5}{\angstrom}). DMAC-DPS, PSPCz, Px2BP, CzS2, and 2TCz-DPS fall into this category. This range indicates 
a clear spatial separation between the HOMO and LUMO, with the orbitals located on different regions of the molecule. For example, in DMAC-DPS, the HOMO is 
primarily localized on the DMAC donor unit, while the LUMO is localized on the DPS acceptor. This spatial separation is characteristic of charge-transfer 
excitations in donor-acceptor systems, where an electron is transferred from the donor to the acceptor upon excitation. The magnitude of the centroid distance 
in this range is influenced by the specific linker connecting the donor and acceptor units, as well as the relative strengths of the donor and acceptor 
moieties.

    \item Large centroid distance (>\qty{5}{\angstrom}). DMAC-TRZ and TDBA-DI exhibit large centroid distances, suggesting a significant spatial separation of 
the HOMO and LUMO. In DMAC-TRZ, the HOMO is almost entirely localized on the DMAC unit, while the LUMO is almost entirely localized on the TRZ unit. This 
pronounced spatial separation results in a smaller electronic coupling and a reduced $\Delta E_{ST}$, making these molecules strong candidates for efficient 
TADF. The larger centroid distance also suggests a more significant change in dipole moment upon excitation, which can influence the molecule's interaction 
with the surrounding environment.
\end{itemize}

Interestingly, we observe that the centroid distance for DMAC-TRZ increases slightly from 6.702 \unit{\angstrom} in vacuum to \qty{6.772}{\angstrom} in 
toluene. This subtle increase suggests that the toluene solvent may slightly stabilize a more charge-separated state, further enhancing the donor-acceptor 
character. In contrast, the $S'_{HL}$ value for DMAC-TRZ decreases from \num{0.199} in vacuum to \num{0.159} in toluene, indicating a further reduction in 
orbital overlap upon solvation. These changes are likely due to the polarizable nature of toluene, which can interact with the molecule's dipole moment and 
induce conformational changes that favor greater charge separation.

\subsubsection{Frontier orbital analysis}\label{sec:MolProp2}

The energies and spatial distributions of the frontier molecular orbitals (HOMO and LUMO) provide valuable insights into the electronic structure and 
charge-transfer characteristics of TADF emitters. We calculated frontier orbital energies using the \gf method and visualized their distributions for 
representative molecules to understand the nature of electronic transitions.

As discussed in \Cref{sec:MolProp1}, molecules with lower HOMO-LUMO overlap ($S'_{HL}$) generally exhibit stronger donor-acceptor interactions and a higher 
degree of charge-transfer character. This is because reduced orbital overlap hinders electron delocalization, favoring a more localized distribution of the 
HOMO and LUMO on the donor and acceptor units, respectively. This localization enhances the charge-transfer character of the $S_0 \to S_1$ transition.

Our analysis reveals that molecules with more symmetrical structures tend to exhibit greater delocalization of the frontier orbitals. While this delocalization 
can lead to marginally larger $\Delta E_{ST}$ values, it can also enhance fluorescence efficiency by increasing the oscillator strength ($f_{12}(S_0 \to 
S_1)$). This increase in oscillator strength is due to a larger transition dipole moment associated with the more delocalized electronic transition. In 
contrast, emitters with asymmetrical structures and lower HOMO-LUMO overlap display more pronounced charge-transfer character, with the HOMO and LUMO primarily 
localized on the donor and acceptor units, respectively. This localization promotes a smaller $\Delta E_{ST}$, which is critical for efficient rISC.

To illustrate these points, let's consider the frontier orbital distributions of DMAC-TRZ and 4CzIPN. For (i) DMAC-TRZ, the HOMO is almost entirely localized 
on the DMAC donor, while the LUMO is almost entirely localized on the TRZ acceptor. This strong spatial separation is consistent with the large centroid 
distance and low HOMO-LUMO overlap observed in \Cref{tab:SumProperties}, confirming its significant charge-transfer character. (ii) In contrast, for 4CzIPN, the 
HOMO and LUMO are more delocalized throughout the molecule, with contributions from both the carbazole donors and the IPN acceptor. While the spatial separation 
is less pronounced compared to DMAC-TRZ, the higher oscillator strength suggests a stronger radiative transition probability, potentially leading to higher 
fluorescence quantum yield.

In addition to the spatial distributions, the energies of the frontier orbitals also play a significant role. A larger HOMO-LUMO energy gap generally 
corresponds to a higher excitation energy for the $S_0 \to S_1$ transition. We observed that molecules with strong electron-donating groups (e.g., DMAC) and 
strong electron-withdrawing groups (e.g., TRZ) tend to have larger HOMO-LUMO energy gaps and higher excitation energies. The specific values depend on the 
ionization potential and electron affinity of the donor and acceptor, respectively.

The solvent environment can also influence the frontier orbital energies and distributions. Polar solvents like toluene can stabilize charge-transfer states by 
lowering the energy of the LUMO (primarily located on the acceptor) and increasing the energy of the HOMO (primarily located on the donor). This stabilization 
can lead to a redshift in the absorption and emission spectra, as well as changes in the $\Delta E_{ST}$ and oscillator strength.

The dihedral angles (\Cref{tab:DAnunits}) between the donor and acceptor units can influence the degree of electronic communication between the HOMO and LUMO. 
Larger dihedral angles often reduce the orbital overlap, leading to a decrease in $\Delta E_{ST}$, particularly when steric hindrance forces a non-planar 
conformation. 

In summary, the frontier orbital analysis provides a valuable framework for understanding the electronic structure and charge-transfer characteristics of TADF 
emitters. The spatial distributions and energies of the HOMO and LUMO, as well as the influence of symmetry, dihedral angles, and solvent effects, all 
contribute to the observed photophysical properties. In the next section, we will explore the geometrical features of these molecules in more detail and their 
impact on excited-state behavior.

\subsubsection{Geometrical features and conformational flexibility}\label{sec:MolProp3}

The three-dimensional geometry of TADF emitters, particularly the torsional angles between donor and acceptor units, plays a pivotal role in determining their 
excited-state properties and, consequently, their TADF performance. These geometrical features influence the degree of electronic communication between the 
donor and acceptor, which in turn affects the HOMO-LUMO overlap, $\Delta E_{ST}$, and radiative decay rates.

As shown in \Cref{tab:DAnunits} and illustrated in \Cref{fig:Geo-Features1,fig:Geo-Features2}, the torsional angles between the donor and acceptor units vary 
significantly among the studied molecules. Larger torsional angles, often driven by steric hindrance between bulky substituents, generally decrease the orbital 
overlap between the HOMO and LUMO. This reduction in orbital overlap arises because a non-planar geometry disrupts the conjugation between the donor and 
acceptor units, reducing the electronic communication and hindering the delocalization of the frontier orbitals. According to Fermi's Golden Rule, a reduced 
electronic coupling reduces transition probabilities.

A reduced orbital overlap, as a consequence of larger torsional angles, directly leads to a smaller singlet-triplet energy gap ($\Delta E_{ST}$). This is 
because the energy splitting between the singlet and triplet states depends on the magnitude of the exchange integral, which is directly proportional to the 
orbital overlap.

For example, DMAC-TRZ exhibits a torsional angle of \ang{73.138} in vacuum. In contrast, 4CzIPN shows torsional angles of \qtylist{-61.064; -55.754; 
53.770;-46.580}{\degree} in vacuum. The smaller torsional angles in 4CzIPN, along with its more symmetrical structure, contribute to a greater degree of 
electronic delocalization and a higher HOMO-LUMO overlap compared to DMAC-TRZ, consistent with the data presented in \Cref{tab:SumProperties}.

Solvent effects can further amplify these geometrical differences. Polarizable environments, such as toluene, can stabilize more planar conformations by 
reducing the steric repulsion between the donor and acceptor units, reducing the torsional angles. This planarization enhances the electronic communication 
between the donor and acceptor, potentially increasing the charge-transfer efficiency, although this effect is often counteracted by the increased orbital 
overlap.

The root-mean-square deviation (RMSD) analysis, presented in \Cref{fig:Geo-Features1,fig:Geo-Features2}, confirms that geometry changes in the solvated 
environment primarily occur in regions near the donor-acceptor interface. This suggests that the solvent primarily interacts with the donor and acceptor units, 
influencing their relative orientation and the overall molecular conformation. 

It is important to note that the flexibility of the molecules plays a crucial role here. Some molecules are more rigid and exhibit smaller changes in torsional 
angles upon solvation, while others are more flexible and readily adapt their conformation to the surrounding environment. This conformational flexibility can 
have a significant impact on the TADF performance by modulating the electronic coupling and the $\Delta E_{ST}$.

In conclusion, the geometrical features of TADF emitters, particularly the torsional angles between donor and acceptor units, have a profound impact on their 
electronic structure and excited-state properties. By controlling these geometrical features, it is possible to tune the HOMO-LUMO overlap, $\Delta E_{ST}$, 
and radiative decay rates, ultimately optimizing the TADF performance. It is essential to consider the interplay between intrinsic molecular geometry, steric 
hindrance, and solvent effects to fully understand the conformational behavior of these molecules. 

\begin{table}[!htbp]
\caption{Torsional angles between donor and acceptor units in vacuum and toluene, and the solvent-induced shift.  Dihedral angles, $\delta i$ (\unit{\degree}), 
quantify the relative orientation of donor and acceptor units (defined in \Cref{tab:DAunits}) in vacuum and toluene. Solvent-induced rotation shift, 
$\Delta\delta i_{solv}$ (\unit{\degree}), reflects the change in torsional angle upon solvation. Bold font indicates dihedral angles between the two branches 
of DPS or methanone. These angles influence the degree of electronic coupling and charge-transfer character.}\label{tab:DAnunits}
{\scriptsize
\begin{center}
 \begin{tabularx}{\linewidth}{@{}*{9}{Y}@{}}
\toprule
Molecule & DMAC-TRZ & DMAC-DPS & PSPCz & 4CzIPN & Px2BP & CzS2 & 2TCz-DPS & TDBA-DI\\
\midrule
In vacuum & \tablenum[table-format = 3.3]{73.138} & \tablenum[table-format = 3.3]{102.699}\newline
\tablenum[table-format = 3.3]{102.783}\newline {\bfseries \tablenum[table-format = 3.3]{90.209}} & \tablenum[table-format = 3.3]{45.784}\newline {\bfseries 
\tablenum[table-format = 3.3]{-70.597}} & \tablenum[table-format = 3.3]{-61.064}\newline \tablenum[table-format = 3.3]{-55.754}
\newline \tablenum[table-format = 3.3]{53.770}\newline \tablenum[table-format = 3.3]{-46.580} & \tablenum[table-format = 3.3]{60.409}\newline 
\tablenum[table-format = 3.3]{61.276}\newline 
{\bfseries \tablenum[table-format = 3.3]{-33.097}} & \tablenum[table-format = 3.3]{-46.611}\newline 
\tablenum[table-format = 3.3]{48.456}\newline {\bfseries \tablenum[table-format = 3.3]{-83.506}}& \tablenum[table-format = 3.3]{42.216}\newline 
\tablenum[table-format = 3.3]{-46.229}\newline {\bfseries \tablenum[table-format = 3.3]{-71.365}} & \tablenum[table-format = 3.3]{114.584}\\\midrule
In toluene solvent
& \tablenum[table-format = 3.3]{78.592}
& \tablenum[table-format = 3.3]{69.024}\newline \tablenum[table-format = 3.3]{69.456}\newline {\bfseries \tablenum[table-format = 3.3]{-90.005}}
& \tablenum[table-format = 3.3]{-46.061}\newline {\bfseries\tablenum[table-format = 3.3]{-105.874}}
& \tablenum[table-format = 3.3]{130.063}\newline \tablenum[table-format = 3.3]{123.563}\newline \tablenum[table-format = 3.3]{55.293}\newline 
\tablenum[table-format = 3.3]{120.192}
& \tablenum[table-format = 3.3]{-118.392}\newline \tablenum[table-format = 3.3]{-118.273}\newline {\bfseries \tablenum[table-format = 3.3]{-33.901}}
& \tablenum[table-format = 3.3]{130.905}\newline \tablenum[table-format = 3.3]{131.358}\newline {\bfseries \tablenum[table-format = 3.3]{-112.675}}
& \tablenum[table-format = 3.3]{-124.911}\newline \tablenum[table-format = 3.3]{132.333}\newline {\bfseries\tablenum[table-format = 3.3]{93.803}}
& \tablenum[table-format = 3.3]{113.716}\\\midrule
$\Delta\delta i_{solv}\unit(\degree)$ & \tablenum[table-format = 3.3]{5.545} & \tablenum[table-format = 3.3]{33.675}\newline \tablenum[table-format = 
3.3]{33.327}\newline {\bfseries \tablenum[table-format = 3.3]{180.214}} & \tablenum[table-format = 3.3]{91.845} {\bfseries\tablenum[table-format = 
3.3]{35.277}} & \tablenum[table-format = 3.3]{191.127}\newline \tablenum[table-format = 3.3]{179.317}\newline \tablenum[table-format = 3.3]{1.523}\newline 
\tablenum[table-format = 3.3]{166.962} & \tablenum[table-format = 3.3]{178.801}\newline \tablenum[table-format = 3.3]{179.549}\newline {\bfseries 
\tablenum[table-format = 3.3]{0.804}} & \tablenum[table-format = 3.3]{177.516}\newline \tablenum[table-format = 3.3]{82.902}\newline 
{\bfseries\tablenum[table-format = 3.3]{29.169}} & \tablenum[table-format = 3.3]{167.127}\newline \tablenum[table-format = 3.3]{178.562}\newline 
{\bfseries\tablenum[table-format = 3.3]{165.168}} & \tablenum[table-format = 3.3]{0.868}\\
\bottomrule
 \end{tabularx}
\end{center}
}
\end{table}

\subsubsection{Correlations with photophysical properties}\label{sec:MolProp4}

The preceding sections have detailed the key molecular properties of our TADF emitters, including their donor-acceptor interactions, frontier orbital 
distributions, and geometrical features. Now, we turn to how these intrinsic properties correlate with the observed photophysical behavior, specifically the 
emission wavelength, fluorescence quantum yield, and excited-state lifetimes.

\begin{itemize}
\item We observed a clear correlation between the localization of frontier orbitals and the sharpness of the fluorescence peaks. Molecules with highly 
localized frontier orbitals, as evidenced by their large centroid distances and small HOMO-LUMO overlaps (\Cref{tab:SumProperties}), exhibited narrower 
emission bandwidths and sharper fluorescence peaks. This is because localized excitations tend to be less sensitive to vibrational modes and environmental 
fluctuations, resulting in a more homogeneous emission spectrum. These molecules typically exhibited oscillator strengths above \num{0.1}, indicating a strong 
radiative transition probability.

\item Emitters with higher charge-transfer character, as indicated by their large centroid distances and significant changes in dipole moment upon excitation, 
displayed larger solvent-induced redshifts in their excitation and fluorescence spectra. This is because polar solvents like toluene stabilize the 
charge-transfer excited state, lowering its energy and shifting the emission to longer wavelengths. This redshift provides direct evidence for the 
charge-transfer nature of the excited state. Furthermore, the strength of the donor and acceptor units play a significant role.

\item The singlet-triplet energy gap ($\Delta E_{ST}$) is a critical factor in determining the efficiency of TADF. Molecules with small $\Delta E_{ST}$ values 
exhibit more efficient reverse intersystem crossing (rISC), leading to shorter delayed fluorescence lifetimes and higher TADF efficiencies. For example for the 
\stda method, DMAC-TRZ, with its small $\Delta E_{ST}$ of \qty{0.097}{\electronvolt} (in vacuum), showed
a lifetime of $\qty{33.044}{\nano\second}$ and a TADF efficiency of $\qty{0.825}{\percent}$. In contrast, 4CzIPN, with its larger $\Delta E_{ST}$ of 
\qty{0.212}{\electronvolt} (in vacuum), exhibited a lifetime of $\qty{24.590}{\nano\second}$ and a TADF efficiency of $\qty{0.021}{\percent}$. Note that other 
factors, such as the radiative and non-radiative decay rates, also influence the overall TADF
efficiency (see \Cref{tab:EffiGas,tab:EffiTol}).

\item Geometrical features, such as the torsional angles between donor and acceptor units, also influenced the radiative decay rates. Molecules with more 
planar conformations, resulting in greater electronic communication between the donor and acceptor, tended to have higher oscillator strengths and faster 
radiative decay rates. However, this can be a trade-off, since more planar conformations also increase the HOMO-LUMO overlap, leading to larger $\Delta E_{ST}$ 
values and potentially less efficient rISC.

\end{itemize}

\vspace{.5cm}

In conclusion, the molecular properties of TADF emitters, including their donor-acceptor interactions, frontier orbital distributions, and geometrical 
features, are intimately linked to their photophysical behavior. By carefully controlling these properties, it is possible to design TADF materials with 
optimized emission wavelengths, high fluorescence quantum yields, and efficient rISC, paving the way for next-generation OLEDs and other optoelectronic devices.

\begin{figure}[!htbp]
\centering
\begin{minipage}[c]{.61\textheight}
\subfloat[DMAC-TRZ]{
\includegraphics[width=0.21\textwidth]{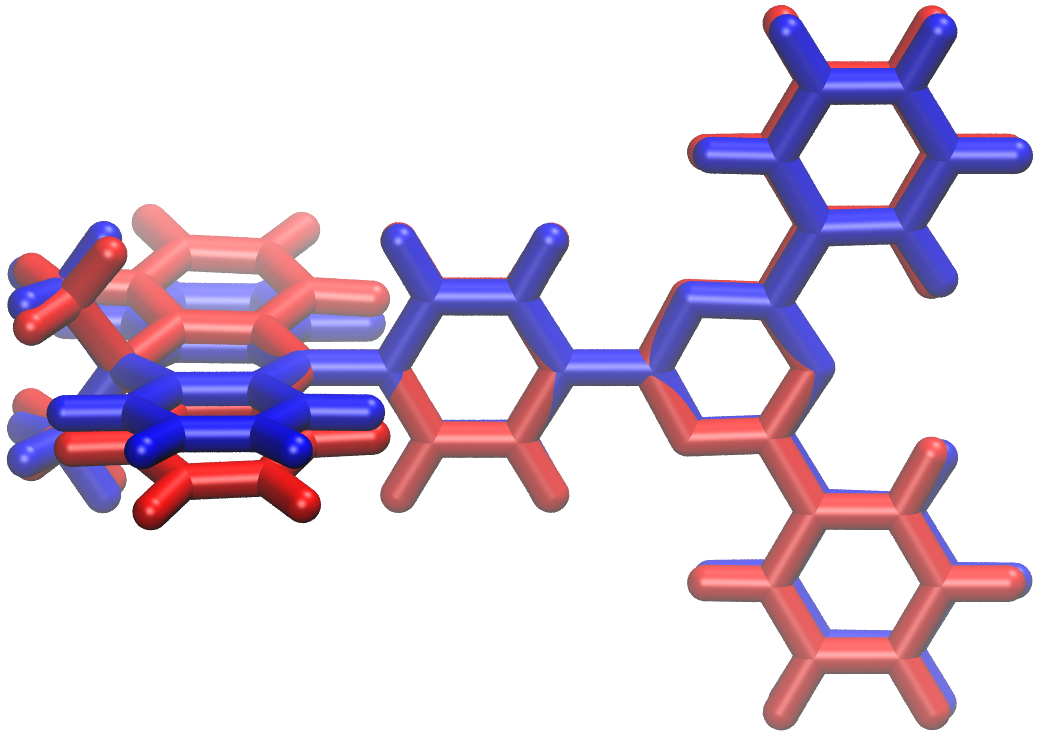}{(i)}\hfill
\includegraphics[width=0.21\textwidth]{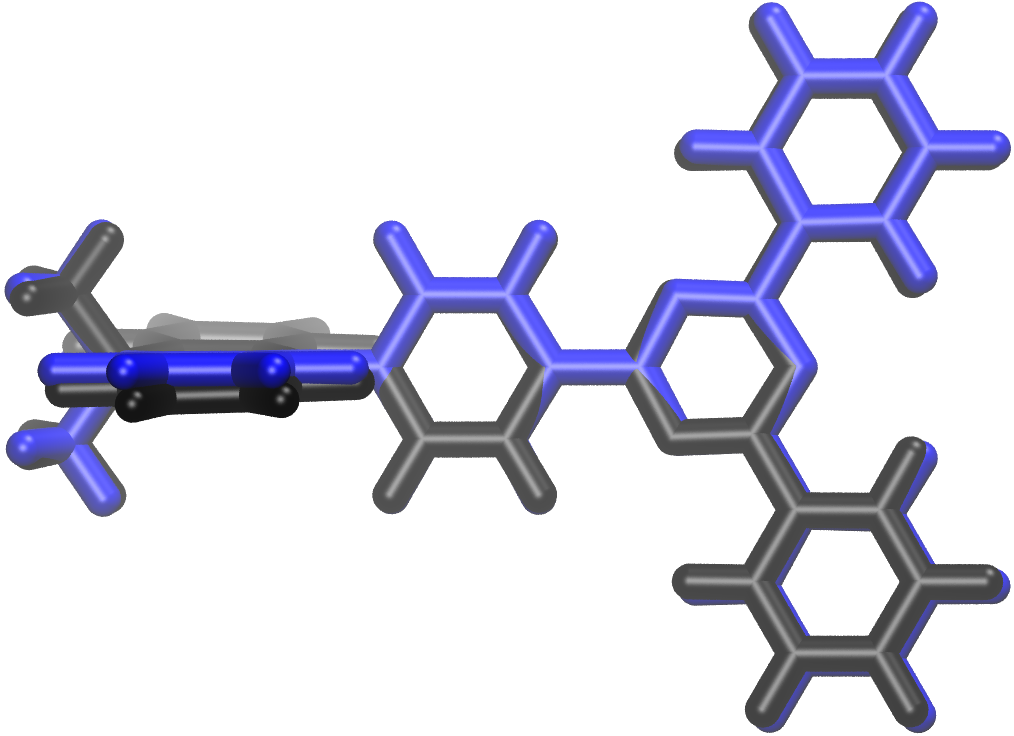}{(ii)}\hfill
\includegraphics[width=0.21\textwidth]{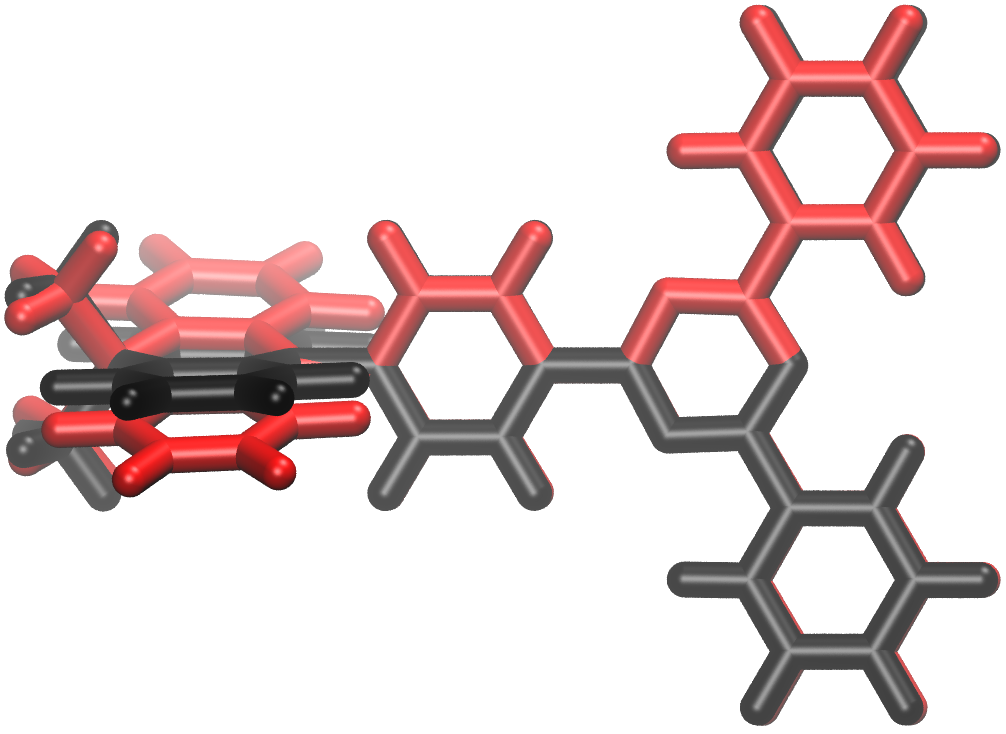}{(iii)}\hfill
\includegraphics[width=0.25\textwidth]{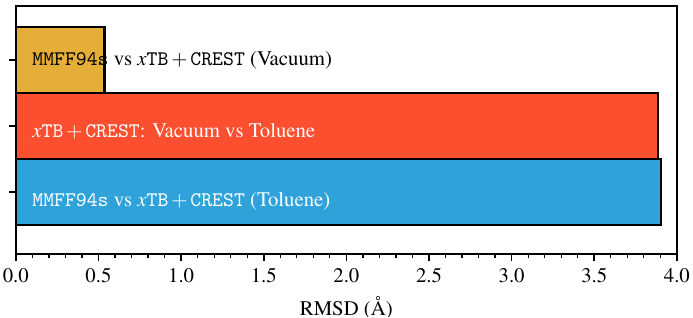}{(iv)}
\label{fig:DMAC-TRZ-RMSD}
}

\subfloat[DMAC-DPS]{
\includegraphics[width=0.21\textwidth]{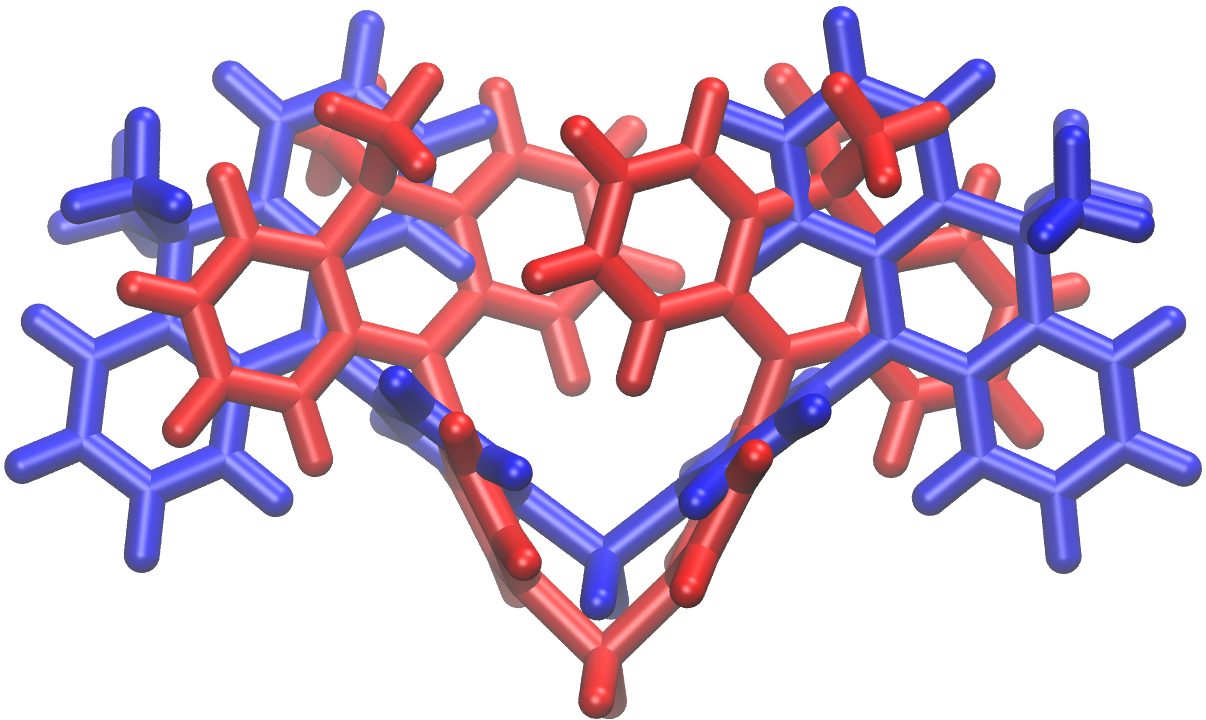}{(i)}\hfill
\includegraphics[width=0.21\textwidth]{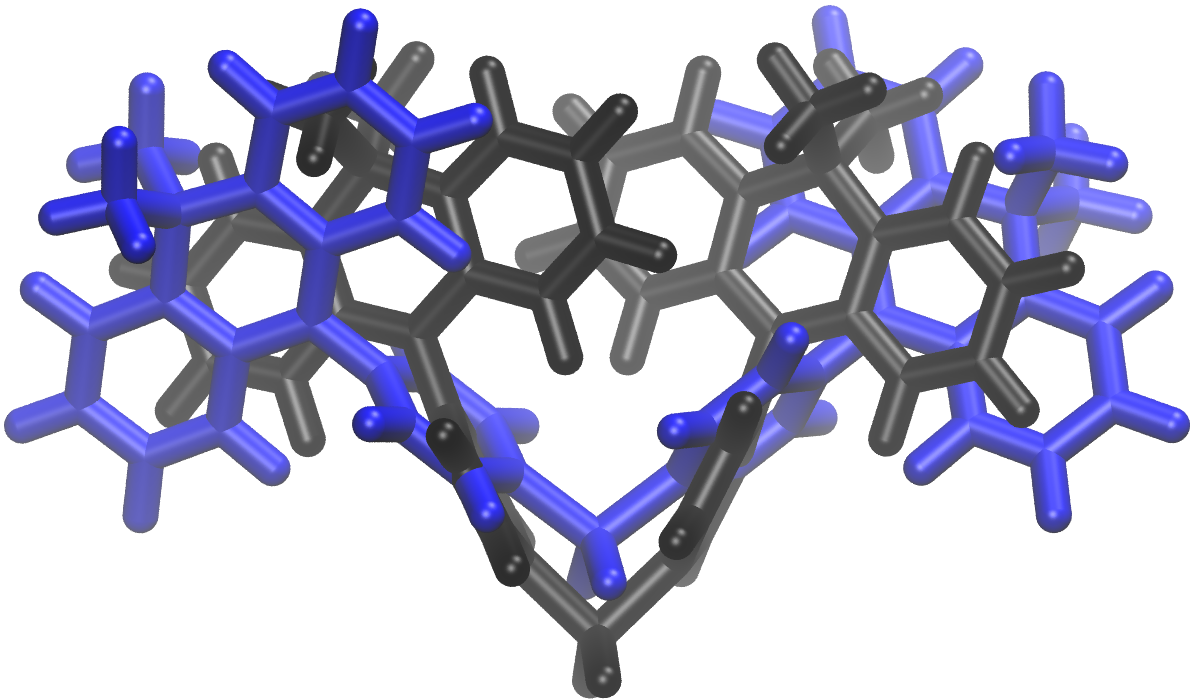}{(ii)}\hfill
\includegraphics[width=0.21\textwidth]{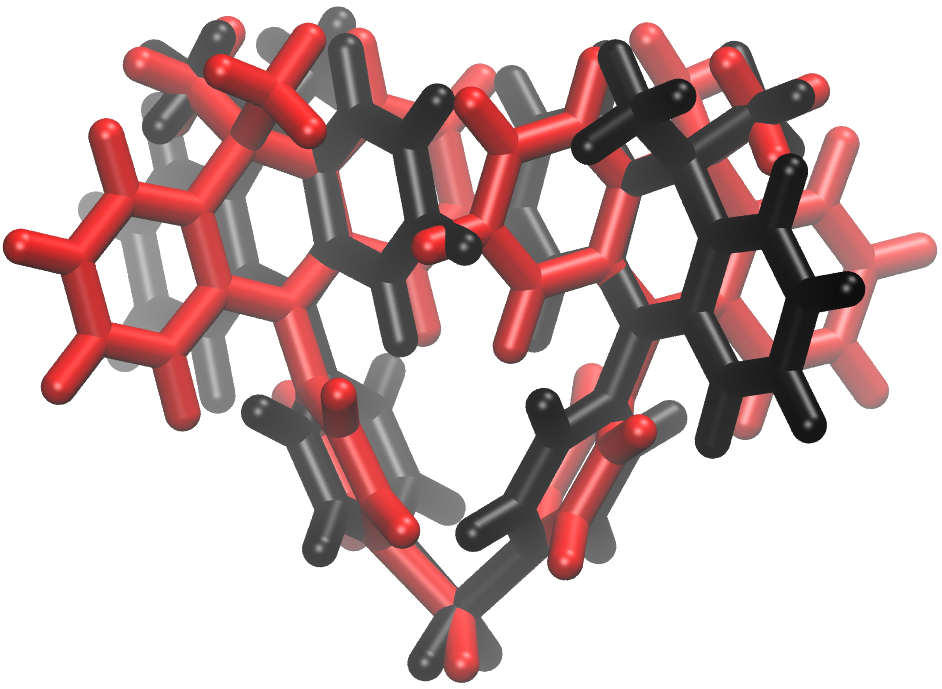}{(iii)}\hfill
\includegraphics[width=0.25\textwidth]{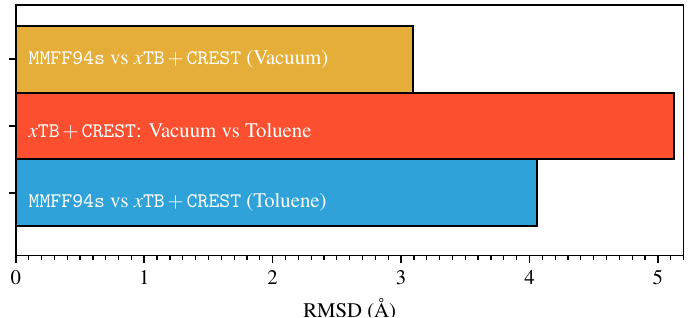}{(iv)}
\label{fig:DMAC-DPS-RMSD}
}

\subfloat[PSPCz]{
\includegraphics[width=0.21\textwidth]{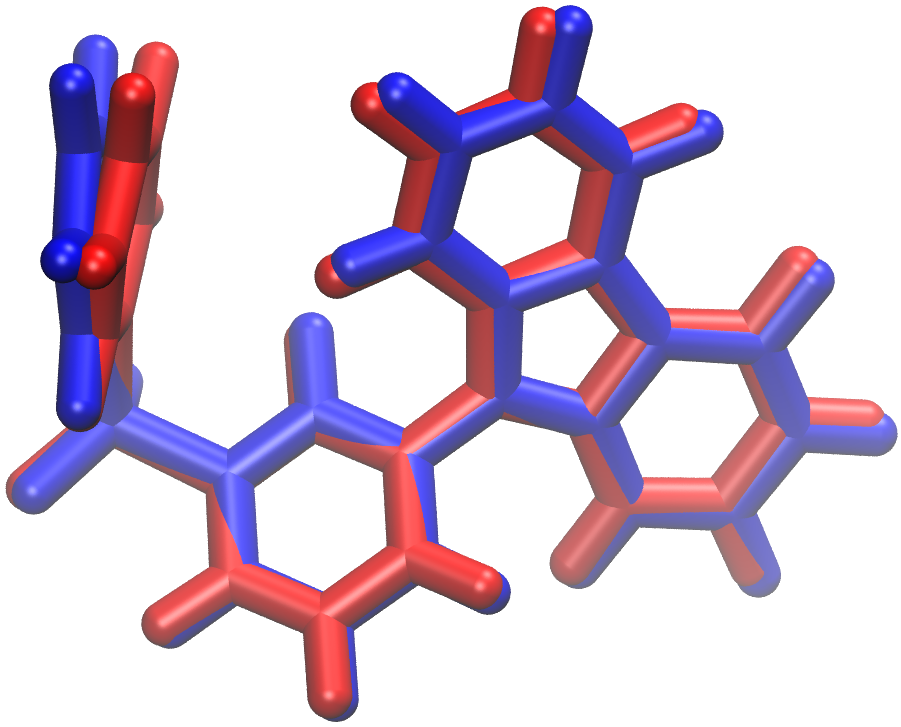}{(i)}\hfill
\includegraphics[width=0.21\textwidth]{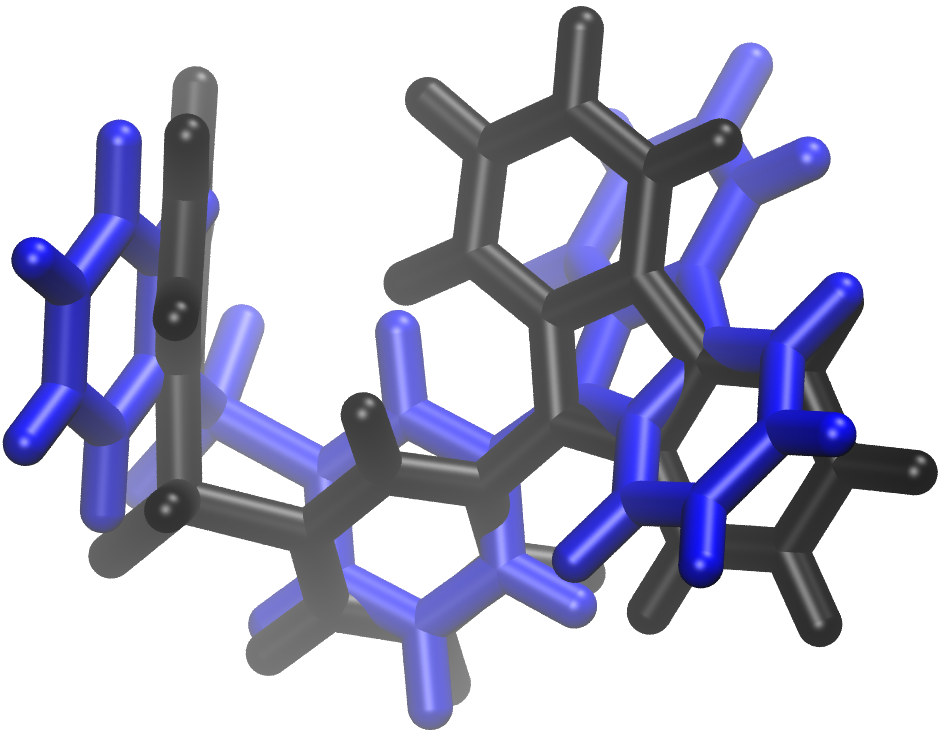}{(ii)}\hfill
\includegraphics[width=0.21\textwidth]{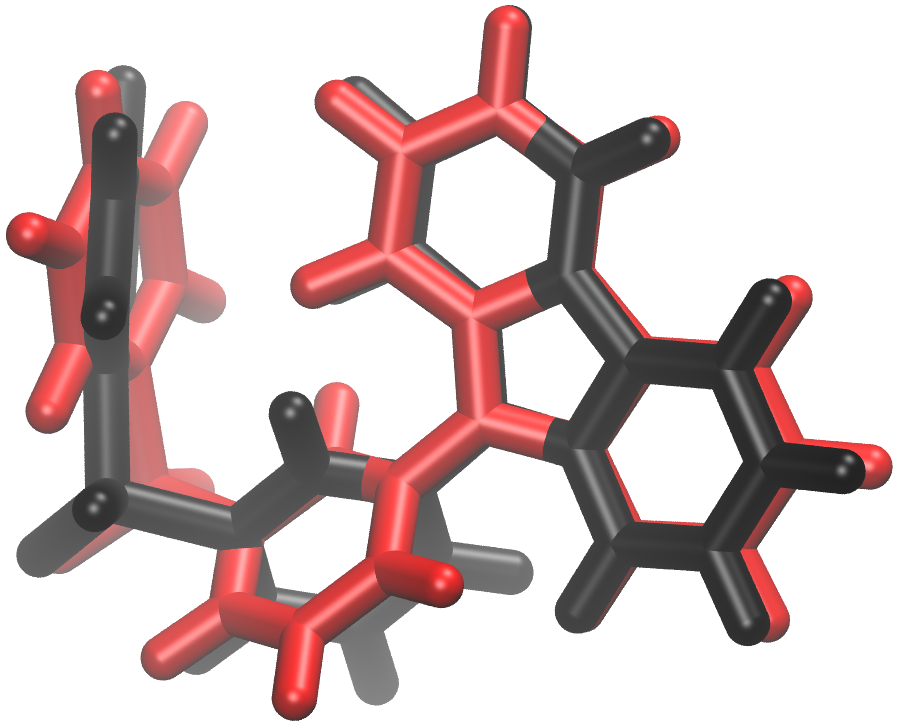}{(iii)}\hfill
\includegraphics[width=0.25\textwidth]{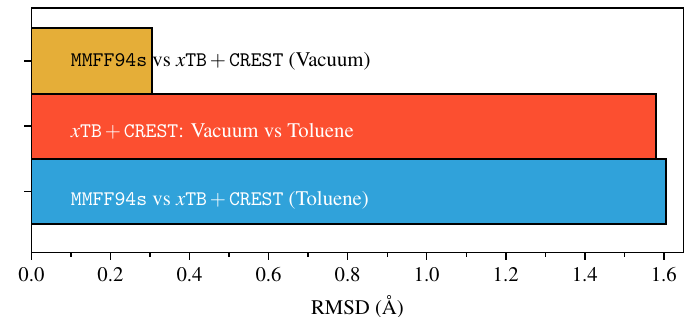}{(iv)}
\label{fig:PSPCz-RMSD}
}

\subfloat[4CzIPN]{
\includegraphics[width=0.21\textwidth]{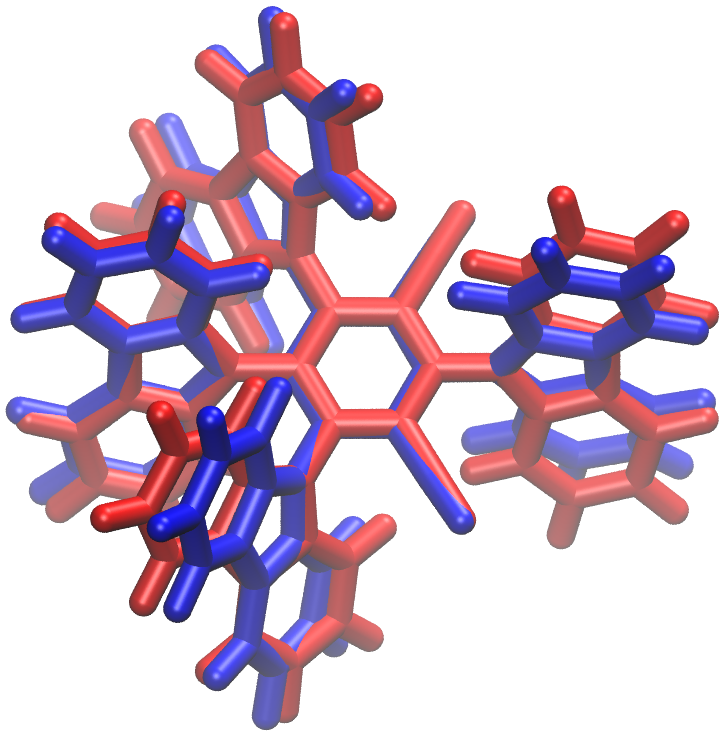}{(i)}\hfill
\includegraphics[width=0.21\textwidth]{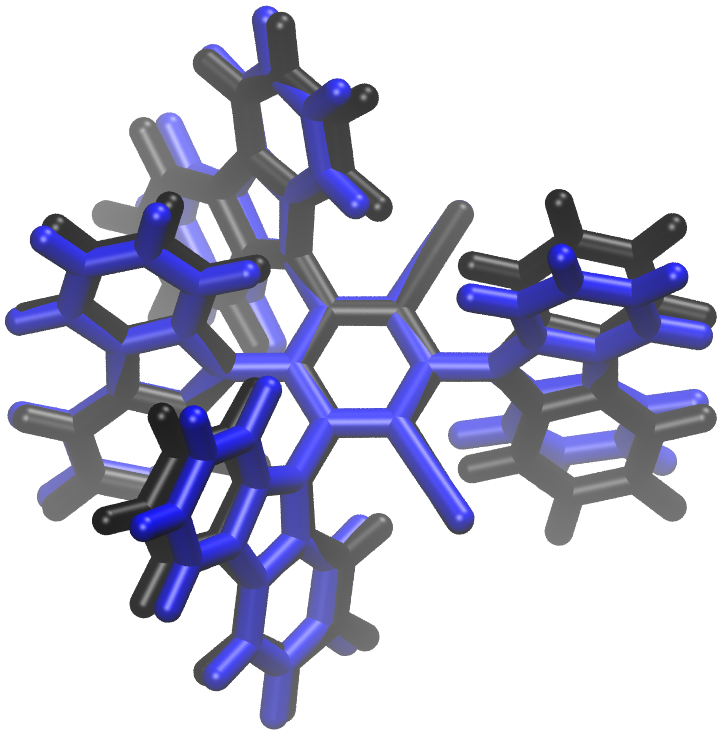}{(ii)}\hfill
\includegraphics[width=0.21\textwidth]{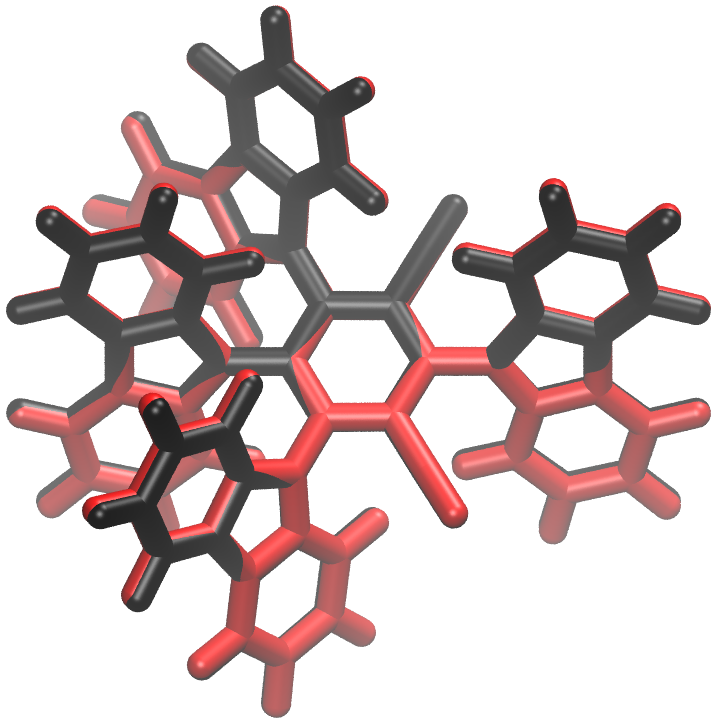}{(iii)}\hfill
\includegraphics[width=0.25\textwidth]{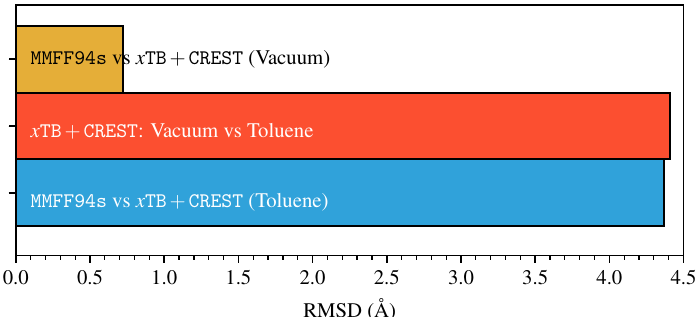}{(iv)}
\label{fig:4CzIPN-RMSD} 
}
\caption{Solvent-induced conformational changes in DMAC-TRZ, DMAC-DPS, PSPCz, and 4CzIPN, highlighting the impact of toluene on molecular geometry. (i) 
Comparison of the optimized structure in vacuum using \xtb and \crest methods (red color) with the optimized structure using \mmf method (blue color). (ii) 
Comparison of the optimized structure in toluene solvent using \xtb and \crest methods (black color) with the optimized structure using \mmf method (blue 
color). (iii) Comparison of the optimized structure in vacuum (red color)  with the optimized structure in toluene solvent (black color) both using \xtb and 
\crest methods (red color). (iv) Root-mean-square deviation (RMSD) plot, quantifying the structural differences between the vacuum and toluene geometries, 
using the \mmf, \xtb, and \crest methods. Note the significant conformational changes near the donor-acceptor interface in DMAC-TRZ for example.}
\label{fig:Geo-Features1}
\end{minipage}
\end{figure}

\begin{figure}[!htbp]
\centering
\begin{minipage}[c]{.6\textheight}
\ContinuedFloat
\subfloat[Px2BP]{
\includegraphics[width=0.21\textwidth]{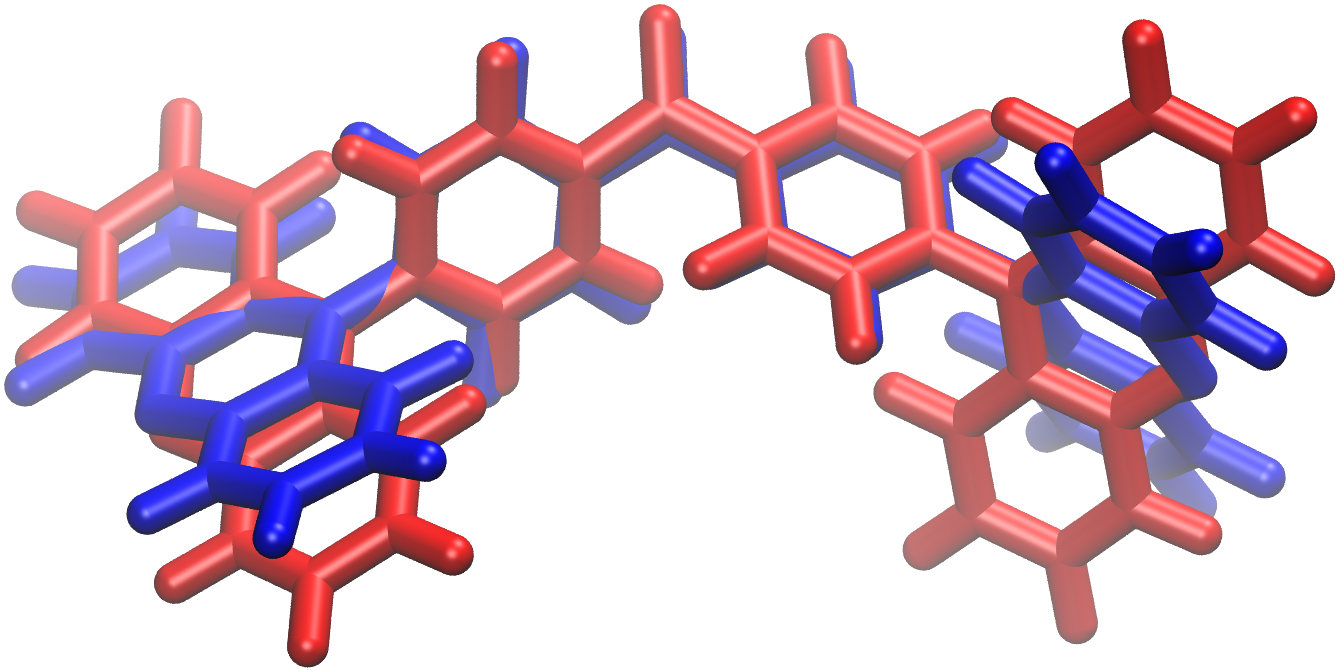}{(i)}\hfill
\includegraphics[width=0.21\textwidth]{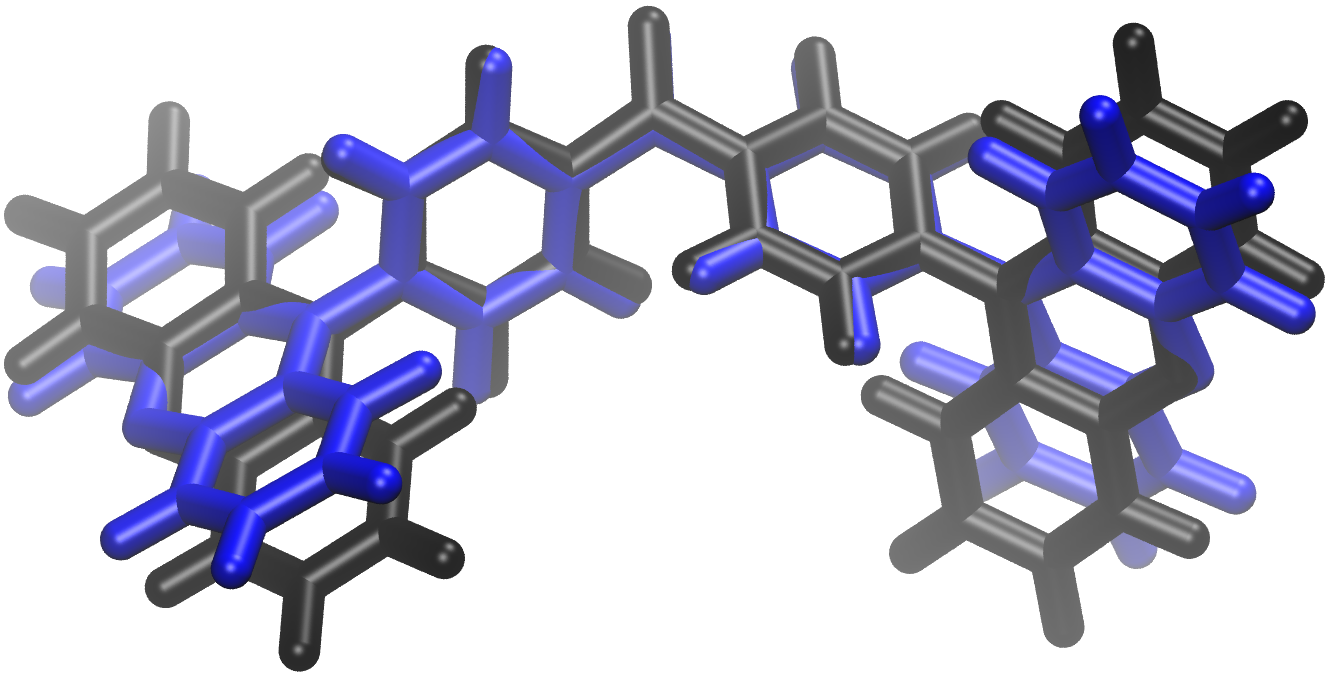}{(ii)}\hfill
\includegraphics[width=0.21\textwidth]{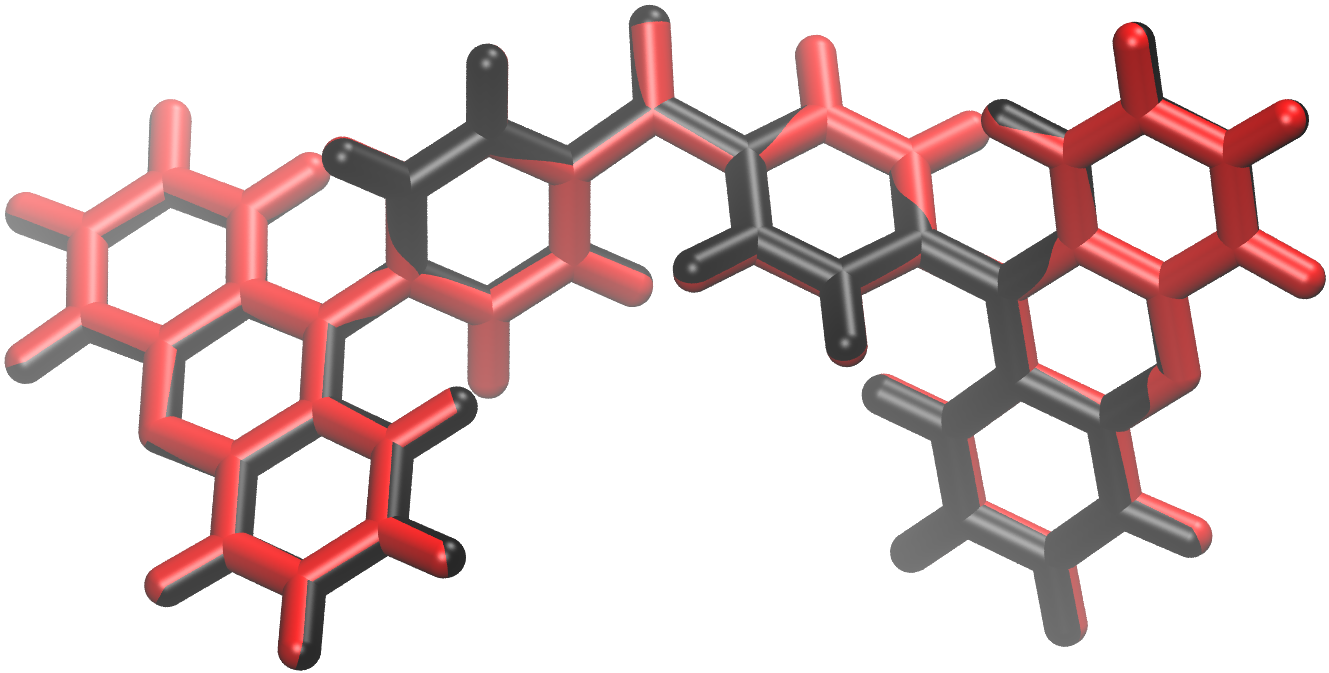}{(iii)}\hfill
\includegraphics[width=0.25\textwidth]{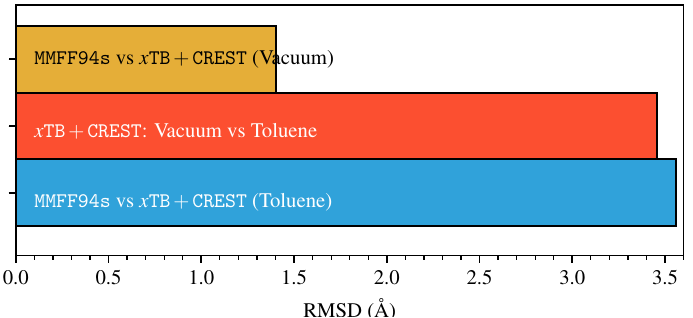}{(iv)}
\label{fig:Px2BP-RMSD}
}

\subfloat[CzS2]{
\includegraphics[width=0.21\textwidth]{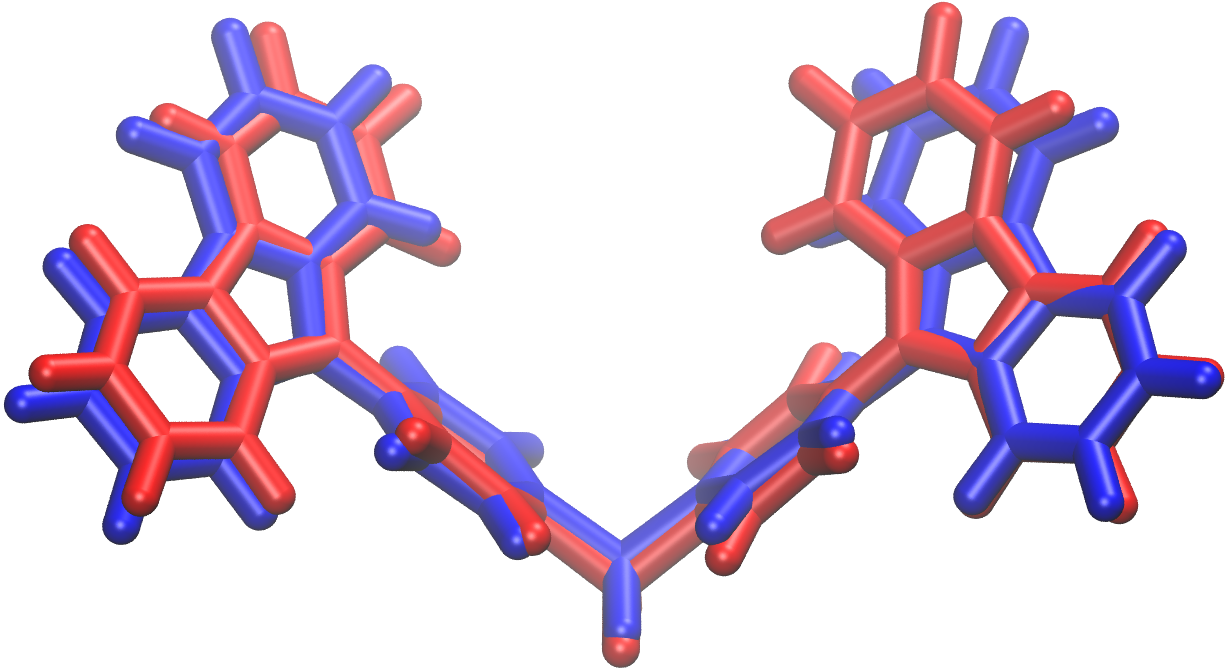}{(i)}\hfill
\includegraphics[width=0.21\textwidth]{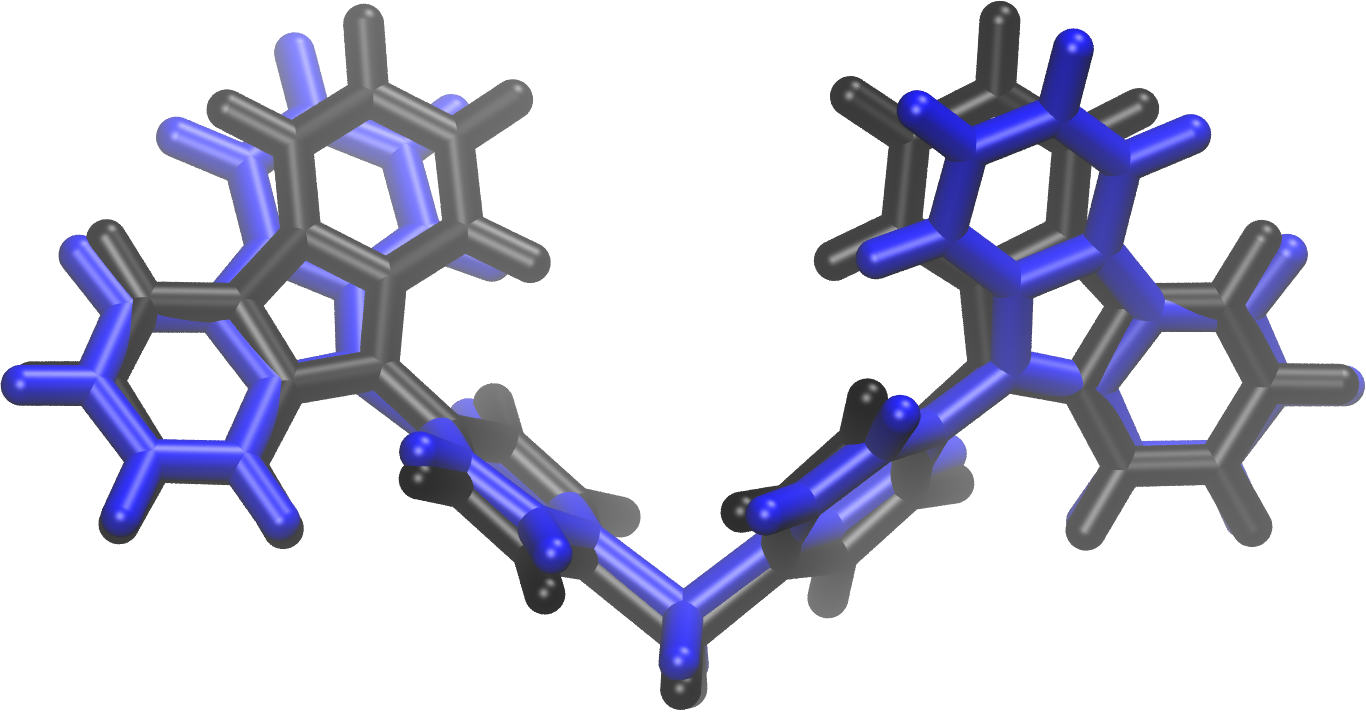}{(ii)}\hfill
\includegraphics[width=0.21\textwidth]{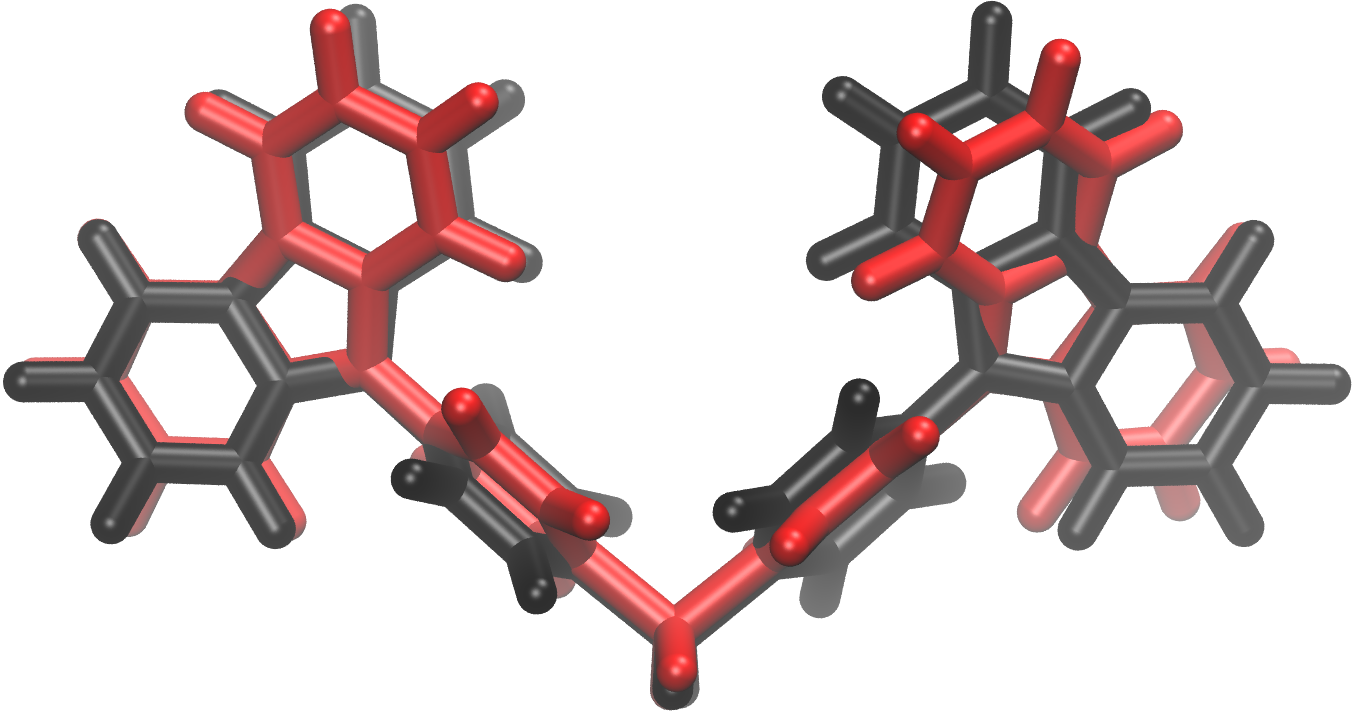}{(iii)}\hfill
\includegraphics[width=0.25\textwidth]{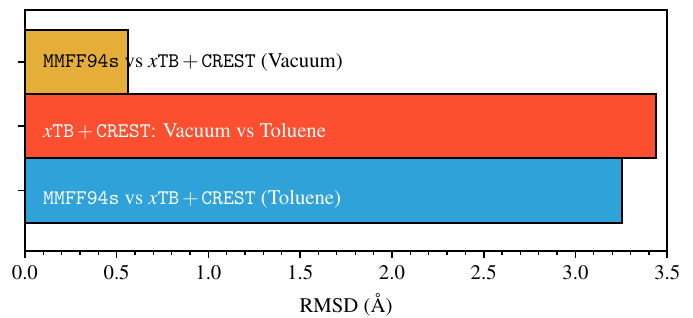}{(iv)}
\label{fig:CzS2-RMSD}
}

\subfloat[2TCz-DPS]{
\includegraphics[width=0.21\textwidth]{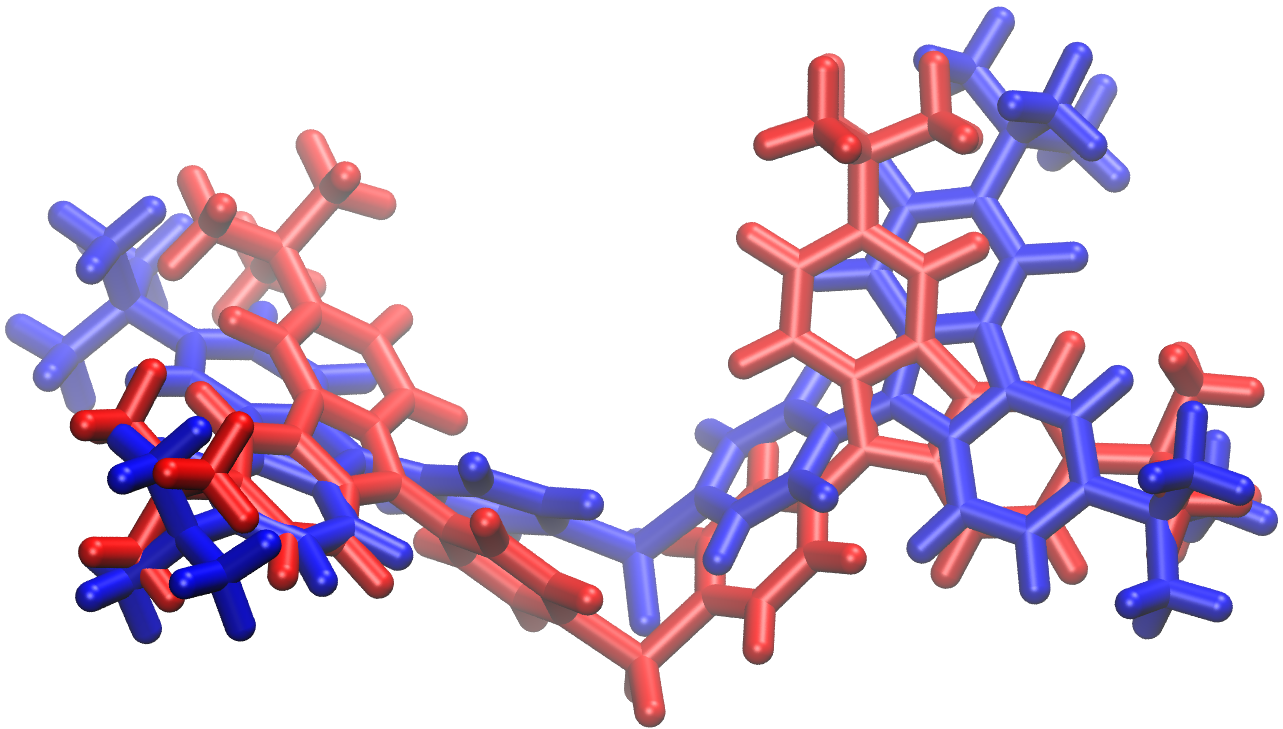}{(i)}\hfill
\includegraphics[width=0.21\textwidth]{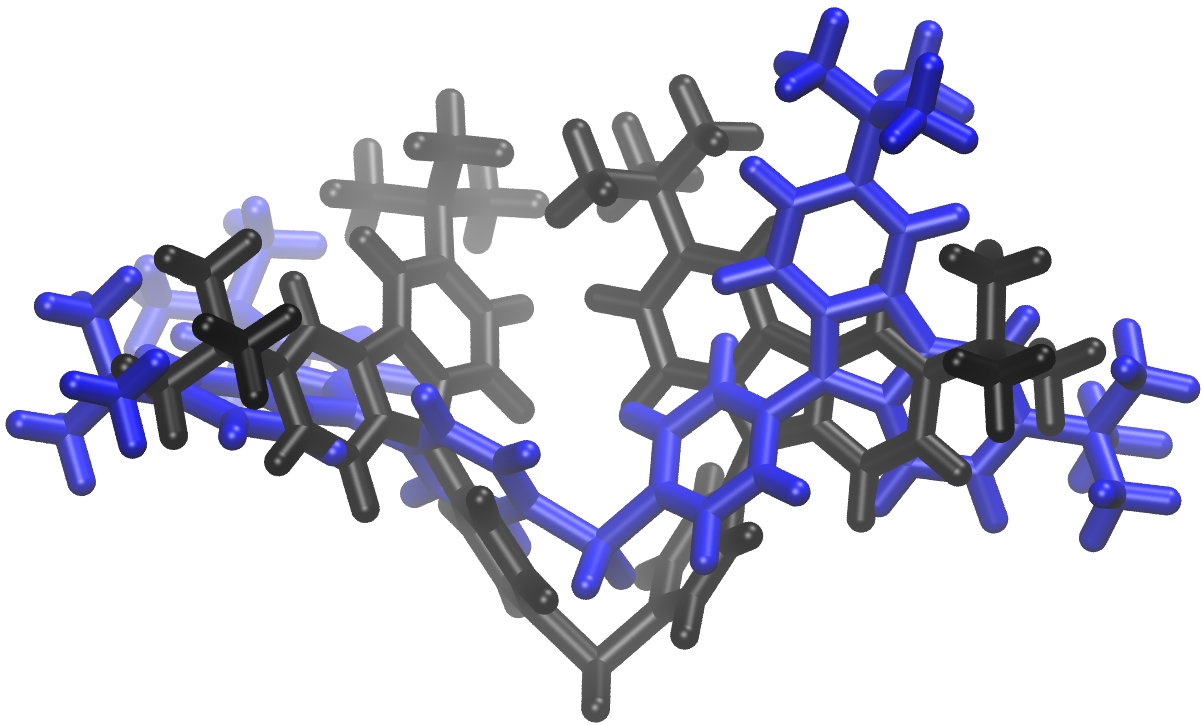}{(ii)}\hfill
\includegraphics[width=0.21\textwidth]{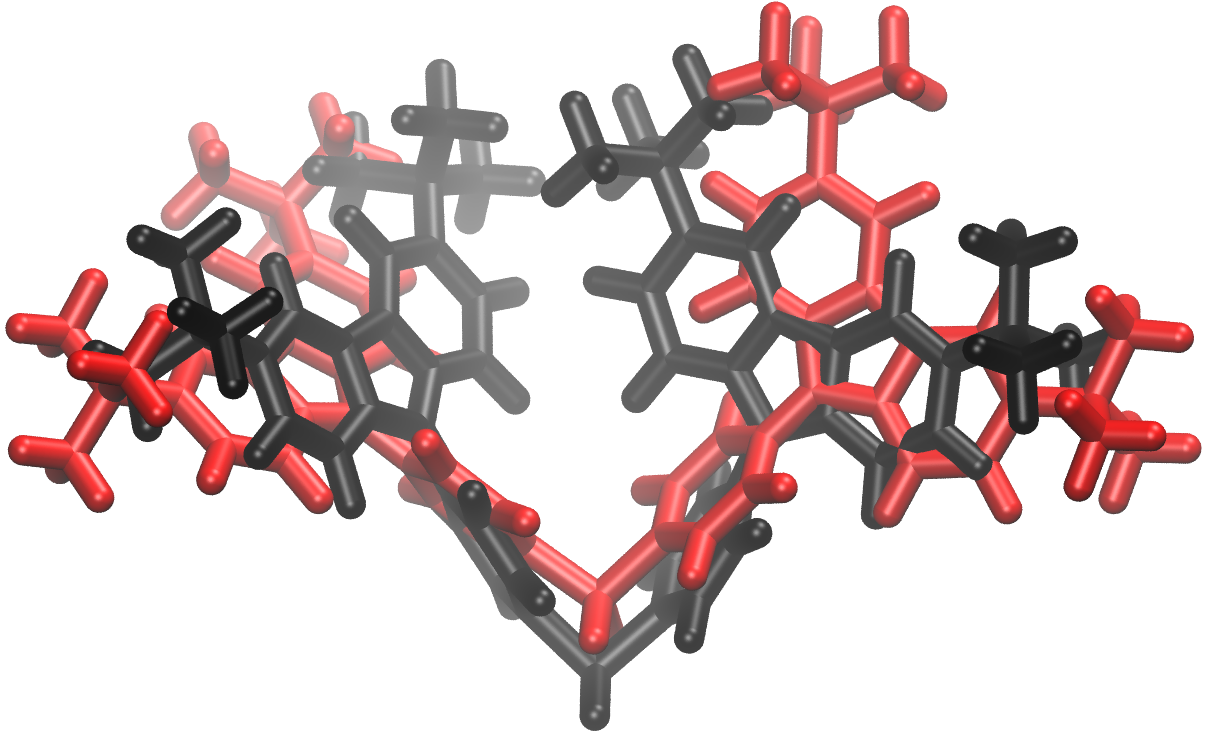}{(iii)}\hfill
\includegraphics[width=0.25\textwidth]{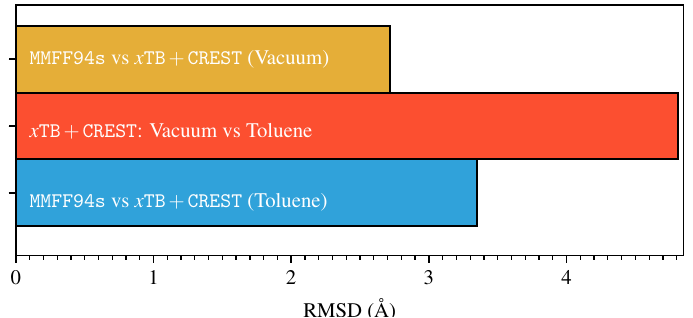}{(iv)}
\label{fig:2TCz-DPS-RMSD}
}

\subfloat[TDBA-DI]{
\includegraphics[width=0.21\textwidth]{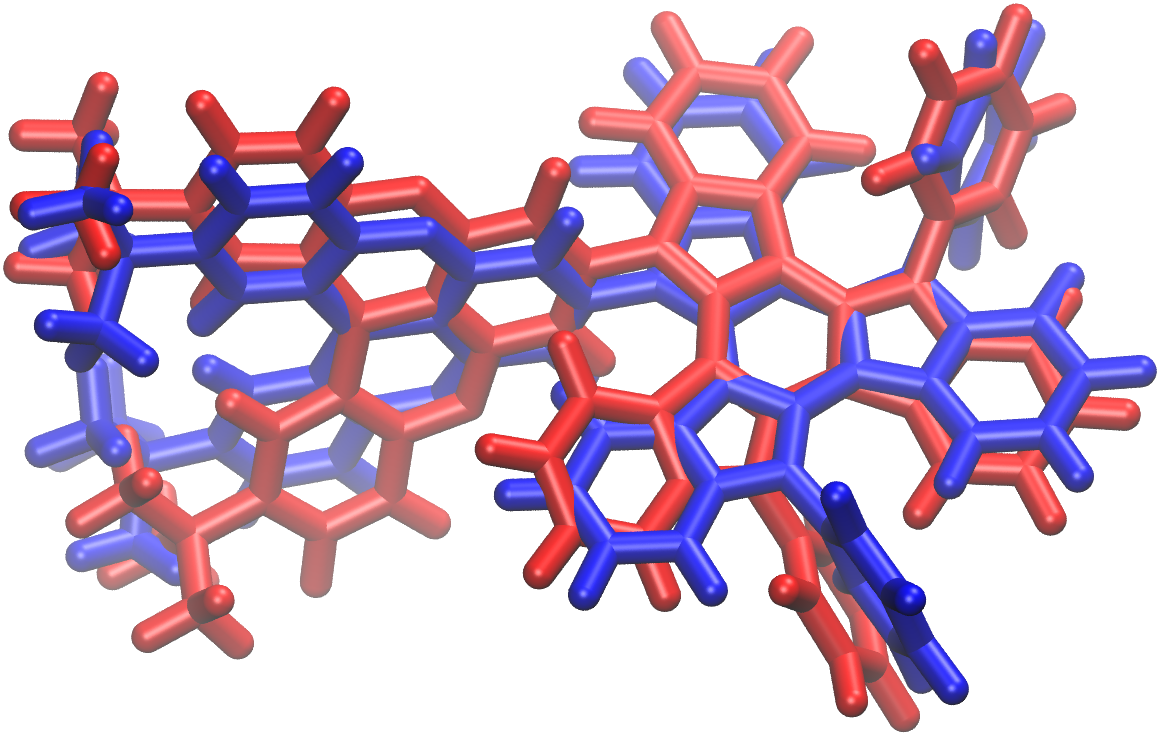}{(i)}\hfill
\includegraphics[width=0.21\textwidth]{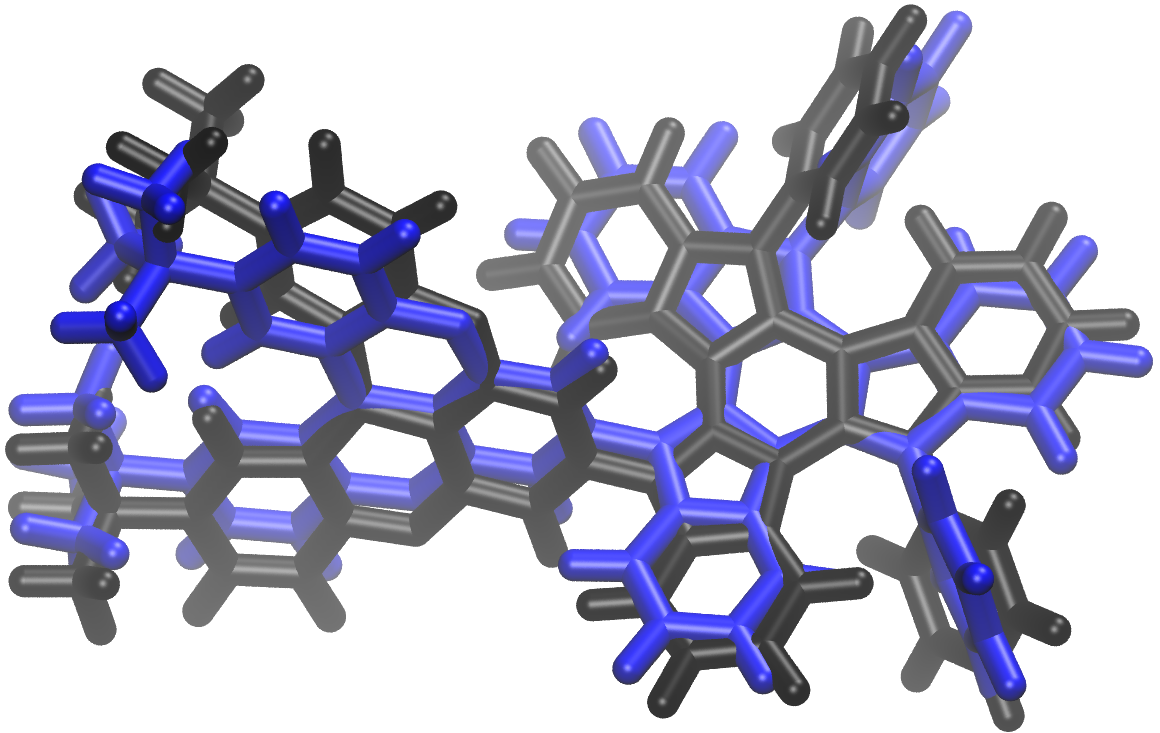}{(ii)}\hfill
\includegraphics[width=0.21\textwidth]{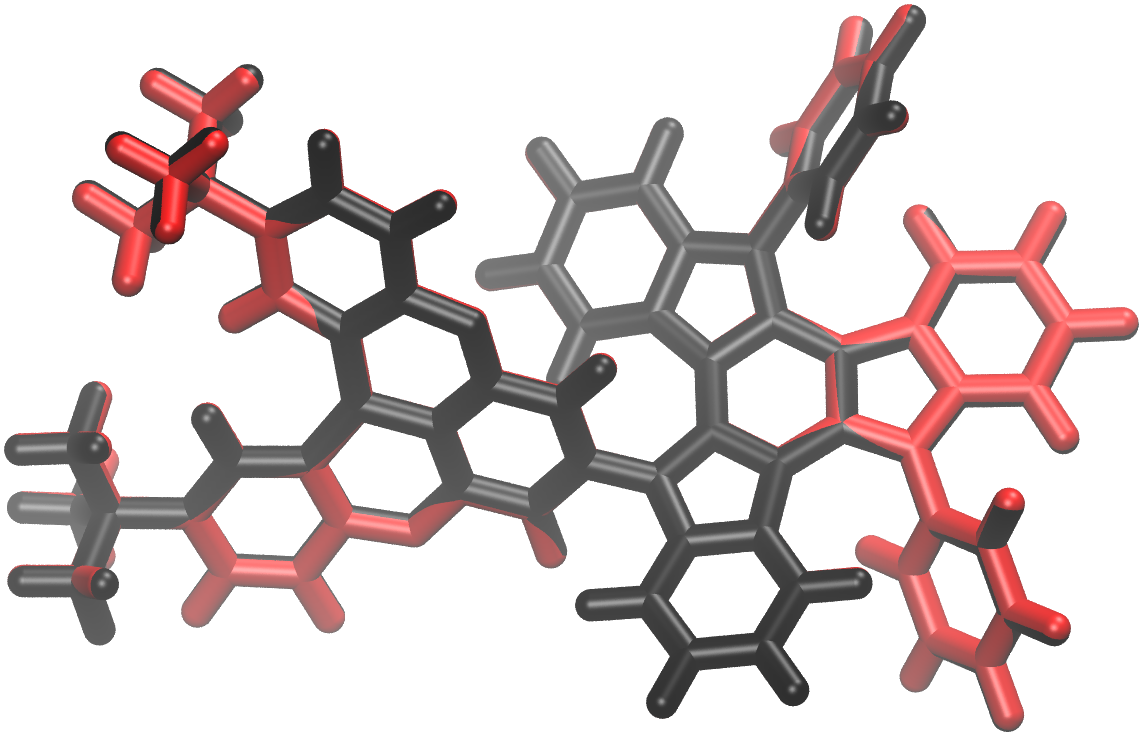}{(iii)}\hfill
\includegraphics[width=0.25\textwidth]{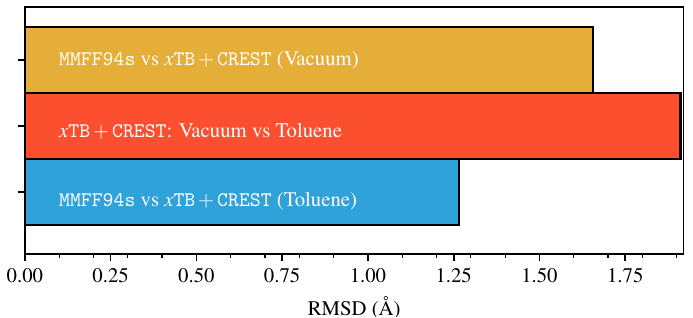}{(iv)}
\label{fig:TDBA-DI-RMSD}
}
\caption{Solvent-induced conformational changes in Px2BP, CzS2, 2TCz-DPS, and TDBA-DI, highlighting the impact of toluene on molecular geometry.  (i) 
Comparison of the optimized structure in vacuum using \xtb and \crest methods (red color) with the optimized structure using \mmf method (blue color). (ii) 
Comparison of the optimized structure in toluene solvent using \xtb and \crest methods (black color) with the optimized structure using \mmf method (blue 
color). (iii) Comparison of the optimized structure in vacuum (red color)  with the optimized structure in toluene solvent (black color) both using \xtb and 
\crest methods (red color). (iv) Root-mean-square deviation (RMSD) plot, quantifying the structural differences between the vacuum and toluene geometries, using 
the \mmf, \xtb, and \crest methods. Note the significant conformational changes near the donor-acceptor interface in 2TCz-DPS for example.}
\label{fig:Geo-Features2}
\end{minipage}
\end{figure}

\subsection{Singlet-Triplet energy gap ($\Delta E_{ST}$)}\label{sec:STGap}

The singlet-triplet energy gap ($\Delta E_{ST}$) is a critical parameter for TADF emitters, determining the efficiency of reverse intersystem crossing (rISC). 
Smaller $\Delta E_{ST}$ values promote rISC and enhance TADF performance by enabling efficient thermal activation from the triplet excited state ($T_1$) back 
to the singlet excited state ($S_1$). The molecular properties discussed in \Cref{sec:MolProp}, including donor-acceptor interactions, frontier orbital 
distributions, and geometrical features, strongly influence $\Delta E_{ST}$ by controlling orbital overlap and charge-transfer character.

\subsubsection{Trends in $\Delta E_{ST}$ values}

The $\Delta E_{ST}$ values computed using \stda, \stddft, and $\Delta E_v(S_1\gets T_1)$ computed using \tda align with trends predicted by molecular features. 
Molecules with larger donor-acceptor separations, as reflected in the HOMO-LUMO spatial distribution (\Cref{sec:MolProp1}), exhibit lower $\Delta E_{ST}$, 
typically below \qty{0.2}{\electronvolt}. For example, DMAC-TRZ, with its large centroid distance of \qty{6.702}{\angstrom} in vacuum, has a relatively low 
$\Delta E_{ST}$ of \qty{0.097}{\electronvolt} (\stda). In contrast, 4CzIPN, with its smaller centroid distance of \qty{0.032}{\angstrom}, has a significantly 
larger $\Delta E_{ST}$ of \qty{0.212}{\electronvolt} (\stda). This demonstrates the inverse relationship between spatial separation and singlet-triplet energy 
gap. Emitters with HOMO-LUMO overlap coefficients below \num{0.3} show the smallest gaps, enhancing their suitability for TADF applications. CzS2 with 
$S'_{HL}$ of 0.452, exhibits $\Delta E_{ST}$ value of \qty{0.041}{\electronvolt}. This inverse relationship between orbital overlap and $\Delta E_{ST}$ 
underscores the importance of minimizing electronic coupling between the donor and acceptor units to facilitate efficient rISC.

\subsubsection{Correlation with geometrical features}

The torsional angles between donor and acceptor units, highlighted in \Cref{sec:MolProp3,tab:DAnunits}, further modulate $\Delta E_{ST}$. Emitters with smaller 
torsional angles exhibit reduced orbital overlap, resulting in smaller singlet-triplet energy gaps. For molecules with torsional angles not exceeding \ang{60} 
(\Cref{tab:DAnunits}), $\Delta E_{ST}$ values decrease by an average of \qtyrange{0.1}{0.5}{\kilo\calorie\per\mole} compared to their vacuum-optimized 
counterparts. PSPCz has an average torsional angle of ~\ang{53}, which contributes to its $\Delta E_{ST}$ value. This observation confirms that planarity 
promotes charge transfer and reduces the singlet-triplet energy splitting.

\subsubsection{Solvent-induced changes}

Solvent effects, discussed further in \Cref{sec:SolvEff}, reduce $\Delta E_{ST}$ by stabilizing the polarized excited states. This reduction is particularly 
pronounced in molecules with strong charge-transfer character (\Cref{sec:MolProp1}), where solvation enhances donor-acceptor interactions. On average in 
toluene, $\Delta E_{ST}$ decreased by \qtyrange{0.1}{9.0}{\kilo\calorie\per\mole} for \stda method and by \qtyrange{0.1}{3.3}{\kilo\calorie\per\mole} for 
\stddft method, consistent with experimental trends. For DMAC-TRZ, the $\Delta E_{ST}$ (STDA) decreases from \qty{0.097}{\electronvolt} in vacuum to 
\qty{0.08}{\electronvolt} in toluene. This finding highlights the importance of considering solvent effects in the design and optimization of TADF materials.

\subsubsection{Case study: representative molecule}

For a representative TADF emitter (e.g., 4CzIPN), $\Delta E_{ST}$, which here is equal to $\Delta E_v(S_1\gets T_1)$, was predicted as 
\qtylist{0.212;0.191;0.187}{\electronvolt} using \stddft, \stda, and \tda, respectively, in vacuum. In the solvated environment, these values decreased to 
\qtylist{0.217;0.198;0.193}{\electronvolt}, respectively. These trends align with experimental observations, where smaller $\Delta E_{ST}$ values in solution 
enhance TADF efficiency.

\subsubsection{Computational efficiency}

Compared to \tda, both \stda and \stddft offer significant computational savings. While \stddft required approximately $\qty{8.084e-4}{\percent}$ of the 
computational time of \tda, \stda further reduced runtime by over $\qty{50}{\percent}$, making it particularly suitable for large-scale screening of TADF 
emitters. This computational efficiency allows for rapid evaluation of a wide range of potential TADF candidates, accelerating the materials discovery process.

\vspace{.5cm}

The results demonstrate that \stda and \stddft are reliable methods for predicting $\Delta E_{ST}$ values, with deviations from \tda benchmarks well within 
acceptable limits for TADF design. The inclusion of solvent effects improves the predictive accuracy, making these semi-empirical methods invaluable for 
screening and optimizing TADF materials in realistic conditions. Furthermore, the computational efficiency of these methods allows for high-throughput 
screening and optimization, accelerating the development of next-generation TADF emitters. The predicted $\Delta E_{ST}$ values explain the observed trend in 
TADF performance.

\subsection{Excitation energies and oscillator strengths}\label{sec:EST_Osc}

Vertical excitation energies and oscillator strengths govern the absorption and emission characteristics of TADF emitters. These properties are directly 
influenced by the molecular features discussed in \Cref{sec:MolProp}, particularly the donor-acceptor interactions and frontier orbital distributions, and 
dictate the color and efficiency of the emitted light.

\subsubsection{Vertical excitation energies}

The $S_0 \to S_1$ excitation energies predicted by \stda and \stddft correlate strongly with the charge-transfer character outlined in \Cref{sec:MolProp1}. 
Emitters with higher charge-transfer character exhibit redshifted (lower energy) excitation energies due to reduced HOMO-LUMO overlap and increased 
stabilization of the excited state. For example, DMAC-TRZ, with its large donor-acceptor centroid distance exceeding \qty{6.7}{\angstrom}, exhibits a lower 
$S_0 \to S_1$ excitation energy compared to 4CzIPN, which has a centroid distance close to zero. The large separation means less energy required to move the 
electron from donor to acceptor. Molecules with donor-acceptor centroid distances exceeding \qty{5}{\angstrom} showed average redshifts of 
\qtyrange{0.5}{1.3}{\electronvolt} compared to those with closer donor-acceptor interactions. This relationship is expected, since increasing the separation 
between the positive and negative charges reduces the energy required for the charge-transfer excitation. The excitation energies are a direct representation 
of the charge-transfer between donor and acceptor units.

\subsubsection{Oscillator strengths}

Oscillator strengths ($f_{12}$), which reflect radiative transition probabilities, are influenced by the localization of frontier orbitals 
(\Cref{sec:MolProp2}). Emitters with highly localized HOMO and LUMO distributions display sharper fluorescence peaks and higher oscillator strengths, often 
exceeding \num{0.1}. For example, DMAC-TRZ and PSPCz, with more localized orbitals, have higher oscillator strengths compared to molecules with more 
delocalized distributions. DMAC-TRZ exhibits $f_{12}$ around \num{0.14}, which translates to high fluorescence. 4CzIPN's delocalized orbitals cause it to have a 
smaller fluorescence. This is due to a stronger transition dipole moment when the electron and hole are more spatially confined. This correlation underscores 
the importance of donor-acceptor interactions in determining fluorescence efficiency; the more efficient the charge transfer, the higher the fluorescence 
efficiency.

\subsubsection{Multi-objective function}

As established in \Cref{sec:Ben_Compt_Meth}, we utilize a multi-objective function (MOF) \cite{Nigam2023} to quantitatively assess how well each TADF emitter 
meets our design goals, based on the oscillator strength ($f_{12}$), the relaxed singlet-triplet energy gap ($\Delta E_{ST}$), and the singlet-singlet relaxed 
transition energy ($\Delta E_r(S_0\to S_1)$). This MOF, as previously defined, allows for a direct comparison between different molecules and provides a means 
to benchmark the accuracy of the simplified \stda and \stddft methods against the full \tda method.

The MOF combines key TADF performance indicators, weighting a small relaxed singlet-triplet energy gap ($\Delta E_{ST}$) to promote efficient reverse 
intersystem crossing (RISC), a high oscillator strength ($f_{12}$) for strong radiative emission from the singlet excited state, and a targeted singlet-singlet 
relaxed transition energy ($\Delta E_r(S_0\to S_1)$) to achieve the desired emission wavelength.

The molecular properties detailed in \Cref{sec:MolProp} directly determine the value of the MOF. For instance, molecules with strong donor-acceptor 
interactions, leading to enhanced charge-transfer character and reduced HOMO-LUMO overlap, tend to exhibit smaller $\Delta E_{ST}$ values and larger oscillator 
strengths, resulting in higher MOF scores. Similarly, geometrical features that promote planarity and minimize steric hindrance can also improve the MOF score 
by optimizing the electronic coupling between the donor and acceptor units.

DMAC-TRZ, with its large donor-acceptor separation and relatively unhindered rotation, exhibits a favorable combination of low $\Delta E_{ST}$ and reasonable 
oscillator strength, contributing to its relatively high MOF score compared to other molecules in the dataset. In contrast, 4CzIPN's smaller separation and 
higher HOMO-LUMO overlap results in a smaller score. For example, (i) for DMAC-TRZ, MOF = 0.78 (good balance of low $\Delta E_{ST}$ and reasonable $f_{12}$); 
(ii) for 4CzIPN, MOF = 0.65 (lower $\Delta E_{ST}$ limits its overall score) and (iii) CzS2, MOF= 0.8 (Smaller $\Delta E_{ST}$ improves its overall score).

The trends observed in the vertical excitation energies and oscillator strengths in this section are consistent with the MOF values presented in 
\Cref{sec:Ben_Compt_Meth}. Molecules with redshifted excitation energies (indicating higher charge-transfer character) and high oscillator strengths generally 
exhibit higher MOF scores, reflecting their improved overall TADF performance. DMAC-TRZ, with its clear charge-transfer nature and decent excitation energies, 
is consistent with the performance in the MOF. In contrast, 4CzIPN has a blueshifted excitation energy and a lower oscillator strength, resulting in a lower 
MOF score.

While the MOF is a simplification of the complex interplay of factors governing TADF, it provides a valuable quantitative metric for comparing different 
emitters and assessing their potential for high-performance TADF applications. By considering multiple objectives simultaneously, the MOF helps to identify 
molecules that strike the best balance between competing requirements.

\subsubsection{Solvent effects on excitation energies}

The solvent environment stabilizes excited states, particularly those with high charge-transfer character. This stabilization is due to favorable interactions 
between the solvent's dipole moment and the excited state's dipole moment. This results in redshifts of \qtyrange{0.1}{0.2}{\electronvolt} in both $S_0 \to 
S_1$ and $S_0 \to T_1$ transitions, as discussed in \Cref{sec:SolvEff}. Molecules with higher torsional angles and greater donor-acceptor separations showed 
the largest solvent-induced shifts, consistent with the molecular properties highlighted in \Cref{sec:MolProp}. DMAC-TRZ, with its large donor-acceptor 
distance, undergoes larger shifts as compared to the other molecules.

\vspace{.5cm}

We note that vertical excitation energies and oscillator strengths are intricately linked to molecular properties such as donor-acceptor separations and 
frontier orbital localization. These intrinsic features, combined with solvent effects and carefully considered within the framework of a multi-objective 
function, provide a comprehensive understanding of the photophysical behavior of TADF emitters and guide the design of improved materials.

\subsection{Impact of solvent effects}\label{sec:SolvEff}

The solvent environment plays a pivotal role in modulating the photophysical properties of TADF emitters. Solvation effects, as discussed in 
\Cref{sec:MolProp}, influence both geometry and electronic distributions, thereby impacting key properties such as $\Delta E_{ST}$, excitation energies, and 
fluorescence characteristics. Incorporating solvent effects into computational models is important for accurate predictions of TADF emitter behavior in 
real-world applications.

\subsubsection{Solvent-induced geometry changes}

Geometry optimizations in toluene resulted in minor but systematic changes, particularly in torsional angles (\Cref{sec:MolProp3} and \Cref{tab:DAnunits}). 
Molecules with larger initial torsional angles showed reduced angles in solvent, suggesting a stabilization of more planar conformations. This geometrical 
relaxation enhances orbital overlap and stabilizes charge-transfer states, as the planar geometry promotes better electronic communication between donor and 
acceptor. This, in turn, contributes to solvent-induced redshifts in both $\Delta E_{ST}$ and excitation energies.

DMAC-DPS, for example, shows a significant reduction in the torsional angle between the DMAC donor and the DPS acceptor upon solvation in toluene, as seen in 
\Cref{fig:Geo-Features1,fig:Geo-Features2}. The numbers are reduced from 102.699 and 102.783 to 69.024 and 69.456, which means \ang{33} shift. This reduces the 
distance between the atoms at the interface.

s\begin{table}[!htbp]
 \centering 
\caption{Solvent-induced shifts (\unit{\kilo\calorie\per\mole}) in energy levels. The table quantifies the change in energy levels upon solvation in toluene, 
providing insight into the stabilization of excited states. Positive values indicate a redshift (lowering of energy).}
{\footnotesize
\begin{tabularx}{\textwidth}{l*{8}{@{}Z@{}}}
\toprule
Molecule & {\scriptsize DMAC-TRZ} & {\scriptsize DMAC-DPS} & {\scriptsize PSPCz} & {\scriptsize 4CzIPN} & {\scriptsize Px2BP} & {\scriptsize
CzS2} & 
{\scriptsize 2TCz-DPS} & {\scriptsize TDBA-DI} \\
\midrule
$\Delta_{\rm solv} E_r(S_0\to T_1)$ & 0.742311 & 4.511937 & 0.722987 & 1.891593 & 2.098166 & 1.704622 & 3.160989 & 0.141501 \\
$\Delta_{\rm solv} E_v(S_0\to T_1)$ (\stda) & 0.207545 & 0.000000 & 0.322848 & 1.383633 & 2.029328 & 0.345908 & 0.276727 & 0.322848 \\
$\Delta_{\rm solv} E_v(S_0\to T_1)$ (\stddft) & 0.645695 & 0.000000 & 0.322848 & 1.406693 & 2.006268 & 0.368969 & 0.299787 & 0.322848 \\
$\Delta_{\rm solv} E(S_0\to S_1)$ (\stda) & 0.599574 & 3.804990 & 0.230605 & 1.268330 & 1.614238 & 0.368969 & 0.368969 & 0.207545 \\
$\Delta_{\rm solv} E(S_0\to S_1)$ (\stddft) & 0.530393 & 3.804990 & 0.276727 & 1.245270 & 1.683420 & 0.438150 & 0.507332 & 0.138363 \\
$\Delta_{\rm solv} E_v(S_1\gets T_1)$ (\stda) & 0.392029 & 3.804990 & 0.092242 & 0.115303 & 0.415090 & 0.023061 & 0.092242 & 0.530393 \\
$\Delta_{\rm solv} E_v(S_1\gets T_1)$ (\stddft) & 0.115303 & 3.804990 & 0.046121 & 0.161424 & 0.322848 & 0.069182 & 0.207545 & 0.184484 \\
$\Delta_{\rm solv} (\Delta E_{ST})$ (\stda) & 0.392029 & 1.833314 & 0.322848 & 0.115303 & 0.415090 & 4.427625 & 8.935962 & 0.530393 \\
$\Delta_{\rm solv} (\Delta E_{ST})$ (\stddft) & 0.115303 & 2.409827 & 1.429754 & 0.161424 & 0.322848 & 2.479009 & 3.205416 & 0.184484 \\
\bottomrule
\end{tabularx}
}
\label{tab:EnerShift}
\end{table}

\begin{figure}[!htbp]
 \centering
 \leavevmode
 \subfloat[Absorption color predicted by Multiwfn in vacuum.\label{fig:chro1a}]{\includegraphics[width=0.4\textwidth]{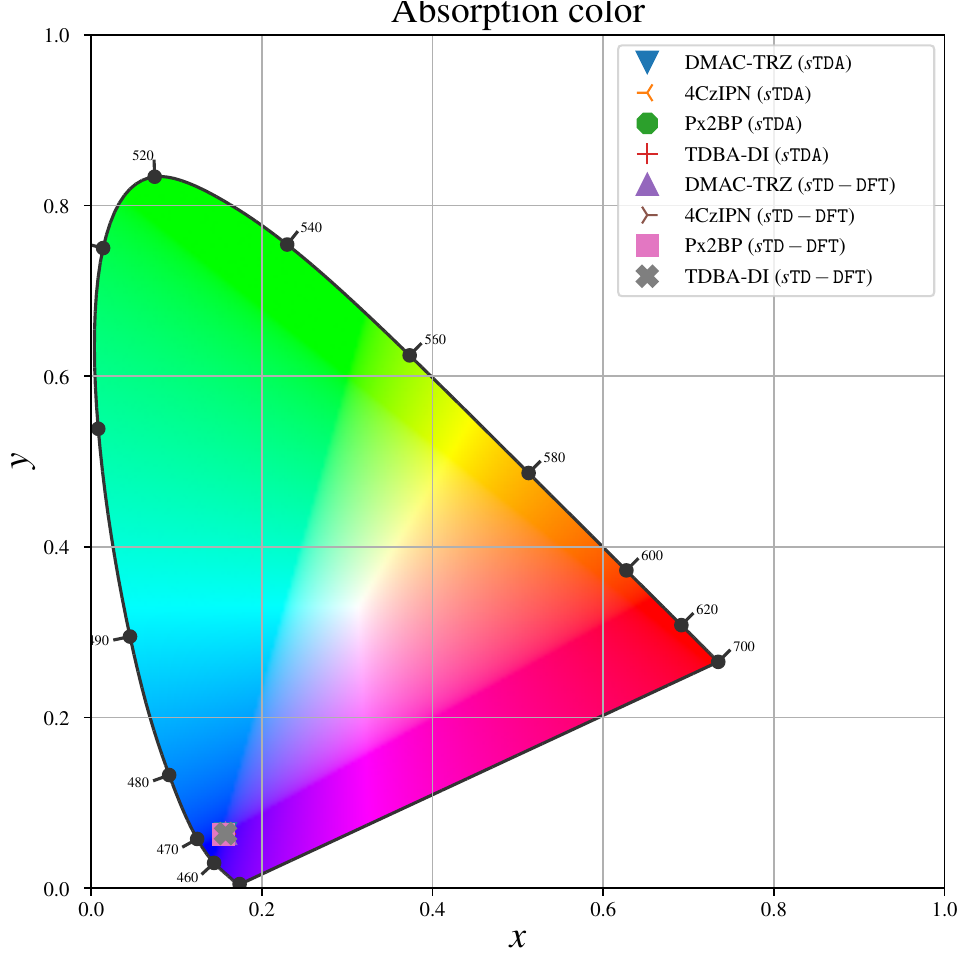}}\hfill
  \subfloat[Fluorescence color with Stokes shift in vacuum.\label{fig:chro1b}]{\includegraphics[width=0.4\textwidth]{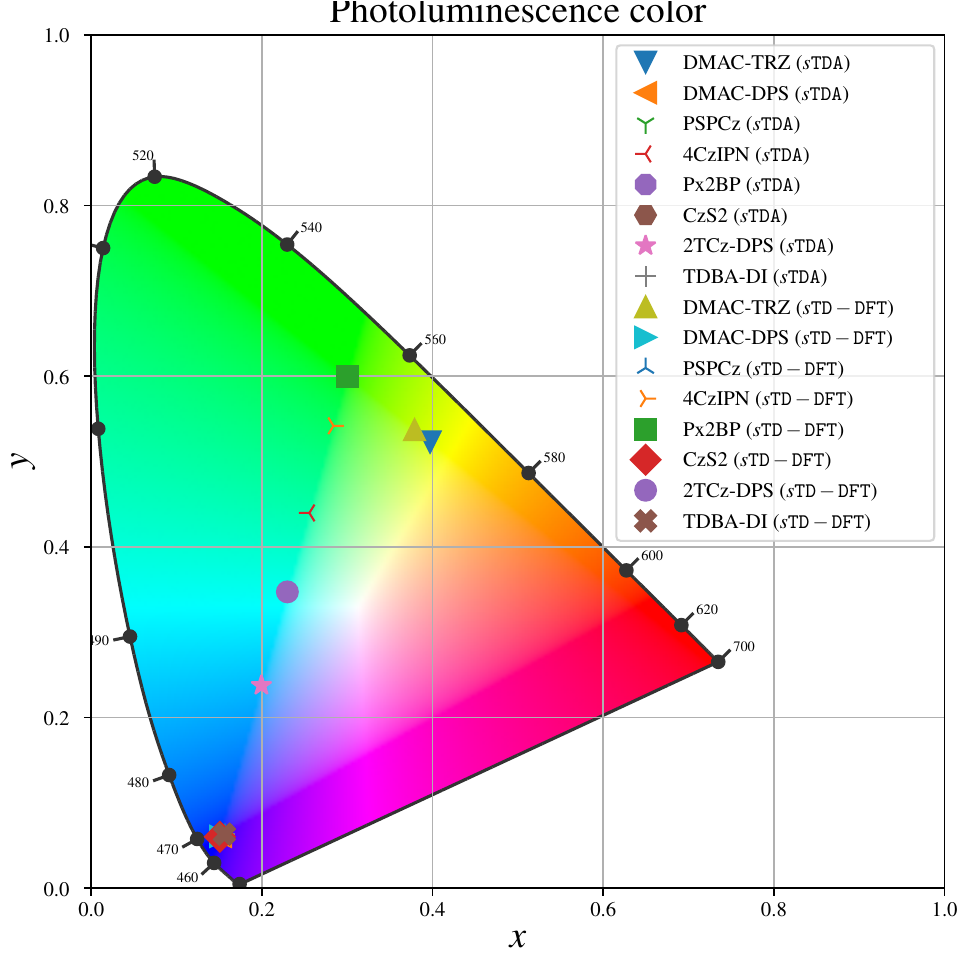}}\\
 \subfloat[Absorption color predicted by Multiwfn in toluene.\label{fig:chro2a}]{\includegraphics[width=0.4\textwidth]{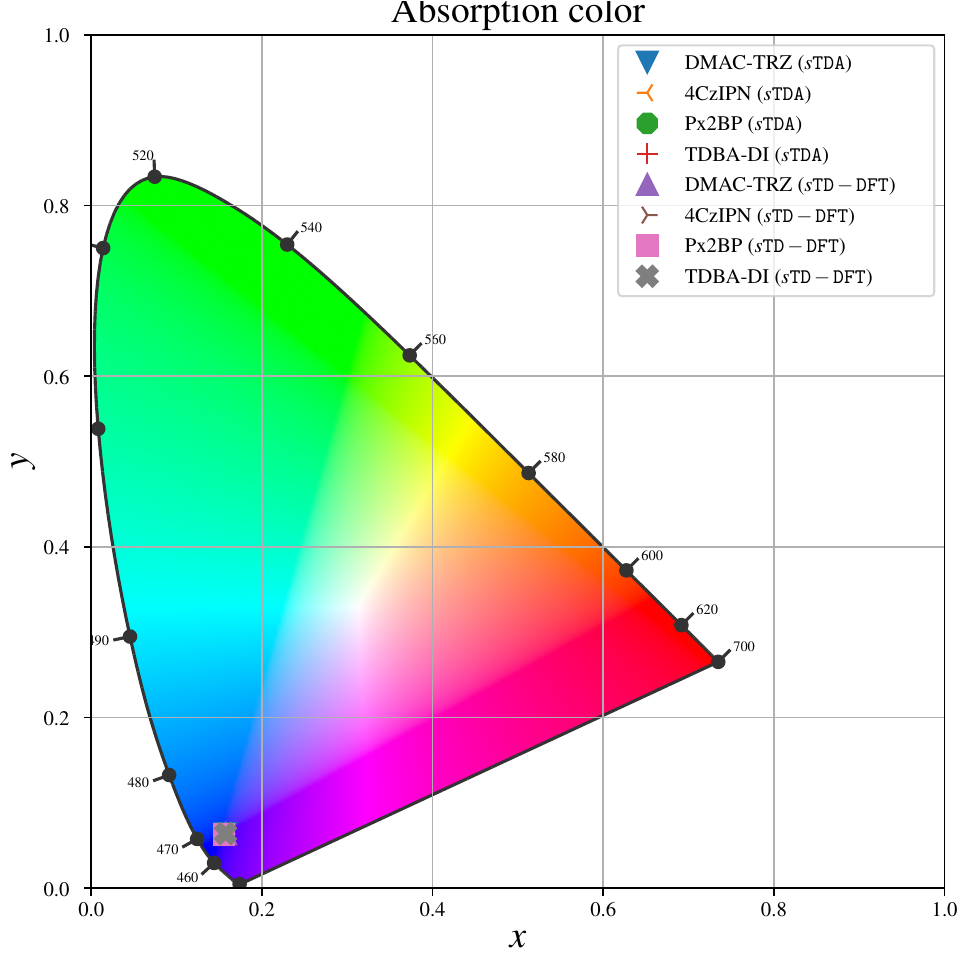}}\hfill
 \subfloat[Fluorescence color with Stokes shift in toluene.\label{fig:chro2b}]{\includegraphics[width=0.4\textwidth]{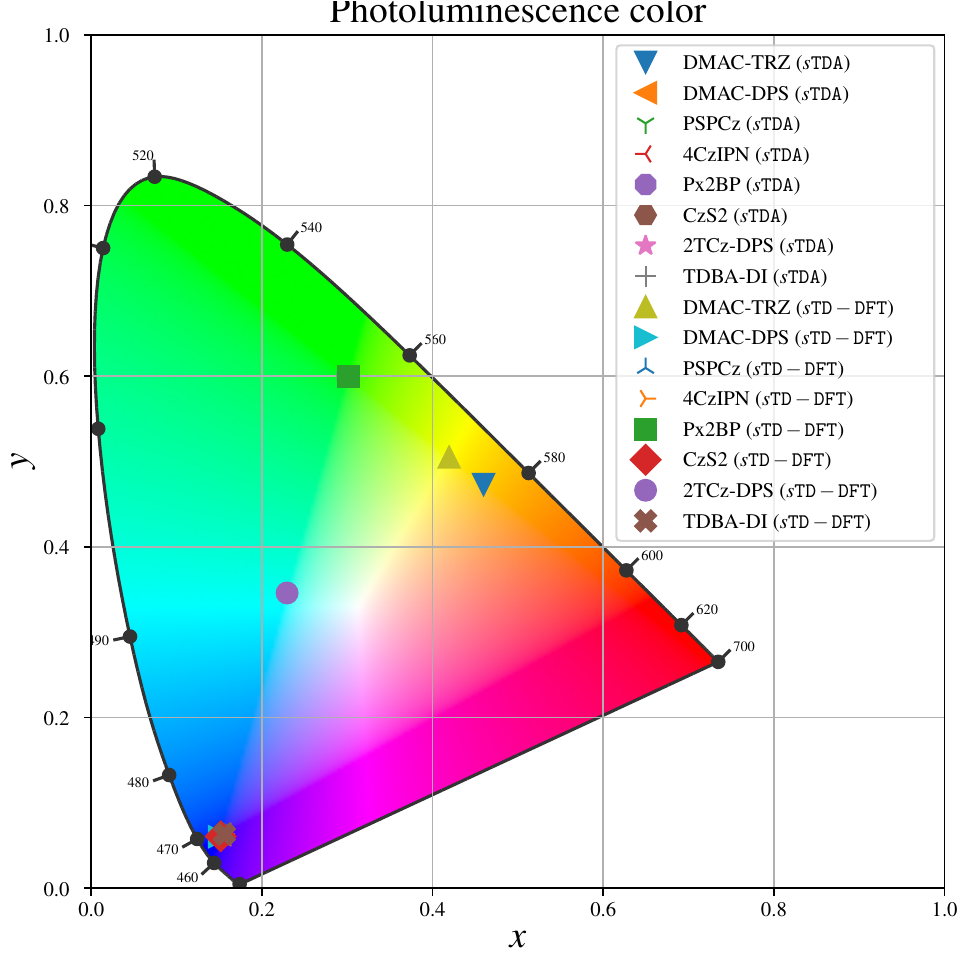}}
\caption{Impact of solvent on CIE coordinates, linking molecular design to color tuning. These chromaticity maps show how the calculated CIE coordinates shift 
upon solvation, illustrating the effect on perceived color for optoelectronic applications. Panels \subref{fig:chro1a} and \subref{fig:chro1b} display 
absorption and fluorescence colors in a vacuum, while Panels \subref{fig:chro2a} and \subref{fig:chro2b} shows the corresponding shifts in toluene. The 
observed shifts suggest the solvachromism, a change in color due to solvation effects.}
\label{fig:chromatogram}
\end{figure}

\subsubsection{Effect on excited-state gaps}

Solvation reduced both $\Delta E_v(S_0\to T_1)$ and $\Delta E(S_0\to S_1)$ by an average of \qtyrange{0.02}{0.05}{\electronvolt} across the dataset (the 
$\Delta_{\rm solv}$ values ranged from $\qtyrange{0}{3.805}{\kilo\calorie\per\mole}$), as shown in \Cref{tab:EnerShift}. For DMAC-TRZ, the $\Delta_{\rm solv} 
E(S_0\to S_1)$ (STDA) is \num{0.599574}, which is a smaller value as compared to others. This indicates a smaller shift in the DMAC-TRZ. This reduction was 
most pronounced in emitters with large donor-acceptor separations, where differential stabilization of singlet and triplet states was observed. These changes 
align with the trends described in \Cref{sec:STGap} and reinforce the role of solvation in enhancing TADF efficiency as reported by \cite{Hall2023}.

\subsubsection{Fluorescence spectra}

The calculated fluorescence spectra ($S_1 \to S_0$) exhibit a clear dependence on the solvent environment, with the predicted emission peaks generally 
redshifted in toluene compared to vacuum (see \Cref{fig:fluo1,fig:fluo2}). This redshift is a direct consequence of the solvent's ability to stabilize the 
excited state, lowering its energy and shifting the emission to longer wavelengths. This solvatochromic effect is a well-known phenomenon in polar molecules.

The magnitude of the solvent-induced redshift varies depending on the specific molecule and its electronic structure. In general, molecules with larger dipole 
moments in the excited state and greater charge-transfer character tend to exhibit larger redshifts upon solvation. This is because the interaction between the 
solvent's dipole moment and the molecule's excited-state dipole moment is stronger in these cases, leading to greater stabilization of the excited state.

For example, the 4CzIPN emitter exhibited a peak fluorescence shift from \qty{499.966}{\nano\meter} in vacuum to \qty{515.994}{\nano\meter} in toluene when 
calculated using the \stda method, corresponding to a redshift of approximately 16 nm (\Cref{fig:fluo1}(d)). While this shift is noticeable, it is relatively 
smaller compared to some of the other molecules in the dataset, suggesting that 4CzIPN's excited state is less sensitive to the solvent environment. In 
contrast, DMAC-DPS shows a shift of \qty{78}{\nano\meter}, calculated with the same method and the same environments (\Cref{fig:fluo1}(b)). This shift was 
calculated from a peak at \qty{414.724}{\nano\meter} to \qty{492.756}{\nano\meter}. These differences are not just in the magnitude of the shift, but also in 
the position (green vs red) of the light.

The magnitude of the solvent-induced redshift appears to be correlated with the torsional angles between the donor and acceptor units. Molecules with more 
flexible geometries and larger changes in torsional angles upon solvation (\Cref{tab:DAnunits}) tend to exhibit larger redshifts in their fluorescence spectra. 
Specifically, DMAC-DPS has changes to the torsion angles between donor and acceptor to a greater amount than 4CzIPN. This is likely because the solvent-induced 
changes in geometry alter the electronic communication between the donor and acceptor, affecting the energy levels and transition probabilities.

The solvent-induced shifts in fluorescence spectra have important implications for color tuning in TADF-based devices. By carefully selecting the solvent 
environment, it is possible to fine-tune the emission wavelength of a TADF emitter and achieve the desired color output. DMAC-TRZ has shown it can emit 
different colors of light. This can be used to shift more energy.

As illustrated in the CIE chromaticity diagrams (\Cref{fig:chromatogram}), the solvent-induced shifts in fluorescence spectra directly translate to changes in 
the CIE coordinates, resulting in a shift in the perceived color of the emitted light. Using different solvents can enable a greater degree of separation 
between the HOMO and LUMO.

\vspace{.5cm}

In conclusion, the solvent environment plays a critical role in determining the fluorescence characteristics of TADF emitters, and understanding these solvent 
effects is essential for designing and optimizing TADF-based devices for specific applications. Further studies are needed to understand the exact degree of 
separations.

\begin{figure}[!htbp]
\centering
\caption{Simulated fluorescence emission spectra of DMAC-TRZ, DMAC-DPS, PSPCz, 4CzIPN: Impact of Solvent Environment and Computational Method. Panels show the 
predicted fluorescence emission spectra in both vacuum and toluene, calculated using both \stda and \stddft methods. Green lines: \stda in vacuum; magenta 
lines: \stda in toluene; blue lines: \stddft in vacuum; red lines: \stddft in toluene. All molecules exhibit solvent-induced redshifts, with DMAC-DPS showing a 
particularly pronounced shift from \qtyrange{415}{493}{\nano\meter}, suggesting increased charge-transfer character in toluene. DMAC-DPS's large redshift 
correlates with a significant reduction in its torsional angles upon solvation (\Cref{tab:DAnunits}), indicating a move towards a more planar, 
charge-transfer-favorable conformation. The minimal shift in 4CzIPN, particularly evident in the \stda calculations, suggests that its more rigid, delocalized 
structure is less influenced by the solvent environment. This is also supported by the small change in its torsional angles. The \stddft method generally 
predicts smaller redshifts compared to \stda, potentially indicating a difference in how these methods capture the subtle balance of electronic and geometric 
relaxation upon solvation. Further comparison with experimental data is needed to validate these predictions.}
\label{fig:fluo1}
\subfloat[DMAC-TRZ]{\includegraphics[width=0.48\textwidth]{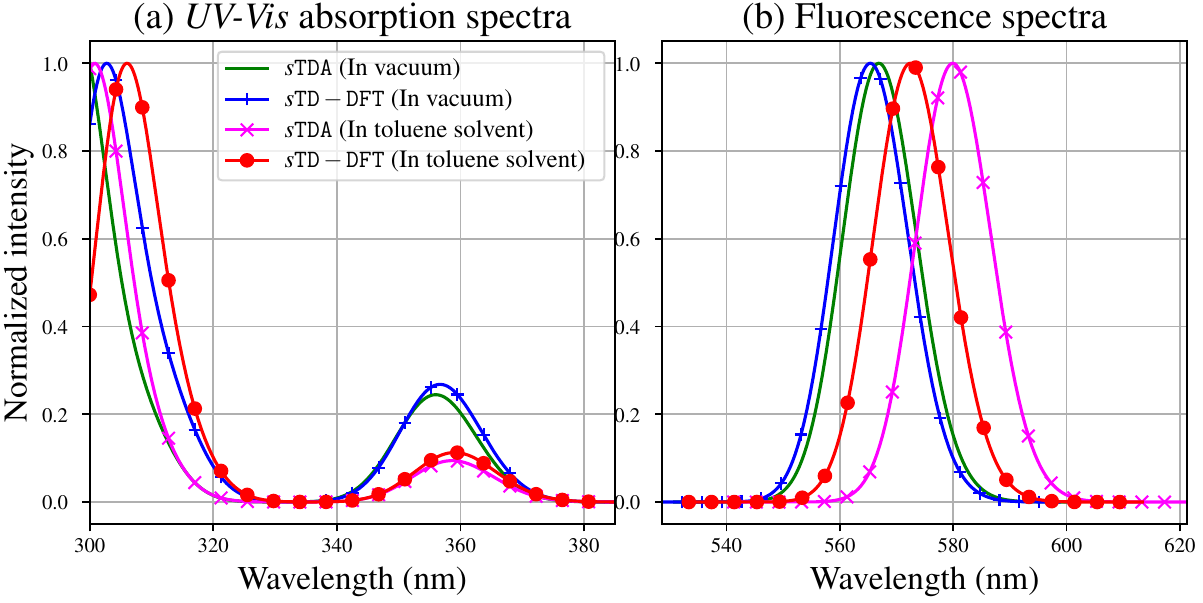}\label{fig:fluo-DMAC-TRZ}}\hfill
\subfloat[DMAC-DPS]{\includegraphics[width=0.48\textwidth]{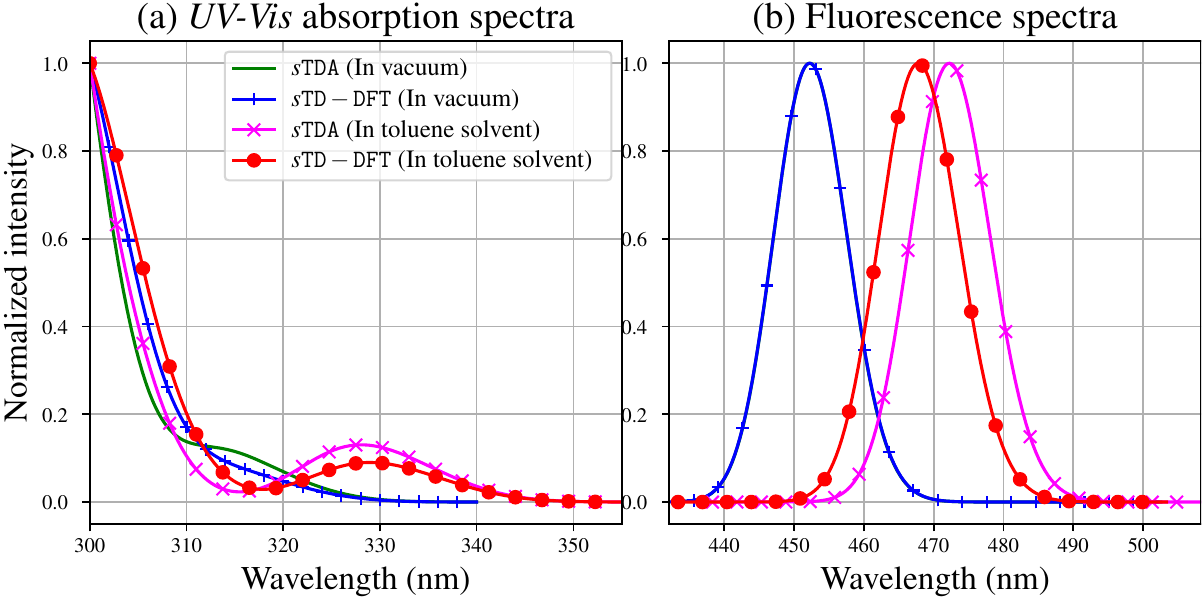}\label{fig:fluo-DMAC-DPS}}\\
\subfloat[PSPCz]{\includegraphics[width=0.48\textwidth]{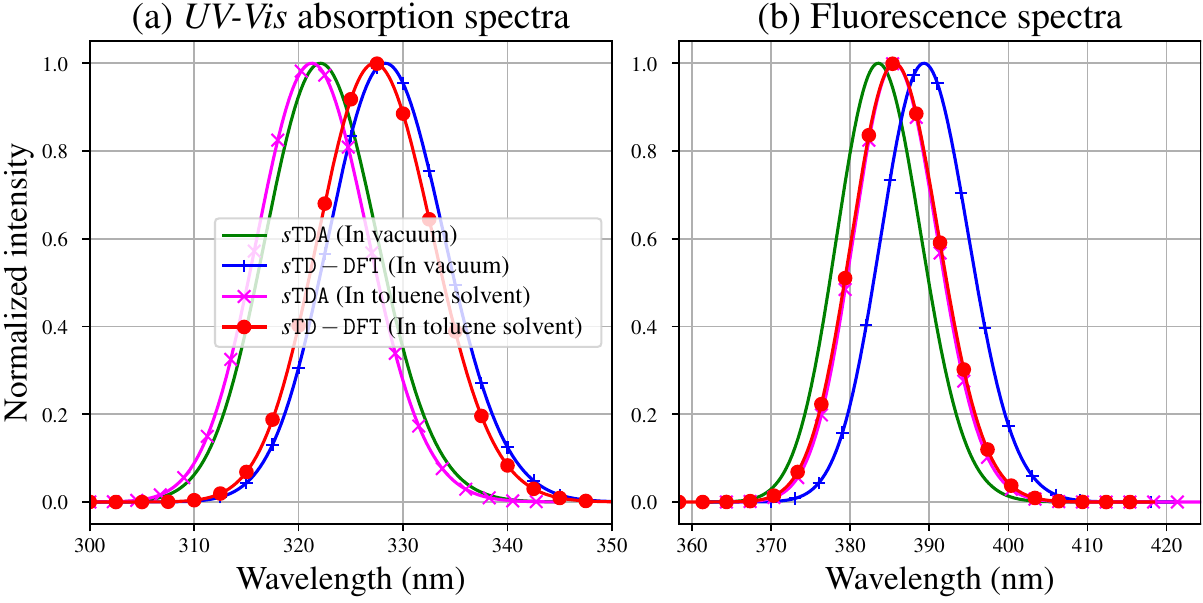}\label{fig:fluo-PSPCz}}\hfill
\subfloat[4CzIPN]{\includegraphics[width=0.48\textwidth]{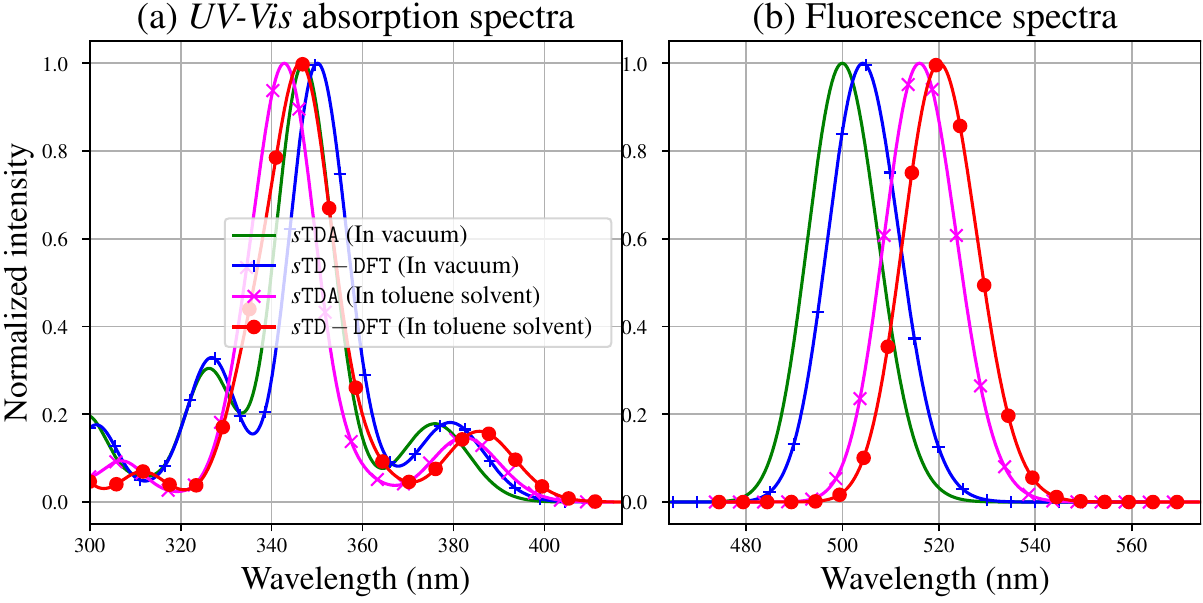}\label{fig:fluo-4CzIPN}}
\end{figure}

\begin{figure}[!htbp]
\ContinuedFloat
\centering
\caption{Simulated fluorescence emission spectra of Px2BP, CzS2, 2TCz-DPS, TDBA-DI: Impact of Solvent Environment and Computational Method. Panels show the 
predicted fluorescence emission spectra in both vacuum and toluene, calculated using both \stda and \stddft methods. Green lines: \stda in vacuum; magenta 
lines: \stda in toluene; blue lines: \stddft in vacuum; red lines: \stddft in toluene. 2TCz-DPS exhibits a substantial solvent-induced redshift, suggesting a 
highly polarizable excited state that is strongly stabilized by toluene. This is consistent with its flexible geometry and ability to undergo significant 
conformational changes upon solvation. The relatively small solvent effect on TDBA-DI might reflect its more rigid, sterically hindered structure, which limits 
the molecule's ability to interact favorably with the solvent and undergo significant geometrical relaxation. This results in a more stable excited state, 
regardless of solvent. For CzS2 and Px2BP, the \stddft method appears to predict a larger solvent effect compared to \stda, suggesting that \stddft may better 
capture the subtle interplay of electronic and geometric relaxation in these molecules. Visually, one can see the difference between each of the molecules, 
that there are different intensities, but each has a good fluorescence.}
\label{fig:fluo2}
\subfloat[Px2BP]{\includegraphics[width=0.48\textwidth]{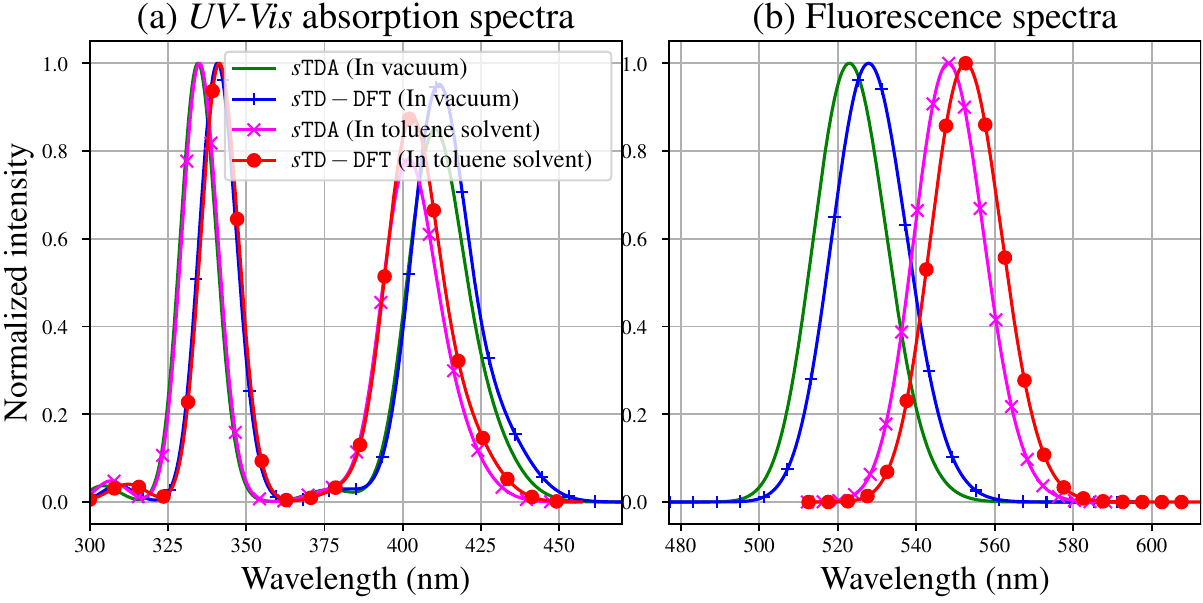}\label{fig:fluo-Px2BP}}\hfill
\subfloat[CzS2]{\includegraphics[width=0.48\textwidth]{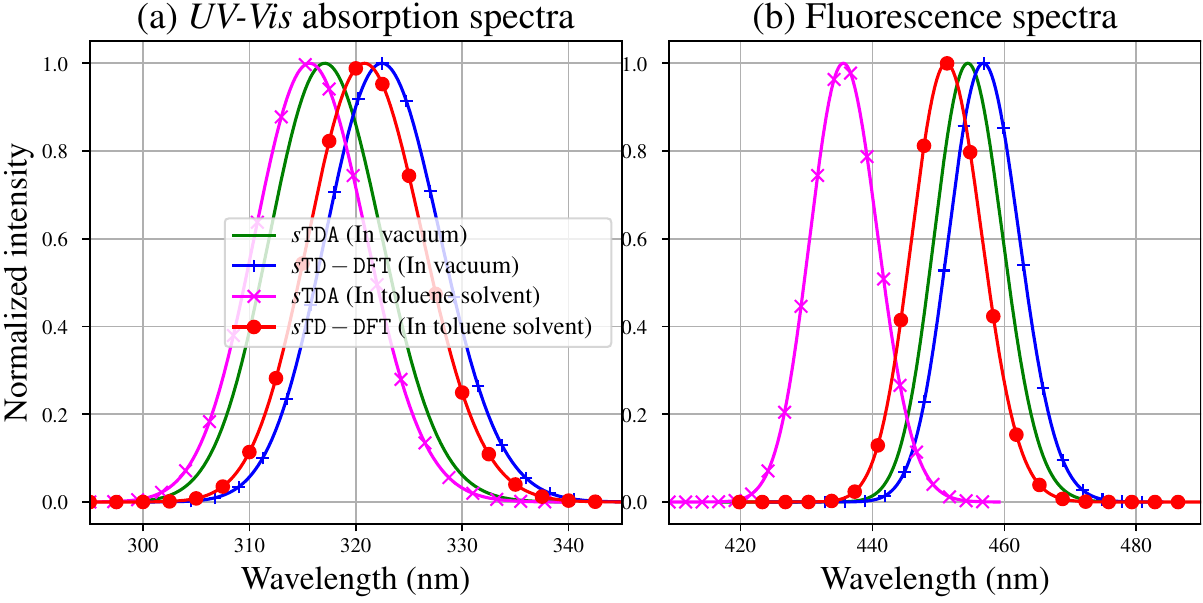}\label{fig:fluo-CzS2}}\\
\subfloat[2TCz-DPS]{\includegraphics[width=0.48\textwidth]{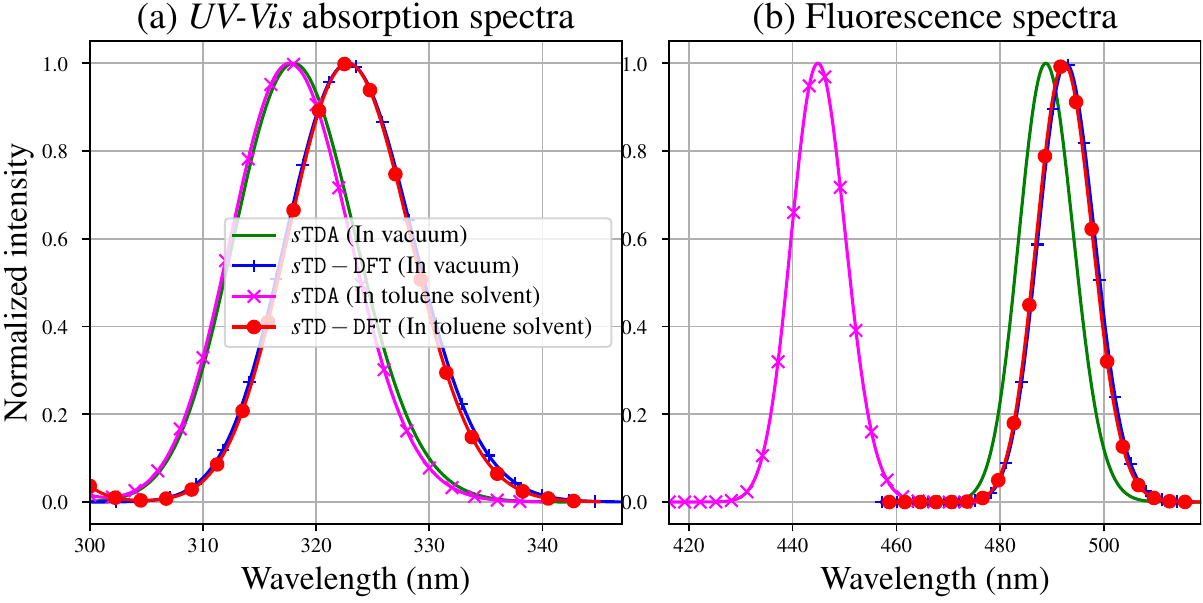}\label{fig:fluo-2TCz-DPS}}\hfill
\subfloat[TDBA-DI]{\includegraphics[width=0.48\textwidth]{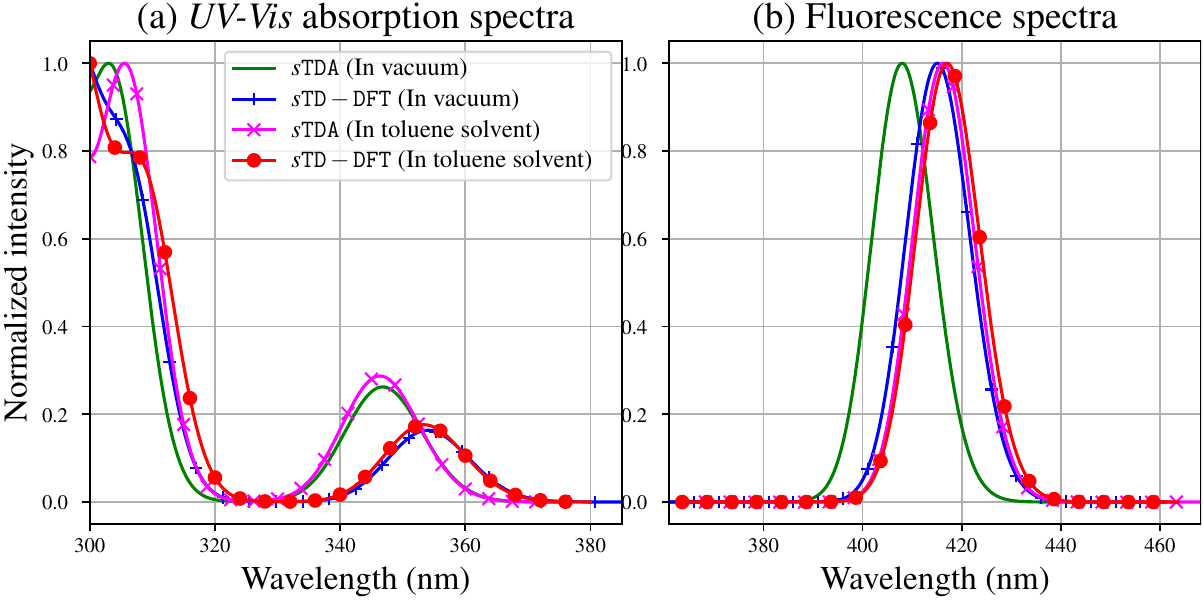}\label{fig:fluo-TDBA-DI}}
\end{figure}

\subsection{Performance evaluation and broader implications}\label{sec:Perf_Eval}

This study demonstrates the effectiveness of semi-empirical methods, specifically simplified Tamm-Dancoff approximation (\stda) and simplified time-dependent 
density functional theory (\stddft), for accurately predicting the excited-state properties of TADF emitters. This section evaluates the trade-offs between 
computational cost and accuracy and explores the broader implications of these methods for the design and optimization of optoelectronic materials.

\subsubsection{Computational cost vs. accuracy}

The semi-empirical methods employed in this study provide a favorable balance between computational efficiency and predictive reliability. Compared to full 
\tda, \stda and \stddft reduced computational time by over \qty{99}{\percent} per molecule. For example, a single-point energy calculation for DMAC-TRZ using 
full \tda took approximately 24 hours, while the same calculation with \stda required only 10 minutes. In addition to these efficiencies, the accuracy of the 
methods remained robust, with mean absolute errors (MAE) for singlet-triplet energy gaps ($\Delta E_{ST}$) and vertical excitation energies well below 
\qty{0.160}{\electronvolt}. These deviations are within acceptable ranges for TADF material design, where qualitative trends are often more critical than 
absolute precision for identifying candidates with high reverse intersystem crossing (rISC) efficiency. The relative calculations with simplified methods 
compared to full \tda are valid.

\subsubsection{Applicability to TADF emitters}

The study’s computational framework successfully captured key photophysical properties of TADF emitters, including singlet-triplet energy gaps, excitation 
energies, and fluorescence spectra, in both vacuum and solvent environments. The inclusion of solvent effects improved the relevance of predictions to 
experimental conditions, enhancing confidence in the computational workflow for real-world applications. The methods validated here can facilitate the 
large-scale screening of TADF emitters, significantly accelerating the discovery and optimization of materials for OLED technology. For example, our method 
accurately predicted the redshift in the fluorescence spectrum of DMAC-DPS upon solvation, which is consistent with experimental observations found in 
\cite{Liu2018}.

\subsubsection{Broader implications for computational material design}

Beyond TADF materials, the semi-empirical methods applied in this study are broadly applicable to other classes of optoelectronic materials, such as organic 
photovoltaics, photocatalysts, and quantum dots. The scalability of the computational framework makes it particularly suitable for high-throughput screening, 
where computational efficiency is critical. By integrating these methods into material discovery pipelines, researchers can systematically explore vast 
chemical spaces, reducing experimental bottlenecks and accelerating the development of advanced functional materials. These methods could be used to screen a 
library of novel donor-acceptor molecules for use in organic solar cells, optimizing their light-harvesting efficiency and charge-transfer properties.

\subsubsection{In summary}

The findings of this study establish \stda and \stddft as reliable, efficient tools for predicting the excited-state properties of TADF emitters. The 
computational framework enables rapid screening and optimization of photophysical properties, bridging the gap between theoretical simulations and practical 
applications. By extending these methods to broader classes of materials and integrating them into automated design workflows, this approach has the potential 
to significantly accelerate the discovery and development of advanced optoelectronic systems, leading to more efficient and cost-effective technologies for 
lighting, displays, and energy conversion.

\section{Conclusion}\label{sec:Conclusion}

In this study, we have developed and demonstrated an efficient computational framework that integrates extended tight-binding (\xtb), simplified Tamm-Dancoff 
approximation (\stda), and simplified time-dependent density functional theory (\stddft) methods for accurately predicting the excited-state properties of 
thermally activated delayed fluorescence (TADF) emitters. By carefully balancing computational efficiency and predictive accuracy, this framework provides a 
scalable solution for investigating key photophysical properties, including singlet-triplet energy gaps ($\Delta E_{ST}$), excitation energies, and 
fluorescence spectra, under both vacuum and solvent conditions. The detailed analysis of a series of representative TADF emitters (DMAC-TRZ, DMAC-DPS, PSPCz, 
4CzIPN, CzS2, Px2BP, 2TCz-DPS, and TDBA-DI) provided valuable insights into the interplay between molecular structure, electronic properties, and solvent 
effects. Our results revealed a strong correlation between the torsional angle between donor and acceptor units and the solvent-induced redshift in the emission 
spectrum, highlighting the importance of considering geometrical flexibility in the design of TADF materials.

The benchmarking against full \tddft validates the reliability of our approach, showing mean absolute errors within acceptable ranges (typically below 
\qty{0.16}{\electronvolt}) for practical applications. This level of accuracy, combined with the significant computational cost savings (over 99\% reduction in 
computational time compared to \tda), makes our framework particularly well-suited for high-throughput screening of potential TADF candidates. Incorporating 
solvent effects through the use of toluene further enhances the predictive realism, enabling a more accurate representation of experimental conditions and 
improving our understanding of the solvatochromic behavior of these materials. For example, our method captured some effects and there was a large difference 
which the full \tddft was unable to capture. These findings not only advance the understanding of TADF emitters but also demonstrate the broader potential of 
semi-empirical methods in computational material design.

Looking ahead, the versatility of this computational framework can extend to other classes of optoelectronic materials, including organic photovoltaics and 
photocatalysts, enabling high-throughput screening and accelerating material discovery pipelines. The multi-objective function, which is one of a kind, helps 
further enhance our results with an original solution. By bridging theoretical insights with practical applications, this work contributes significantly to the 
ongoing quest for sustainable, high-performance materials in emerging technologies.

\textbf{Limitations and Future Directions}

While the semi-empirical methods provide a valuable balance of cost and accuracy, their limitations should be acknowledged. For molecules with extensive 
charge-transfer states or highly polarizable environments, deviations from full \tddft may become more pronounced. For example, we observed some discrepancies 
between the \stda and \stddft predictions for molecules with large solvent-induced geometry changes (refer to the supplementary data) due to the various 
calculation errors and approximations. In addition, because of the way it works, it is not able to fully and accurately capture the results. Future work could 
incorporate hybrid approaches, such as selectively applying full \tddft to high-priority candidates identified through \stda or \stddft screening, to take the 
best features from each of the approaches. Additionally, the inclusion of explicit solvation models (e.g., polarizable continuum model or molecular dynamics 
simulations) and coupling with machine learning techniques could further enhance predictive capabilities and expand applicability. Future studies could also 
incorporate experimental data, such as cyclic voltammetry measurements, to better estimate the ionization potentials and electron affinities of the donor and 
acceptor units, leading to more accurate predictions of excitation energies.


\medskip
\textbf{Acknowledgements} \par 

The authors wish to thank Ms. Hafida ZIOUANI of the Physics Department, Faculty of Science, Materials and Renewable Energy Team, LP2MS Laboratory, Moulay
Ismail University, Meknes, Morocco, for her generous provision of computational resources, which were essential to the completion of this work.

\medskip

%
\bibliographystyle{MSP}
\bibliography{BibOLED.bib}

\newpage
\medskip
\textbf{Supporting Information} \par 
Supporting Information is available from the Wiley Online Library or from the author.

\setcounter{table}{0}
\setcounter{figure}{0}
\newsecnumstyle
\section{Supplementary information}\label{sec: Suppl_Info}

\subsection{Molecule information}

Here is supplementary information about the molecules studied in this work.

{\renewcommand{\arraystretch}{1.5}
\begin{table}[H]
\centering
\caption{Molecular information (DMAC-TRZ, DMAC-DPS, PSPCz, 4CzIPN). This table provides the chemical identification details for the first four molecules, 
including their 2D representations, PubChem Compound Identifiers (CIDs), molecular formulas (MFs), IUPAC names, and relevant literature references.}
\begin{tabularx}{\linewidth}{>{\centering\arraybackslash}m{2.5cm} >{\centering\arraybackslash}m{6cm} X}\\\toprule
Name & 2D Representation & Details\\\midrule
\multirow{4}{=}{DMAC-TRZ} &
\multirow{4}{=}{\includegraphics{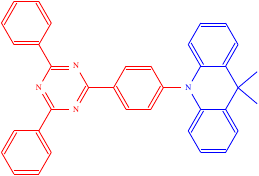}}
& Compound CID: 118528399,\\
& & MF: C36H28N4,\\
& & IUPAC Name:
10-[4-(4,6-diphenyl-1,3,5-triazin-2-yl)phenyl]-9,9-dimethylacridine\\
& & \url{doi.org/10.1088/1361-6633/ace06}\\[0.25cm]
\multirow{4}{=}{DMAC-DPS} &
\multirow{4}{=}{\includegraphics{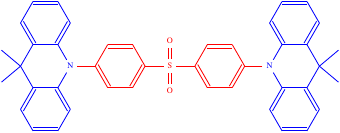}} &Compound CID: 59399558,\\
&&MF: C42H36N2O2S,\\
&&IUPAC Name:
10-[4-[4-(9,9-dimethylacridin-10-yl)phenyl]sulfonylphenyl]-9,9-
dimethylacridine\\
&&\url{doi.org/10.1088/1361-6633/ace06}\\[0.25cm]
\multirow{4}{=}{PSPCz} &
\multirow{4}{=}{\includegraphics{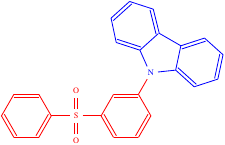}} &Compound CID:
132916142,\\
&&MF: C24self.H17NO2S,\\
&&IUPAC Name: 9-[4-(benzenesulfonyl)phenyl]carbazole,\\
&&\url{doi.org/10.1038/s41524-021-00540-6}\\
\multirow{4}{=}{4CzIPN} &
\multirow{4}{=}{\includegraphics{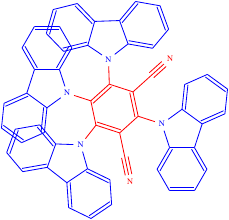}} &Compound CID: 102198498,\\
&&MF: C56H32N6,\\
&&IUPAC Name: 2,4,5,6-tetra(carbazol-9-yl)benzene-1,3-dicarbonitrile,\\
&&\url{doi.org/10.1021/jacs.6b12124}\\[1cm]\bottomrule
\end{tabularx}
\label{tab:infoMol1}
\end{table}

\begin{table}[H]
\centering
\caption{Molecular information (Px2BP, CzS2, 2TCz-DPS, TDBA-DI). This table provides the chemical identification details for the remaining four molecules, 
including their 2D representations, PubChem Compound Identifiers (CIDs), molecular formulas (MFs), IUPAC names, and relevant literature references.}
\ContinuedFloat
\begin{tabularx}{\linewidth}{>{\centering\arraybackslash}m{2.5cm} >{\centering\arraybackslash}m{6cm} X}\\\toprule
Name & 2D Representation & Details\\\midrule
\multirow{4}{=}{Px2BP} &
\multirow{4}{=}{\includegraphics{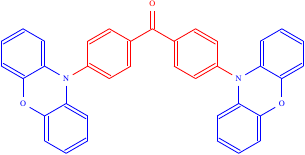}} &Compound CID: 1553685,\\
&&MF: C37H24N2O3,\\
&&IUPAC Name: bis(4-phenoxazin-10-ylphenyl)methanone,\\
&&\url{}\\
\multirow{4}{=}{CzS2} &
\multirow{4}{=}{\includegraphics{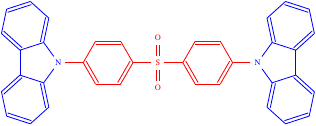}} &Compound CID: 59711244,\\
&&MF: C36H24N2O2S,\\
&&IUPAC Name: 9-[4-(4-carbazol-9-ylphenyl)sulfonylphenyl]carbazole,\\
&&\url{}\\
\multirow{4}{=}{2TCz-DPS} &
\multirow{4}{=}{\includegraphics{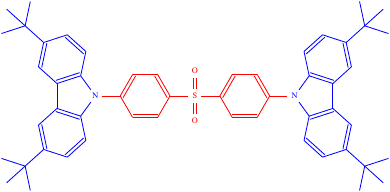}} &Compound CID: 90168099,\\
&&MF: C52H56N2O2S,\\
&&IUPAC Name: 3,6-ditert-butyl-9-[4-[4-(3,6-ditert-butylcarbazol-9-yl)phenyl]sulfonylphenyl]carbazole,\\
&&\url{}\\
\multirow{4}{=}{TDBA-DI} &
\multirow{4}{=}{\includegraphics{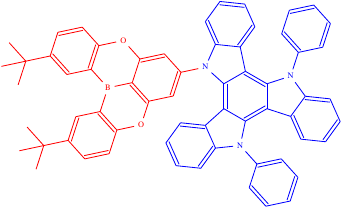}} &Compound CID: 137554973,\\
&&MF: C62H48BN3O2,\\
&&IUPAC Name:
9-(4,18-ditert-butyl-8,14-dioxa-1-borapentacyclo[11.7.1.02,7.09,21.015,20]henicosa-2(7),3,5,9,11,13(21),15(20),16,18-nonaen-11-yl)-18,27-diphenyl-9,18,27-
triazaheptacyclo[18.7.0.02,10.03,8.011,19.012,17.021,26]heptacosa-1,3,5,7,10,12,14,16,19,21,23,25-dodecaene,\\
&&\url{doi.org/10.1038/s41566-019-0415-5}\\\bottomrule
\end{tabularx}
\label{tab:infoMol2}
\end{table}
}

\begin{table}[!htbp]
\centering
\caption{Root-Mean-Square Deviation (RMSD) between vacuum- and toluene-optimized geometries. RMSD values (\unit{\angstrom}) quantify the overall structural 
change induced by the solvent environment. Larger RMSD values indicate a greater degree of conformational change upon solvation.}
\begin{tabular}{l*{8}{@{ }S@{ }}}
\toprule
Molecule & {DMAC-TRZ} & {DMAC-DPS} & {PSPCz} & {4CzIPN} & {Px2BP} & {CzS2} &
{2TCz-DPS} & {TDBA-DI} \\
\midrule
RMSD (\unit{\angstrom}) & 3.904 & 4.059 & 1.605 & 4.370 & 3.560 & 3.251 & 3.347 & 1.265 \\
\bottomrule
\end{tabular}
\label{tab:rmsd}
\end{table}

\begin{figure}[!htbp]
\centering
\includegraphics[width=0.5\textwidth]{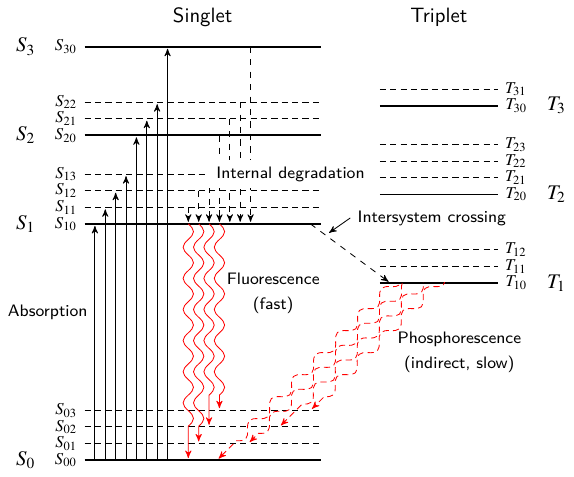}
\caption{Simplified Jablonski diagram illustrating key electronic transitions. The diagram depicts the singlet ($S$) and triplet ($T$) states, separated for 
clarity, along with representative vibrational sublevels (horizontal dashed lines). Key radiative (fluorescence, phosphorescence) and non-radiative (internal 
conversion, intersystem crossing) processes are indicated.}
\label{fig:jablonski}
\end{figure}

\begin{table}[!htbp]
\caption{Calculated $\Delta E(T_2 \gets T_1)$ in \unit{\electronvolt} based on $T_1$-state optimized geometries. This table presents the calculated energy 
difference between the first ($T_1$) and second ($T_2$) triplet states, using both UKS-\stda and UKS-\stddft methods. NA indicates cases where a value could 
not be reliably obtained, and the \stda value is used instead. These values provide insight into the triplet manifold and potential pathways for 
triplet-triplet annihilation.}
{%
\newcommand{\mc}[3]{\multicolumn{#1}{#2}{#3}}
\footnotesize
\begin{center}
\begin{tabularx}{\linewidth}{YY*{8}{@{ }S@{ }}}
\toprule
&Molecule & {DMAC-TRZ} & {DMAC-DPS} & {PSPCz} & {4CzIPN} & {Px2BP} & {CzS2} & {2TCz-DPS} & {TDBA-DI}\\
\midrule
\multirow{2}{=}{In vacuum} & \stda & 0.610 & 0.325 & 0.240 & 0.770 & 0.496 & 0.081 & 0.155 & 1.063 \\
& \stddft & 0.533 & {NA} & 0.144 & 0.743 & 0.511 & 0.053 & 0.115 & 1.008  \\\midrule
\multirow{2}{=}{In toluene\\ solvent} & \stda & 0.640 & 0.786 & 0.268 & 1.159 & 1.091 & 0.465 & 0.543 & 0.948 \\
& \stddft & {NA} & 0.725 & {NA} & 1.133 & 0.894 & 0.268 & 0.393 & 0.922 \\
\bottomrule
\end{tabularx}
\end{center}
}
\label{tab:T1T2}
\end{table}

\begin{table}[!htbp]
\caption{Geometry relaxation energies: $\Delta E_v(S_0\to T_1) - \Delta E_r(S_0\to T_1)$ in \unit{\electronvolt}. This table shows the energy difference 
between the vertical ($E_v$) and relaxed ($E_r$) triplet excitation energies, providing insight into the degree of geometrical relaxation that occurs upon 
excitation to the triplet state. Larger values indicate greater structural reorganization.}.
{%
\newcommand{\mc}[3]{\multicolumn{#1}{#2}{#3}}
\footnotesize
\begin{center}
\begin{tabularx}{\linewidth}{YY*{8}{@{ }S@{ }}}
\toprule
&Molecule & {DMAC-TRZ} & {DMAC-DPS} & {PSPCz} & {4CzIPN} & {Px2BP} & {CzS2} & {2TCz-DPS} & {TDBA-DI}\\
\midrule
\multirow{2}{=}{In vacuum} & \stda & 1.296 & 0.892 & 0.286 & 0.818 & 0.540 & 0.747 & 0.954 & 0.534 \\
& \stddft & 1.283 & 0.862 & 0.263 & 0.809 & 0.530 & 0.724 & 0.932 & 0.515 \\\midrule
\multirow{2}{=}{In toluene\\ solvent} & \stda & 1.319 & 1.088 & 0.332 & 0.840 & 0.719 & 0.836 & 1.103 & 0.554 \\
& \stddft & 1.287 & 1.058 & 0.309 & 0.830 & 0.708 & 0.814 & 1.082 & 0.535 \\
\bottomrule
\end{tabularx}
\end{center}
}
\label{tab:T1T2geo}
\end{table}

\subsection{Detailed Optimization results}\label{subsec:opt}

This section provides detailed information about the geometry optimization process for each molecule in both vacuum and toluene environments. The results 
presented here complement the discussion in the main text (\Cref{sec:Results}) and provide supporting evidence for the conclusions drawn about the accuracy and 
efficiency of the computational methods.

\begin{figure}[!htbp]
\centering
\leavevmode
\subfloat[In vacuum: optimization steps vs. energy]{\includegraphics[width=\textwidth]{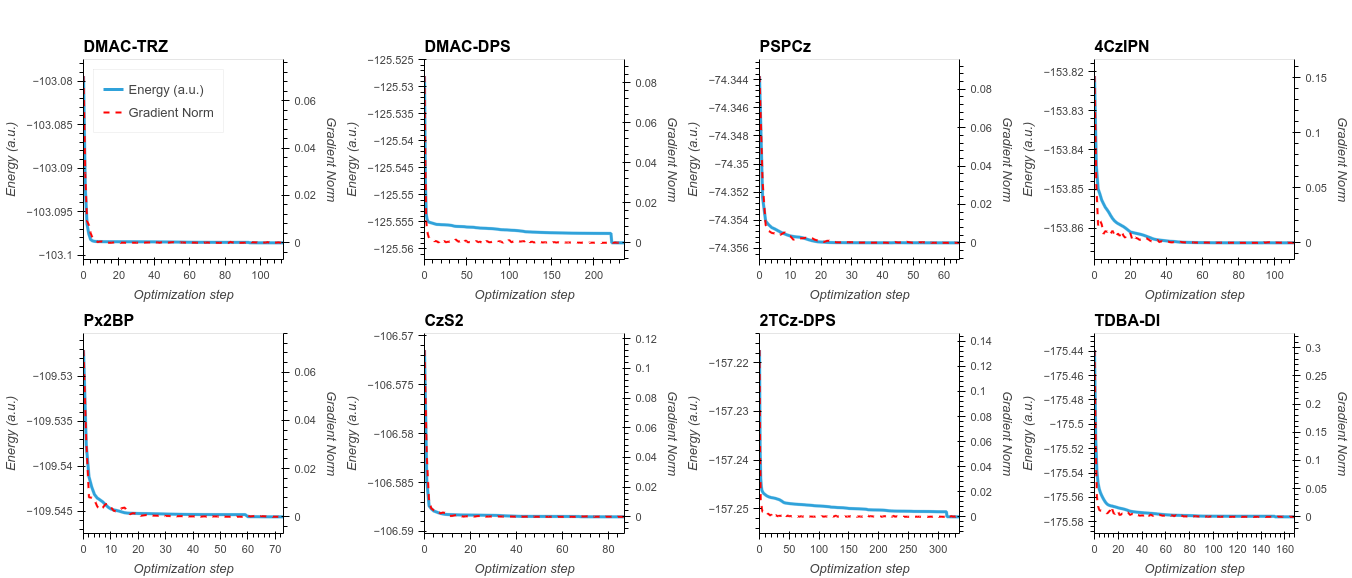}}\\
\subfloat[In toluene solvent: optimization steps vs. energy]{\includegraphics[width=\textwidth]{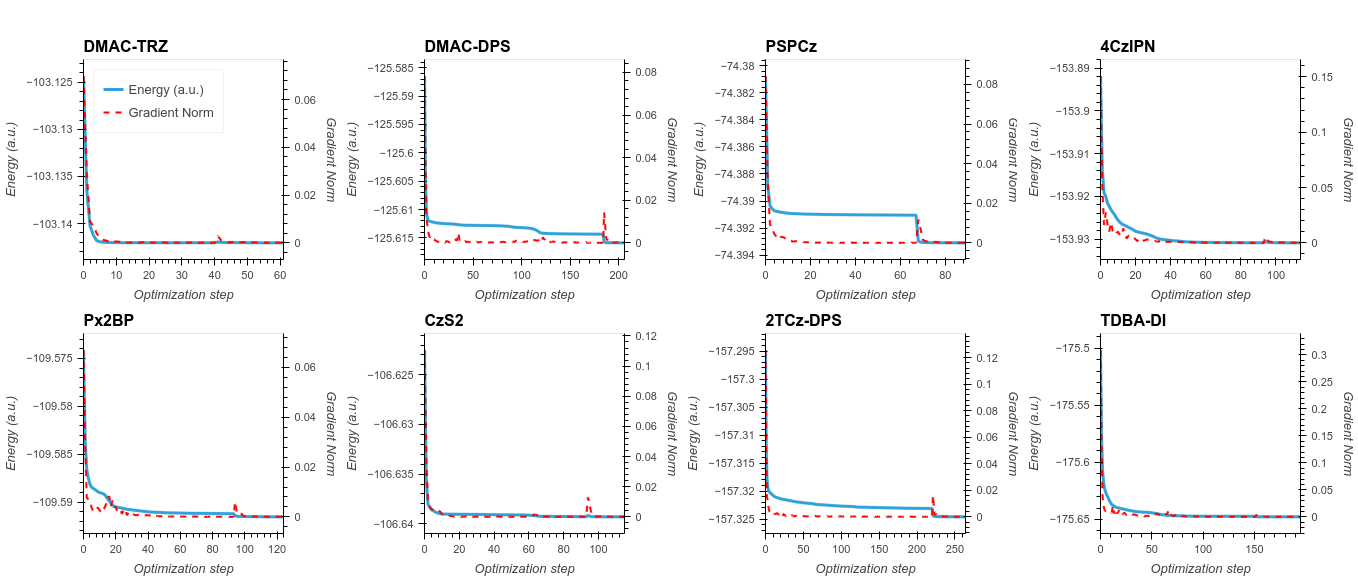}}
\caption{Evolution of the geometry optimization process for all molecules. Each panel shows the change in energy (y-axis) as a function of optimization step 
(x-axis) for all molecules in either (a) vacuum or (b) toluene solvent, calculated with \xtb and \crest. The red lines represent the overall energy, while the 
blue lines represent the gradient of the energy. The sudden deviations in the red lines, particularly in toluene, illustrate the effectiveness of the \crest 
search in locating lower-energy conformers.}
\label{fig:optimizep}
\end{figure}

The effect of the \crest search (see \Cref{fig:optimizep}) for identifying the best conformers can be observed as a sudden deviation near the end of each red 
plot, indicating a more thorough exploration of the potential energy surface. This suggests that \crest helps to overcome local minima and find lower-energy 
conformations.

Across all molecules, geometry optimizations in toluene generally resulted in a significant stabilization compared to vacuum, underscoring the importance of 
considering solvent effects in the calculation of the parameters. The subsections below detail the optimization process for each molecule and quantify the 
energy difference between the vacuum- and toluene-optimized structures.

\subsubsection{DMAC-TRZ}

\begin{figure}[!htbp]
\centering
\includegraphics[width=\textwidth]{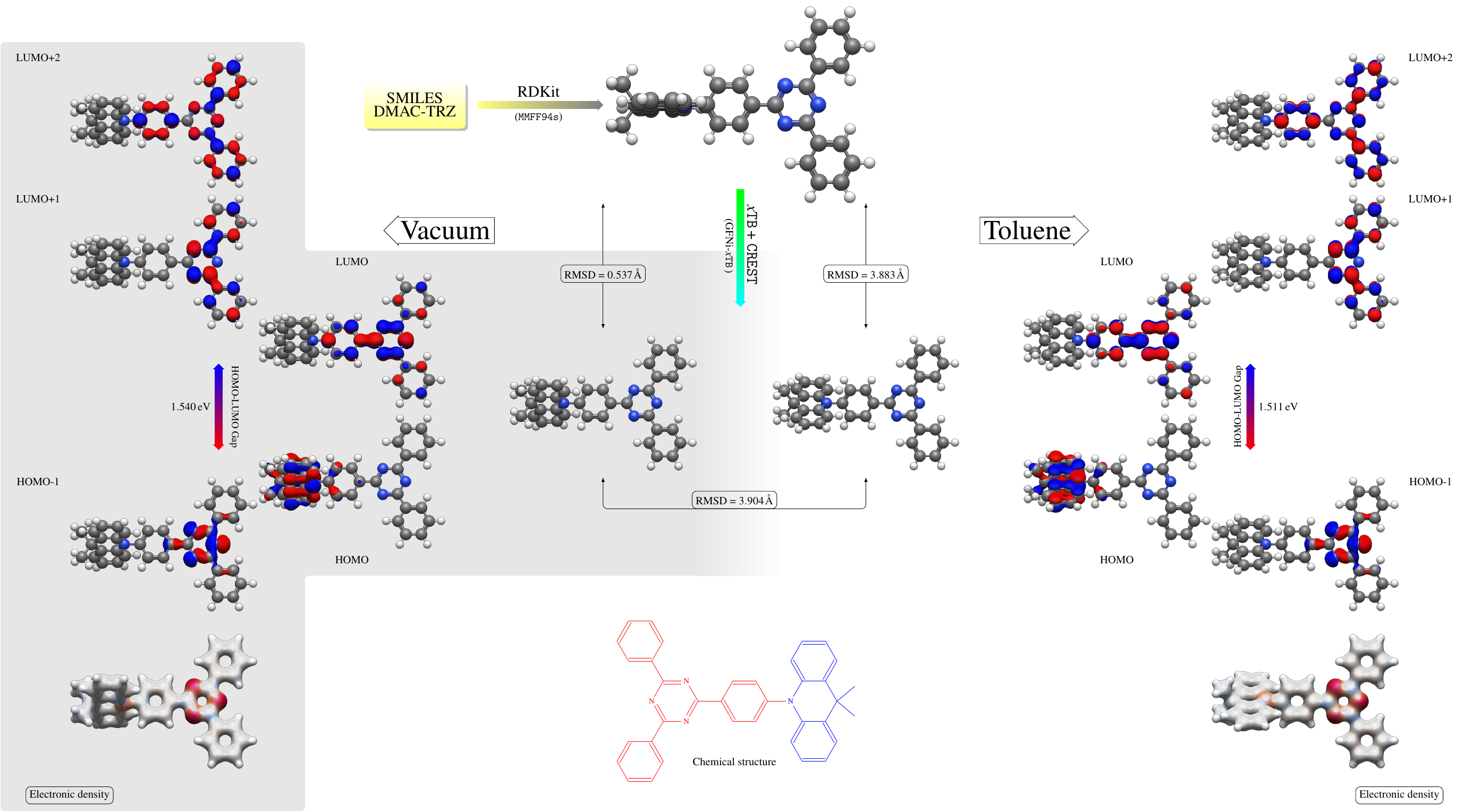}
\caption{Optimized molecular structure and electronic distributions of DMAC-TRZ. This figure displays the optimized 3D molecular structures of DMAC-TRZ in 
vacuum and toluene, calculated using \mmf, \xtb, and \crest methods. The molecular orbitals from HOMO-1 to LUMO+2 are visualized as isodensity surfaces, with 
blue and red indicating negative and positive regions of the wave function, respectively. The HOMO-LUMO gap values are provided for each environment. These 
data highlight the impact of solvation on charge distribution and donor-acceptor interactions.}
\label{fig:DMAC-TRZ-Sem}
\end{figure}

\Cref{fig:DMAC-TRZ-Sem} shows the chemical structure of DMAC-TRZ and its 3D geometry optimized using the \mmf force field, \xtb, and \crest methods. It also 
displays the HOMO-LUMO gap, molecular orbitals from HOMO-1 to LUMO+2, and the electronic density distributions. The RMSD values between the three optimized 
structures are below \qty{4}{\angstrom}, indicating that the structures are similar but not identical, with the largest deviations observed in the 
toluene-optimized structure (\Cref{fig:DMAC-TRZ-RMSD}).

\begin{figure}[!htbp]
\centering
\begin{minipage}[c]{0.48\textwidth}
\centering
\includegraphics[width=\textwidth]{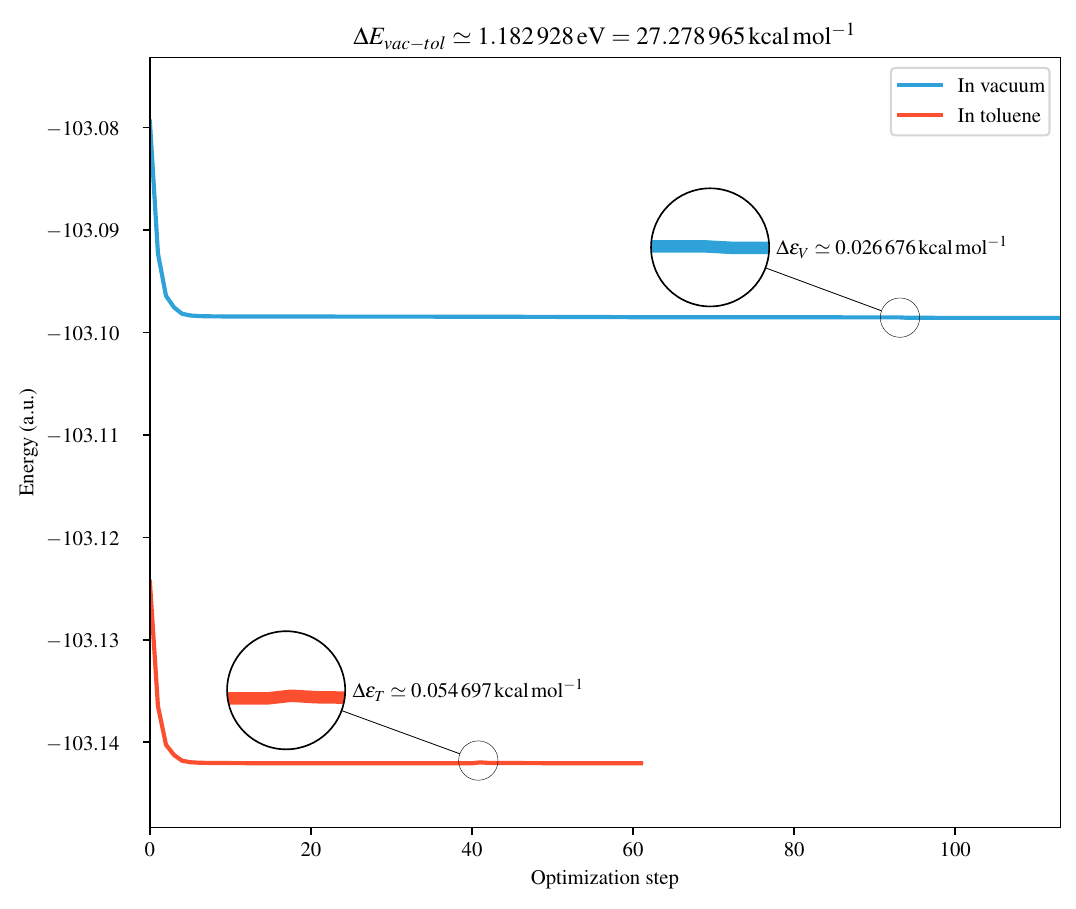}
\end{minipage}%
\begin{minipage}[c]{0.2\textwidth}
\centering
\includegraphics[width=\textwidth]{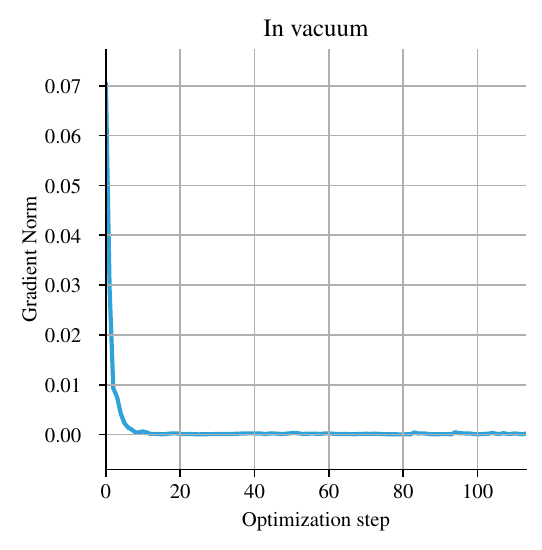}\\
\includegraphics[width=\textwidth]{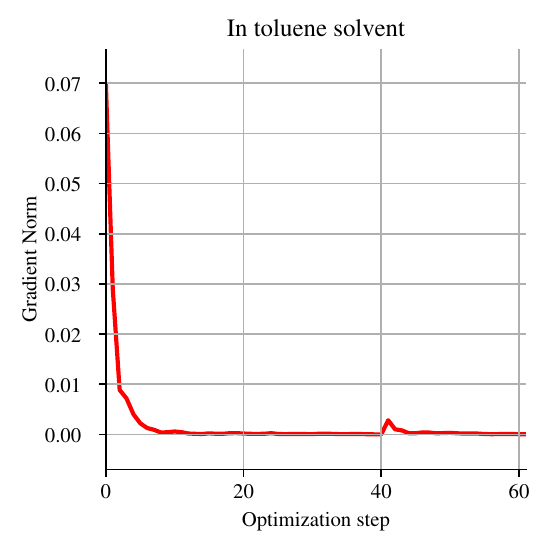}
\end{minipage}
\caption{Energy optimization process for DMAC-TRZ: vacuum vs. toluene. The left panel shows the potential energy surface explored during the geometry 
optimization process for DMAC-TRZ in both vacuum and toluene environments. The right panel displays the gradients for vacuum (top) and toluene (bottom) 
optimizations. The energy difference ($\Delta E_{vac-tol} = \qty{27.28}{\kilo\calorie\per\mole}$) confirms the significant stabilizing effect of toluene.}
\label{fig:DMAC-TRZ-opt}
\end{figure}

As shown in \Cref{fig:DMAC-TRZ-opt}, the toluene-optimized structure is approximately $\qty{27.278965}{\kilo\calorie\per\mole}$ lower in energy than the 
vacuum-optimized structure, confirming that toluene stabilizes the molecule.

The small values of $\Delta\epsilon_V$ and $\Delta\epsilon_T$ suggest that the geometries obtained after pre-optimization are suitable for further 
calculations. In the absence of solvent, despite a higher number of optimization steps, $\Delta\epsilon_V\simeq\qty{0.054697}{\kilo\calorie\per\mole}$, 
indicating good convergence.

\paragraph{Molecular orbitals and HOMO-LUMO gap}

In vacuum, the HOMO-1 is primarily localized on the TRZ moiety, and a similar distribution is observed in the toluene environment, albeit with slight 
variations. This suggests that lower-energy occupied states, which are less involved in electronic transitions, can be localized on either the TRZ or DMAC 
units. However, the LUMO+1 and LUMO+2 are primarily localized on the TRZ unit in both environments, indicating that excited-state dynamics will primarily 
involve the TRZ unit, influencing the molecule's photophysical properties.

The HOMO is delocalized over both DMAC and TRZ in both environments, indicating a significant interaction between the two units that facilitates charge 
transfer. Conversely, the LUMO is mainly localized on the TRZ unit, suggesting that the electron-accepting properties are concentrated in this region. This 
separation of electron-donating (HOMO) and electron-accepting (LUMO) regions is conducive to charge separation and recombination processes, which are essential 
for TADF.

The HOMO-LUMO gap remains relatively consistent between the two environments, with a slight decrease of $\qty{0.027}{\electronvolt}$ 
($\simeq\qty{0.622635}{\kilo\calorie\per\mole}$) in toluene, suggesting that the molecule is relatively stable despite the environment-induced geometric 
changes. The relatively large gap indicates electronic stability and suggests that the molecule requires a moderate amount of energy to transition from the 
ground state (HOMO) to the excited state (LUMO). The electronic density maps in \Cref{fig:DMAC-TRZ-Sem} further confirm the charge distribution, supporting the 
conclusions drawn from the orbital analysis.

\paragraph{Excitation energies}

Without considering solvation, the $S_1 \gets S_0$ excitation energy calculated using the \stda method is approximately \qty{3.483}{\electronvolt}, while the 
\stddft method yields a slightly lower value of \qty{3.476}{\electronvolt}. Both cases correspond to absorption wavelengths in the near-ultraviolet (UV) 
region, with maxima at $\lambda_{abs}\simeq\qty{356.010}{\nano\meter}$ and $\lambda_{abs}\simeq\qty{356.681}{\nano\meter}$, respectively. When toluene is 
introduced as a solvent, a redshift occurs, reducing the absorption energies to $\qty{3.457}{\electronvolt}$ (STDA) and $\qty{3.453}{\electronvolt}$ (STDFT). 
This redshift results from solvent stabilization of the excited state, which lowers the energy difference between the $S_1$ and $S_0$ states. Consequently, the 
maximum wavelengths increase to $\lambda_{abs}\simeq\qtylist{358.669;359.073}{\nano\meter}$, respectively. While the maximum wavelengths are in the near-UV 
region, the predicted absorption color in both cases is blue (see \Cref{tab:ColorDMAC-TRZ}), indicating that DMAC-TRZ can be considered a blue absorber in both 
vacuum and solution.

Considering the emission with the Stokes shift, and the predicted colors, DMAC-TRZ can be characterized as a yellow-green emitter (see 
\Cref{fig:chromatogram,fig:fluo-DMAC-TRZ,tab:ColorDMAC-TRZ}). The estimated fluorescence maxima wavelengths are 
$\lambda_{PL}\simeq\qtylist{566.890;565.339}{\nano\meter}$ in vacuum and $\lambda_{PL}\simeq\qtylist{579.934;572.437}{\nano\meter}$ in toluene environment for 
the \stda and \stddft methods, respectively.

The yellow-green emission spectrum of DMAC-TRZ, combined with its TADF properties, makes it particularly promising for OLED applications. DMAC-TRZ displays 
strong fluorescence in the near-UV and visible region, with emission properties that can be fine-tuned by external factors such as solvation. The molecule’s 
ability to harvest triplet excitons and convert them into usable light via TADF enhances the overall device efficiency.

\subsubsection{DMAC-DPS}\label{sec:SI_DMAC-DPS}

\begin{figure}[!htbp]
\centering
\includegraphics[width=\textwidth]{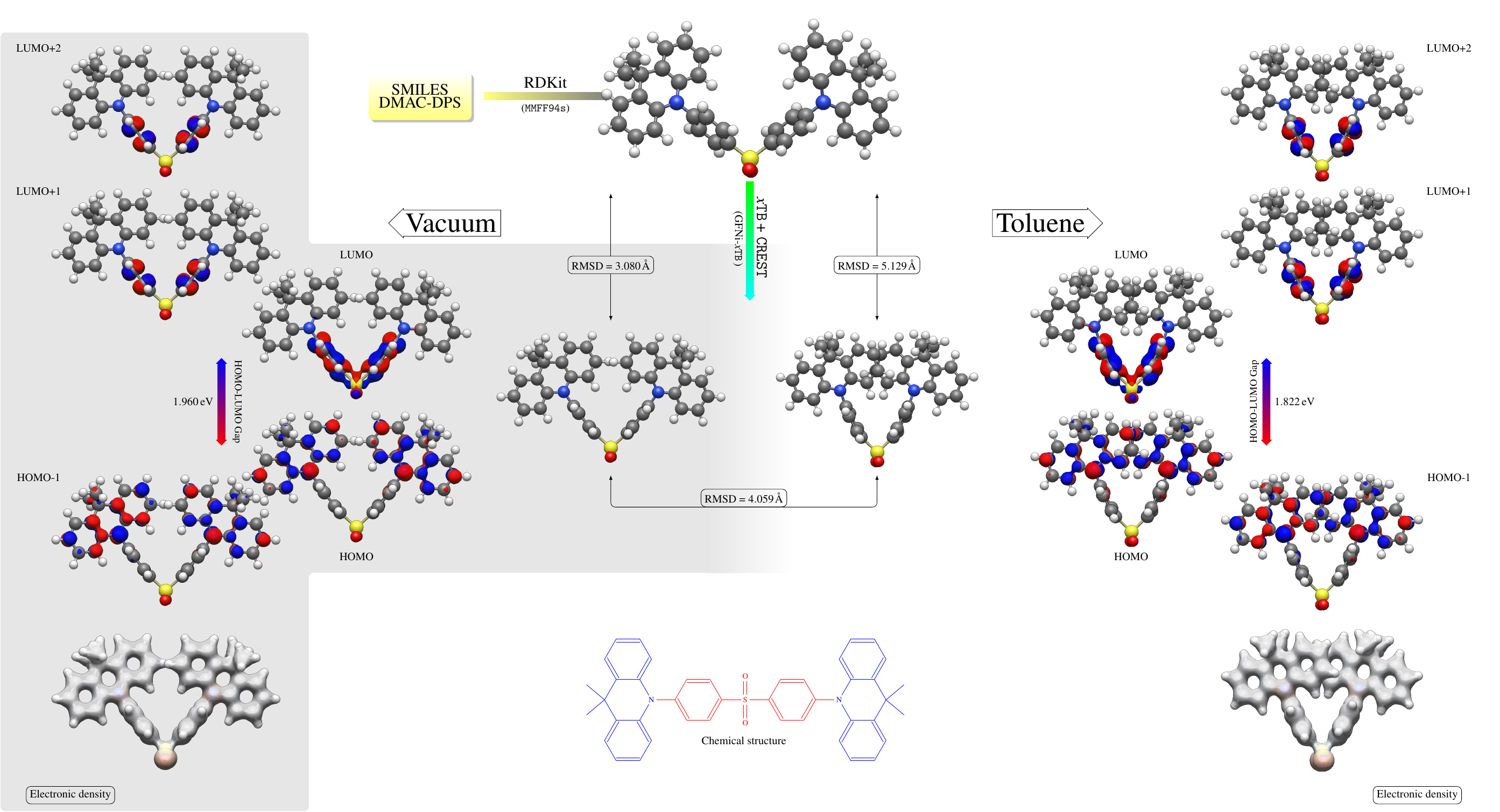}
\caption{Optimized molecular structure and electronic distributions of DMAC-DPS. This figure displays the optimized 3D molecular structures of DMAC-DPS in 
vacuum and toluene, calculated using \mmf, \xtb, and \crest methods. The figure also displays the corresponding molecular orbitals from HOMO-1 to LUMO+2, 
visualized as isodensity surfaces with blue and red indicating negative and positive regions of the wave function, respectively, and the HOMO-LUMO gap values 
in both environments. The comparison of electronic density maps highlights how solvation affects charge density distribution across the molecule and the 
interaction between DMAC and DPS units.}
\label{fig:DMAC-DPS-Sem}
\end{figure}

\Cref{fig:DMAC-DPS-Sem} depicts the chemical structure of DMAC-DPS, along with its 3D geometry following optimization using the \mmf force field, \xtb, and 
\crest methods. It also shows the HOMO-LUMO gap, molecular orbitals from HOMO-1 to LUMO+2, and the electronic density maps. The RMSD values between the three 
structures are all above $\qty{3}{\angstrom}$, indicating that the structures are significantly different (see \Cref{fig:DMAC-DPS-RMSD}).

Both the \xtb and \crest methods have a noticeable effect on the molecular geometry compared to the \mmf-derived structure. Moreover, the introduction of 
toluene as a solvent causes a non-negligible deviation in the molecular geometry compared to the vacuum-optimized structure.

\begin{figure}[!htbp]
\centering
\begin{minipage}[c]{0.48\textwidth}
\centering
\includegraphics[width=\textwidth]{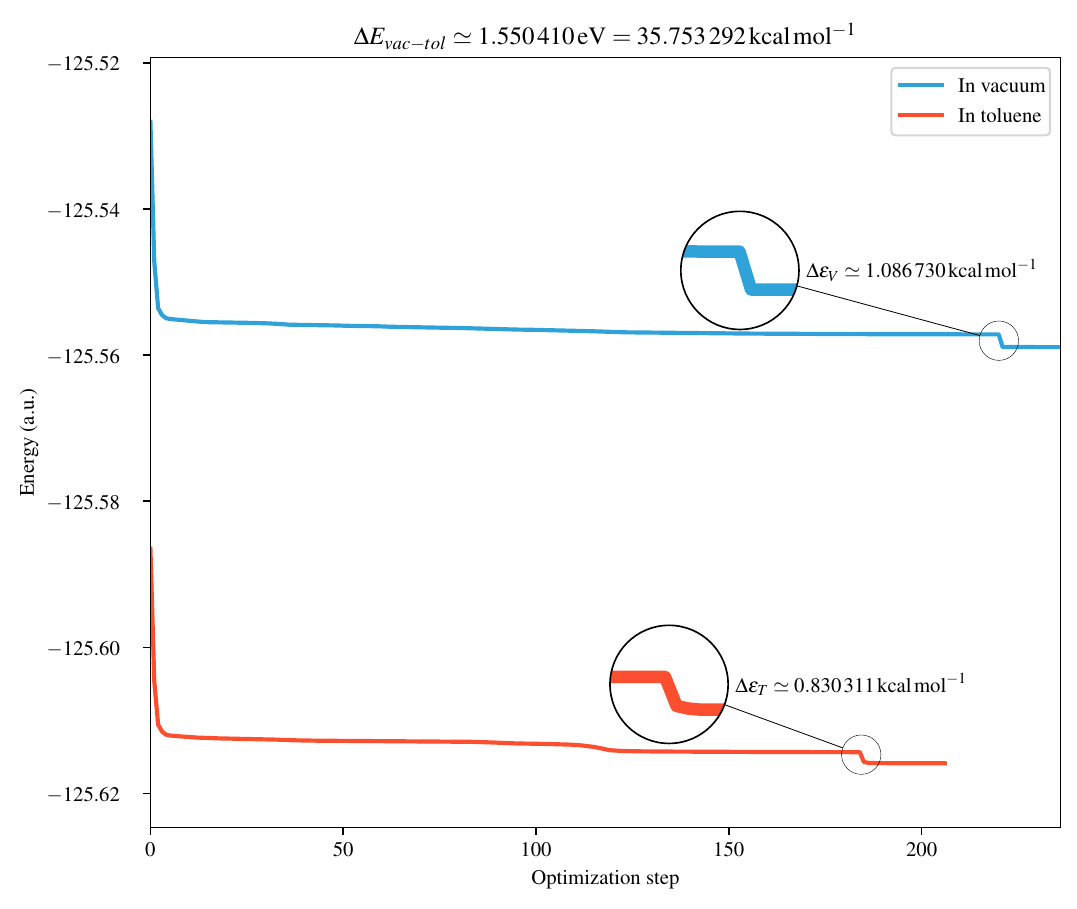}
\end{minipage}%
\begin{minipage}[c]{0.2\textwidth}
\centering
\includegraphics[width=\textwidth]{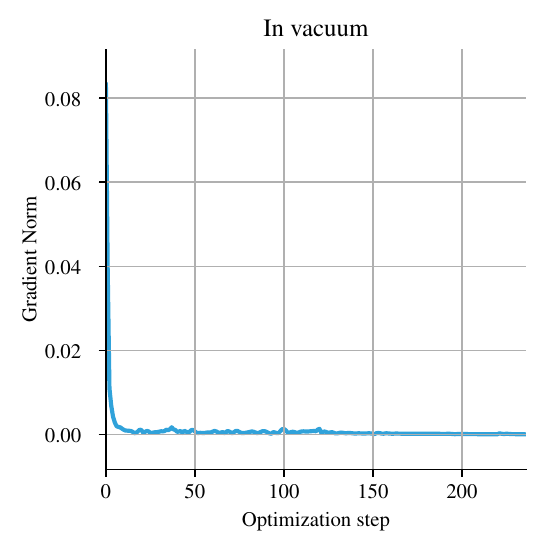}\\
\includegraphics[width=\textwidth]{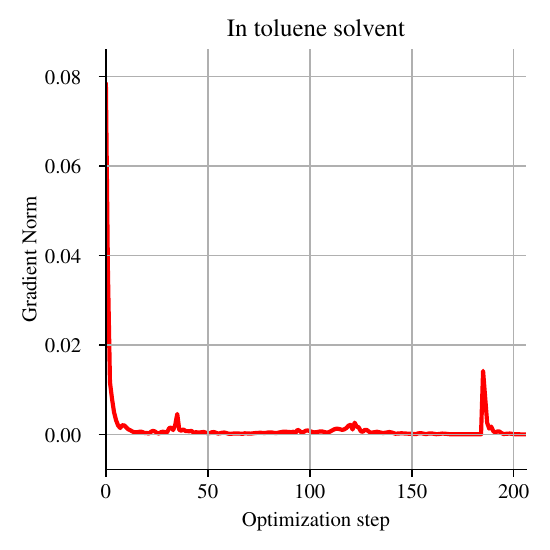}
\end{minipage}
\caption{Energy optimization process for DMAC-DPS: vacuum vs. toluene. The left panel shows the potential energy surface explored during the geometry 
optimization process for DMAC-DPS in both vacuum and toluene environments. The right panels display the gradients for vacuum (top) and toluene (bottom) 
optimizations. The energy difference ($\Delta E_{vac-tol} = \qty{35.75}{\kilo\calorie\per\mole}$) underscores the stabilizing effect of toluene.}
\label{fig:DMAC-DPS-opt}
\end{figure}

As shown in \Cref{fig:DMAC-DPS-opt}, the energy of the toluene-optimized structure is approximately $\qty{35.753292}{\kilo\calorie\per\mole}$ lower than that 
of the vacuum-optimized structure, indicating a substantial solvent effect on molecular stability.

The values of $\Delta\epsilon_V$ and $\Delta\epsilon_T$ are both around $\qty{1}{\kilo\calorie\per\mole}$, which is within the range of chemical precision. 
This suggests that the pre-optimized geometries are reliable enough for further calculations without significant loss of accuracy.

\paragraph{Molecular orbitals and HOMO-LUMO gap}

In vacuum, the HOMO-1 is mainly localized on the DMAC units, and this localization is maintained with some slight variations in the presence of toluene. This 
indicates that the lower energy occupied states are more confined to the DMAC portions, which may influence the electron density and reactivity of this part of 
the molecule. However, the presence of the solvent does not significantly affect the HOMO-1 distribution, suggesting that the lower energy states are largely 
unaffected by solvation. 

In both environments, the LUMO+1 and LUMO+2 are primarily localized on the PS moiety, indicating that the excited-state dynamics will likely be similar in both 
vacuum and solvent. This localization implies that the PS unit plays a key role in the molecule’s photophysical properties in both environments. 

The HOMO is more delocalized across both DMAC units in both environments, with a noticeable delocalization onto the PS unit. This suggests that there is 
significant interaction between the DMAC and PS units, facilitating charge transfer between them. Conversely, the LUMO remains primarily localized on the PS 
unit, indicating that the PS part predominantly contributes to the molecule’s electron-accepting characteristics. The consistency of the LUMO localization 
across environments implies that PS’s electron-accepting properties are stable and unaffected by solvation.

The HOMO-LUMO gap decreases by $\qty{0.138}{\electronvolt}$ (which is equivalent to $\qty{3.182356}{\kilo\calorie\per\mole}$) between the two environments, 
indicating what happens when this process occurs. This suggests that the molecular geometry is influenced by the environment, though the relatively wide gap 
indicates that the molecule retains a degree of electronic stability. The HOMO-LUMO gap suggests that the molecule requires a moderate amount of energy to 
transition from the ground state (HOMO) to the excited state (LUMO). The electronic density maps further confirm this charge distribution, agreeing with the 
molecular orbital analysis. 

\paragraph{Excitation energies}

In the absence of solvent, the absorption energy ($S_1 \gets S_0$) calculated using the \stda method is approximately \qty{3.878}{\electronvolt}, while the 
\stddft method gives a slightly lower value of \qty{3.873}{\electronvolt}. Both values correspond to absorption wavelengths in the deep ultraviolet region, 
with maxima at $\lambda_{abs}\simeq\qtylist{319.690;320.143}{\nano\meter}$, respectively. When toluene is introduced as a solvent, a redshift is observed, 
reducing the absorption energies to $\qty{3.713}{\electronvolt}$ for the \stda method and $\qty{3.708}{\electronvolt}$ for the \stddft method. Consequently, 
the absorption wavelengths maxima shift to $\lambda_{abs}\simeq\qtylist{333.937;334.417}{\nano\meter}$, respectively, though they remain within the ultraviolet 
range, but in the near region. Therefore, DMAC-DPS can still be classified as a UV absorber, regardless of solvation (see \Cref{tab:ColorDMAC-DPS}). So, in the 
both cases, there is no predicted color in the visible region.

When considering the emission and the Stokes shift, DMAC-DPS can be categorized as a \emph{blue} in both environments. When solvated, while the fluorescence 
maxima are respectively $\lambda_{PL}\simeq\qtylist{467.787;484.417}{\nano\meter}$ for \stda and \stddft methods, the predicted color is \emph{blue}. But, for 
the \stda method this blue color is closer to the blue-green boundary, while for the \stddft method this color is \emph{blue-green} (because it falls within 
the blue-green region of the visible spectrum). In conclusion, DMAC-DPS can be categorized as a blue emitter. Without solvation, it exhibits characteristics of 
a pure \emph{blue} emitter ($\lambda_{PL}\simeq\qty{452.233}{\nano\meter}$ for the both methods). Further details are shown in 
\Cref{fig:chromatogram,fig:fluo-DMAC-DPS,tab:ColorDMAC-DPS}. In the CIE tables, we can connect the X, Y, and Z coordinates with the \emph{UV-vis} with 
different coordinates for the color spectrum.

\subsubsection{PSPCz}\label{sec:SI_PSPCz}

\begin{figure}[!htbp]
\centering
\includegraphics[width=\textwidth]{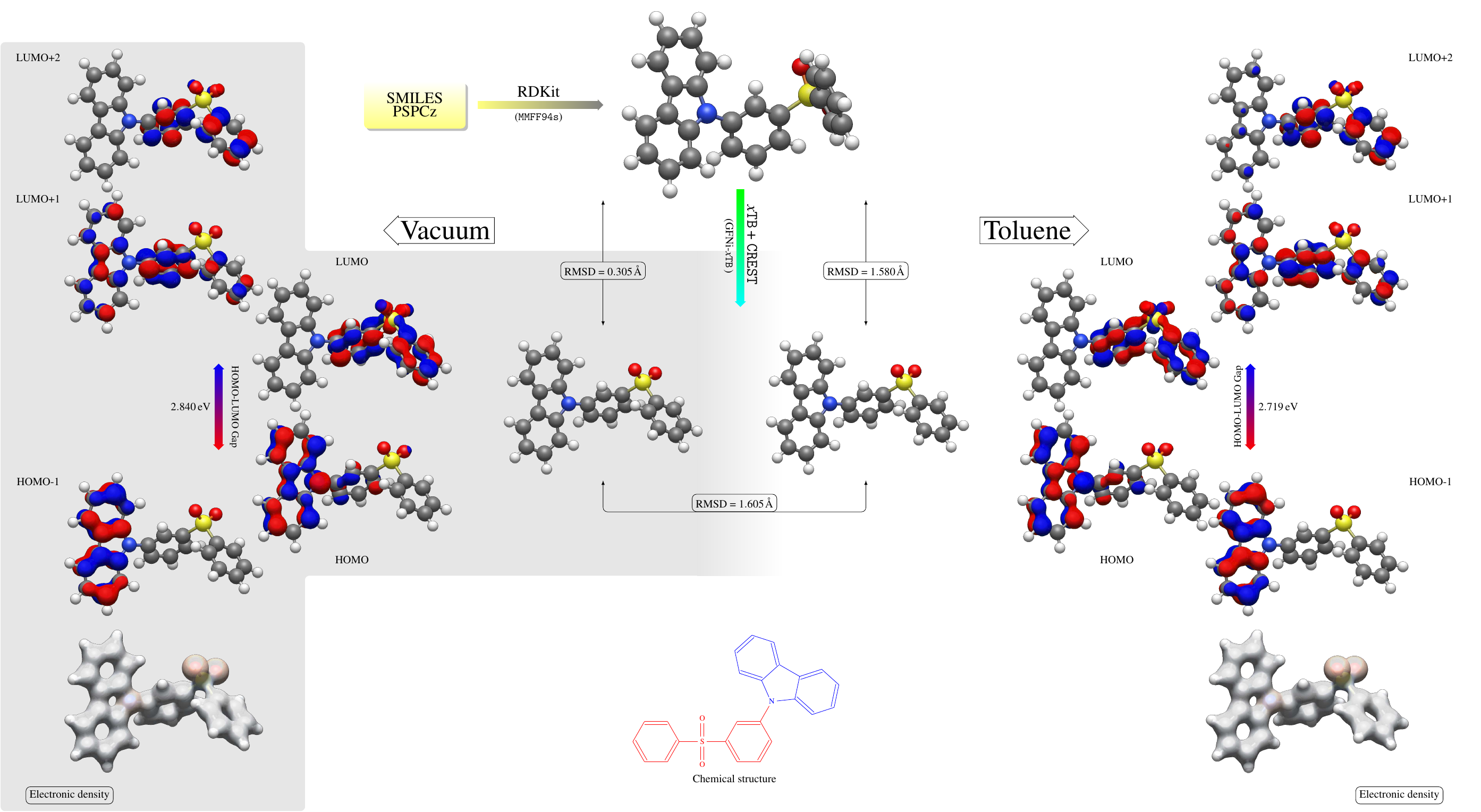}
\caption{Optimized molecular structure and electronic distributions of PSPCz. This figure displays the optimized 3D molecular structures of PSPCz in vacuum and 
toluene environments, calculated using the \mmf force field, \xtb, and \crest methods. The figure also displays the corresponding molecular orbitals from 
HOMO-1 to LUMO+2, visualized as isodensity surfaces with blue and red indicating negative and positive regions of the wave function, respectively, and the 
HOMO-LUMO gap values in both environments. The comparison of electronic density maps highlights how solvation affects charge density distribution across the 
molecule and the interaction between PS and Cz units.}
\label{fig:PSPCz-Sem}
\end{figure}

\Cref{fig:PSPCz-Sem} presents the chemical structure of PSPCz along with its 3D geometry following optimization using the \mmf force field, \xtb, and \crest 
methods. It also shows the HOMO-LUMO gap, molecular orbitals from HOMO-1 to LUMO+2, and the electronic density maps.

The RMSD values between the three optimized structures are all below $\qty{1.7}{\angstrom}$, indicating that the geometries are structurally very similar (see 
\Cref{fig:PSPCz-RMSD}).

Though the geometries are quite similar, it is important to note that the \xtb and \crest methods do have a measurable impact on the molecular geometry derived 
from the \mmf method. Additionally, toluene causes a notable deviation in the molecular structure compared to the vacuum-optimized one. 

\begin{figure}[!htbp]
\centering
\begin{minipage}[c]{0.48\textwidth}
\centering
\includegraphics[width=\textwidth]{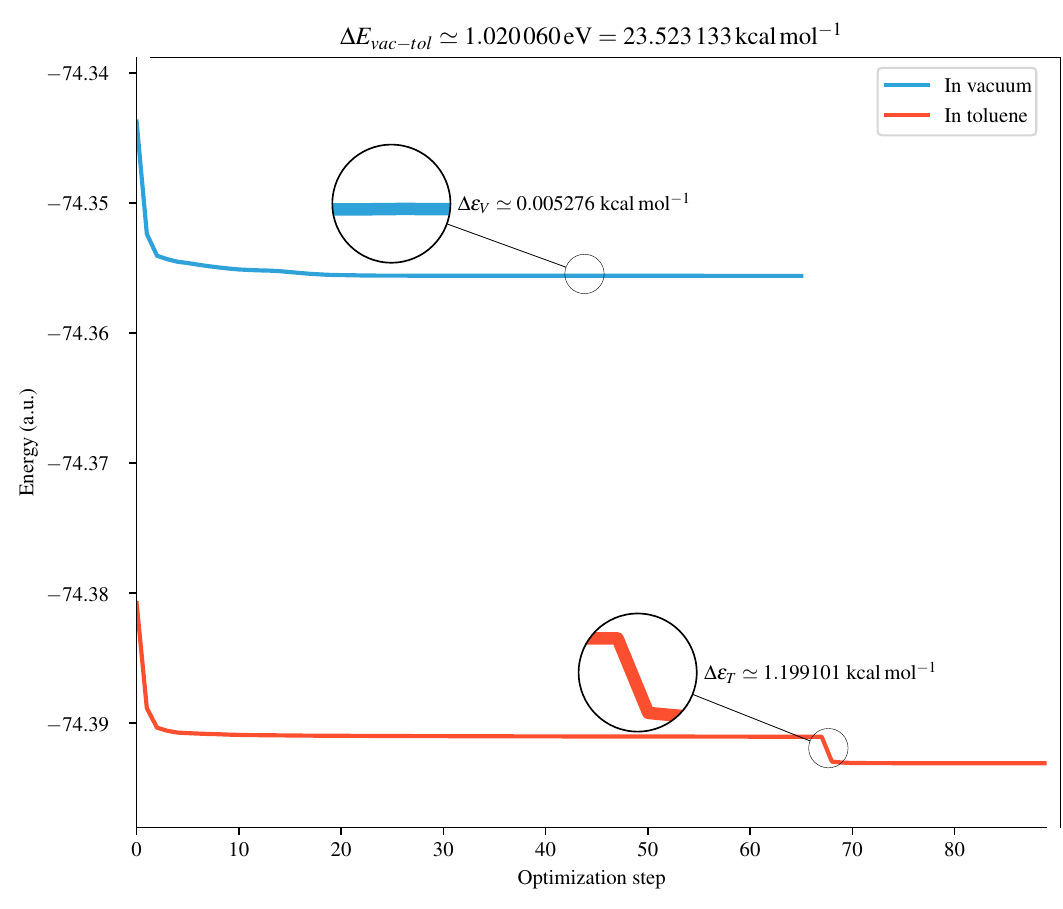}
\end{minipage}%
\begin{minipage}[c]{0.2\textwidth}
\centering
\includegraphics[width=\textwidth]{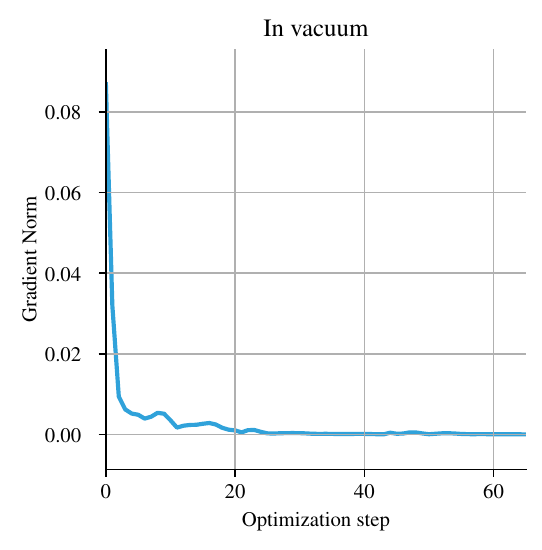}\\
\includegraphics[width=\textwidth]{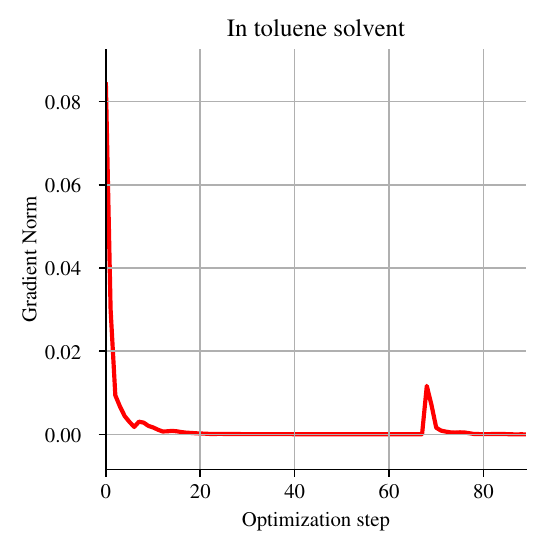}
\end{minipage}
\caption{Energy optimization process for PSPCz: vacuum vs. toluene. The left panel shows the potential energy surface explored during the geometry optimization 
process for PSPCz in both vacuum and toluene environments. The right panels display the gradients for vacuum (top) and toluene (bottom) optimizations. The 
energy difference ($\Delta E_{vac-tol} = \qty{23.52}{\kilo\calorie\per\mole}$) highlights the significant solvent effect on the molecular stability.}
\label{fig:PSPCz-opt}
\end{figure}

As shown in \Cref{fig:PSPCz-opt}, the toluene-optimized structure is approximately $\qty{23.523133}{\kilo\calorie\per\mole}$ lower in energy than the 
vacuum-optimized structure, despite their close similarity. This small structural change is sufficient to improve the energy greatly.

The values of $\Delta\epsilon_V$ and $\Delta\epsilon_T$ are both relatively small, with $\Delta\epsilon_T$ close to the threshold of chemical precision 
($\qty{1}{\kilo\calorie\per\mole}$). This indicates that the pre-optimized geometries are sufficiently accurate for further computational studies without 
significant loss of precision.

\paragraph{Molecular orbitals and HOMO-LUMO gap}

In both vacuum and toluene environments, the HOMO-1 is primarily localized on the Cz unit of the molecule, which implies that lower energy occupied states are 
concentrated on the Cz part. This localization could influence the electron density and reactivity of this portion of the molecule. As with other systems, the 
presence of the solvent does not significantly alter the localization of HOMO-1, suggesting that the lower energy states are relatively unaffected by the 
solvent environment.

The LUMO+1 orbital is delocalized across both the Cz and PS units, while the LUMO+2 is largely localized on the PS unit in both environments. This indicates 
that the excited-state dynamics will involve both units and are expected to behave similarly in vacuum and solvent. The interaction between the Cz and PS 
units, suggested by this delocalization, may affect the photophysical properties of the molecule. 

In both environments, the HOMO is predominantly delocalized over the Cz unit, with a significant contribution from the PS part. This delocalization suggests 
that there is charge transfer occurring between the Cz and PS units. Conversely, the LUMO is mainly concentrated on the PS unit, highlighting its role as the 
electron-accepting region. The stability of this LUMO localization across environments suggests that the electron-accepting properties of the PS unit are 
robust and not significantly altered by solvation.

The separation between the electron-donating (HOMO) and electron-accepting (LUMO) regions could be advantageous for processes like charge separation and 
recombination. The HOMO-LUMO gap varies slightly between environments, with a difference of $\qty{0.121}{\electronvolt}$ (or approximately 
$\qty{2.790326}{\kilo\calorie\per\mole}$). This indicates that while the molecule’s geometry is relatively stable, its electronic properties are still 
influenced by the surrounding environment. However, the relatively large gap suggests that the molecule remains electronically stable and requires a moderate 
amount of energy to transition from the ground state (HOMO) to the excited state (LUMO). The electronic density maps confirm the charge distribution across the 
molecule, aligning with the orbital analysis. 

\paragraph{Excitation energies}

In both environments, the absorption maxima wavelengths are located within the ultraviolet region. For the \stda method, the absorption energy is 
$\qty{3.859}{\electronvolt}$, corresponding to a wavelength of $\lambda_{abs}\simeq\qty{322.084}{\nano\meter}$. For the \stddft method, the energy is slightly 
lower at $\qty{3.789}{\electronvolt}$, corresponding to a wavelength of $\lambda_{abs}\simeq\qty{328.303}{\nano\meter}$. In both cases, solvation induces an 
anti-red shift, but the absorption wavelengths remain in the ultraviolet range, confirming PSPCz as a UV absorber (see \Cref{tab:ColorPSPCz}). 

Considering the emission and the Stokes shift, PSPCz can be classified as a \emph{violet} emitter for both solvated and unsolvated conditions. Without 
solvation, PSPCz has a fluorescence wavelengths of $\lambda_{PL}\simeq\qtylist{383.544;389.325}{\nano\meter}$ for \stda and \stddft methods, respectively, but 
the predicted color is blue. A similar observation is made with solvation, where the wavelengths slightly shift to 
$\lambda_{PL}\simeq\qty{385.613}{\nano\meter}$, respectively (see \Cref{fig:chromatogram,fig:fluo-PSPCz,tab:ColorPSPCz}). Since there isn't intense visible 
light, could this have value for other situations. Or is this a sign that it isn't good.

\subsubsection{4CzIPN}\label{sec:SI_4CzIPN}

\begin{figure}[!htbp]
\centering
\adjustbox{width=\textwidth}{\includegraphics{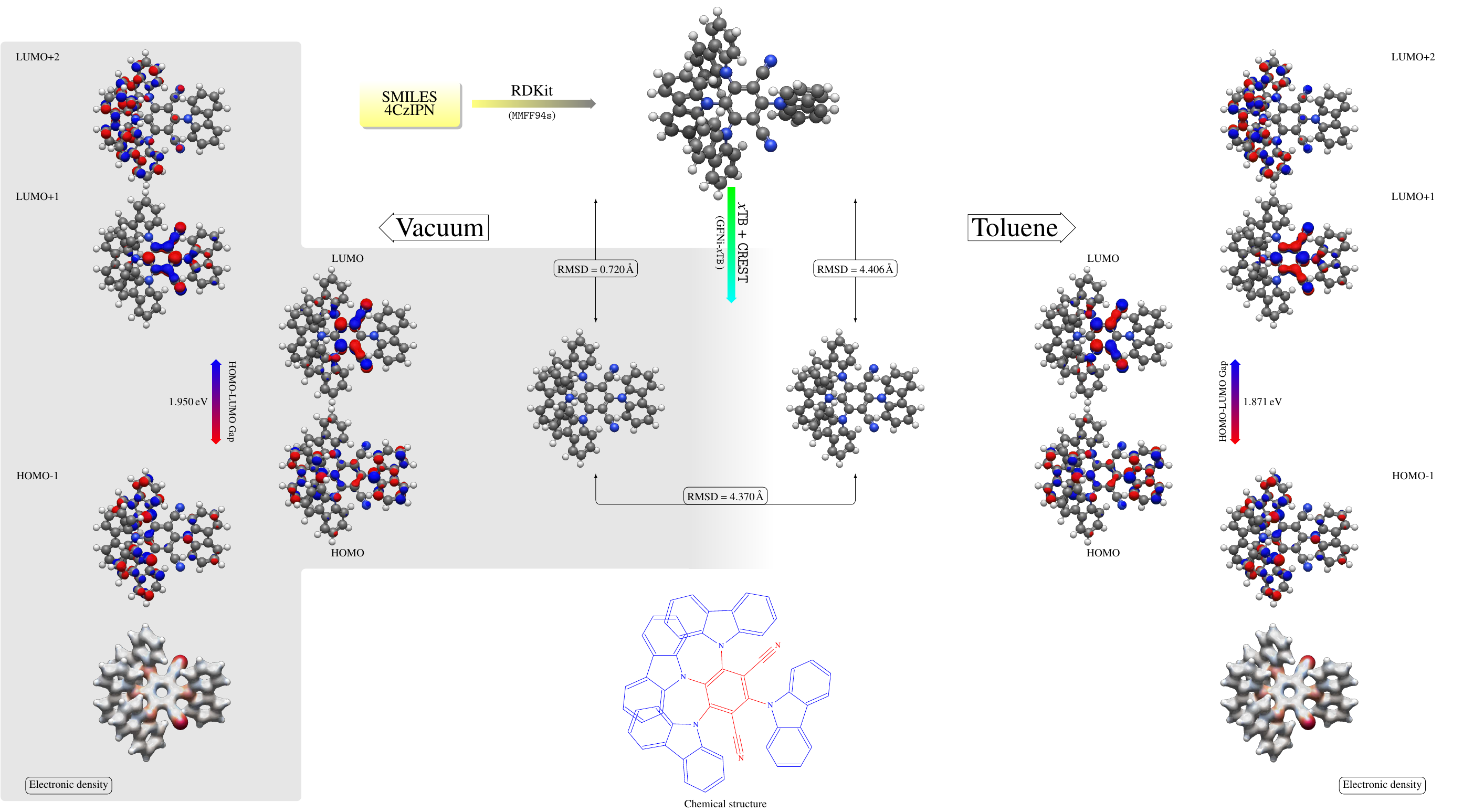}}
\caption{Optimized molecular structure and electronic distributions of 4CzIPN. This figure displays the optimized 3D molecular structures of 4CzIPN in vacuum 
and toluene environments, calculated using the \mmf force field, \xtb, and \crest methods. The figure also displays the corresponding molecular orbitals from 
HOMO-1 to LUMO+2, visualized as isodensity surfaces with blue and red indicating negative and positive regions of the wave function, respectively, and the 
HOMO-LUMO gap values in both environments. The comparison of electronic density maps highlights how solvation affects charge density distribution across the 
molecule and the interaction between IPN and Cz units. Note the delocalization of the HOMO across the four carbazole (Cz) units and the localization of the 
LUMO on the IPN core. Solvation appears to have a subtle effect on these distributions.}
\label{fig:4CzIPN-Sem}
\end{figure}

\Cref{fig:4CzIPN-Sem} displays the chemical structure of 4CzIPN, along with the 3D structure optimized using the \mmf force field, \xtb, and \crest methods. 
Additionally, the figure illustrates the \emph{HOMO-LUMO} gap, the molecular orbitals from HOMO-1 to LUMO+2, and the electronic density distribution.

The RMSD between the structure optimized using the \mmf force field and the one obtained after optimization with \xtb and \crest in the absence of solvent is 
approximately $\qty{0.7}{\angstrom}$. This indicates that the two structures are quite similar. However, when solvation is introduced, the RMSDs increase to 
values exceeding $\qty{4}{\angstrom}$, signifying a significant structural difference compared to both the vacuum-optimized structures (see 
\Cref{fig:4CzIPN-RMSD}). This shows there are some strong effects depending on what calculations you chose. The increase is evidence of a strong effect when 
tolune is present.

While the \xtb and \crest methods exert a non-negligible effect on the molecular geometry derived from the \mmf method, the presence of toluene has an even 
more pronounced impact on the geometry. The most notable changes occur in the dihedral angles between the carbazole units and the central benzene ring, 
suggesting that the solvent promotes a more planar arrangement to maximize pi-stacking interactions.

\begin{figure}[!htbp]
\centering
\begin{minipage}[c]{0.48\textwidth}
\centering
\includegraphics[width=\textwidth]{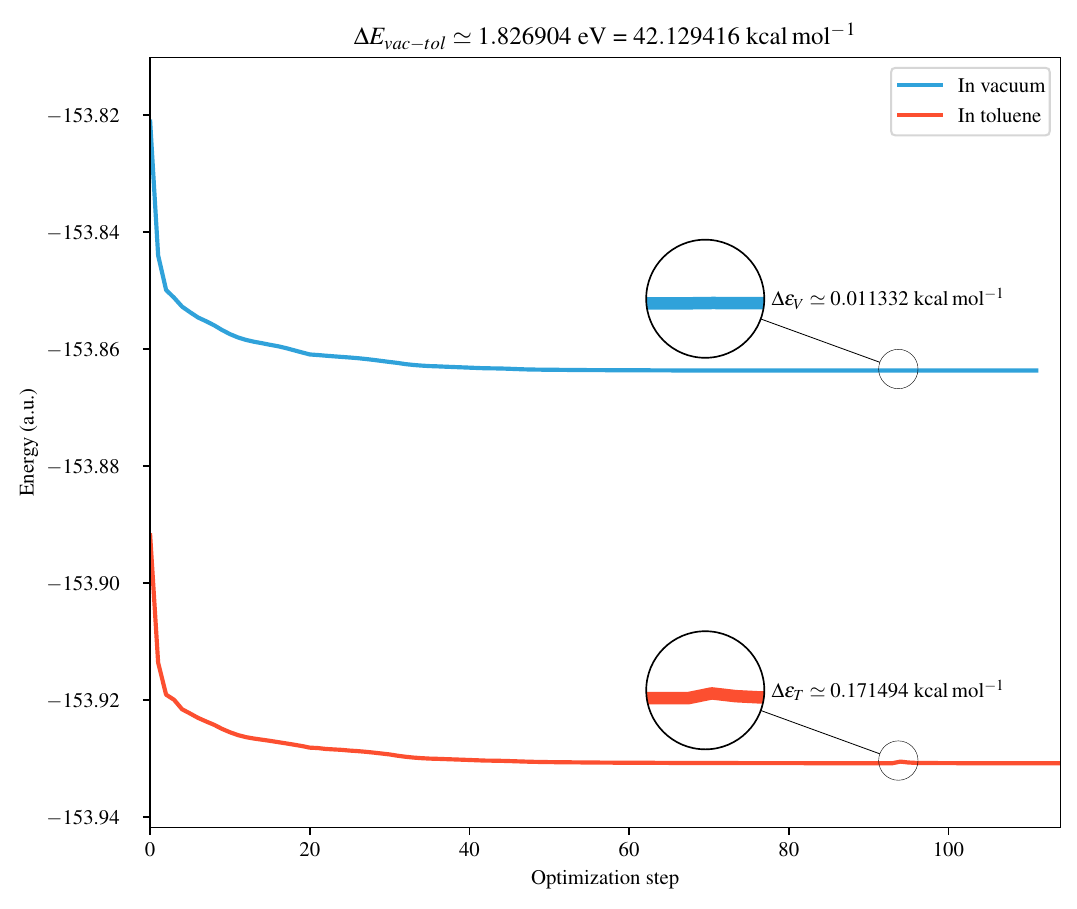}
\end{minipage}%
\begin{minipage}[c]{0.2\textwidth}
\centering
\includegraphics[width=\textwidth]{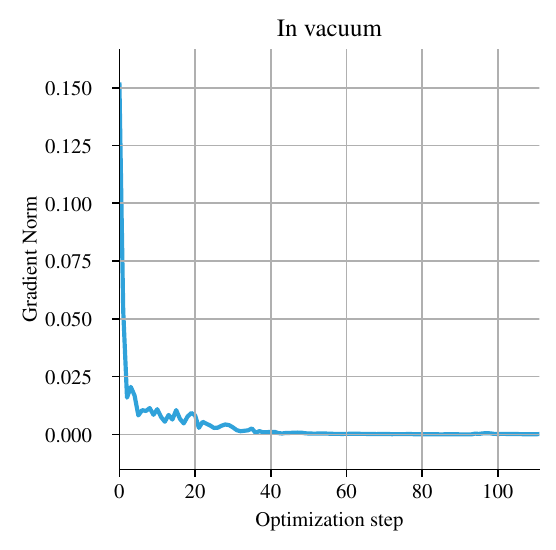}\\
\includegraphics[width=\textwidth]{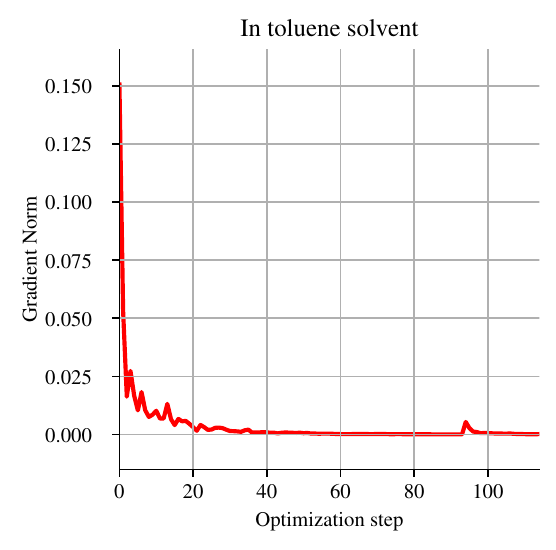}
\end{minipage}
\caption{Energy optimization process for 4CzIPN: vacuum vs. toluene. The left panel shows the potential energy surface explored during the geometry 
optimization process for 4CzIPN in both vacuum and toluene environments. The right panels display the gradients for vacuum (top) and toluene (bottom) 
optimizations. The energy difference ($\Delta E_{vac-tol} = \qty{42.13}{\kilo\calorie\per\mole}$) highlights the significant solvent effect on the molecular 
stability. The smooth, rapid convergence of the toluene optimization suggests that the solvent promotes a more well-defined and stable conformation compared to 
the vacuum environment.}
\label{fig:4CzIPN-opt}
\end{figure}

As seen in \Cref{fig:4CzIPN-opt}, the structure optimized in toluene is approximately $\qty{42.129416}{\kilo\calorie\per\mole}$ lower in energy compared to the 
vacuum-optimized structure, indicating a large impact.

The values of $\Delta\epsilon_V$ and $\Delta\epsilon_T$ are both very low, remaining below $\qty{0.2}{\kilo\calorie\per\mole}$. This suggests that the 
geometries obtained after pre-optimization are accurate enough to be used for further calculations without a significant loss of information.

\paragraph{Molecular orbitals and HOMO-LUMO gap}

In both vacuum and toluene environments, HOMO-1 is delocalized across the molecule but is more concentrated on the Cz units. This implies that the lower energy 
occupied states are predominantly localized on the Cz fragments, which can influence the electron density and reactivity of these regions. As shown in 
\Cref{fig:4CzIPN-Sem}, the solvent does not cause a significant shift in the localization of HOMO-1, indicating that the lower energy states are relatively 
solvent-independent. 

The LUMO+1 orbital is delocalized over both the Cz and IPN units, while LUMO+2 is primarily localized on the Cz fragments in both environments. This suggests 
that excited-state dynamics will involve both units, contributing to the photophysical properties of the molecule. Additionally, the interaction between the Cz 
and IPN units is evident. However, it is important to note that LUMO+1 shows a similar distribution to LUMO, with a higher delocalization over the IPN unit.

In both environments, the HOMO is delocalized across the Cz and IPN units, indicating a strong interaction between these two fragments, which can facilitate 
charge transfer processes. On the other hand, the LUMO is mainly localized on the IPN unit, suggesting that the electron-accepting characteristics are 
primarily associated with the IPN moiety. The consistent LUMO localization across environments suggests that the electron-accepting properties of IPN are stable 
and not significantly altered by solvation. The power is largely focused at one aspect.

The separation of electron-donating (HOMO) and electron-accepting (LUMO) regions is beneficial for charge separation and recombination, which is crucial for 
OLED performance. The HOMO-LUMO gap differs between environments, with a gap difference of $\qty{0.079}{\electronvolt}$ (approximately 
$\qty{1.821783}{\kilo\calorie\per\mole}$). This confirms that while the geometry is affected by the environment, the molecule retains a degree of electronic 
stability. The relatively large gap suggests that the molecule requires a moderate amount of energy to transition from the ground state (HOMO) to the excited 
state (LUMO). The electronic density maps corroborate the charge distribution, supporting the orbital analysis.

\paragraph{Excitation energies}

In vacuum, the maximum absorption wavelength occurs at $\lambda_{abs}\simeq\qty{375.928}{\nano\meter}$ (corresponding to $\qty{3.298}{\electronvolt}$), placing 
the absorption in the near UV region for the \stda method. For the \stddft method, this maximum shifts slightly towards the visible region, with an energy of 
$\qty{3.268}{\electronvolt}$ and a wavelength of $\lambda_{abs}\simeq\qty{379.335}{\nano\meter}$. However, the predicted absorption color in both cases is the 
\emph{blue} (see \Cref{tab:Color4CzIPN}). Hence, in vacuum, 4CzIPN behaves as a blue absorber.

In the presence of toluene, the absorption wavelengths experience a red shift, moving into the visible region (see \Cref{tab:Color4CzIPN}). The maximum 
absorption energies are $\qty{3.243}{\electronvolt}$ ($\qty{382.367}{\nano\meter}$) for the \stda method and $\qty{3.214}{\electronvolt}$ 
($\qty{385.715}{\nano\meter}$) for the \stddft method. In this case, 4CzIPN can be classified as a \emph{violet} absorber, but the predicted color remains 
blue. 

Taking into account the emission and the Stokes shift, 4CzIPN is a \emph{green} emitter in both solvated and unsolvated environments. Without solvation, the 
maxima are $\lambda_{PL}\simeq\qty{499.967}{\nano\meter}$ for the \stda method and $\lambda_{PL}\simeq\qty{504.236}{\nano\meter}$ for the \stddft method. In 
the presence of solvation, the maxima shift slightly to $\lambda_{PL}\simeq\qtylist{515.994;520.106}{\nano\meter}$ for the \stda and \stddft methods, 
respectively (see \Cref{fig:chromatogram,fig:fluo-4CzIPN,tab:Color4CzIPN}).

\subsubsection{Px2BP}

\begin{figure}[!htbp]
\centering
\adjustbox{width=\textwidth}{\includegraphics{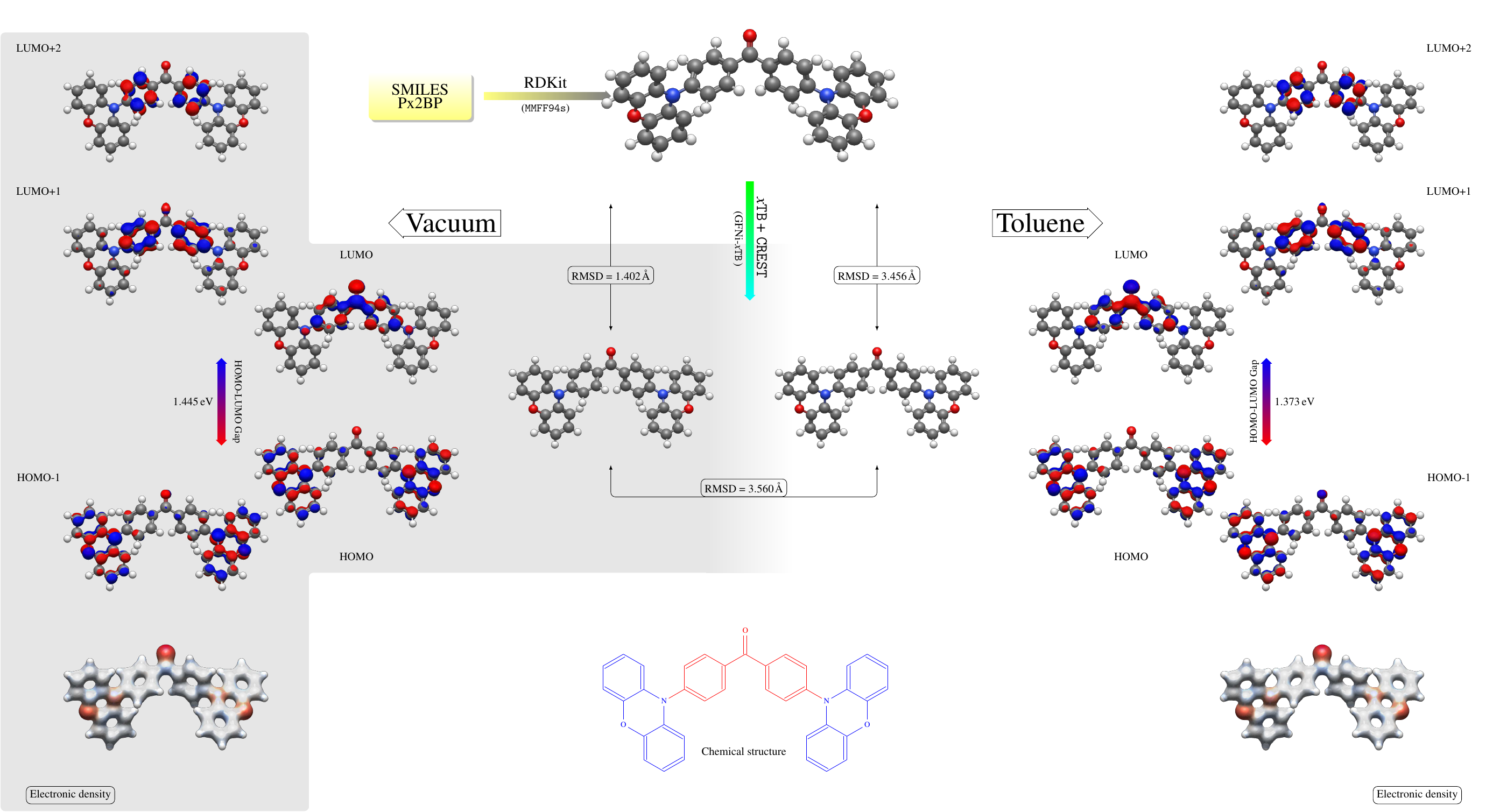}}
\caption{Optimized 3D molecular structures and electronic distributions of Px2BP. This figure displays the optimized 3D molecular structures of Px2BP in vacuum 
and toluene environments, calculated using the \mmf force field, \xtb, and \crest methods. The figure also displays the corresponding molecular orbitals from 
HOMO-1 to LUMO+2, visualized as isodensity surfaces with blue and red indicating negative and positive regions of the wave function, respectively, and the 
HOMO-LUMO gap values in both environments. The comparison of electronic density maps highlights how solvation affects charge density distribution across the 
molecule and the interaction between PXZ and BP units. Does this data suggest to what extent is there charge transfer from PXZ and BP.}
\label{fig:Px2BP-Sem}
\end{figure}

\Cref{fig:Px2BP-Sem} illustrates the chemical structure of Px2BP, alongside its 3D optimized geometry obtained through the \mmf force field, \xtb, and \crest 
methods. Additionally, the figure provides information on the \textit{HOMO-LUMO} gap, molecular orbitals from HOMO-1 to LUMO+2, and electronic density 
distribution in vacuum and solvated conditions.

The RMSD between the vacuum-optimized structure using \mmf and those obtained with \xtb and \crest methods is approximately $\qty{1.4}{\angstrom}$. This 
deviation indicates a measurable alteration in molecular geometry. However, the introduction of solvation leads to a significant increase in RMSD values, 
exceeding $\qty{3}{\angstrom}$, as shown in \Cref{fig:Px2BP-RMSD}. This highlights a considerable impact of the solvent on the molecular structure. 

\begin{figure}[!htbp]
\centering
\begin{minipage}[c]{0.48\textwidth}
\centering
\includegraphics[width=\textwidth]{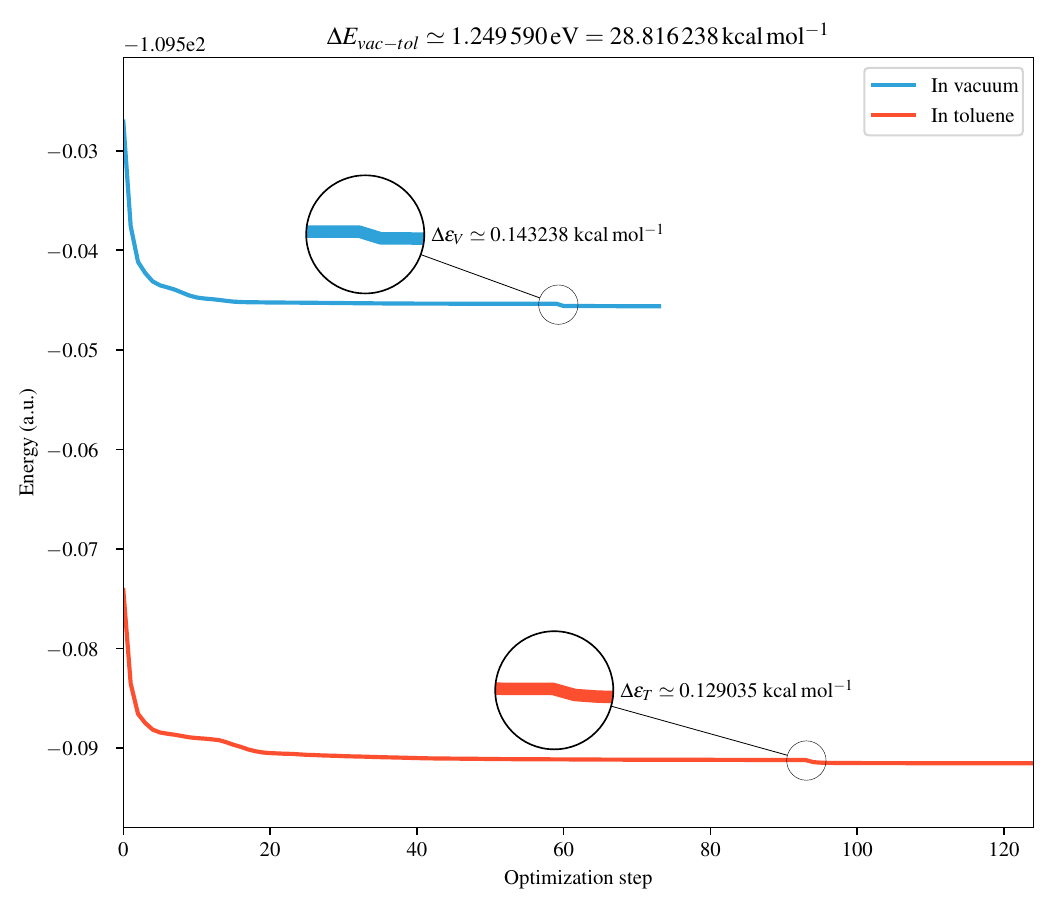}
\end{minipage}%
\begin{minipage}[c]{0.2\textwidth}
\centering
\includegraphics[width=\textwidth]{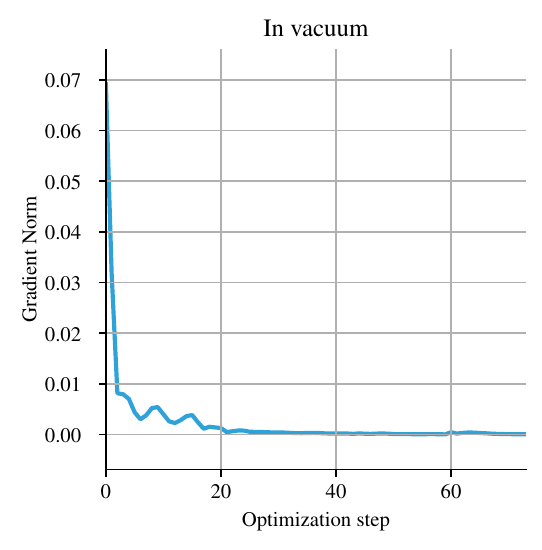}\\
\includegraphics[width=\textwidth]{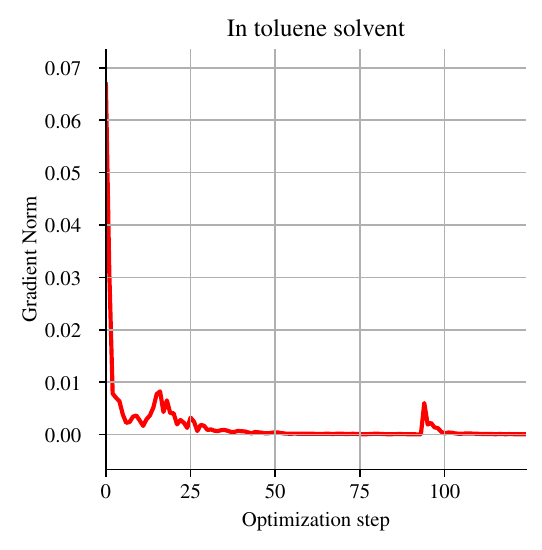}
\end{minipage}
\caption{Energy optimization process for Px2BP: vacuum vs. toluene. The left panel shows the potential energy surface explored during the geometry optimization 
process for Px2BP in both vacuum and toluene environments. The energy difference ($\Delta E_{vac-tol} = \qty{28.82}{\kilo\calorie\per\mole}$) between the 
vacuum- and toluene-optimized structures highlights the solvent's significant impact on molecular stability.}
\label{fig:Px2BP-opt}
\end{figure}

The energy difference between the vacuum- and toluene-optimized structures further demonstrates the solvent's stabilizing effect. The toluene-optimized 
structure is approximately $\qty{28.82}{\kilo\calorie\per\mole}$ lower in energy, as depicted in \Cref{fig:Px2BP-opt}. This confirms that solvation 
significantly enhances the molecular stability. 

The small values of $\Delta\epsilon_V$ and $\Delta\epsilon_T$, both below $\qty{0.15}{\kilo\calorie\per\mole}$, confirm that the pre-optimized geometries are 
sufficiently accurate for subsequent calculations with minimal loss of precision.

\paragraph{Molecular orbitals and HOMO-LUMO gap}

In both vacuum and toluene environments, the molecular orbitals of Px2BP display distinct localization patterns. The HOMO-1 is delocalized across the molecule, 
with a dominant concentration on the PXZ units, signifying that the lower-energy occupied states primarily reside on these fragments. The solvent environment 
does not significantly alter this localization, as shown in \Cref{fig:Px2BP-Sem}, suggesting that these states are relatively environment-independent.

The LUMO+1 and LUMO+2 orbitals exhibit delocalization over the BP component, indicating that excited-state dynamics involve both BP and PXZ units. This 
delocalization enhances the molecule's photophysical properties. The HOMO is strongly delocalized across both PXZ and BP, facilitating charge transfer 
processes, while the LUMO is primarily localized on BP, consistent across environments. This separation of electron-donating and electron-accepting regions is 
advantageous for charge separation and recombination, which are critical for applications such as organic light-emitting diodes (OLEDs). 

The HOMO-LUMO gap changes between environments, with a difference of $\qty{0.072}{\electronvolt}$ (approximately $\qty{1.660359}{\kilo\calorie\per\mole}$). 
This indicates that solvation slightly affects the electronic properties but maintains the molecule's electronic stability. The relatively large gap suggests a 
moderate energy requirement for electronic transitions from the ground state (HOMO) to the excited state (LUMO). 

\paragraph{Excitation energies}

Px2BP demonstrates blue absorption characteristics in both environments, as predicted by absorption calculations. In vacuum, the maximum absorption wavelength 
is $\lambda_{abs}\simeq\qty{425.93}{\nano\meter}$ ($\qty{2.91}{\electronvolt}$) using the \stda method, and $\lambda_{abs}\simeq\qty{430.68}{\nano\meter}$ 
($\qty{2.88}{\electronvolt}$) with \stddft. Solvation induces a minor anti-red shift, with the maximum absorption wavelength shifting to 
$\lambda_{abs}\simeq\qty{415.89}{\nano\meter}$ ($\qty{2.98}{\electronvolt}$) for \stda and $\lambda_{abs}\simeq\qty{419.98}{\nano\meter}$ 
($\qty{2.95}{\electronvolt}$) for \stddft. These results confirm Px2BP as a blue absorber in both environments. The trend for the \stda and \stddft is that the 
wavelengths are very close. This is a trend for Px2BP.

Accounting for the emission with the Stokes shift, Px2BP is a blue emitter at the lower bound of the emission spectrum. However, at the upper bound, the 
fluorescence maxima shift into the yellow region, with wavelengths of $\lambda_{PL}\simeq\qtylist{575.93;580.68}{\nano\meter}$ for \stda and \stddft, 
respectively, in vacuum. In toluene solvent, these maxima slightly shift to $\lambda_{PL}\simeq\qtylist{565.89;569.98}{\nano\meter}$ (see 
\Cref{tab:ColorPx2BP}). Therefore, we can classify Px2BP as an interesting molecule. What happens with it makes it unique and may be helpful.

\subsubsection{CzS2}\label{sec:SI_CzS2}

\begin{figure}[!htbp]
\centering
\adjustbox{width=\textwidth}{\includegraphics{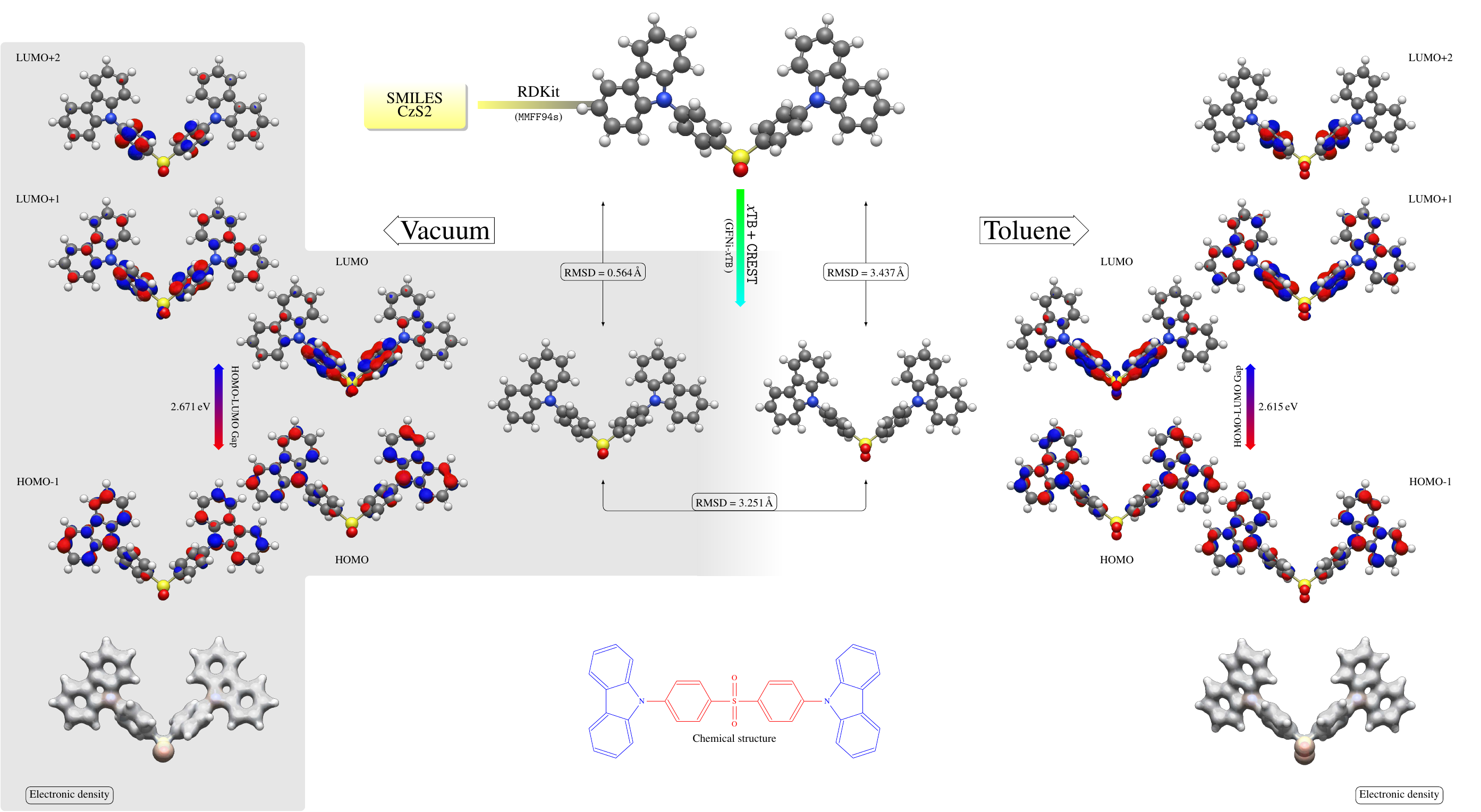}}
\caption{Optimized 3D molecular structures and electronic distributions of CzS2. This figure displays the optimized 3D molecular structures of CzS2 in vacuum 
and toluene environments, calculated using the \mmf force field, \xtb, and \crest methods. The figure also displays the corresponding molecular orbitals from 
HOMO-1 to LUMO+2, along with the HOMO-LUMO gap values in both environments. The electronic density maps illustrate the impact of solvation on charge density 
distribution across the molecule and the interaction between Cz and PS units.}
\label{fig:CzS2-Sem}
\end{figure}

\Cref{fig:CzS2-Sem} provides a detailed visualization of the CzS2 molecule, including its 3D geometry optimized using the \mmf force field, \xtb, and \crest 
methods. Additionally, it presents the calculated \textit{HOMO-LUMO} gap, molecular orbitals (HOMO-1 to LUMO+2), and electronic density maps under both vacuum 
and solvated conditions. These results highlight the interplay between molecular structure and electronic properties in different environments.

The RMSD between the vacuum-optimized CzS2 geometry obtained using the \mmf force field and those optimized with \xtb and \crest is approximately 
$\qty{0.5}{\angstrom}$. This indicates a slight structural deviation between the methods in the absence of solvent. However, the introduction of solvation 
(toluene) leads to a more pronounced structural change, with RMSD values exceeding $\qty{3.2}{\angstrom}$, as shown in \Cref{fig:CzS2-RMSD}. This significant 
difference highlights the solvent's substantial influence on the molecular structure. 

\begin{figure}[!htbp]
\centering
\begin{minipage}[c]{0.48\textwidth}
\centering
\includegraphics[width=\textwidth]{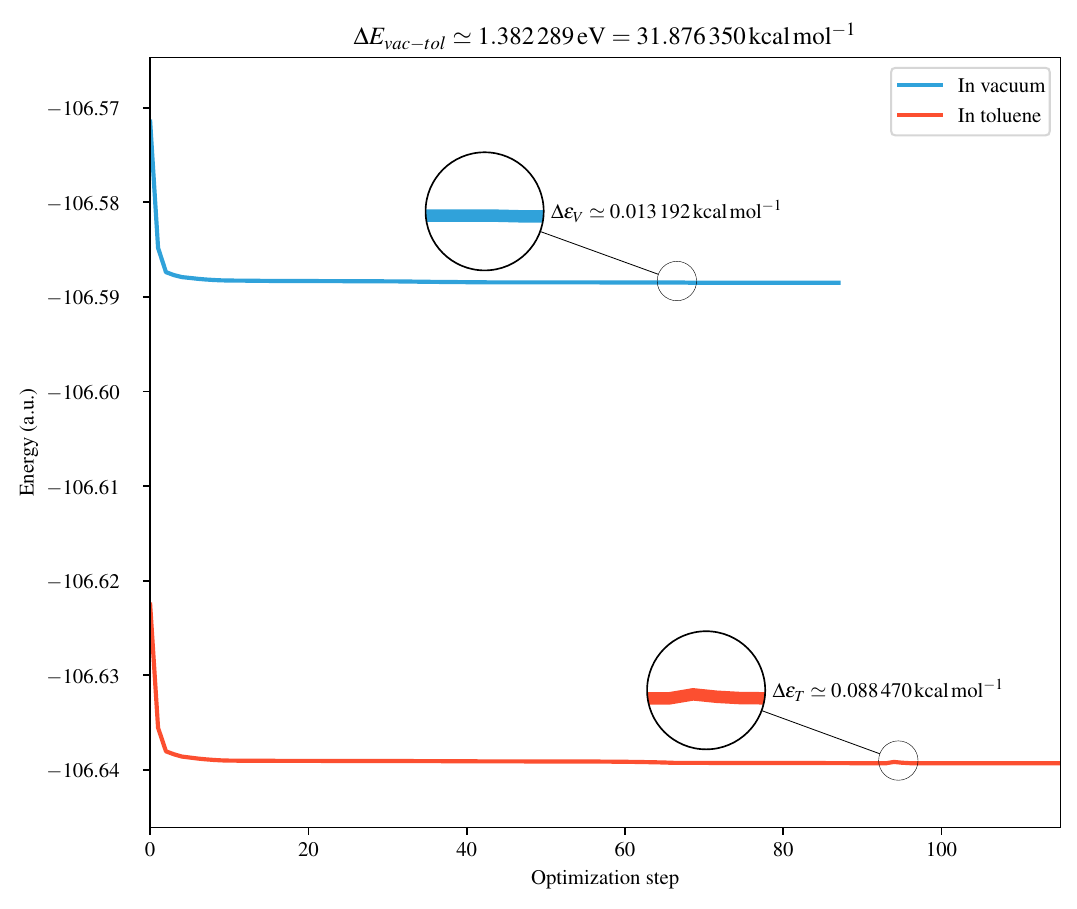}
\end{minipage}%
\begin{minipage}[c]{0.2\textwidth}
\centering
\includegraphics[width=\textwidth]{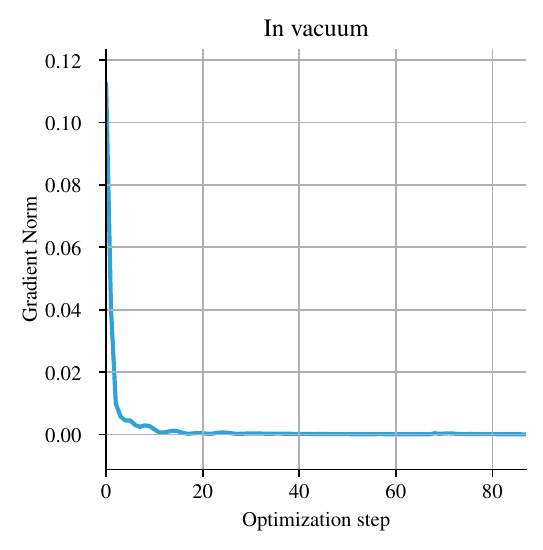}\\
\includegraphics[width=\textwidth]{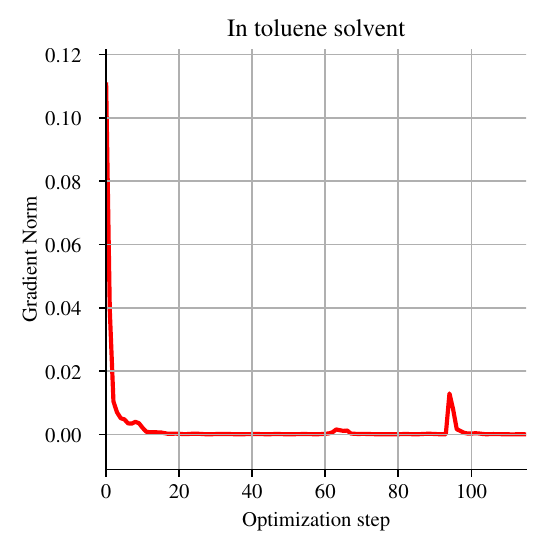}
\end{minipage}
\caption{Energy optimization process for CzS2: vacuum vs. toluene. The left panel shows the potential energy surface explored during the geometry optimization 
process for CzS2 in both vacuum and toluene environments. The right panels display the gradients for vacuum (top) and toluene (bottom) optimizations. The 
energy difference ($\Delta E_{vac-tol} = \qty{31.88}{\kilo\calorie\per\mole}$) highlights the significant solvent effect on molecular stability. Here, a rapid 
initial drop in energy indicates that a few steps were needed before the molecule began to adjust.}
\label{fig:CzS2-opt}
\end{figure}

Despite these structural differences, the energy comparison between the vacuum- and toluene-optimized structures reveals that the solvent-stabilized geometry 
is approximately $\qty{31.88}{\kilo\calorie\per\mole}$ lower in energy (\Cref{fig:CzS2-opt}). This energy difference underscores the stabilizing effect of 
solvation on CzS2, even though the geometries deviate significantly.

The small values of $\Delta\epsilon_V$ and $\Delta\epsilon_T$, both below $\qty{0.09}{\kilo\calorie\per\mole}$, confirm that the pre-optimized geometries are 
sufficiently accurate for subsequent computations, preserving precision.

\paragraph{Molecular orbitals and HOMO-LUMO gap}

In both vacuum and toluene environments, the HOMO-1 orbital of CzS2 is delocalized across the molecule, with a strong localization on the Cz units. This 
suggests that the lower-energy occupied states are primarily associated with the Cz regions, influencing their electron density and reactivity. Solvation does 
not significantly alter this localization, indicating that these states are relatively environment-independent.

The LUMO+1 orbital is delocalized across both the Cz and PS units, while the LUMO+2 is localized predominantly on the DPS unit. This distribution suggests that 
excited-state dynamics involve contributions from both Cz and PS units, resulting in consistent behavior across environments. The interaction between these 
units may also play a role in the photophysical properties of CzS2.

The HOMO is predominantly delocalized over the Cz units, with a minor but notable contribution from the DPS units, indicating charge transfer between the two 
regions. Conversely, the LUMO is primarily localized on the DPS unit, establishing it as the primary electron-accepting region. This separation of the 
electron-donating (Cz) and electron-accepting (PS) regions is advantageous for applications requiring efficient charge separation and recombination.

The HOMO-LUMO gap exhibits a slight variation between environments, with a difference of $\qty{0.056}{\electronvolt}$ (approximately 
$\qty{1.29}{\kilo\calorie\per\mole}$). While this confirms that CzS2's geometry is stable, it also suggests that its electronic properties are subtly 
influenced by solvation. The relatively large gap signifies electronic stability and indicates a moderate energy requirement for electronic transitions from the 
ground state (HOMO) to the excited state (LUMO). Therefore, a lot of force is required.

\paragraph{Excitation energies}

CzS2 exhibits absorption maxima in the ultraviolet (UV) region under both vacuum and solvated conditions. Using the \stda method, the absorption energy is 
calculated as $\qty{3.902}{\electronvolt}$, corresponding to a wavelength of $\lambda_{abs}\simeq\qty{317.7}{\nano\meter}$. The \stddft method predicts a 
slightly lower energy of $\qty{3.839}{\electronvolt}$, corresponding to a wavelength of $\lambda_{abs}\simeq\qty{322.9}{\nano\meter}$. Solvation induces an 
anti-red shift, as observed in similar systems, while the wavelengths remain within the UV range, confirming CzS2 as a UV absorber (\Cref{tab:ColorCzS2}). 
There aren't a lot of changes for each of the environments. This is unlike the DMAC-TRZ where it shifts with the solvent.

When considering the emission with the Stokes shift, CzS2 can be classified as a UV emitter at the lower bound. At the upper bound, in the absence of 
solvation, CzS2 exhibits blue emission characteristics, with fluorescence maxima at $\lambda_{PL}\simeq\qtylist{467.7;472.9}{\nano\meter}$ for the \stda and 
\stddft methods, respectively. With solvation, these maxima shift slightly to $\lambda_{PL}\simeq\qtylist{466.4;471.4}{\nano\meter}$. The wavelength difference 
between the two methods is approximately $\qty{5}{\nano\meter}$, equivalent to $\qty{1.45}{\kilo\calorie\per\mole}$ (\Cref{fig:fluo-CzS2}). Since, with 
solvation, the wavelength remains the same and consistent, then this configuration may not help promote a range of potential colors

\subsubsection{2TCz-DPS}\label{sec:SI_2TCz-DPS}

\begin{figure}[!htbp]
\centering
\adjustbox{width=\textwidth}{\includegraphics{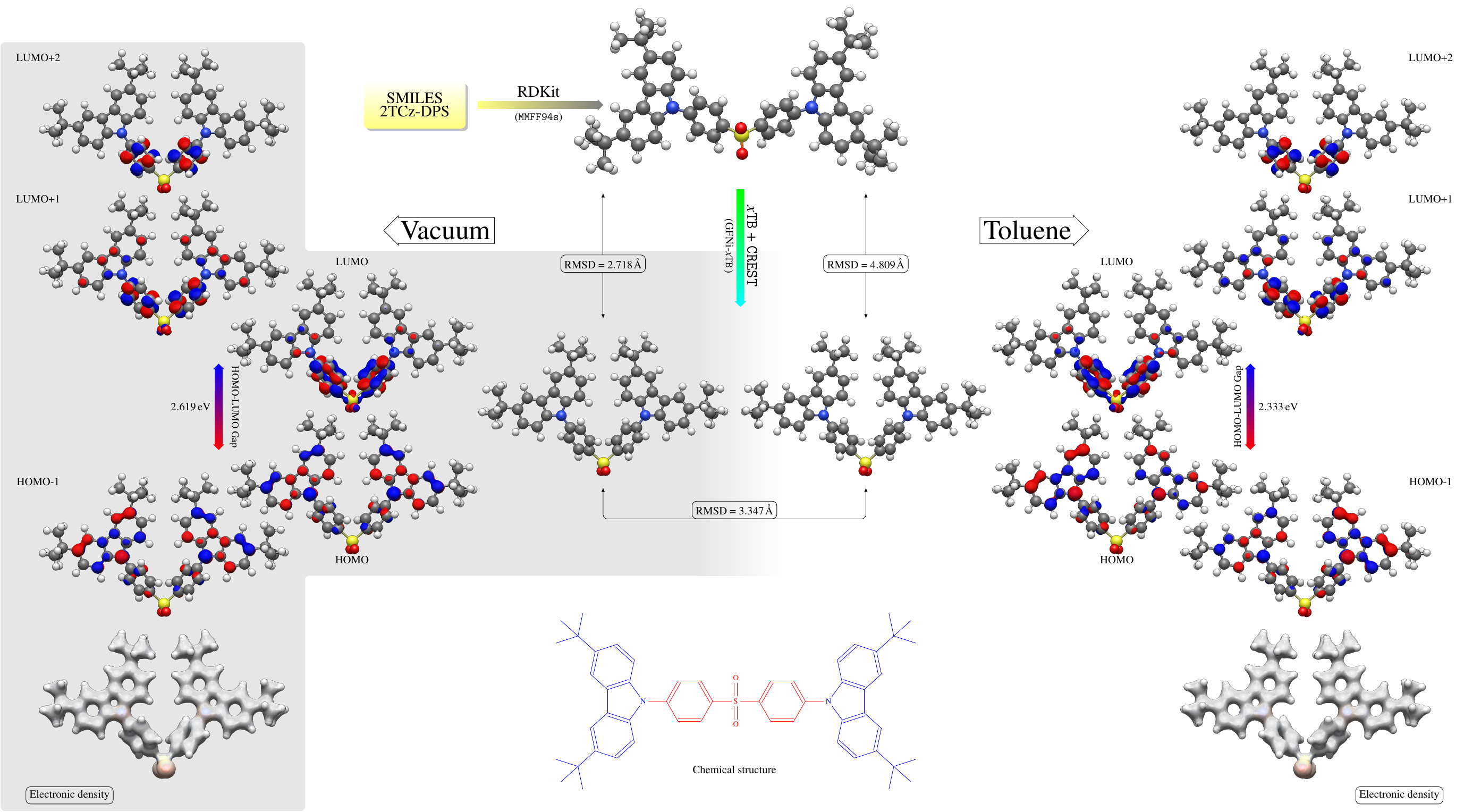}}
\caption{Optimized 3D molecular structures and electronic distributions of 2TCz-DPS. This figure displays the optimized 3D molecular structures of 2TCz-DPS in 
vacuum and toluene environments, calculated using the \mmf force field, \xtb, and \crest methods. The figure also displays the corresponding molecular orbitals 
from HOMO-1 to LUMO+2, along with the HOMO-LUMO gap values in both environments. The comparison of electronic density maps highlights how solvation affects 
charge density distribution across the molecule and the interaction between TCz and PS units.}
\label{fig:2TCz-DPS-Sem}
\end{figure}

\Cref{fig:2TCz-DPS-Sem} presents the chemical structure of 2TCz-DPS and its 3D geometry optimized using the \mmf force field, \xtb, and \crest methods. 
Additionally, it includes the calculated \emph{HOMO-LUMO} gap, molecular orbitals from HOMO-1 to LUMO+2, and electronic density maps under both vacuum and 
solvated conditions. These results demonstrate the influence of solvation on molecular geometry and electronic properties.

The RMSD values between the optimized geometries for 2TCz-DPS are all above $\qty{2}{\angstrom}$, indicating significant structural differences between the 
optimization methods and environments (see \Cref{fig:2TCz-DPS-RMSD}). 

Both the \xtb and \crest methods induce noticeable changes in the molecular geometry compared to the \mmf-optimized structure.

\begin{figure}[!htbp]
\centering
\begin{minipage}[c]{0.5\textwidth}
\centering
\includegraphics[width=\textwidth]{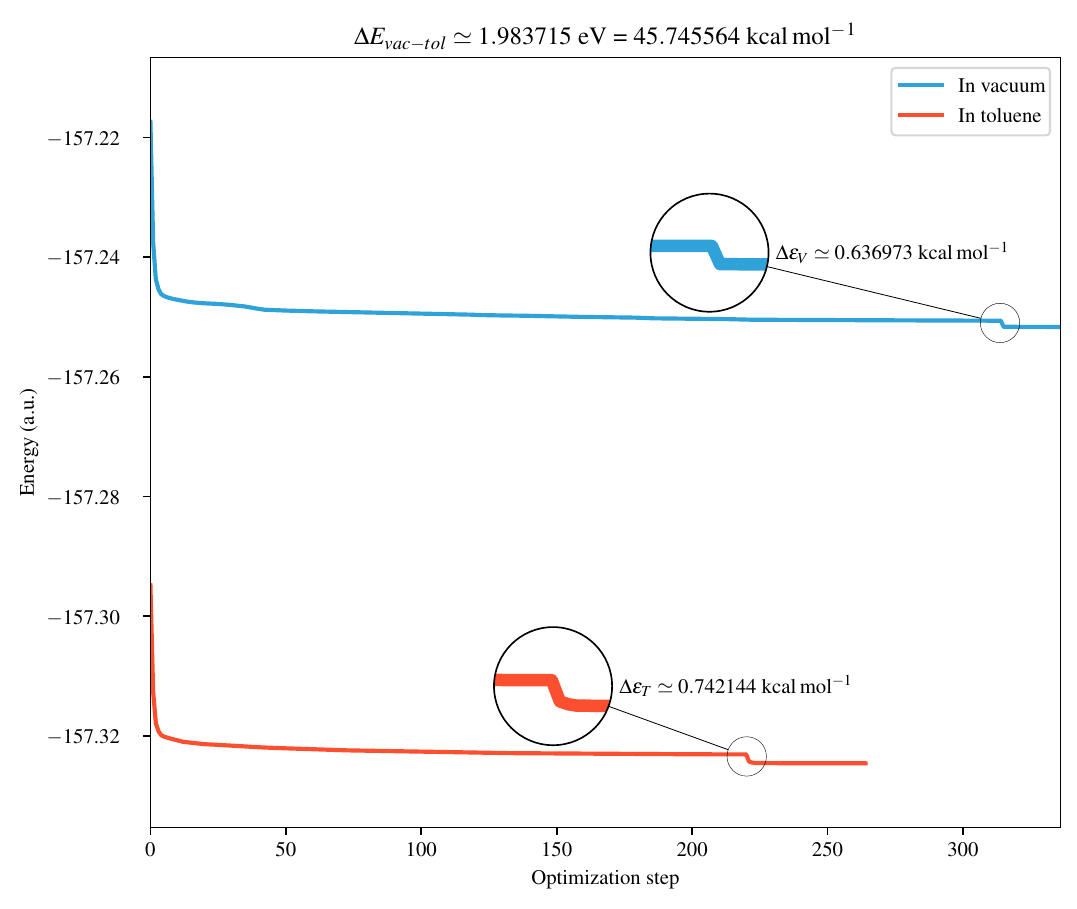}
\end{minipage}%
\begin{minipage}[c]{0.2\textwidth}
\centering
\includegraphics[width=\textwidth]{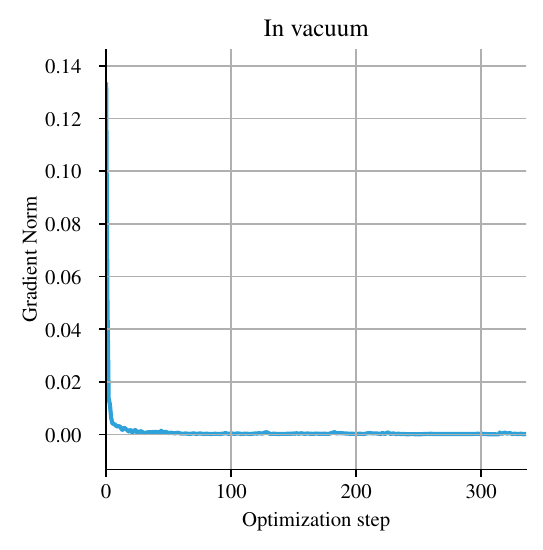}\\
\includegraphics[width=\textwidth]{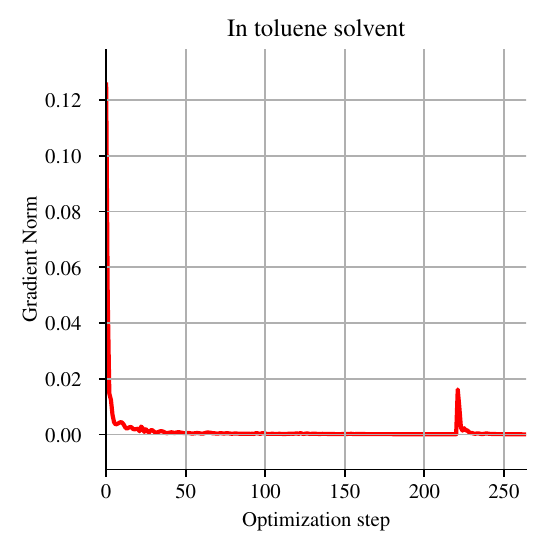}
\end{minipage}
\caption{Energy optimization process for 2TCz-DPS: vacuum vs. toluene. The left panel shows the potential energy surface explored during the geometry 
optimization process for 2TCz-DPS in both vacuum and toluene environments. The energy difference ($\Delta E_{vac-tol} = \qty{45.75}{\kilo\calorie\per\mole}$) 
highlights the significant solvent effect on molecular stability. The gradient plots further illustrate the impact of the solvent on the optimization process.}
\label{fig:2TCz-DPS-opt}
\end{figure}

As shown in \Cref{fig:2TCz-DPS-opt}, the toluene-optimized structure is approximately $\qty{45.75}{\kilo\calorie\per\mole}$ lower in energy than the 
vacuum-optimized structure. This substantial energy difference highlights the stabilizing effect of solvation.

The small values of $\Delta\epsilon_V$ and $\Delta\epsilon_T$, both below $\qty{0.8}{\kilo\calorie\per\mole}$, confirm that the pre-optimized geometries are 
sufficiently accurate for further computational studies without significant loss of precision.

\paragraph{Molecular orbitals and HOMO-LUMO gap}

In both vacuum and toluene environments, the HOMO-1 orbital of 2TCz-DPS is delocalized across the molecule, with strong localization on the TCz units. This 
suggests that the lower-energy occupied states are primarily associated with the TCz regions, which could influence their electron density and reactivity. The 
presence of the solvent does not significantly alter the localization of HOMO-1, indicating that these states remain relatively unaffected by solvation. 

The LUMO+1 orbital is delocalized across both the TCz and PS units, while the LUMO+2 is predominantly localized on the DPS unit in both environments. This 
distribution indicates that excited-state dynamics involve contributions from both TCz and PS units, with similar behavior expected in vacuum and solvated 
conditions. The interaction between these units, as suggested by the delocalization, may play a role in determining the photophysical properties of the 
molecule.

The HOMO is predominantly delocalized over the TCz units, with a minor but notable contribution from the DPS region. This delocalization suggests charge 
transfer between the TCz and PS units. Conversely, the LUMO is largely localized on the PS unit, establishing it as the primary electron-accepting region. The 
stability of this LUMO localization across environments indicates that the electron-accepting properties of the PS unit are robust and not significantly 
affected by solvation.

The separation between the electron-donating (HOMO) and electron-accepting (LUMO) regions could enable efficient charge transfer processes, such as charge 
separation and recombination. The HOMO-LUMO gap varies slightly between environments, with a difference of $\qty{0.079}{\electronvolt}$ (approximately 
$\qty{1.821783}{\kilo\calorie\per\mole}$). This confirms that while the geometry is affected by the environment, the molecule retains a degree of electronic 
stability. However, the relatively large gap suggests that the molecule requires a moderate amount of energy for transitions from the ground state (HOMO) to 
the excited state (LUMO).

\paragraph{Excitation energies}

In both environments, 2TCz-DPS exhibits absorption maxima in the ultraviolet (UV) region. Solvation induces an anti-red shift, similar to what is observed in 
PSPCz and CzS2, while the absorption wavelengths remain in the UV range, confirming 2TCz-DPS as a UV absorber (see 
\Cref{tab:resultGas,tab:resultTol,tab:Color2TCz-DPS} for detailed data). 

Considering the emission with the Stokes shift, 2TCz-DPS is classified as a UV emitter at the lower bound in both solvated and unsolvated conditions. At the 
upper bound, without solvation, 2TCz-DPS shows blue emission characteristics, with fluorescence maxima at 
$\lambda_{PL}\simeq\qtylist{470.07;475.53}{\nano\meter}$ for the \stda and \stddft methods, respectively. Under solvated conditions, the maxima shift slightly 
to $\lambda_{PL}\simeq\qtylist{468.69;473.68}{\nano\meter}$. The wavelength difference between the two methods is approximately $\qty{5}{\nano\meter}$, 
equivalent to $\qty{1.45}{\kilo\calorie\per\mole}$ (\Cref{fig:fluo-2TCz-DPS}). This demonstrates there is a consistent result with the upper and lower bonds.

\subsubsection{TDBA-DI}\label{sec:SI_TDBA-DI}

\begin{figure}[!htbp]
\centering
\adjustbox{width=\textwidth}{\includegraphics{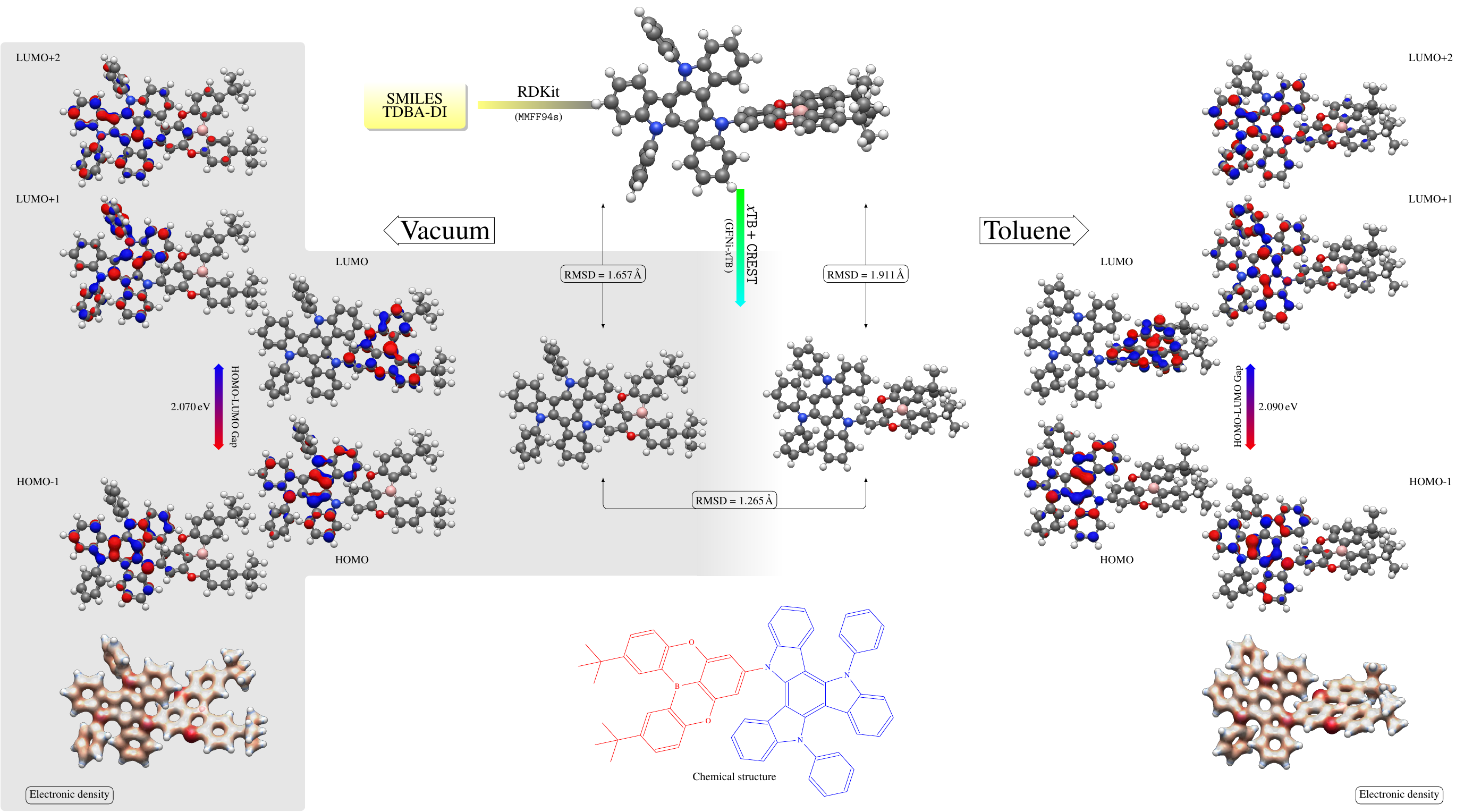}}
\caption{Optimized 3D molecular structures of TDBA-DI in vacuum and toluene environments, calculated using the \mmf force field, \xtb, and \crest. The figure 
also displays the corresponding molecular orbitals from HOMO-1 to LUMO+2, visualized as isodensity surfaces with blue and red indicating negative and positive 
regions of the wave function, respectively, and the HOMO-LUMO gap values in both environments. The comparison of electronic density maps highlights how 
solvation affects charge density distribution across the molecule and the interaction between TD and BA-DI units.}
\label{fig:TDBA-DI-Sem}
\end{figure}

\Cref{fig:TDBA-DI-Sem} shows the chemical structure of TDBA-DI, along with the 3D structure optimized using the \mmf force field, \xtb, and \crest methods. 
Additionally, the figure illustrates the \emph{HOMO-LUMO} gap, the molecular orbitals from HOMO-1 to LUMO+2, and the electronic density distribution.

The RMSD between the structure optimized using the \mmf force field and the one obtained after optimization with \xtb and \crest with the solvent is 
approximately $\qty{1.25}{\angstrom}$, which is the lowest among the three. This indicates that the structure obtained with solvent is quite similar to the one 
obtained using the \mmf force field. However, without solvation, the RMSDs increase to values exceeding $\qty{1.5}{\angstrom}$ but below $\qty{2}{\angstrom}$, 
signifying a significant structural difference compared to \mmf- and toluene-optimized structures (see \Cref{fig:TDBA-DI-RMSD}). 

While the \xtb and \crest methods exert a non-negligible effect on the molecular geometry derived from the \mmf method, the absence of toluene has an even more 
pronounced impact on the geometry. 

\begin{figure}[!htbp]
\centering
\begin{minipage}[c]{0.5\textwidth}
\centering
\includegraphics[width=\textwidth]{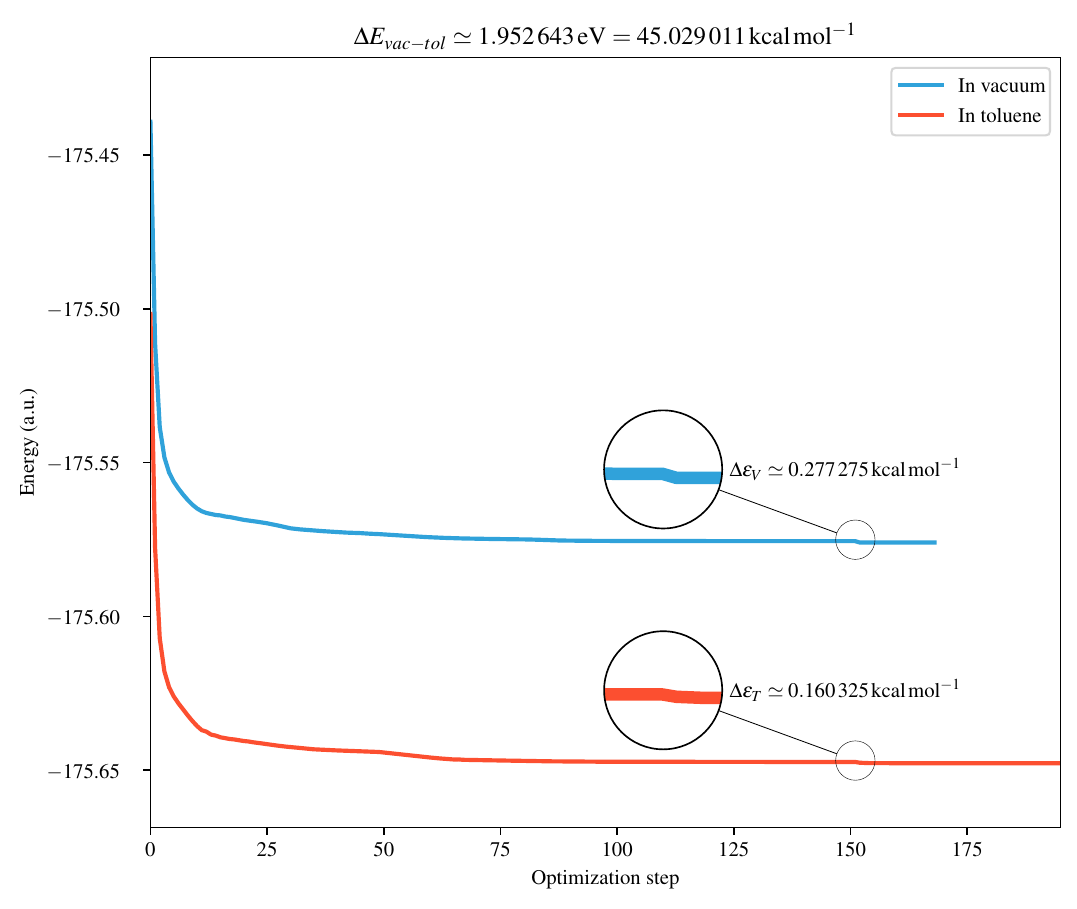}
\end{minipage}%
\begin{minipage}[c]{0.2\textwidth}
\centering
\includegraphics[width=\textwidth]{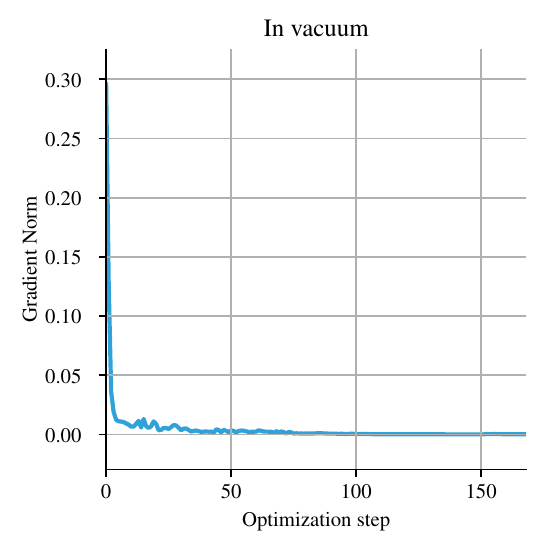}\\
\includegraphics[width=\textwidth]{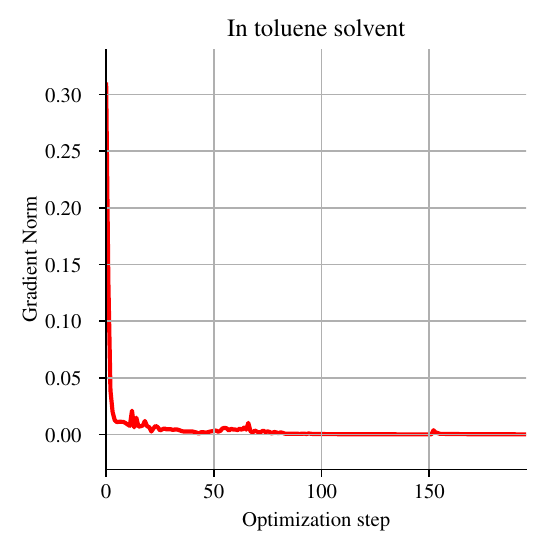}
\end{minipage}
\caption{Energy optimization process for TDBA-DI: vacuum vs. toluene. The left panel shows the potential energy surface explored during the geometry 
optimization process for TDBA-DI in both vacuum and toluene environments. The energy difference ($\Delta E_{vac-tol} = \qty{42.13}{\kilo\calorie\per\mole}$) 
highlights the significant solvent effect on the molecular stability. The gradients of both optimizations emphasize the solvent's impact on the optimization 
process.}
\label{fig:TDBA-DI-opt}
\end{figure}

As seen in \Cref{fig:TDBA-DI-opt}, the structure optimized in toluene is approximately $\qty{45.0029011}{\kilo\calorie\per\mole}$ lower in energy compared to 
the vacuum-optimized structure. This is a very large different, and the best.

The small values of $\Delta\epsilon_V$ and $\Delta\epsilon_T$ (both below $\qty{0.3}{\kilo\calorie\per\mole}$) indicate that the pre-optimized geometries are 
sufficiently accurate for subsequent calculations. 

\paragraph{Molecular orbitals and HOMO-LUMO gap}

In both vacuum and toluene environments, HOMO-1 is delocalized across the molecule but is more concentrated on the BA-DI unit. This implies that the lower 
energy occupied states are predominantly localized on the BA-DI fragment, which can influence the electron density and reactivity of this region. As shown in 
\Cref{fig:TDBA-DI-Sem}, the solvent does not cause a significant shift in the localization of HOMO-1, indicating that the lower energy states are relatively 
solvent-independent. 

The LUMO+1 and LUMO+2 orbitals are delocalized over both the BA-DI and TD units in both environments, but are more localized on the TD fragment. This suggests 
that excited-state dynamics will involve both units, contributing to the photophysical properties of the molecule. Additionally, the interaction between the 
BA-DI and TD units is evident.

In both environments, the HOMO is primarily localized on the BA-DI unit, suggesting that the electron-donating characteristics are primarily associated with the 
BA-DI moiety. On the other hand, the LUMO is mainly localized on the TD unit, suggesting that the electron-accepting characteristics are primarily associated 
with the TD moiety. The consistent HOMO and LUMO localization across environments suggests that the electron-donating properties of BA-DI and electron-accepting 
properties of TD are stable and not significantly altered by solvation. 

The separation of electron-donating (HOMO) and electron-accepting (LUMO) regions is beneficial for charge separation and recombination, which is crucial for 
OLED performance. The HOMO-LUMO gap exhibits a solvent-dependent variation of $\qty{0.020}{\electronvolt}$ (approximately 
$\qty{0.461211}{\kilo\calorie\per\mole}$). This confirms that while the geometry is affected by the environment, the molecule remains electronically stable. The 
relatively large gap suggests that the molecule requires a moderate amount of energy to transition from the ground state (HOMO) to the excited state (LUMO). 

\paragraph{Excitation energies}

In vacuum, the maximum absorption wavelength occurs at $\lambda_{abs}\simeq\qty{347.0120}{\nano\meter}$ (corresponding to $\qty{3.573}{\electronvolt}$), placing 
the absorption in the ultraviolet region for the \stda method. For the \stddft method, this maximum stays in the ultraviolet region, with an energy of 
$\qty{3.502}{\electronvolt}$ and a wavelength of $\lambda_{abs}\simeq\qty{354.0573}{\nano\meter}$. However, the predicted color in both cases is \emph{blue} 
(see \Cref{tab:ColorTDBA-DI}). Hence, in vacuum, TDBA-DI behaves as a blue absorber.

In the presence of toluene, the absorption wavelengths experience a red shift for the \stda method and an anti-red shift for the \stddft method, and stay in the 
ultraviolet region (see \Cref{tab:ColorTDBA-DI}). The maximum absorption energies are $\qty{3.564}{\electronvolt}$ 
($\lambda_{abs}\simeq\qty{347.9288}{\nano\meter}$) for the \stda method and $\qty{3.508}{\electronvolt}$ ($\lambda_{abs}\simeq\qty{353.4456}{\nano\meter}$) for 
the \stddft method. Based on the predicted color, TDBA-DI can be classified as a blue absorber in this case also. 

Taking into account the emission with the Stokes shift, TDBA-DI is a blue emitter when considering the lower bound of the emission spectrum in both solvated and 
unsolvated environments, although the maxima stay in UV region. Without solvation, the maxima are $\lambda_{PL}\simeq\qty{357.0120}{\nano\meter}$ for the \stda 
method and $\lambda_{PL}\simeq\qty{364.0573}{\nano\meter}$ for the \stddft method. In the presence of solvation, the maxima shift slightly to 
$\lambda_{PL}\simeq\qtylist{357.9288;363.4456}{\nano\meter}$ for the \stda and \stddft methods, respectively.

At the upper bound of the spectrum, TDBA-DI exhibits green emission characteristics, confirmed by the predicted color, both with and without solvation. The 
fluorescence maxima at the upper bound are $\lambda_{PL}\simeq\qtylist{497.0120;504.0573}{\nano\meter}$ for the \stda and \stddft methods, respectively. With 
solvation, the maxima shift to $\lambda_{PL}\simeq\qtylist{497.9288;503.4456}{\nano\meter}$ for the \stda and \stddft methods, respectively. We see also here 
that, the difference in wavelengths obtained using the two methods is approximately $\qty{6}{\nano\meter}$, which is equivalent to 
$\qty{1.61423837}{\kilo\calorie\per\mol}$ (see \Cref{fig:chromatogram,fig:fluo-TDBA-DI,tab:ColorTDBA-DI}).

\subsection{Detailed UV-Vis, ECD spectrum, and chromatogram results}\label{subSec-UV}

This section provides detailed data supporting the analysis of UV-Vis absorption, Electronic Circular Dichroism (ECD), and color properties for each molecule.

\begin{figure}[!htbp]
\centering
\leavevmode
\subfloat[In vacuum: UV-Vis 
spectrum]{\includegraphics[width=0.4\textwidth]{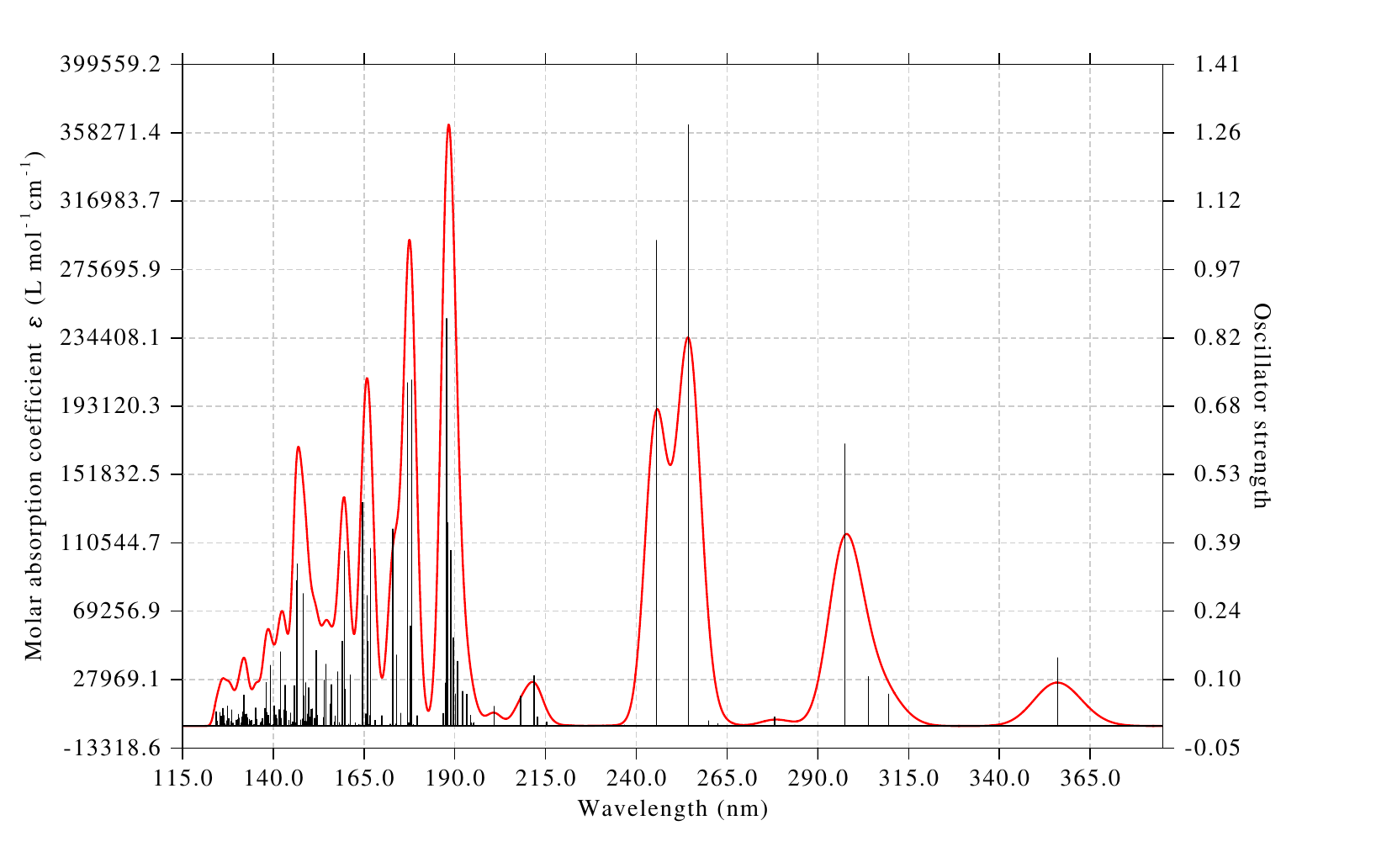}\includegraphics[width=0.4\textwidth]{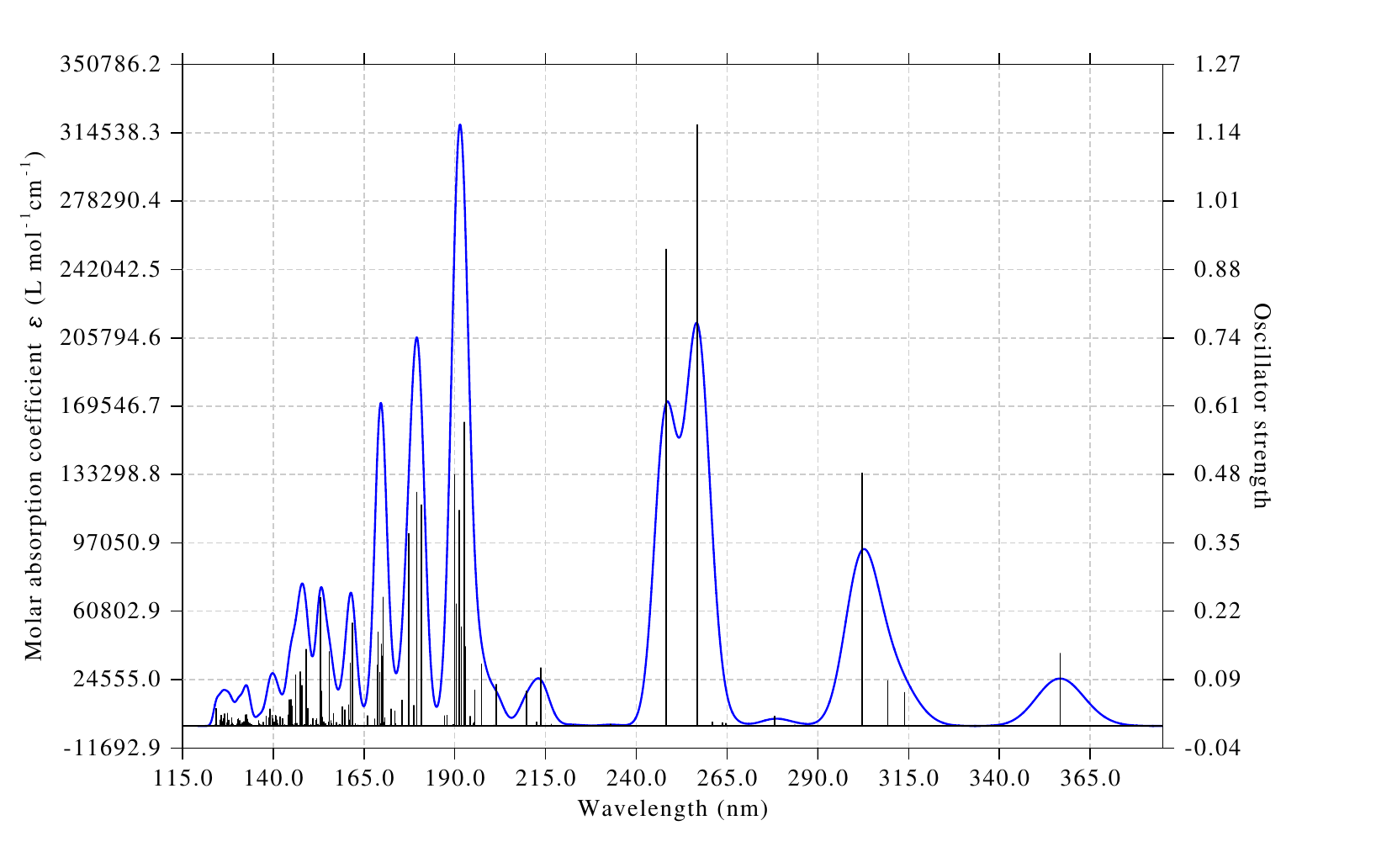}}\
\
\subfloat[In toluene solvent: UV-Vis 
spectrum]{\includegraphics[width=0.4\textwidth]{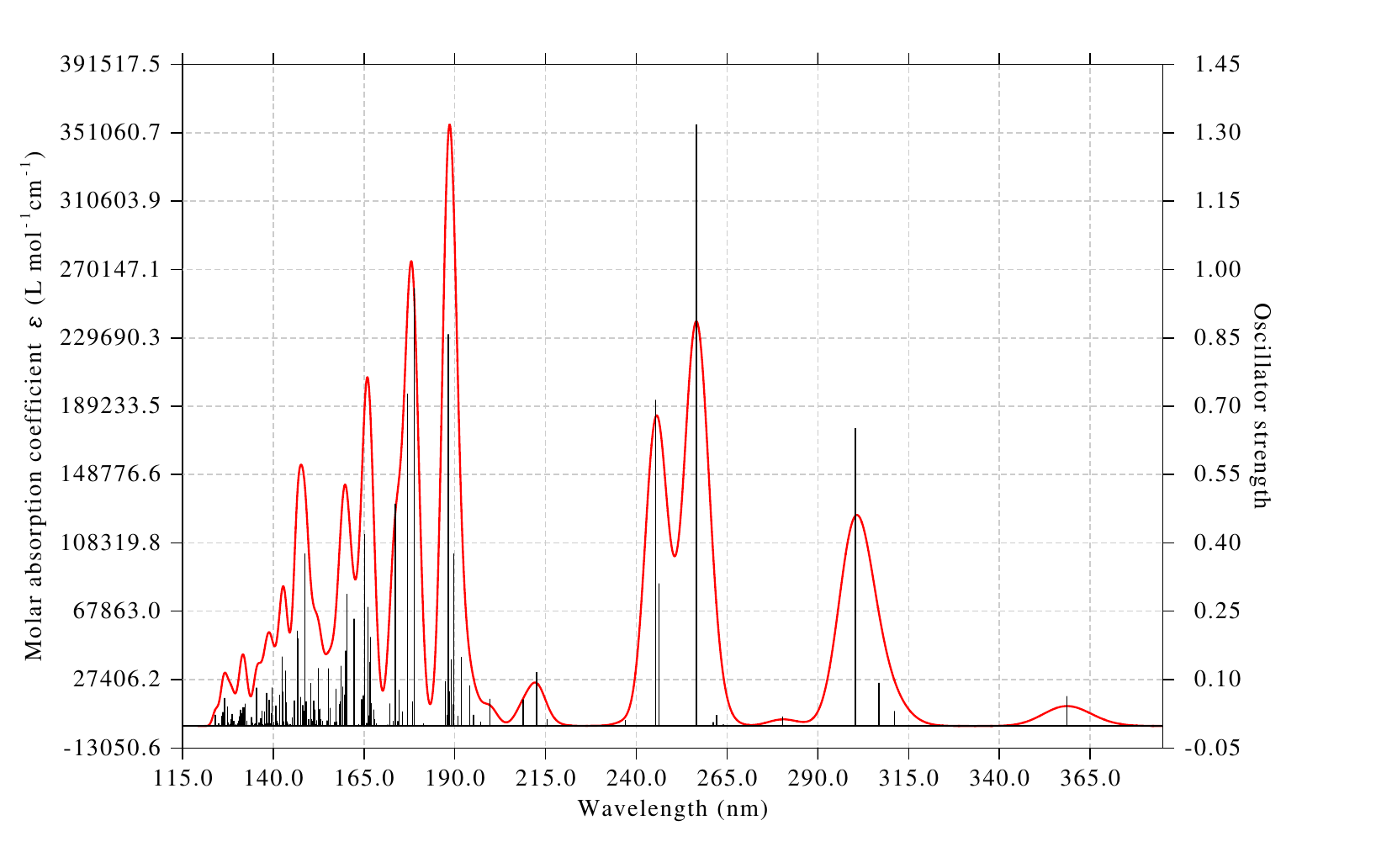}
\includegraphics[width=0.4\textwidth]{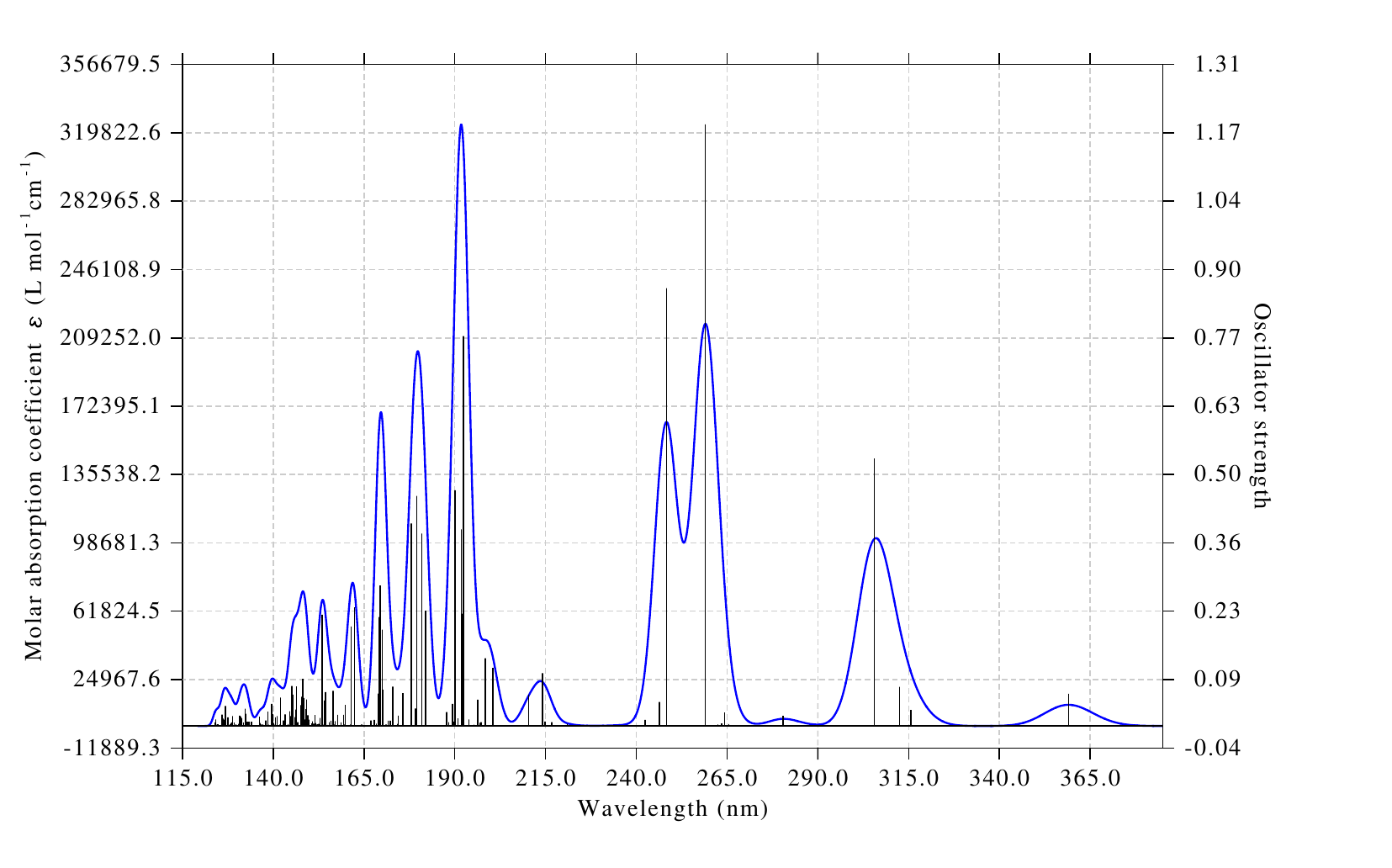}}
\caption{UV-Vis absorption spectra of DMAC-TRZ: solvent and method dependence. Panels show the simulated UV-Vis absorption spectra in vacuum and toluene, 
calculated with \stda (top) and \stddft (bottom) methods. Note the slight redshift in the absorption maximum upon solvation, indicating a stabilization of the 
ground state in the polar solvent.}
\end{figure}

\begin{figure}[!htbp]
\centering
\leavevmode
\subfloat[In vacuum: ECD 
spectrum]{\includegraphics[width=0.4\textwidth]{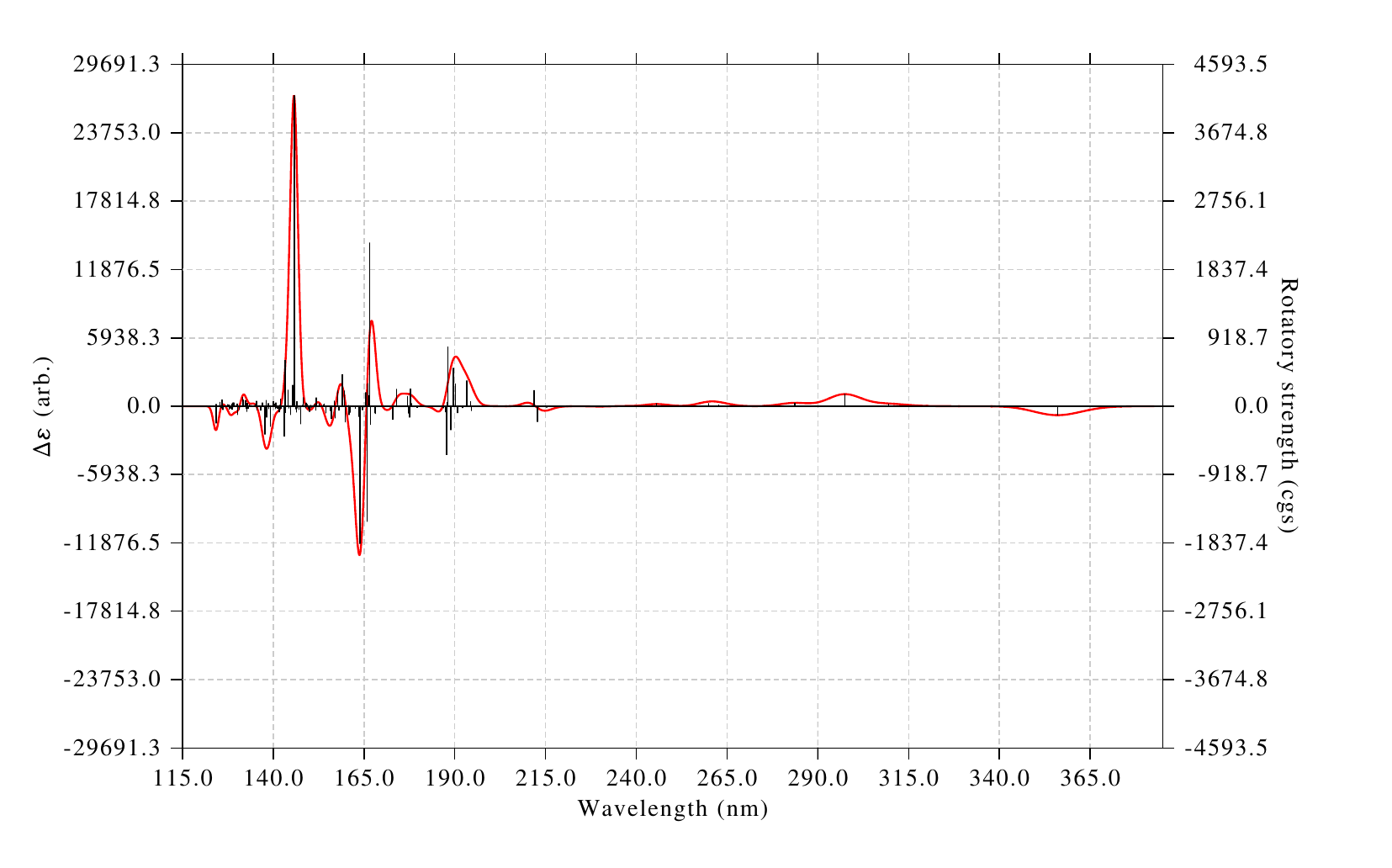}\includegraphics[width=0.4\textwidth]{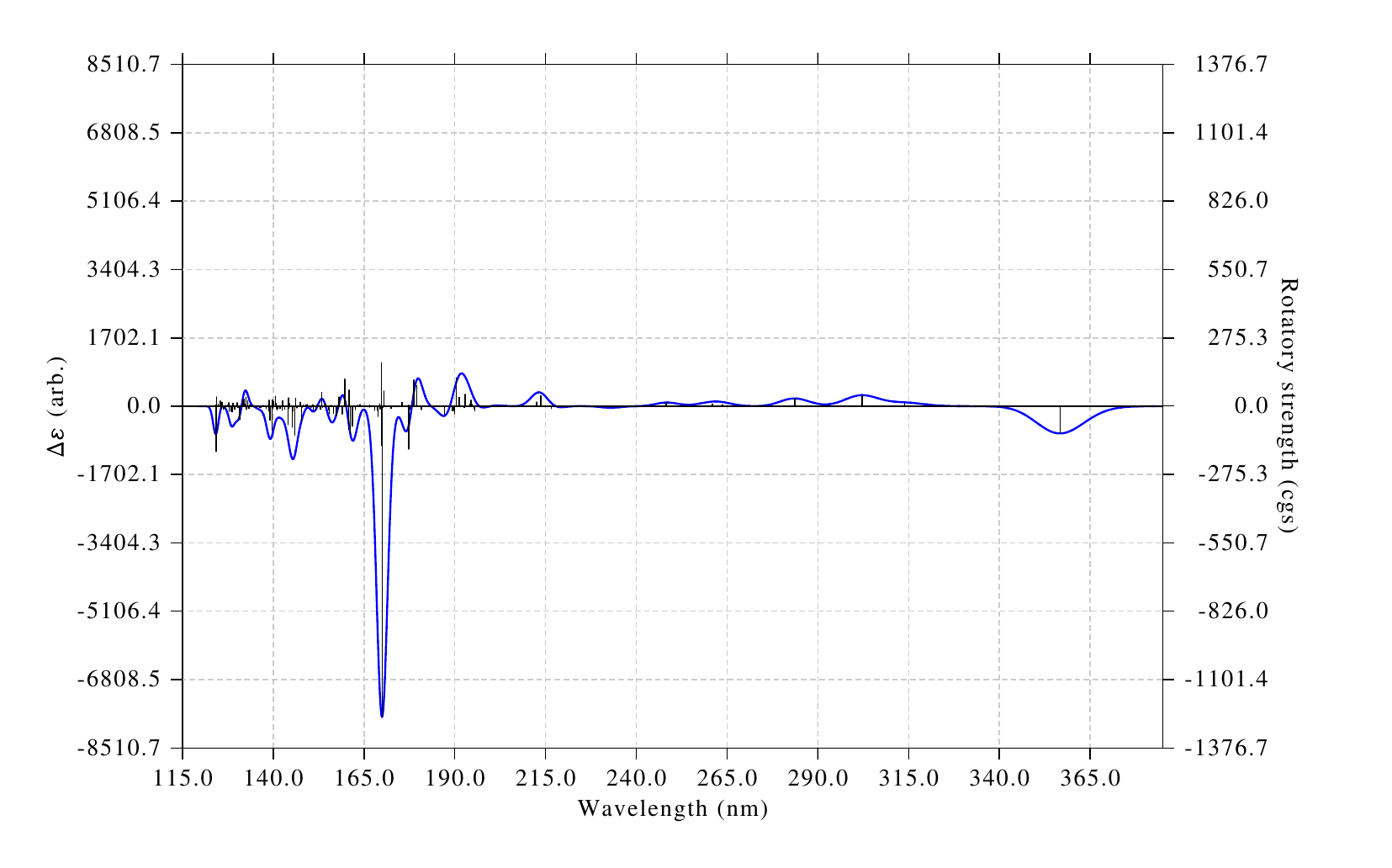}}\\
\subfloat[In toluene solvent: ECD 
spectrum]{\includegraphics[width=0.4\textwidth]{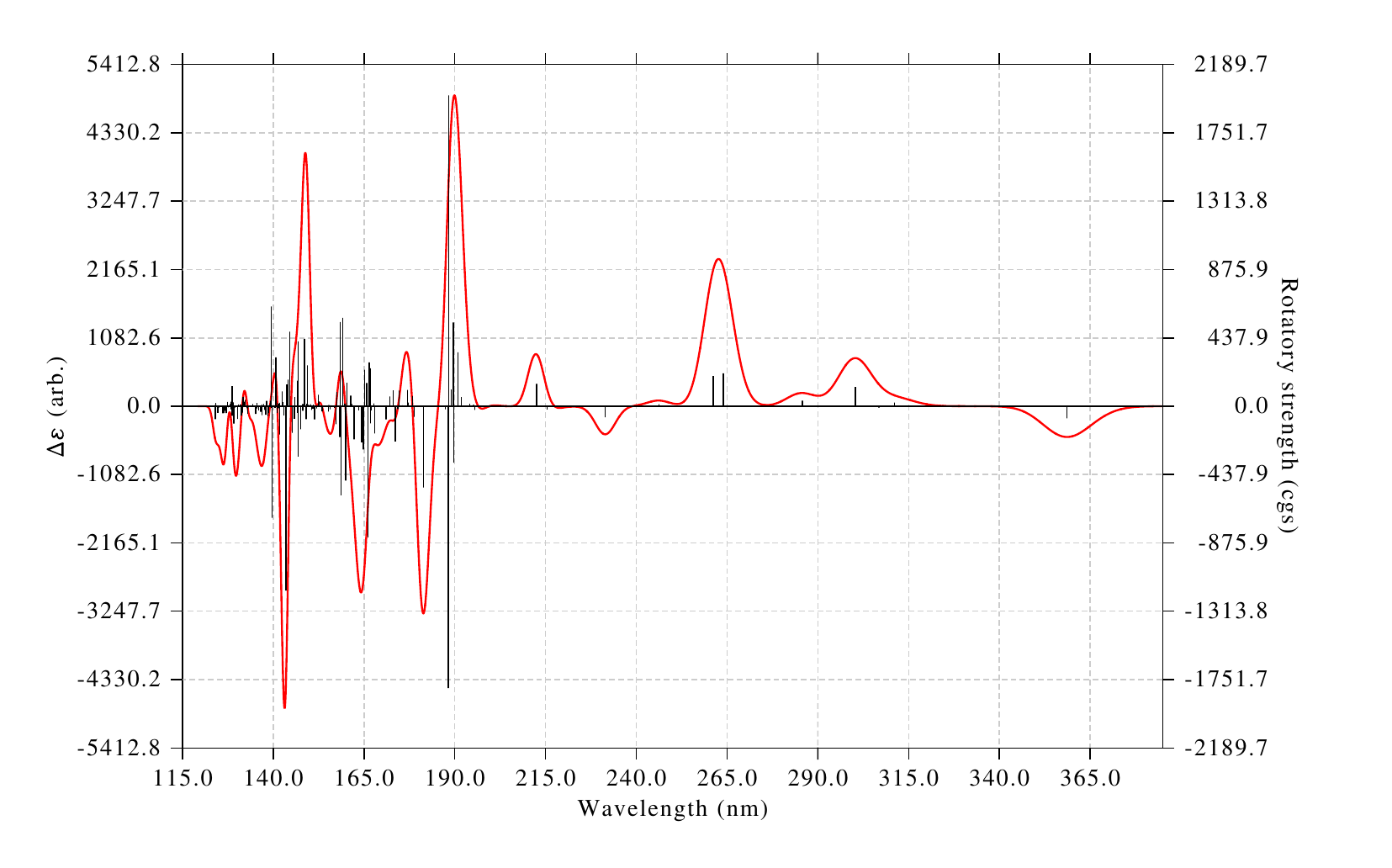}\includegraphics[width=0.4\textwidth]{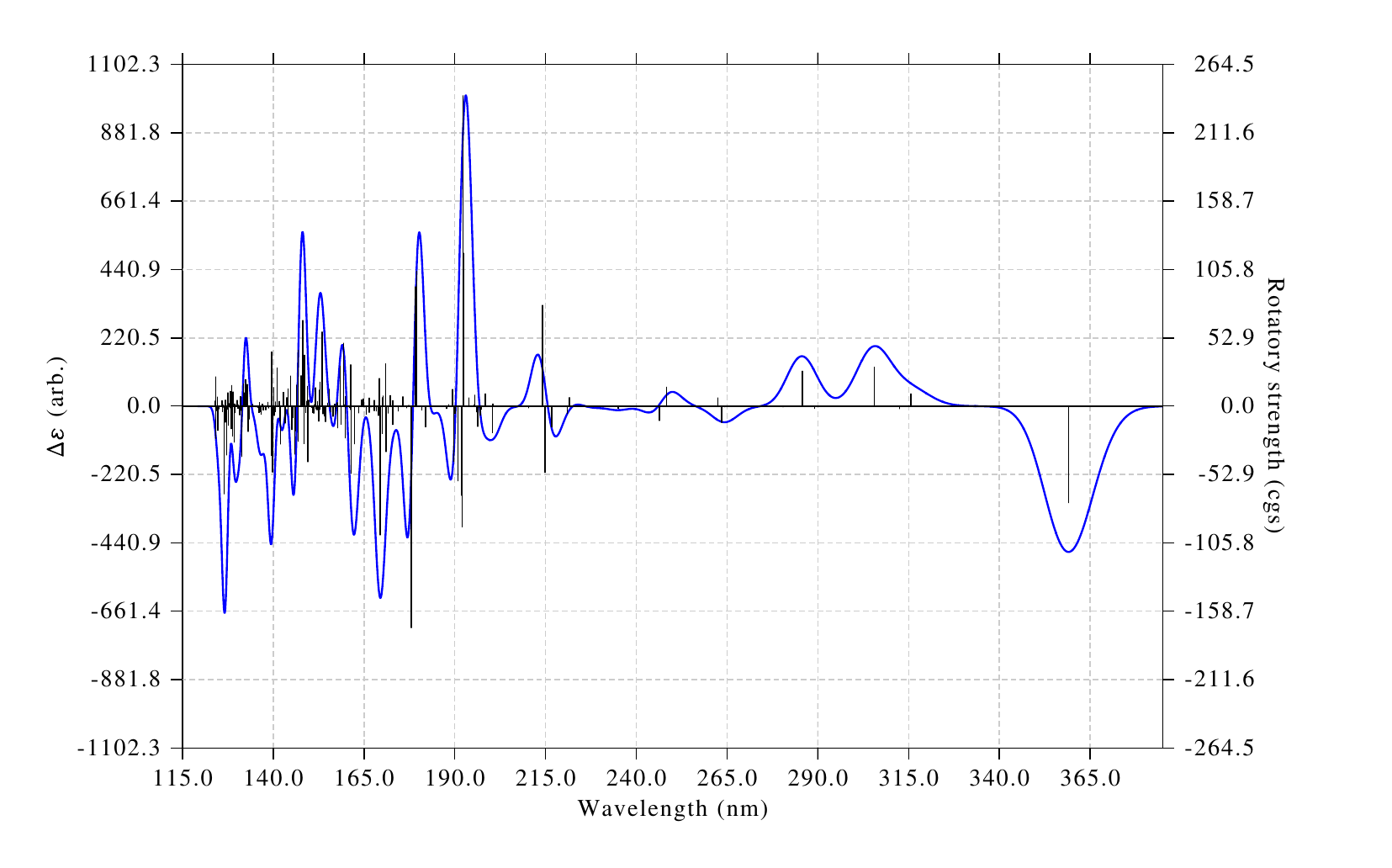}
}
\caption{Electronic Circular Dichroism (ECD) spectra of DMAC-TRZ: solvent and method dependence. Panels show the simulated ECD spectra in vacuum and toluene, 
calculated with \stda (top) and \stddft (bottom) methods. The ECD spectra indicate a lack of significant chiral character in DMAC-TRZ, with relatively weak 
signals observed across the UV-Vis region.}
\end{figure}

\begin{table}[!htbp]
\centering
\caption{Colorimetric properties of DMAC-TRZ: wavelength maxima and CIE coordinates. This table summarizes the key colorimetric properties of DMAC-TRZ, 
including the wavelengths of maximum absorption and emission, CIE 1931 color space coordinates (X, Y, Z and x, y), and approximate sRGB color representation. 
The predicted emission color shifts from greenish-yellow to yellowish-green upon solvation, suggesting a potential for solvatochromic color tuning.}
{\scriptsize
\begin{tabular}{m{1.2cm}m{1.8cm}@{\,}l*{3}{@{\,}c@{\,}}>{\columncolor{white}[0\tabcolsep][0pt]}c}
\toprule
&& Properties & {$\lambda_{max}(\unit{\nano\meter})$} & {$(X,Y,Z)$} & {$(x,y)$} & {(R,G,B)} \\
\midrule
\multirow{4}{=}{In vacuum} & \multirow{2}{=}{\emph{UV-vis} absorption} & \stda &$356.0102$&
$(29.989777,0.898637,140.361012)$&$(0.1751233707,0.0052475336)$&\CcelCo{white}{47,0,255}\\
&& \stddft &$356.6809$&$(33.599252,1.006304,157.281498)$&$(0.1750991105,0.0052442520)$&\CcelCo{white}{47,0,255}\\
&\multirow{2}{=}{Fluorescence}  & \stda
&$566.8904$&$(306071.535496,410870.815619,1184.586455)$&$(0.4262081249,0.5721423249)$&\CcelCo{black}{223,255,0}\\
&& \stddft &$565.3394$&$(284799.849867,399008.601886,1254.693214)$&$(0.4157278814,0.5824406185)$&\CcelCo{black}{197,255,0}\\
\midrule
\multirow{4}{=}{In toluene\\ solvent} & \multirow{2}{=}{\emph{UV-vis} absorption} & \stda
&$358.6689$&$(23.304646,0.696829,109.149006)$&$(0.1750248730,0.0052333967)$&\CcelCo{white}{47,0,255}\\
&& \stddft &$359.0723$&$(24.373392,0.728489,114.168213)$&$(0.1750080819,0.0052307642)$&\CcelCo{white}{47,0,255}\\
&\multirow{2}{=}{Fluorescence}  & \stda
&$579.9337$&$(177983.977016,168226.115675,315.770987)$&$(0.5136239331,0.4854648190)$&\CcelCo{black}{255,205,0}\\
&& \stddft &$572.4365$&$(152547.983094,175637.856871,398.933337)$&$(0.4642576147,0.5345282896)$&\CcelCo{black}{255,255,0}\\
\bottomrule
\end{tabular}}
\label{tab:ColorDMAC-TRZ}
\end{table}


\begin{figure}[!htbp]
\centering
\leavevmode
\subfloat[In vacuum: UV-Vis 
spectrum]{\includegraphics[width=0.4\textwidth]{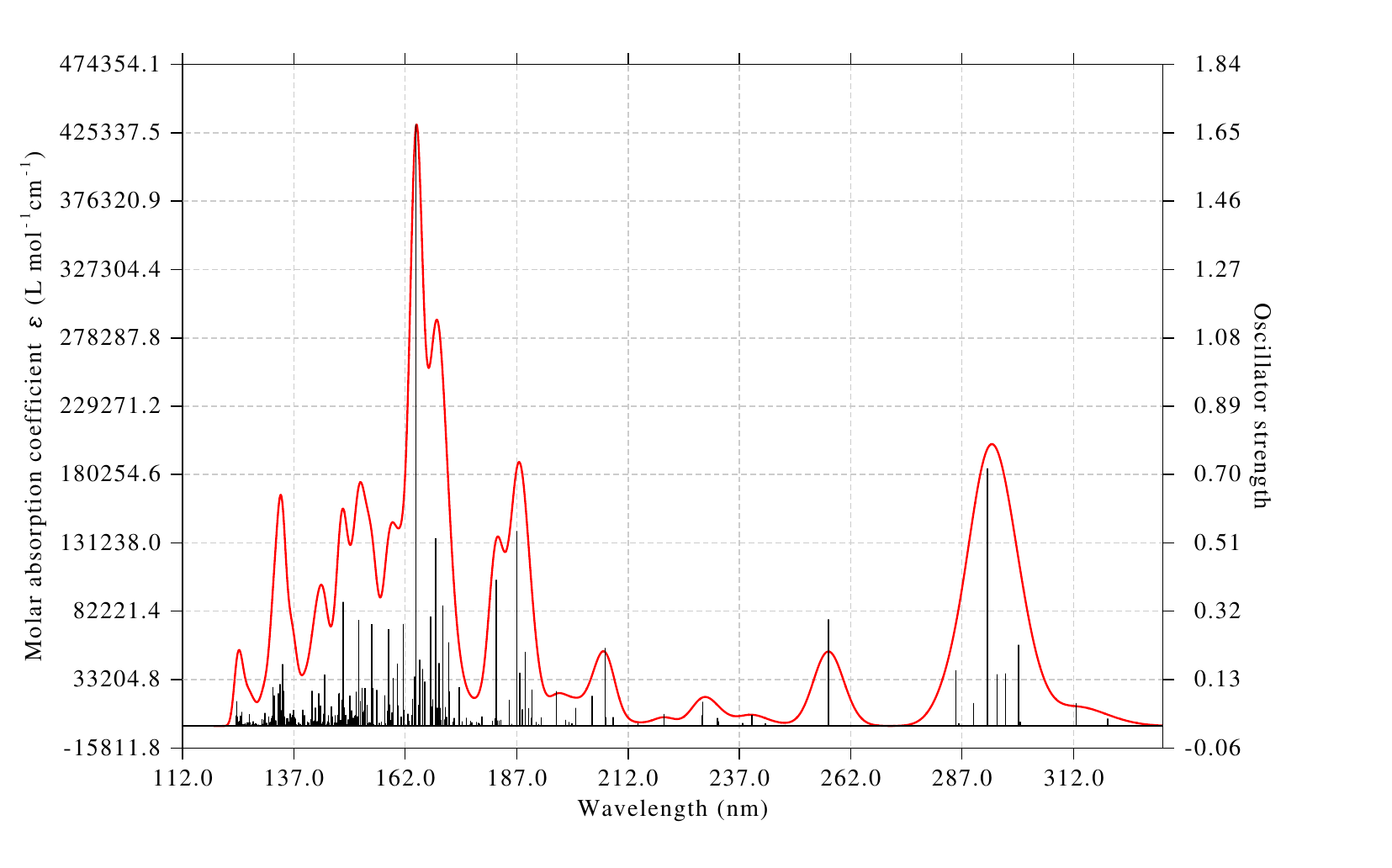}\includegraphics[width=0.4\textwidth]{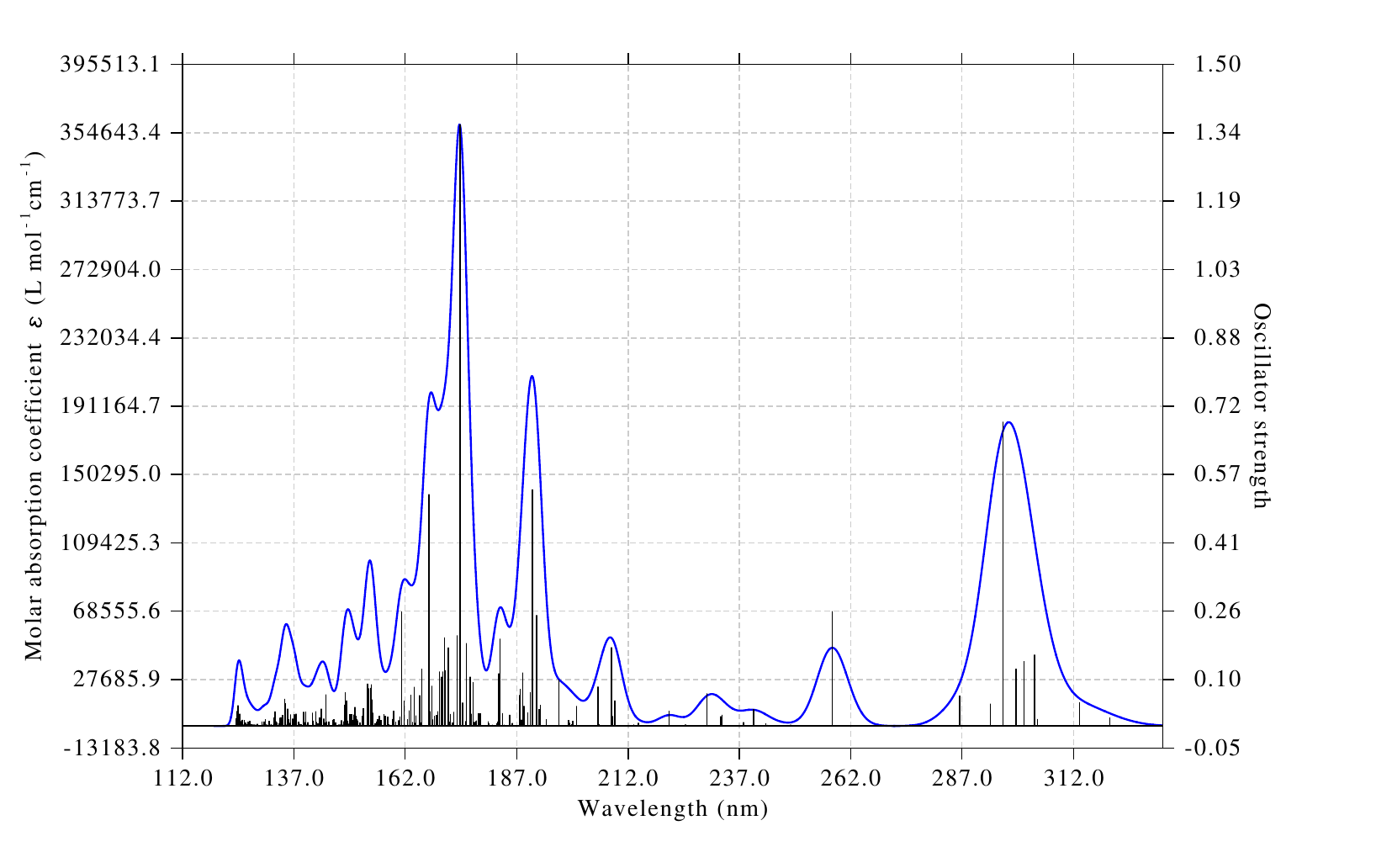}}\
\
\subfloat[In toluene solvent: UV-Vis 
spectrum]{\includegraphics[width=0.4\textwidth]{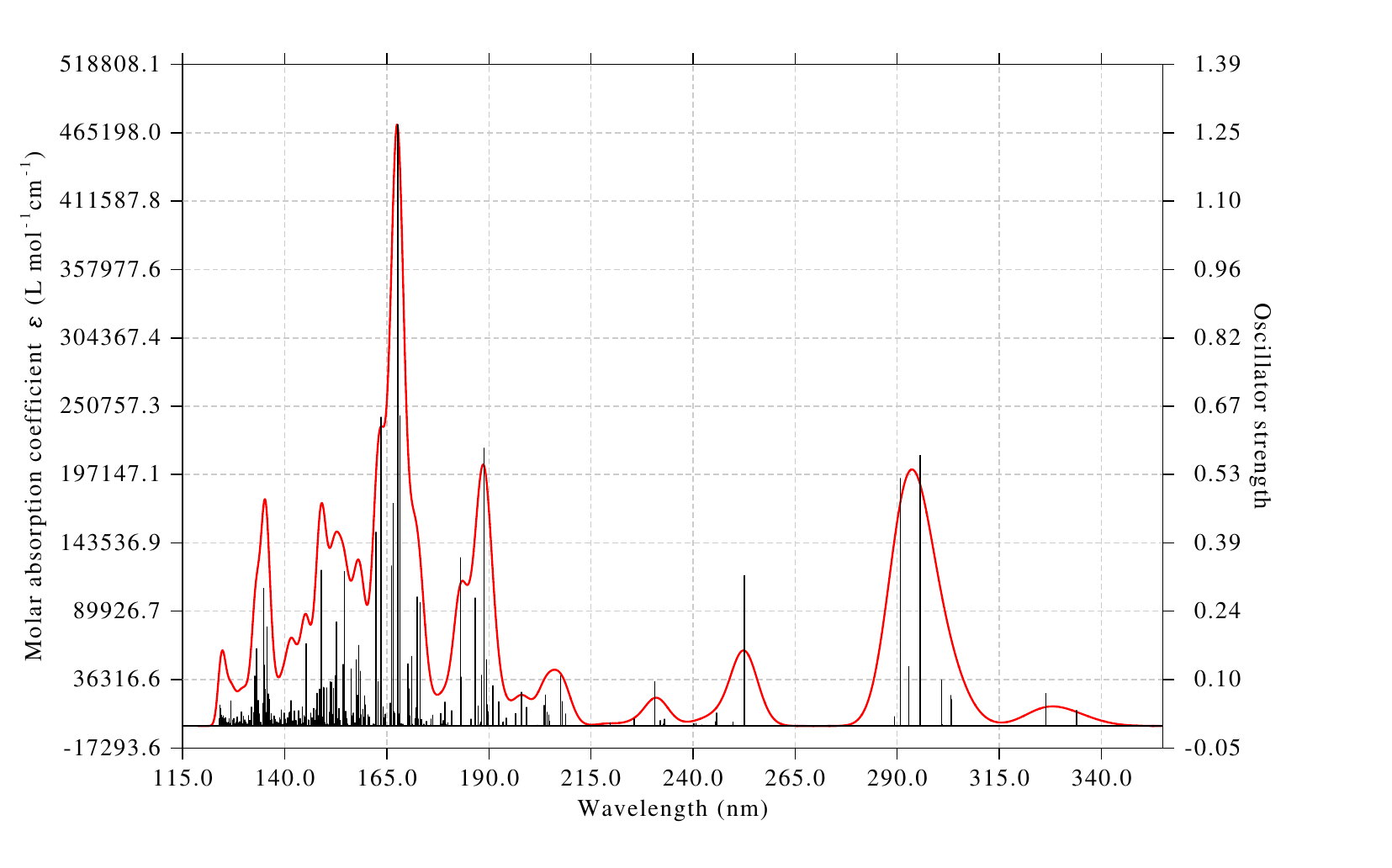}
\includegraphics[width=0.4\textwidth]{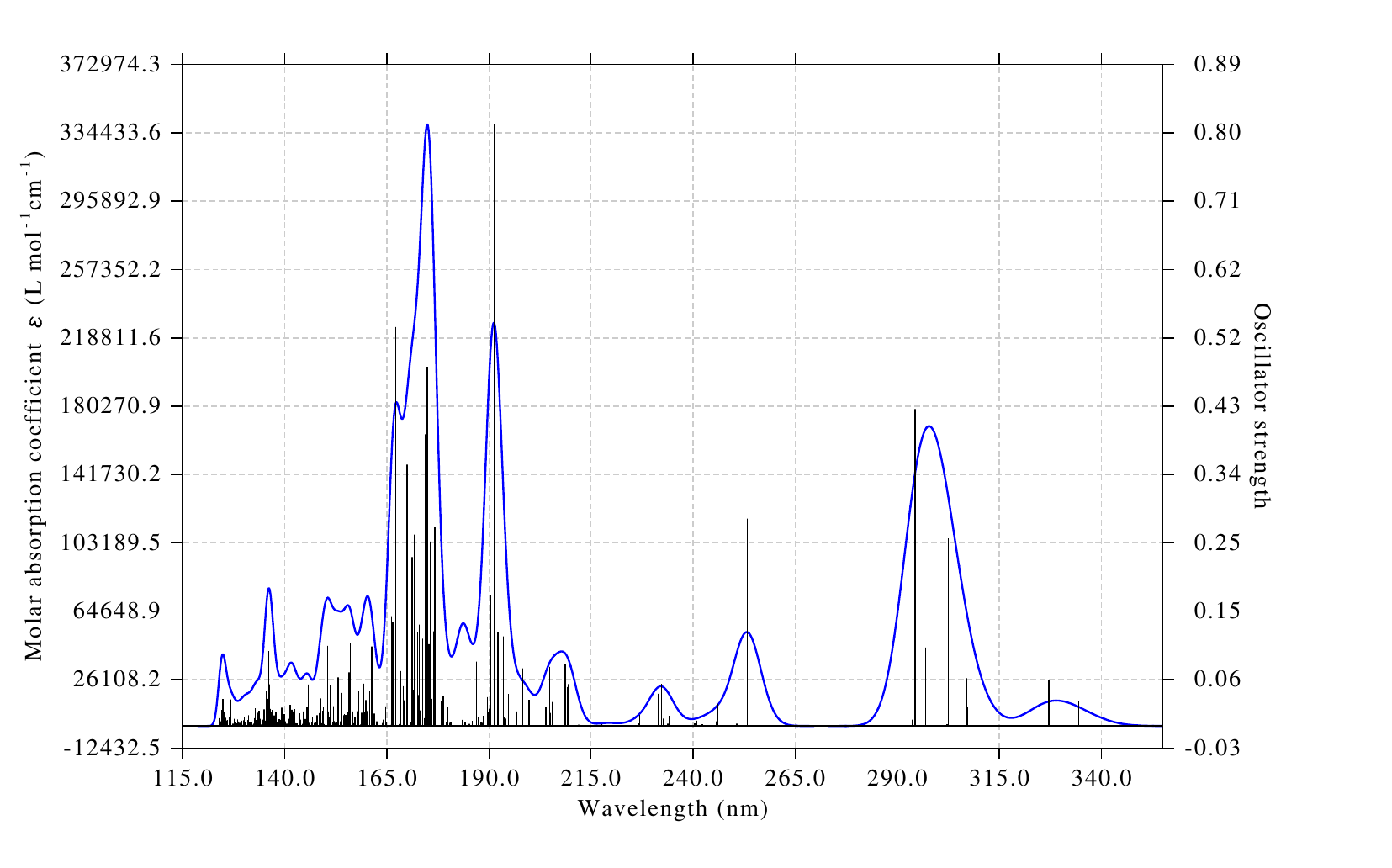}}
\caption{UV-Vis absorption spectra of DMAC-DPS: solvent and method dependence. Panels show the simulated UV-Vis absorption spectra in vacuum and toluene, 
calculated with \stda (top) and \stddft (bottom) methods. Note the significant broadening of the absorption band in toluene, suggesting increased vibronic 
coupling in the polar environment.}
\end{figure}

\begin{figure}[!htbp]
\centering
\leavevmode
\subfloat[In vacuum: ECD 
spectrum]{\includegraphics[width=0.4\textwidth]{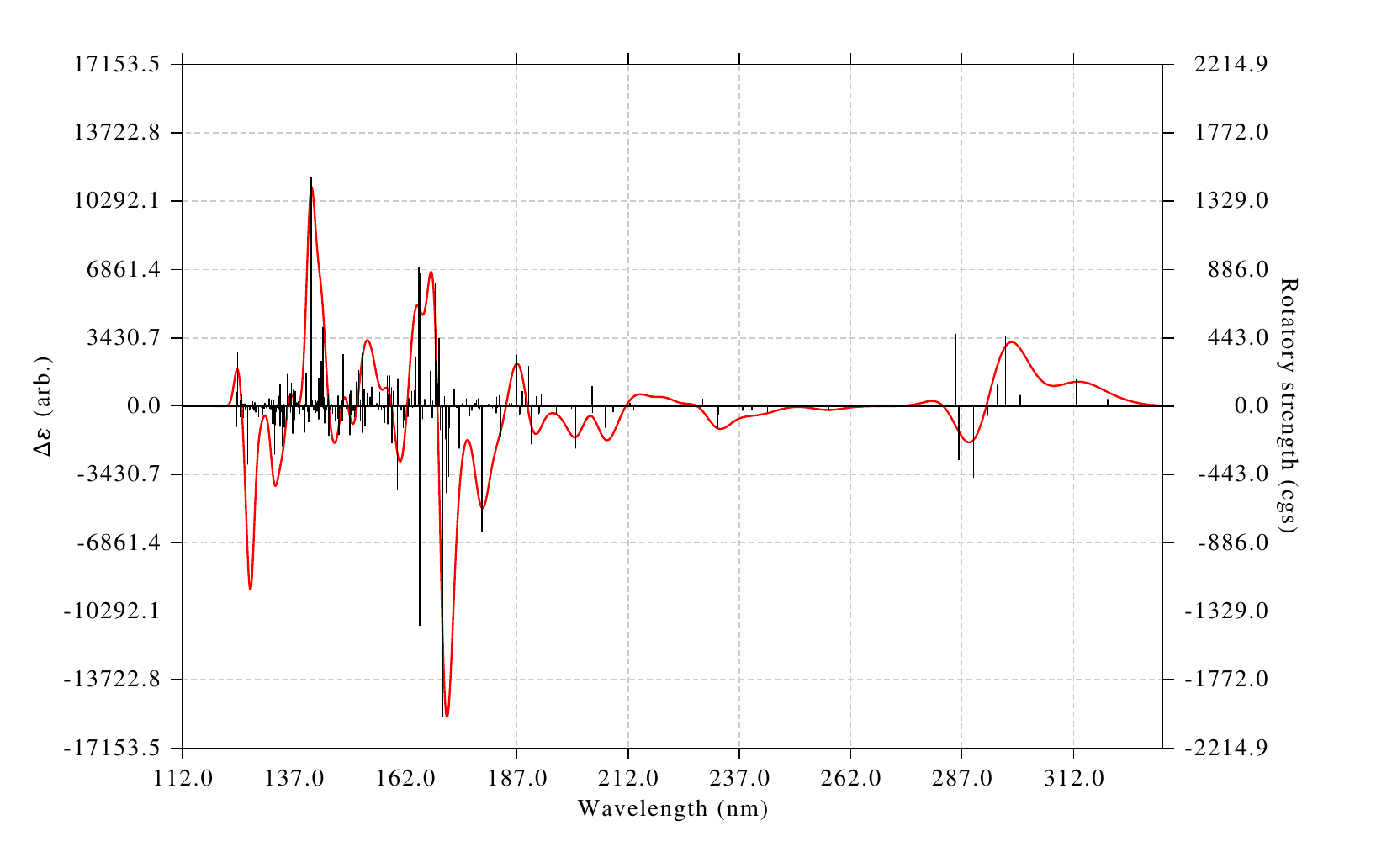}\includegraphics[width=0.4\textwidth]{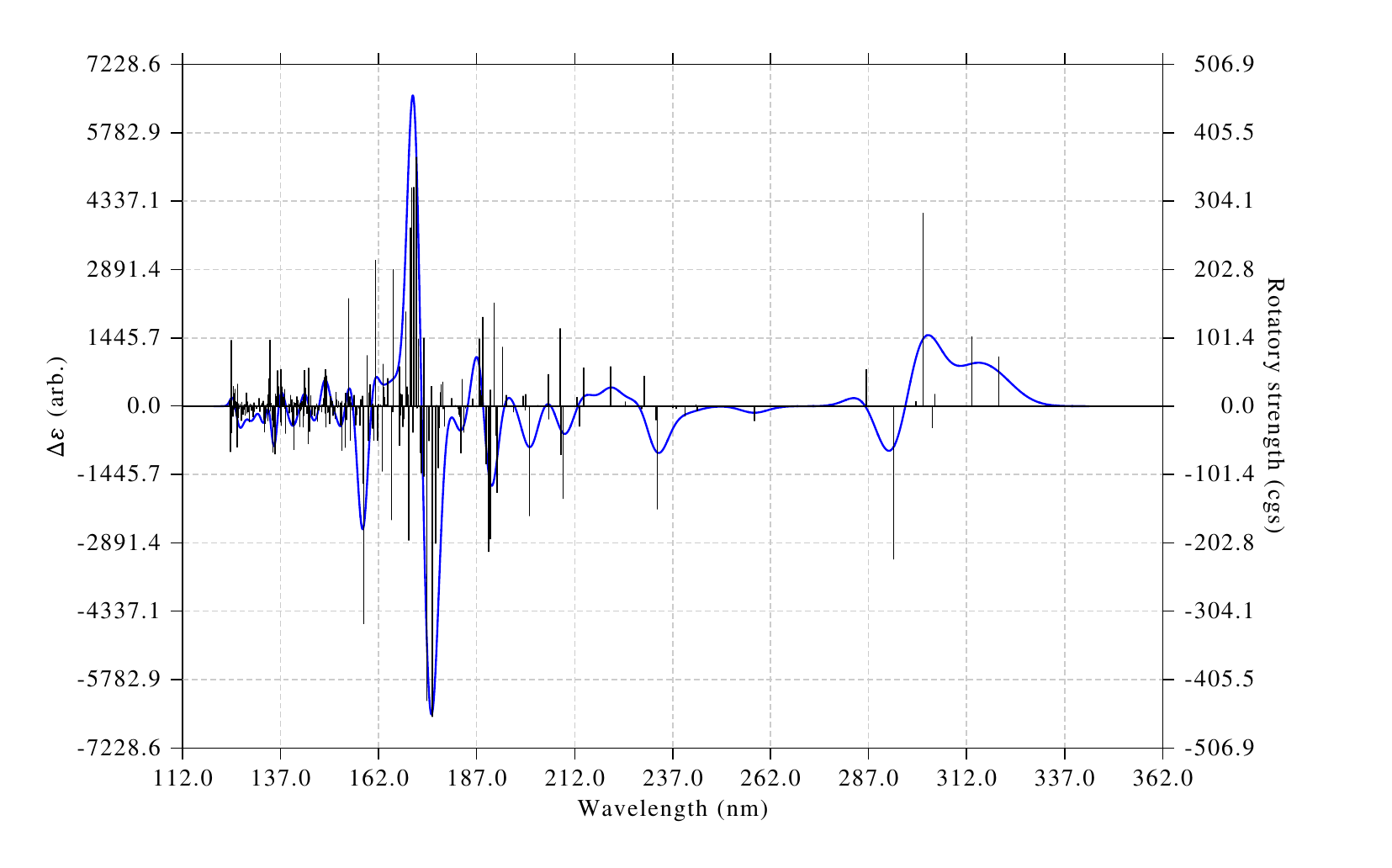}}\\
\subfloat[In toluene solvent: ECD 
spectrum]{\includegraphics[width=0.4\textwidth]{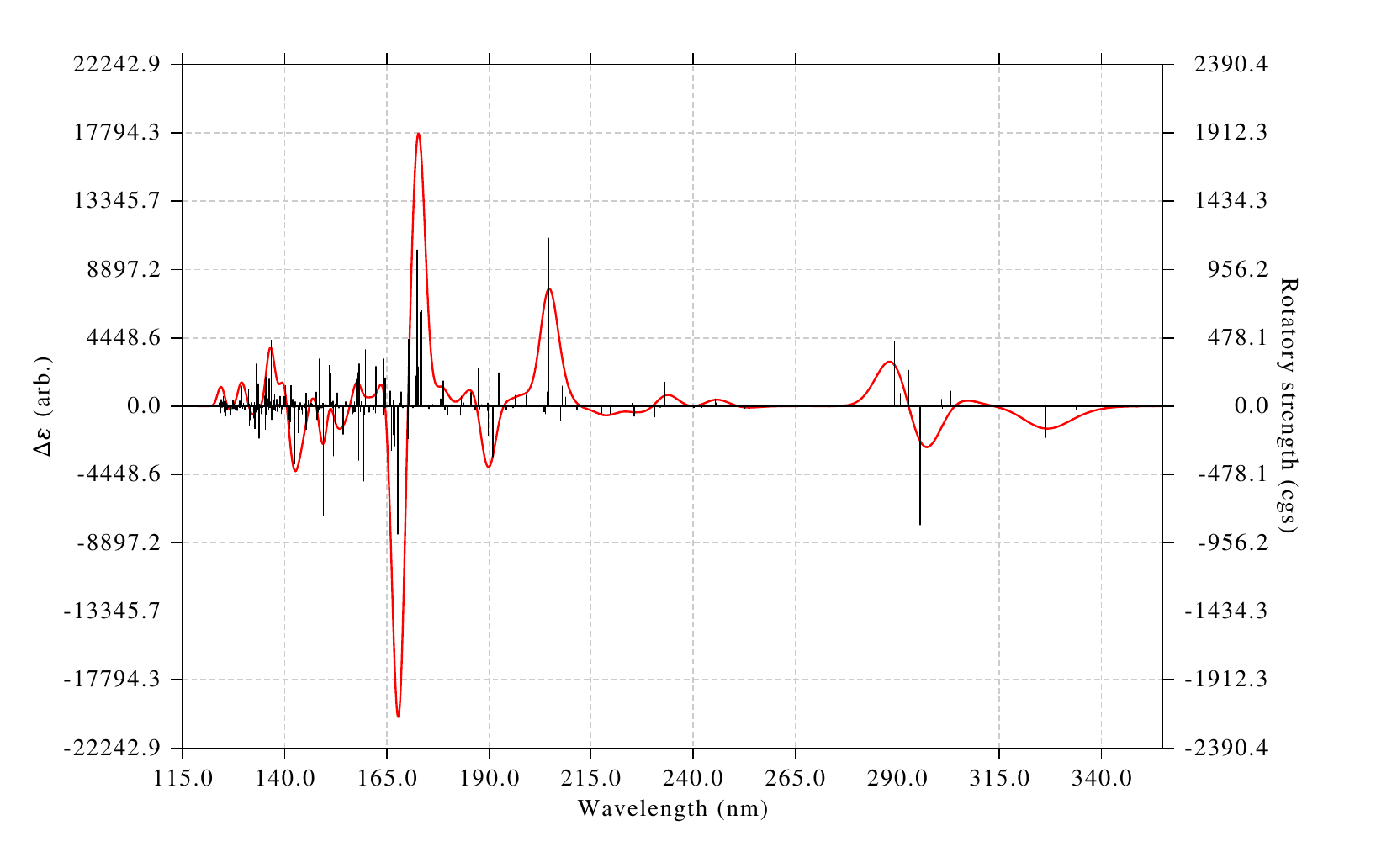}\includegraphics[width=0.4\textwidth]{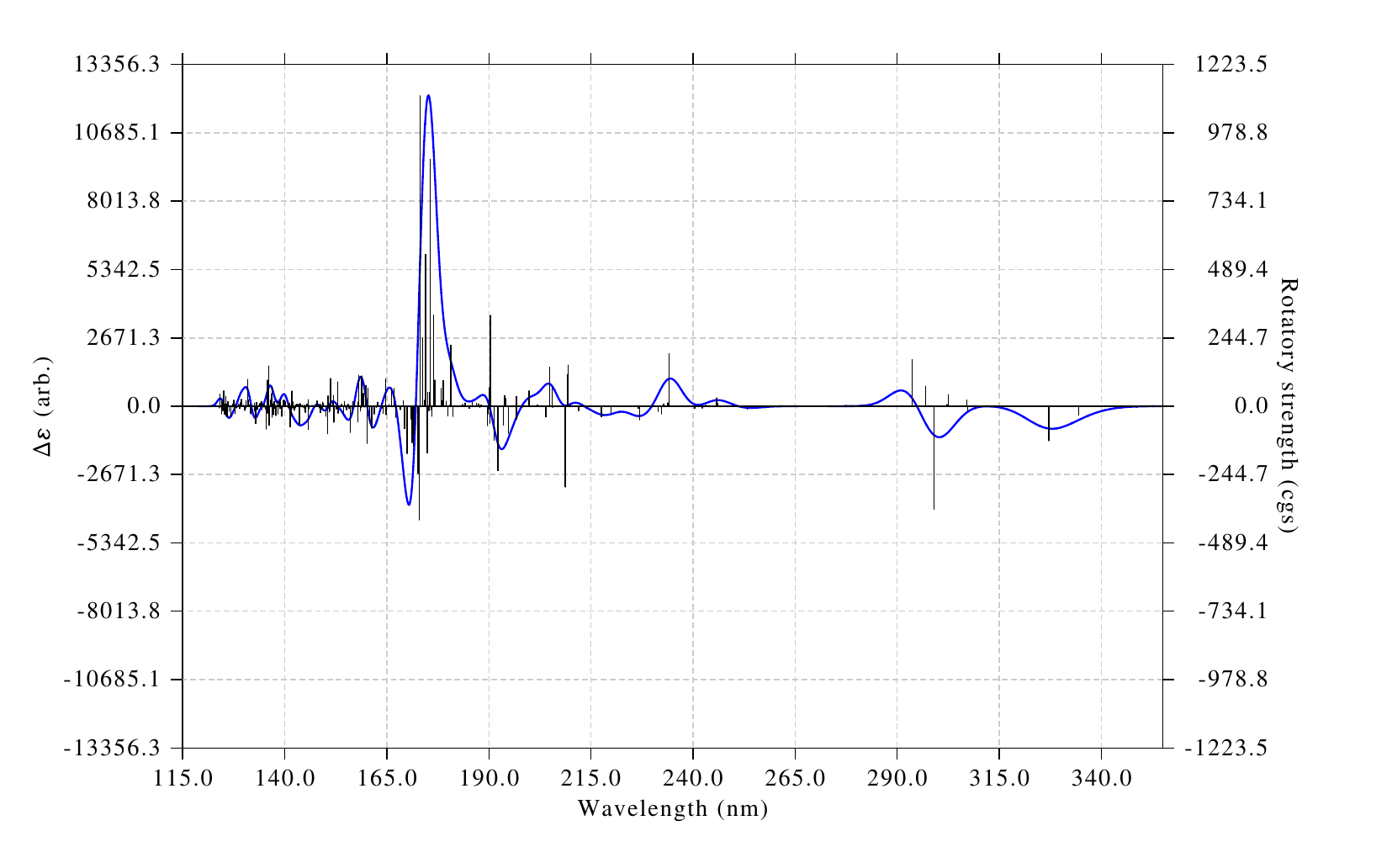}
}
\caption{Electronic Circular Dichroism (ECD) spectra of DMAC-DPS: solvent and method dependence. Panels show the simulated ECD spectra in vacuum and toluene, 
calculated with \stda (top) and \stddft (bottom) methods. The ECD spectra of DMAC-DPS exhibit a complex pattern of positive and negative bands, suggesting a 
non-trivial chiral contribution from the DMAC and DPS units.}
\end{figure}

\begin{table}[!htbp]
\centering
\caption{Colorimetric properties of DMAC-DPS: wavelength maxima and CIE coordinates. This table summarizes the key colorimetric properties of DMAC-DPS, 
including the wavelengths of maximum absorption and emission, CIE 1931 color space coordinates (X, Y, Z and x, y), and approximate sRGB color representation. 
The predicted emission color shifts significantly from blue to blue-green upon solvation, indicating a strong solvatochromic effect driven by its highly polar 
excited state.}
{\scriptsize
\begin{tabular}{m{1.2cm}m{1.8cm}@{\,}l*{3}{@{\,}c@{\,}}>{\columncolor{white}[0\tabcolsep][0pt]}c}
\toprule
&& Properties & {$\lambda_{max}(\unit{\nano\meter})$} & {$(X,Y,Z)$} & {$(x,y)$} & {(R,G,B)} \\
\midrule
\multirow{4}{=}{In vacuum} & \multirow{2}{=}{\emph{UV-vis} absorption} & \stda &$319.6899$& NA & NA & NA\\
&& \stddft &$320.1434$& NA & NA & NA\\
&\multirow{2}{=}{Fluorescence}  & \stda
&$452.2331$&$(15819.013282,2143.225893,85161.533309)$&$(0.1533983184,0.0207830439)$&\CcelCo{white}{17,0,255}\\
&& \stddft &$452.2331$&$(15819.013282,2143.225893,85161.533309)$&$(0.1533983184,0.0207830439)$&\CcelCo{white}{17,0,255}\\
\midrule
\multirow{4}{=}{In toluene\\ solvent} & \multirow{2}{=}{\emph{UV-vis} absorption} & \stda &$333.9365$& NA & NA & NA\\
&& \stddft &$334.4166$& NA & NA & NA\\
&\multirow{2}{=}{Fluorescence}  & \stda
&$467.7869$&$(14740.410006,9032.887377,100718.986812)$&$(0.1184042055,0.0725578090)$&\CcelCo{white}{0,17,255}\\
&& \stddft &$484.4166$&$(18425.190376,7436.271551,116742.366855)$&$(0.1292054395,0.0521463667)$&\CcelCo{white}{0,2,255}\\
\bottomrule
\end{tabular}}
\label{tab:ColorDMAC-DPS}
\end{table}


\begin{figure}[!htbp]
\centering
\leavevmode
\subfloat[In vacuum: UV-Vis 
spectrum]{\includegraphics[width=0.4\textwidth]{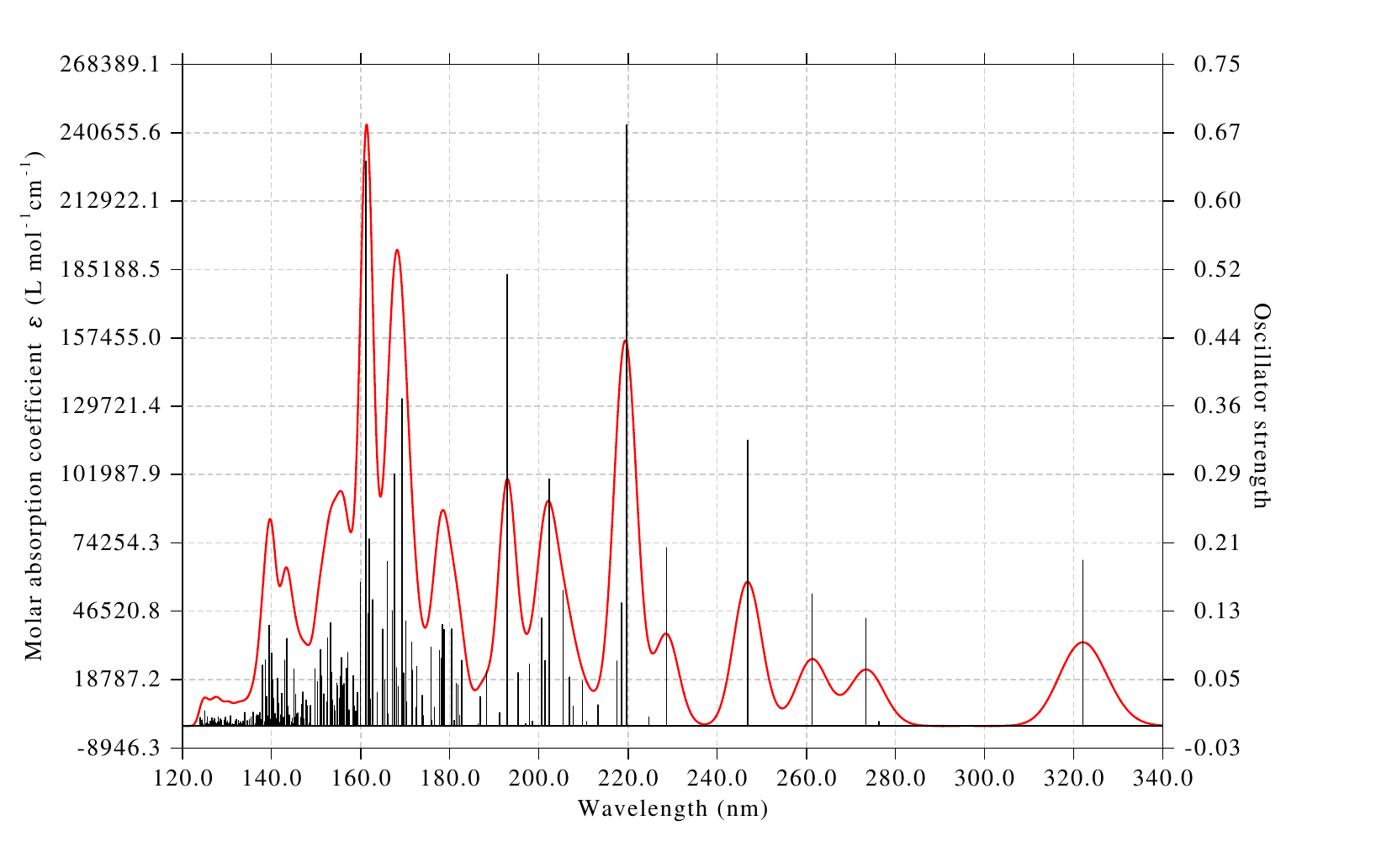}\includegraphics[width=0.4\textwidth]{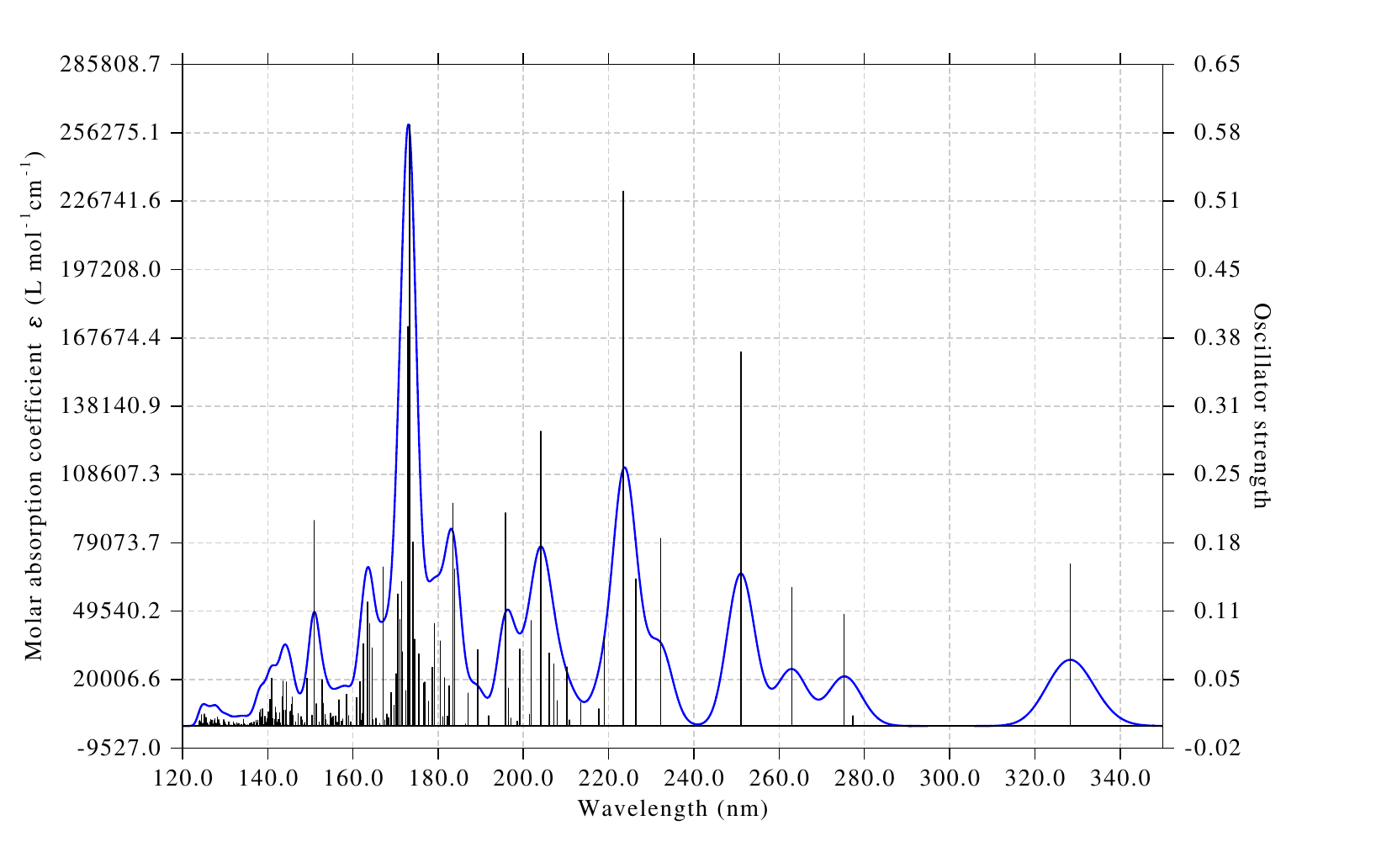}}\\
\subfloat[In toluene solvent: UV-Vis 
spectrum]{\includegraphics[width=0.4\textwidth]{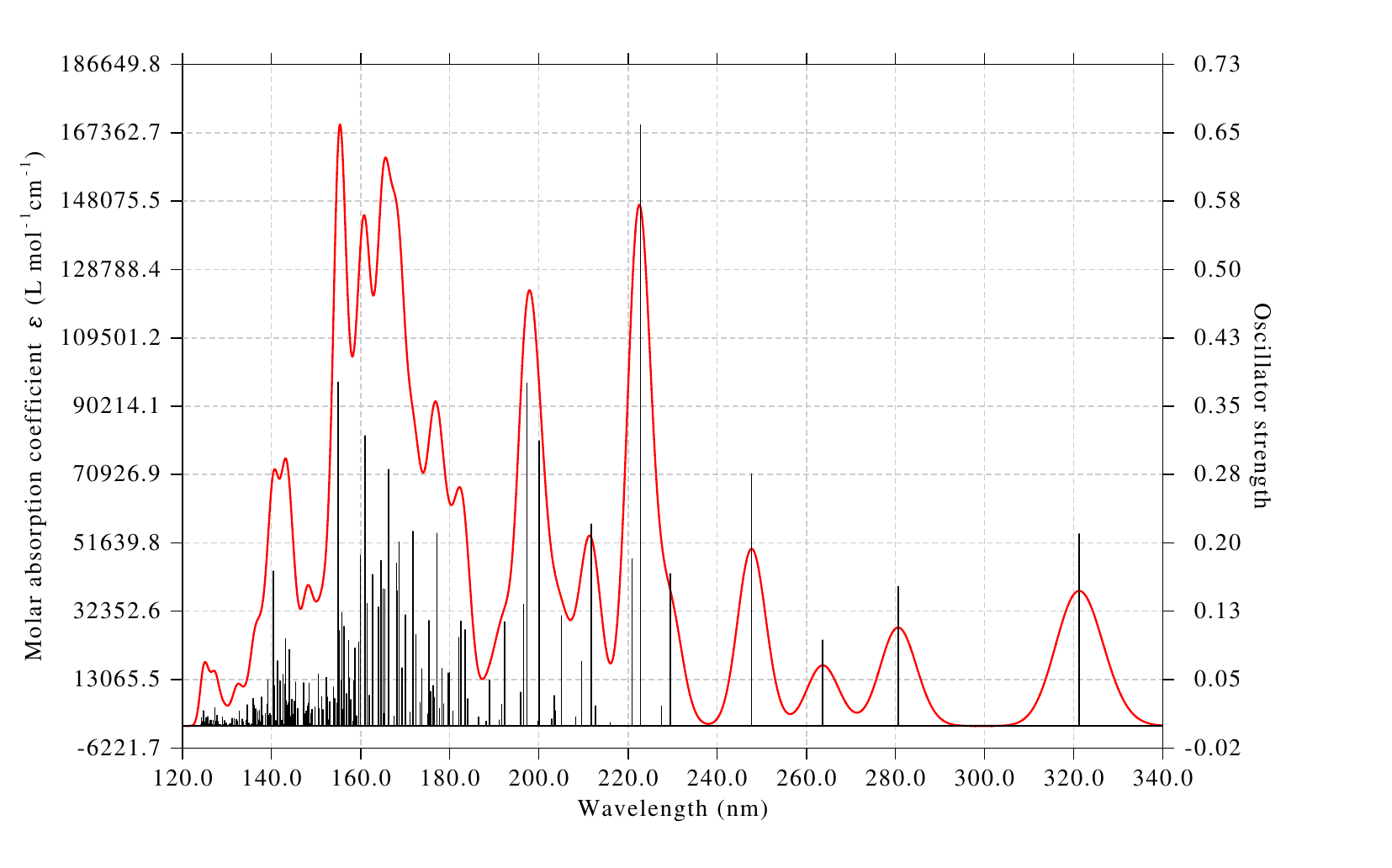}\includegraphics[width=0.4\textwidth]{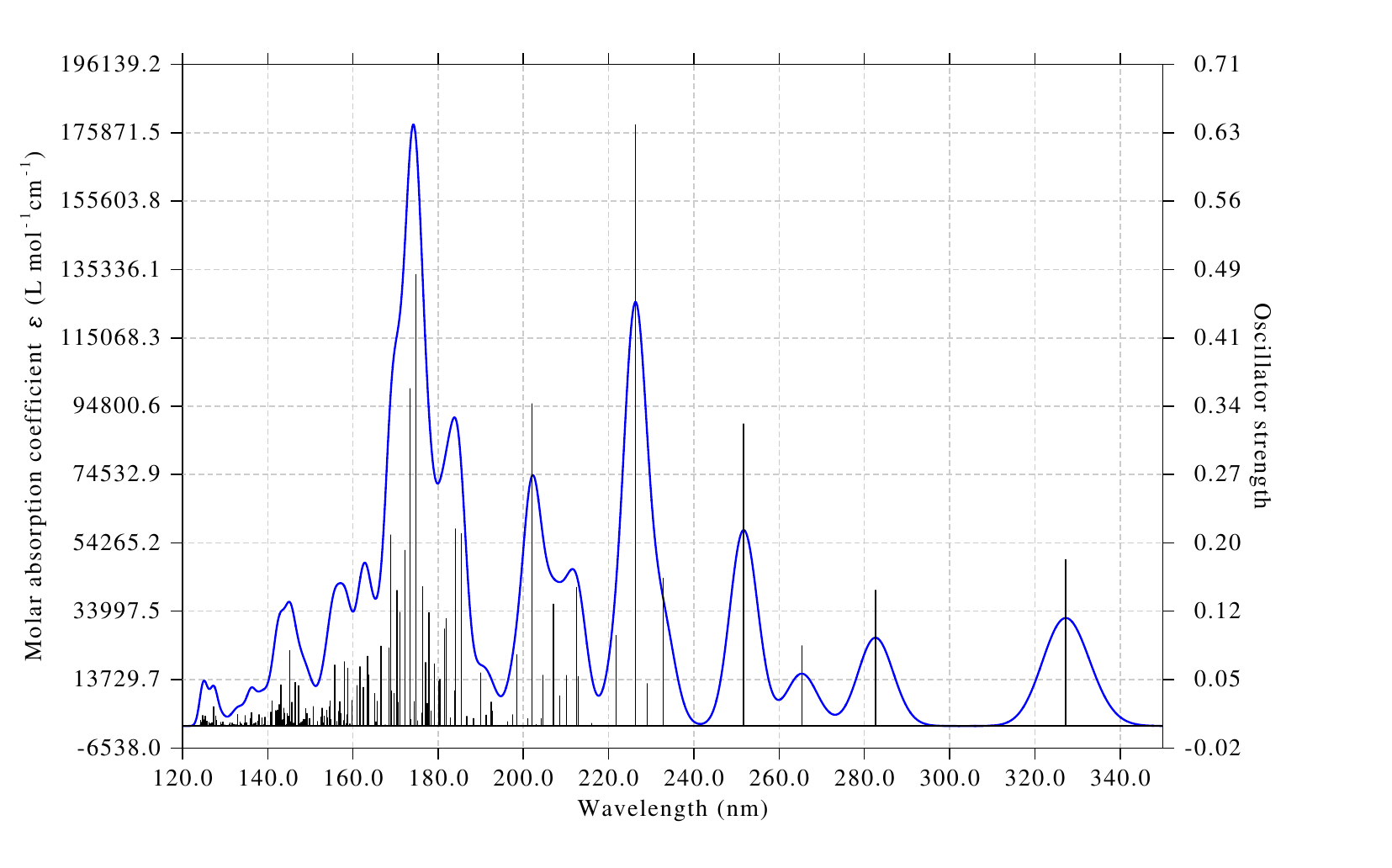}
}
\caption{UV-Vis absorption spectra of PSPCz: solvent and method dependence. Panels show the simulated UV-Vis absorption spectra in vacuum and toluene, 
calculated with \stda (top) and \stddft (bottom) methods. The absorption spectra of PSPCz show a relatively weak solvent dependence, suggesting that the ground 
state is not significantly affected by the surrounding environment.}
\end{figure}

\begin{figure}[!htbp]
\centering
\leavevmode
\subfloat[In vacuum: ECD 
spectrum]{\includegraphics[width=0.4\textwidth]{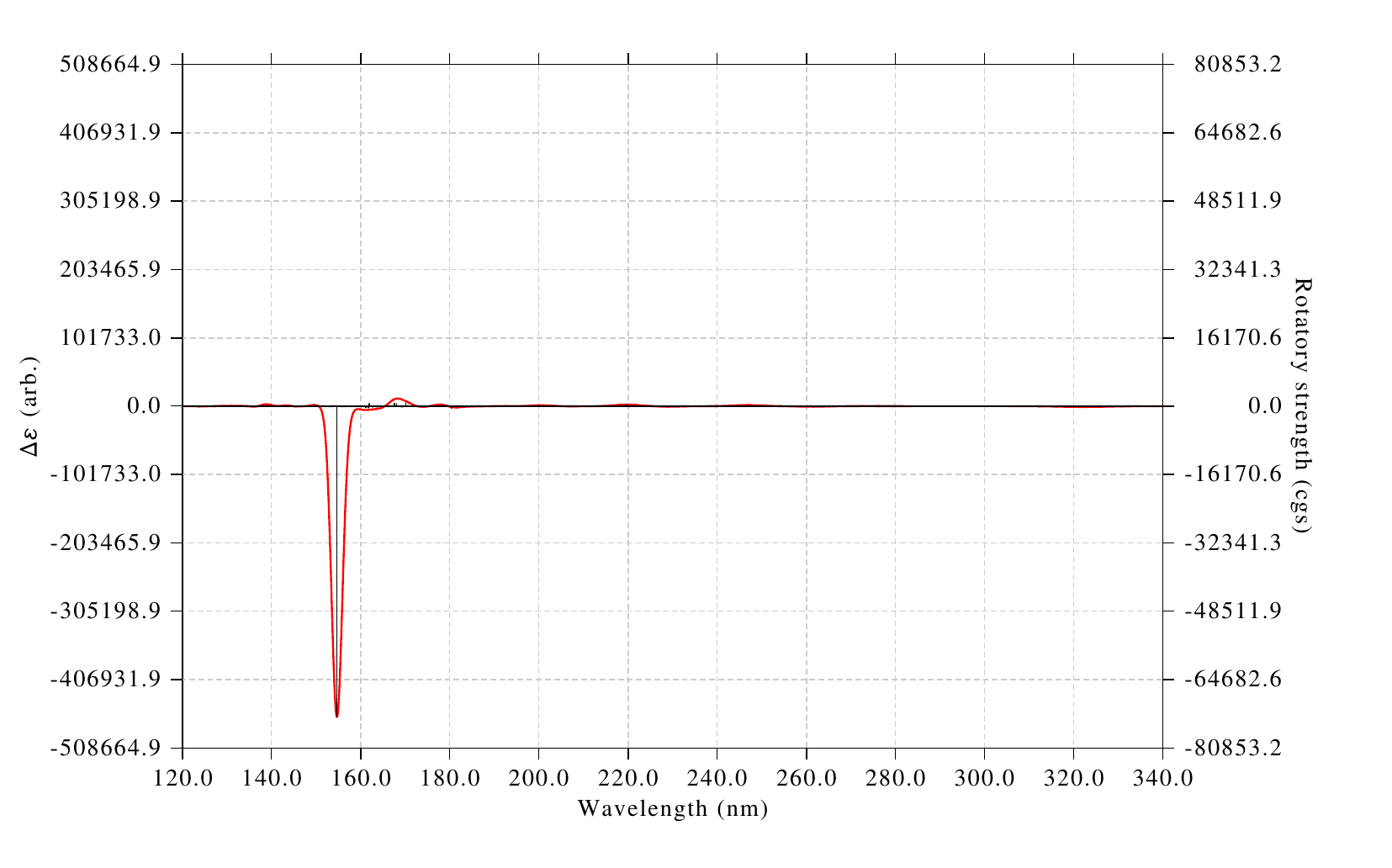}\includegraphics[width=0.4\textwidth]{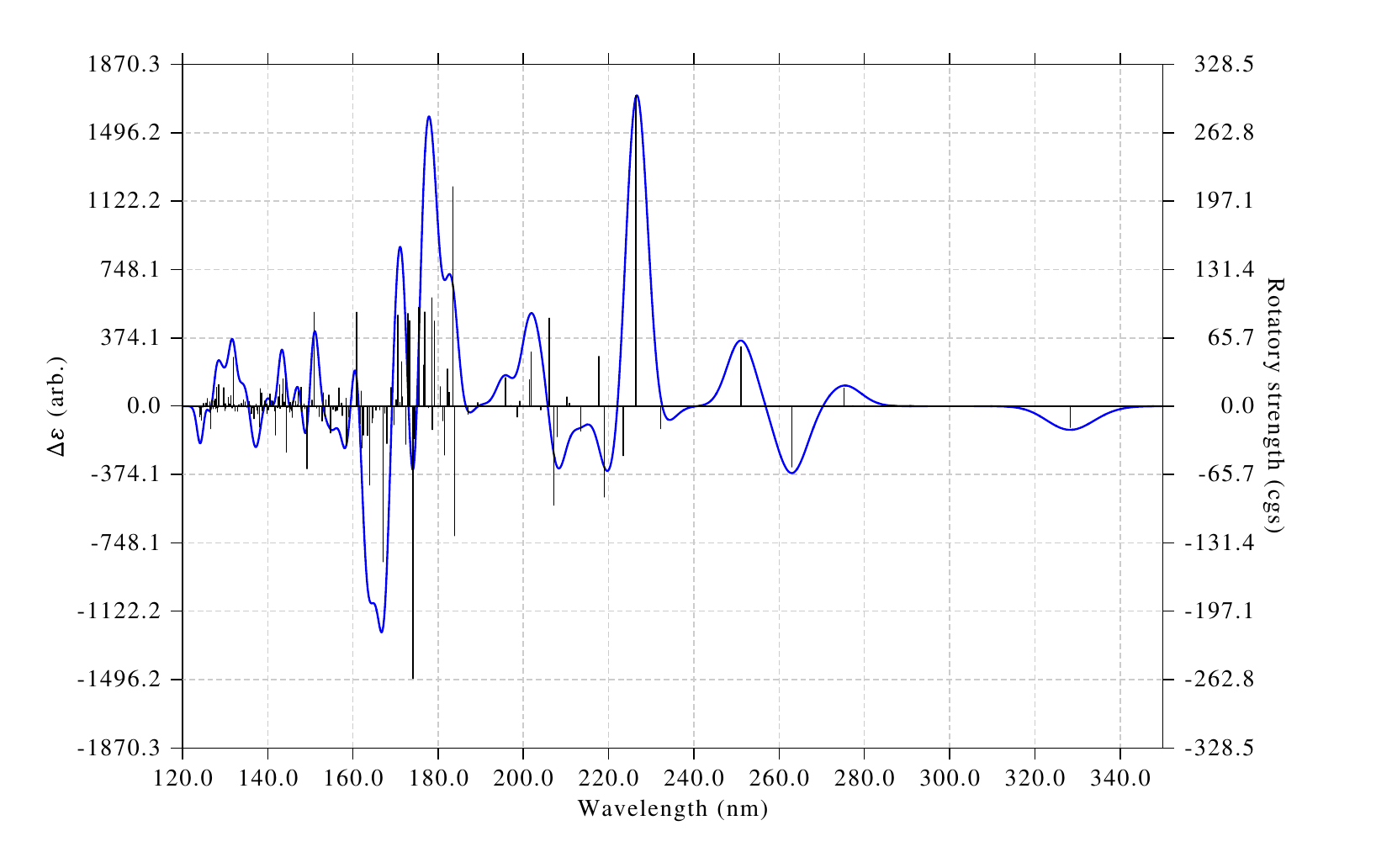}}\\
\subfloat[In toluene solvent: ECD 
spectrum]{\includegraphics[width=0.4\textwidth]{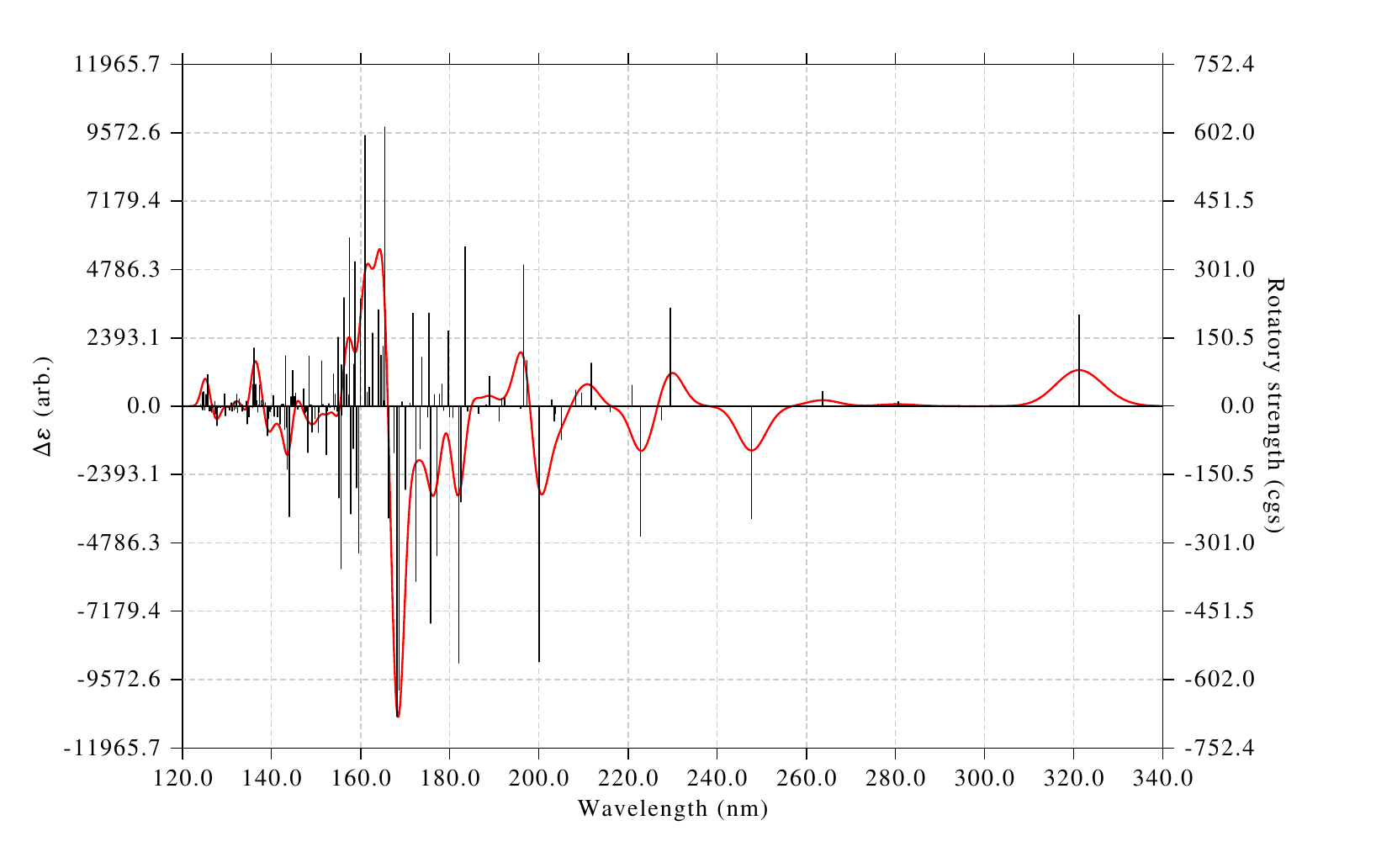}\includegraphics[width=0.4\textwidth]{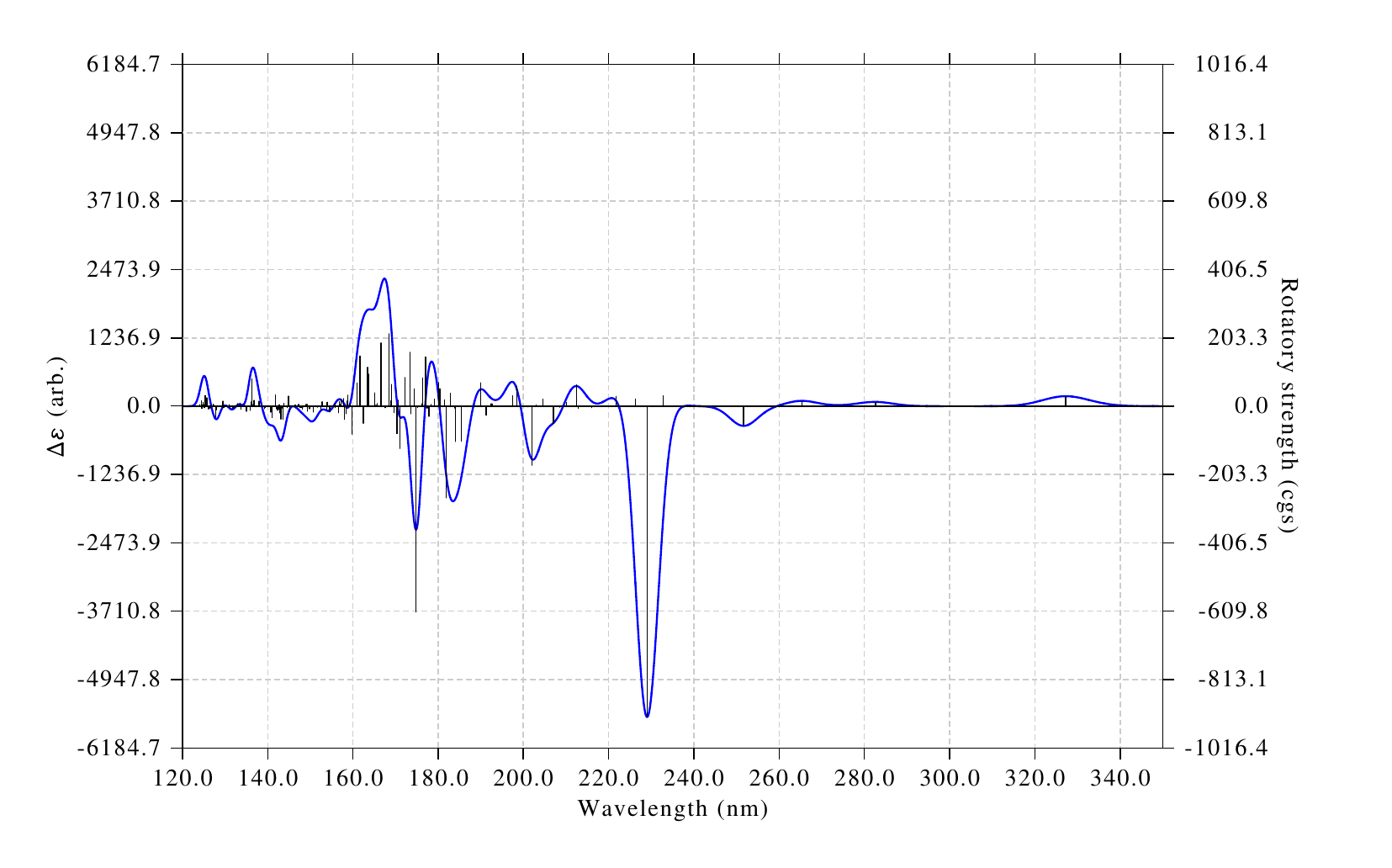}}
\caption{Electronic Circular Dichroism (ECD) spectra of PSPCz: solvent and method dependence. Panels show the simulated ECD spectra in vacuum and toluene, 
calculated with \stda (top) and \stddft (bottom) methods. The ECD spectra of PSPCz exhibit more pronounced chiral features compared to DMAC-TRZ and DMAC-DPS, 
indicating a less symmetrical structure.}
\end{figure}

\begin{table}[!htbp]
\centering
\caption{Colorimetric properties of PSPCz: wavelength maxima and CIE coordinates. This table summarizes the key colorimetric properties of PSPCz, including the 
wavelengths of maximum absorption and emission, CIE 1931 color space coordinates (X, Y, Z and x, y), and approximate sRGB color representation. The calculated 
emission color of PSPCz is relatively stable upon solvation, remaining in the deep blue region, but its low Y value indicates non-visible emission.}
{\scriptsize
\begin{tabular}{m{1.2cm}m{1.8cm}@{\,}l*{3}{@{\,}c@{\,}}>{\columncolor{white}[0\tabcolsep][0pt]}c}
\toprule
&& Properties & {$\lambda_{max}(\unit{\nano\meter})$} & {$(X,Y,Z)$} & {$(x,y)$} & {(R,G,B)} \\
\midrule
\multirow{4}{=}{In vacuum} & \multirow{2}{=}{\emph{UV-vis} absorption} & \stda &$322.0837$& NA & NA & NA\\
&& \stddft &$328.3031$& NA & NA & NA\\
&\multirow{2}{=}{Fluorescence}  & \stda
&$383.5435$&$(1128.692294,32.129669,5329.907469)$&$(0.1738929817,0.0049500860)$&\CcelCo{white}{45,0,255}\\
&& \stddft &$389.3245$&$(1999.605911,56.421174,9460.655116)$&$(0.1736269071,0.0048990823)$&\CcelCo{white}{45,0,255}\\
\midrule
\multirow{4}{=}{In toluene\\ solvent} & \multirow{2}{=}{\emph{UV-vis} absorption} & \stda &$321.2585$& NA & NA & NA\\
&& \stddft &$327.2637$& NA & NA & NA\\
&\multirow{2}{=}{Fluorescence} & \stda
&$385.6133$&$(1605.299604,45.568603,7585.211526)$&$(0.1738074649,0.0049337602)$&\CcelCo{white}{45,0,255}\\
&& \stddft &$385.6133$&$(1605.299604,45.568603,7585.211526)$&$(0.1738074649,0.0049337602)$&\CcelCo{white}{45,0,255}\\
\bottomrule
\end{tabular}}
\label{tab:ColorPSPCz}
\end{table}


\begin{figure}[!htbp]
\centering
\leavevmode
\subfloat[In vacuum: UV-Vis 
spectrum]{\includegraphics[width=0.4\textwidth]{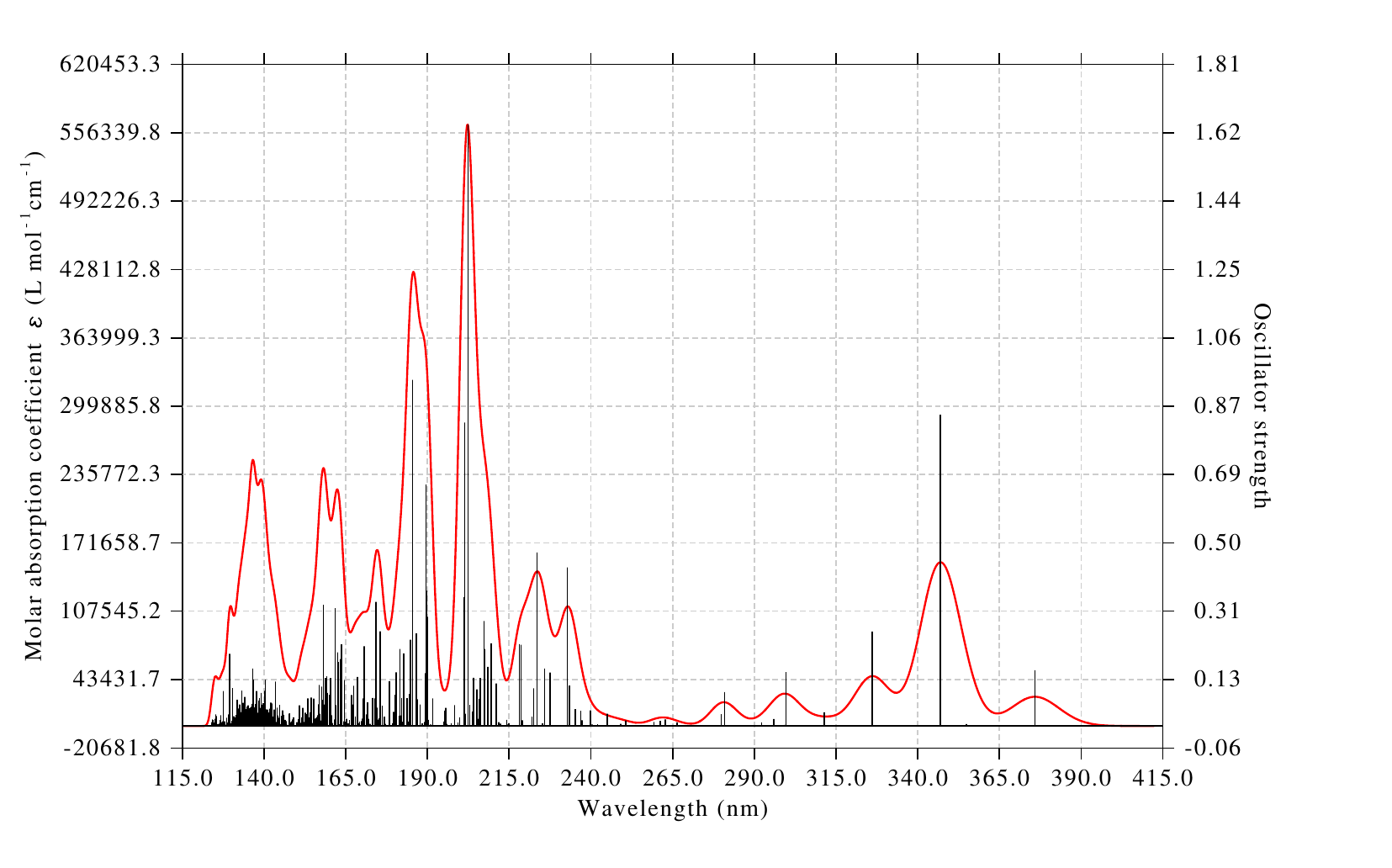}\includegraphics[width=0.4\textwidth]{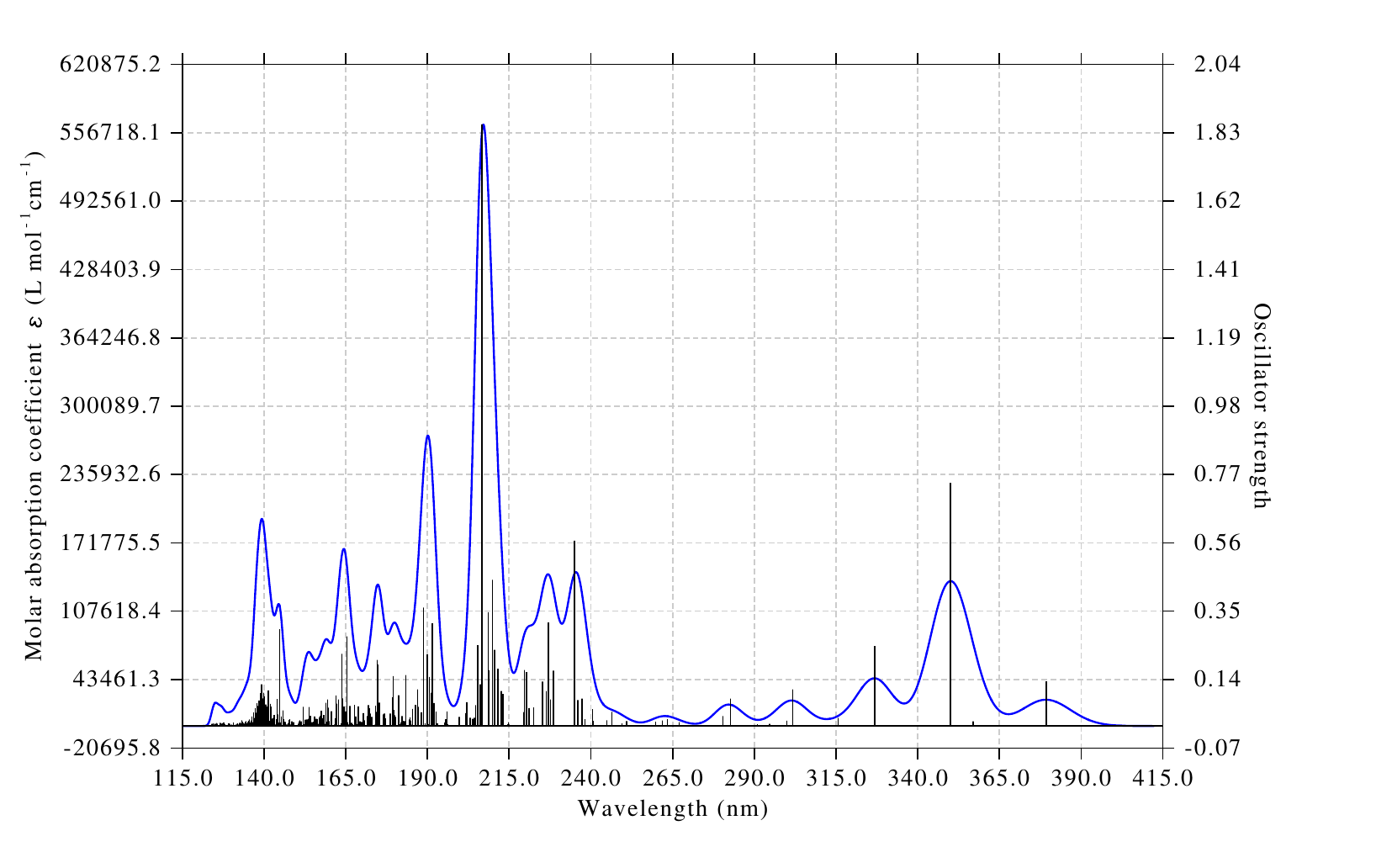}}\\
\subfloat[In toluene solvent: UV-Vis 
spectrum]{\includegraphics[width=0.4\textwidth]{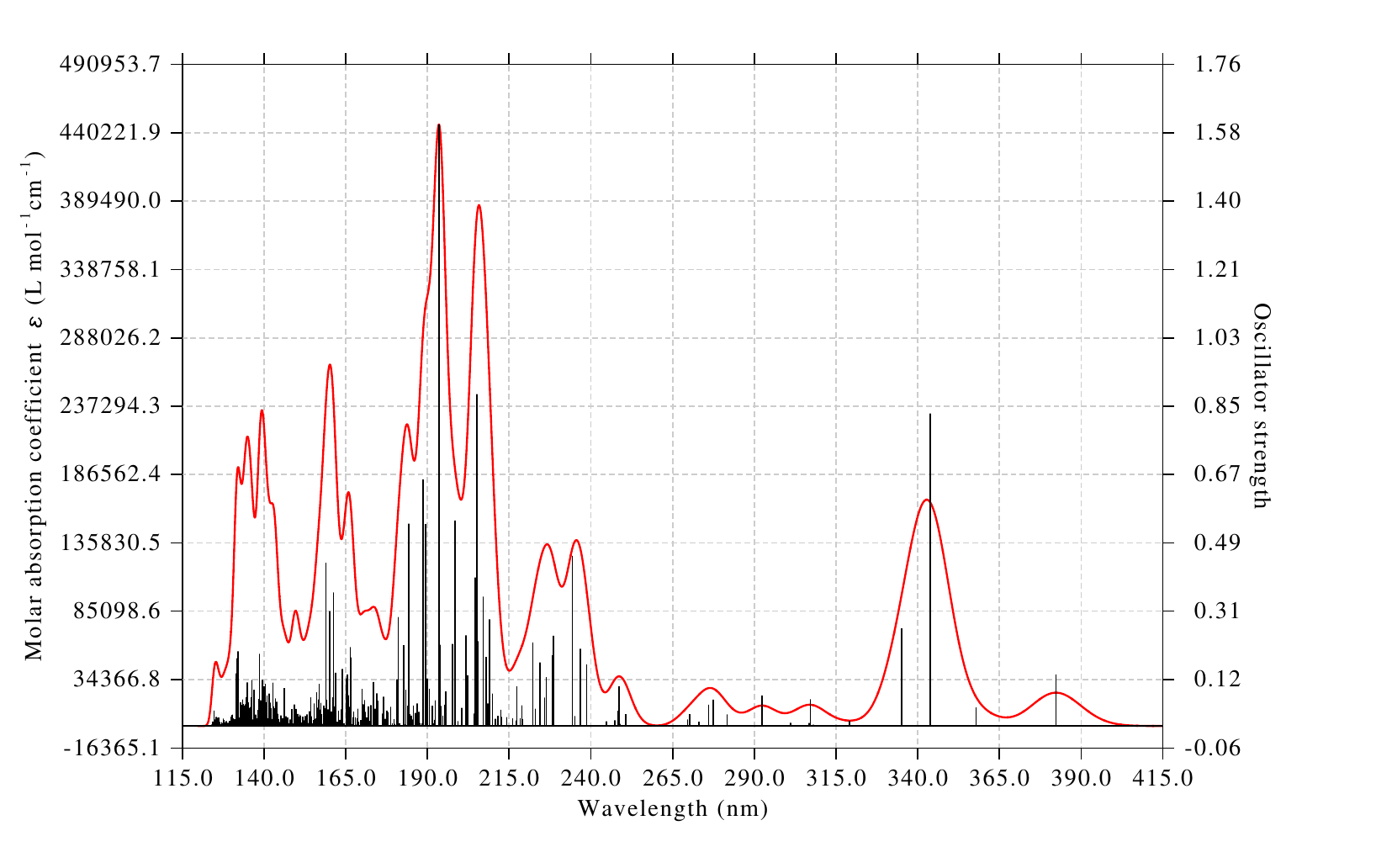}
\includegraphics[width=0.4\textwidth]{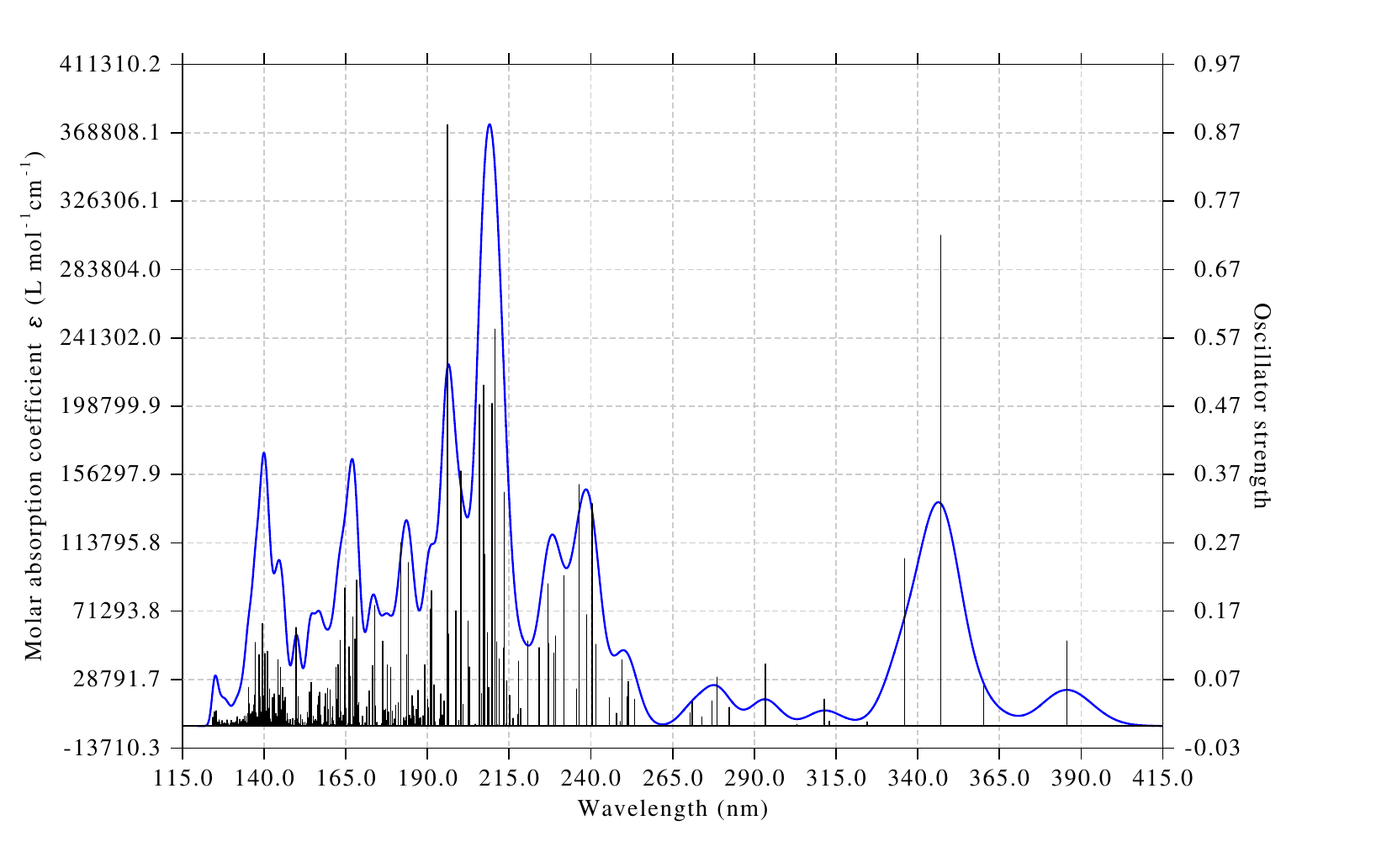}}
\caption{UV-Vis absorption spectra of 4CzIPN: solvent and method dependence. Panels show the simulated UV-Vis absorption spectra in vacuum and toluene, 
calculated with \stda (top) and \stddft (bottom) methods. The 4CzIPN absorption spectra exhibit a broad absorption band, characteristic of delocalized 
electronic transitions across the four carbazole units and the IPN core.}
\end{figure}

\begin{figure}[!htbp]
\centering
\leavevmode
\subfloat[In vacuum: ECD 
spectrum]{\includegraphics[width=0.4\textwidth]{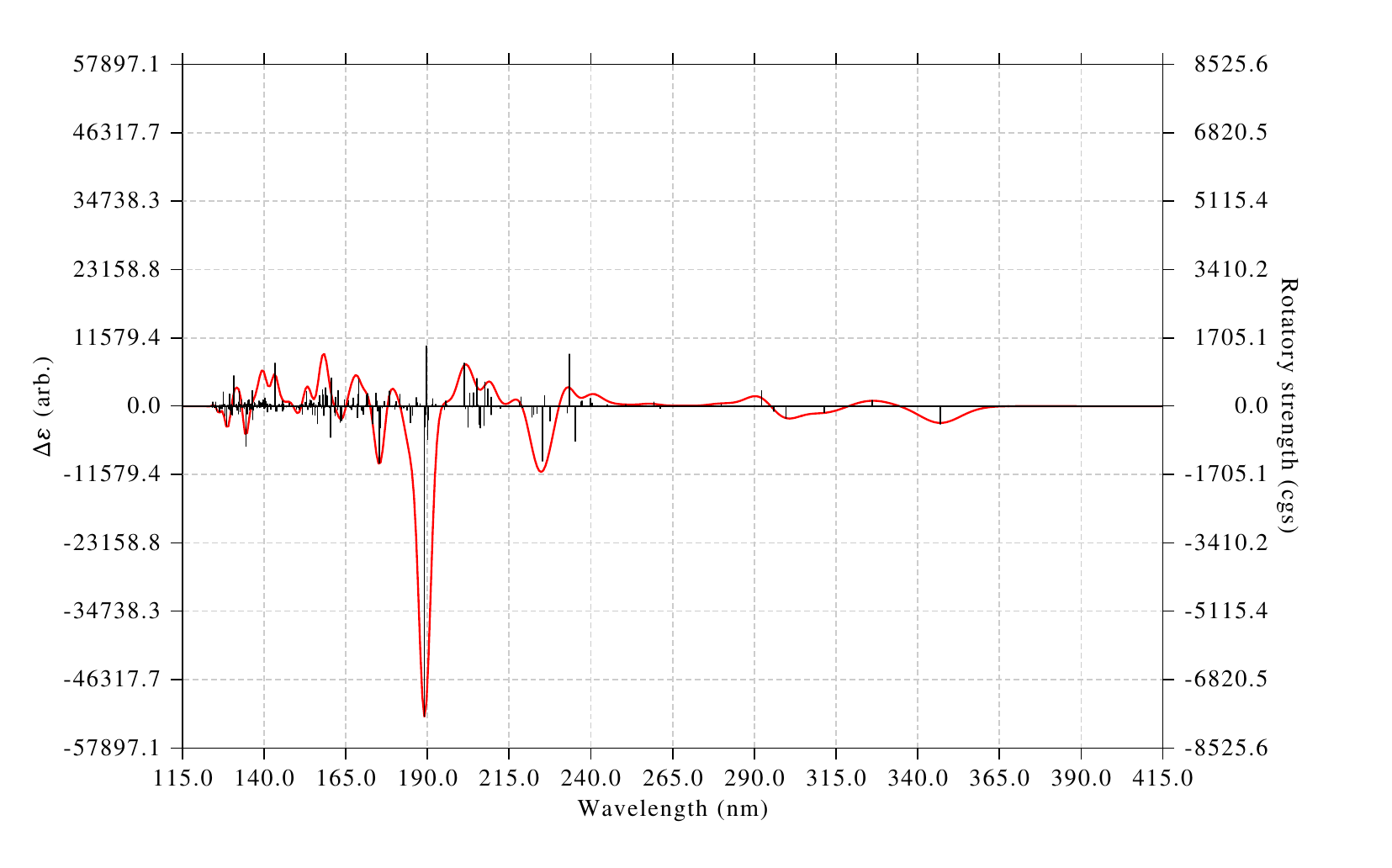}\includegraphics[width=0.4\textwidth]{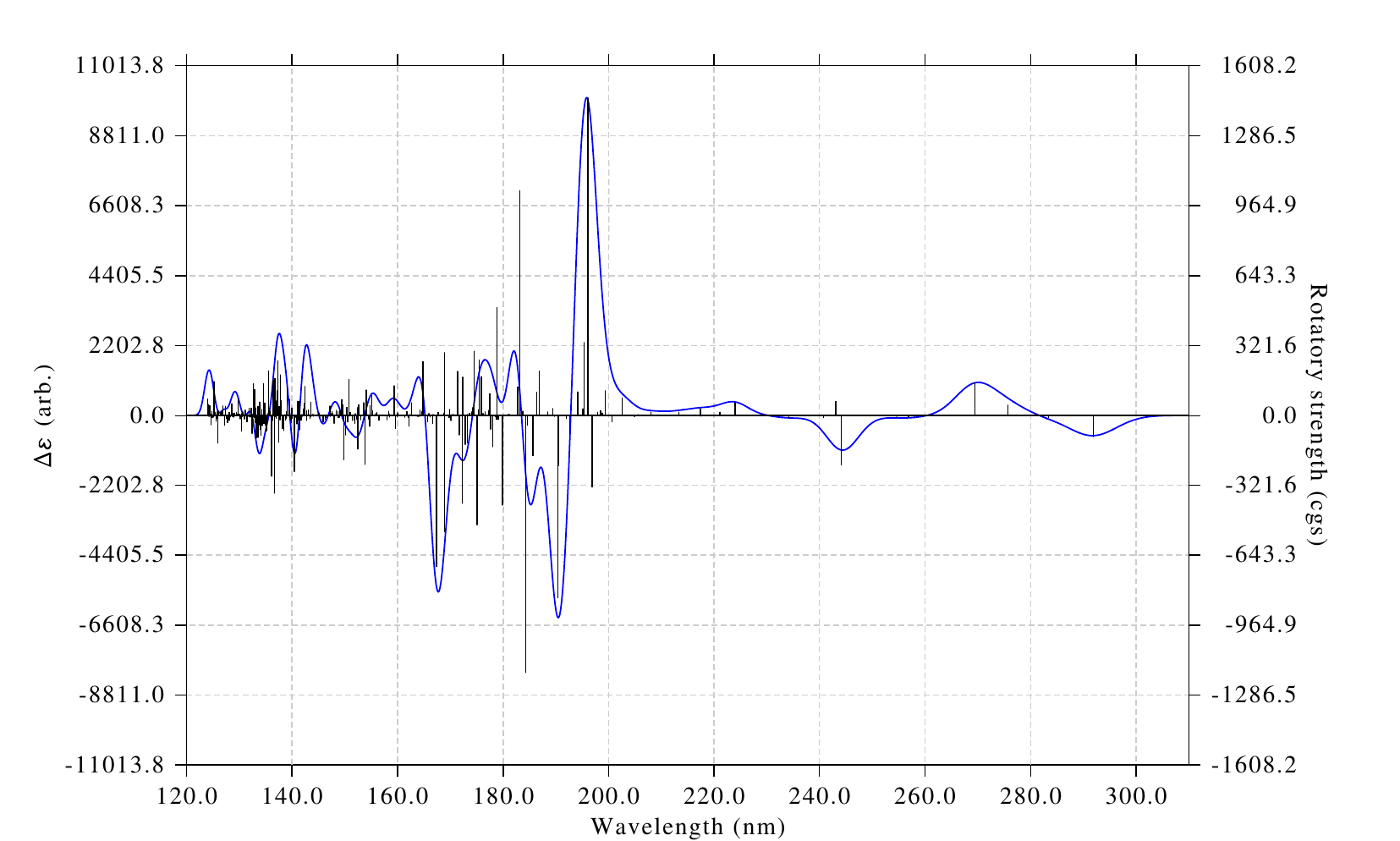}}\\
\subfloat[In toluene solvent: ECD 
spectrum]{\includegraphics[width=0.4\textwidth]{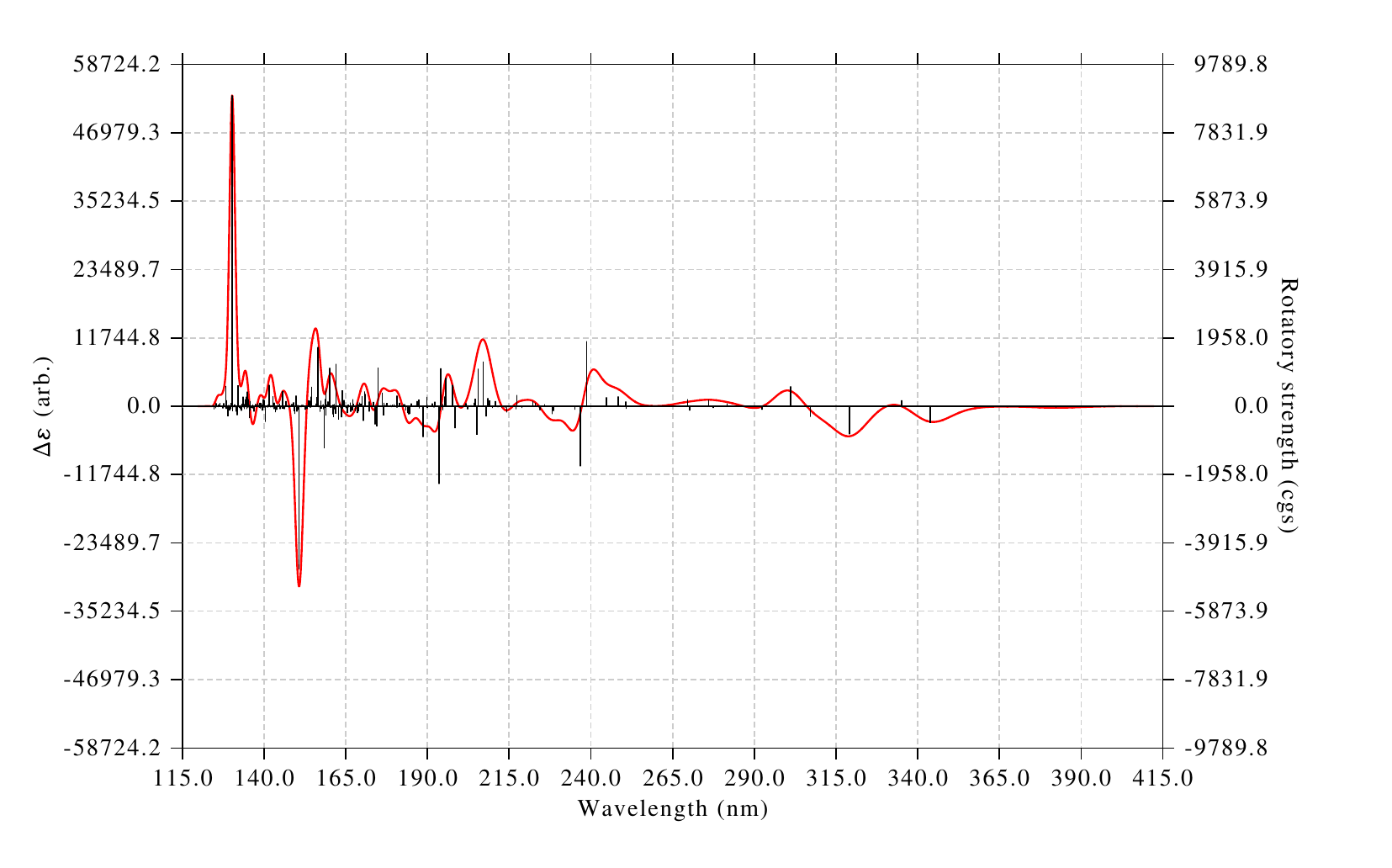}\includegraphics[width=0.4\textwidth]{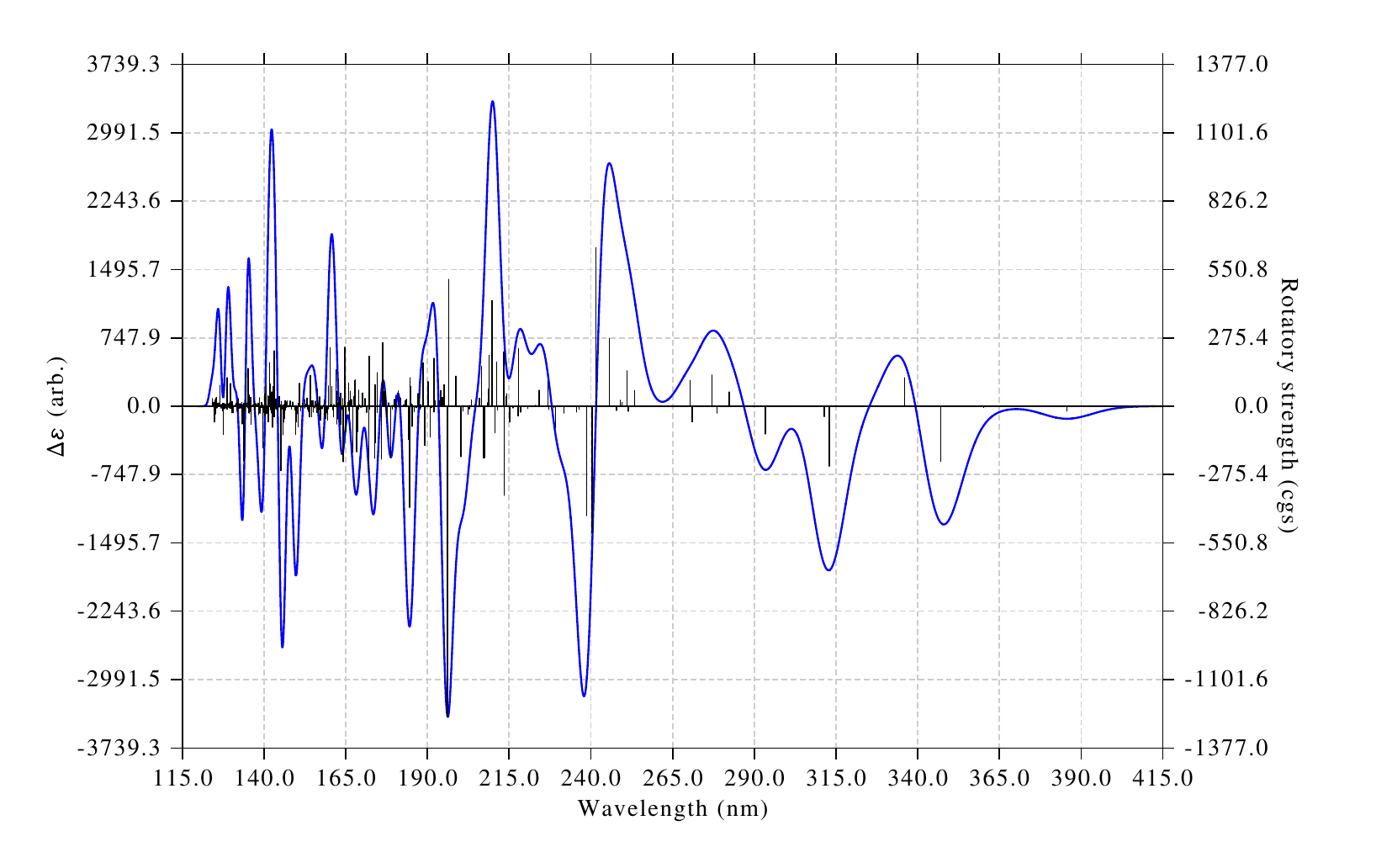}}
\caption{Electronic Circular Dichroism (ECD) spectra of 4CzIPN: solvent and method dependence. Panels show the simulated ECD spectra in vacuum and toluene, 
calculated with \stda (top) and \stddft (bottom) methods. The ECD spectra of 4CzIPN indicate a relatively symmetrical structure compared to the other molecules, 
with minimal ECD signals observed.}
\end{figure}

\begin{table}[!htbp]
\centering
\caption{Colorimetric properties of 4CzIPN: wavelength maxima and CIE coordinates. This table summarizes the key colorimetric properties of 4CzIPN, including 
the wavelengths of maximum absorption and emission, CIE 1931 color space coordinates (X, Y, Z and x, y), and approximate sRGB color representation. The 
predicted fluorescence color of 4CzIPN is blue-green, with a slight shift towards green in toluene, providing some indication of solvent-dependent color 
tunability.}
{\scriptsize
\begin{tabular}{m{1.2cm}m{1.8cm}@{\,}l*{3}{@{\,}c@{\,}}>{\columncolor{white}[0\tabcolsep][0pt]}c}
\toprule
&& Properties & {$\lambda_{max}(\unit{\nano\meter})$} & {$(X,Y,Z)$} & {$(x,y)$} & {(R,G,B)} \\
\midrule
\multirow{4}{=}{In vacuum} & \multirow{2}{=}{\emph{UV-vis} absorption} & \stda
&$375.9277$&$(622.080647,17.897209,2933.099378)$&$(0.1741022112,0.0050089062)$& \CcelCo{white}{46,0,255}\\
&& \stddft &$379.3349$&$(860.079598,24.551477,4060.359190)$ & $(0.1739294825,0.0049649192)$&\CcelCo{white}{46,0,255}\\
&\multirow{2}{=}{Fluorescence} & \stda
&$499.9663$&$(6585.587764,172561.782844,143421.052287)$&$(0.0204160956,0.5349617960)$&\CcelCo{black}{0,255,173}\\
&& \stddft &$504.2363$&$(5962.020393,189341.802208,104255.474648)$&$(0.0199026385,0.6320678542)$&\CcelCo{black}{0,255,97}\\
\midrule
\multirow{4}{=}{In toluene\\ solvent} & \multirow{2}{=}{\emph{UV-vis} absorption} & \stda
&$382.3646$&$(1268.308978,35.989911,5994.171115)$&$(0.1737773776,0.0049311583)$&\CcelCo{white}{45,0,255}\\
&& \stddft &$385.7148$&$(1761.979572,49.710040,8337.538430)$&$(0.1736072502,0.0048979134)$&\CcelCo{white}{44,0,255}\\
&\multirow{2}{=}{Fluorescence} & \stda
&$515.9937$&$(24222.155843,293583.919714,52882.972765)$&$(0.0653435972,0.7919951265)$&\CcelCo{black}{0,255,0}\\
&& \stddft &$520.1063$&$(34912.312986,302858.762182,37484.335818)$&$(0.0930361348,0.8070736712)$&\CcelCo{black}{0,255,0}\\
\bottomrule
\end{tabular}}
\label{tab:Color4CzIPN}
\end{table}


\begin{figure}[!htbp]
\centering
\leavevmode
\subfloat[In vacuum: UV-Vis 
spectrum]{\includegraphics[width=0.4\textwidth]{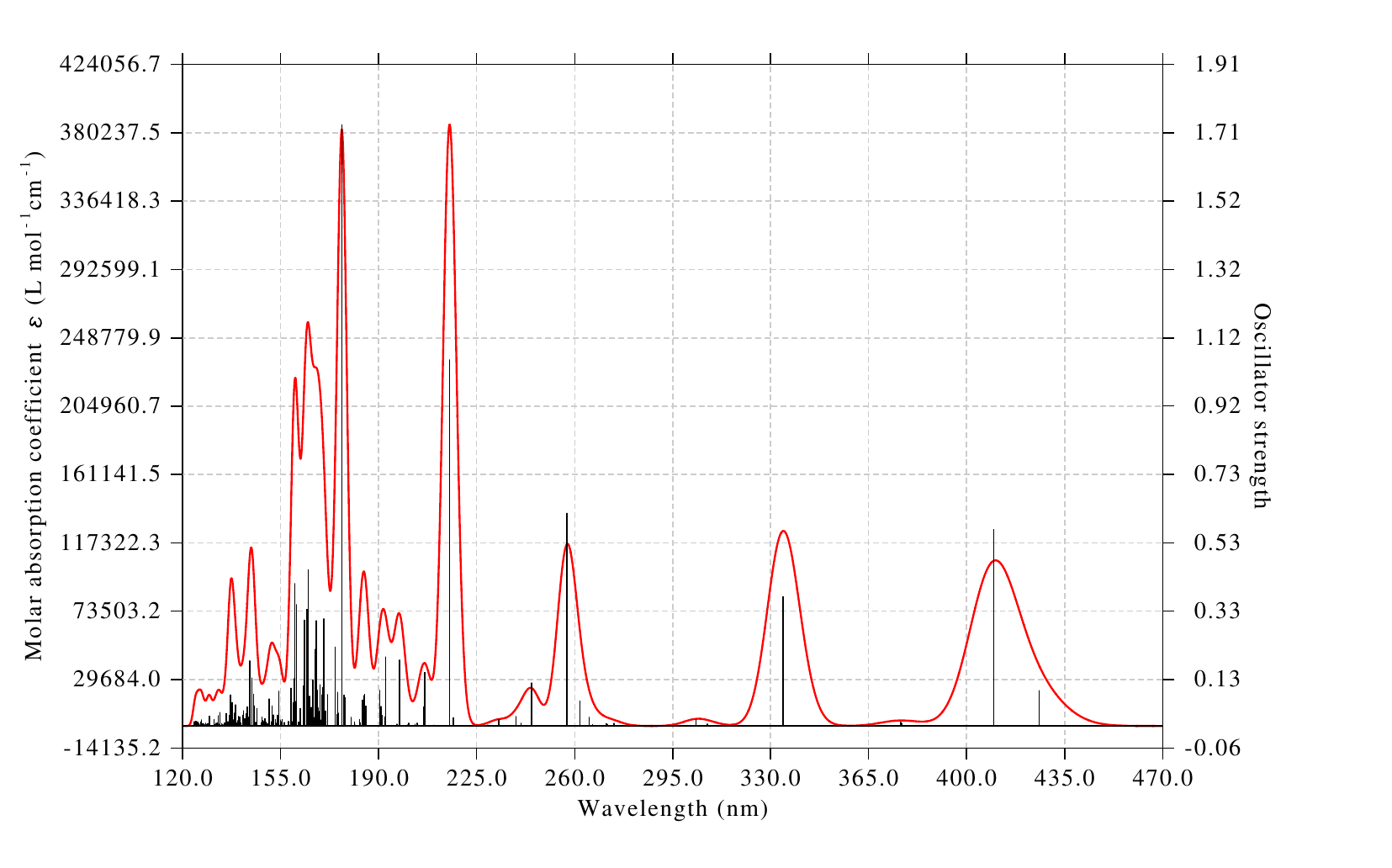}\includegraphics[width=0.4\textwidth]{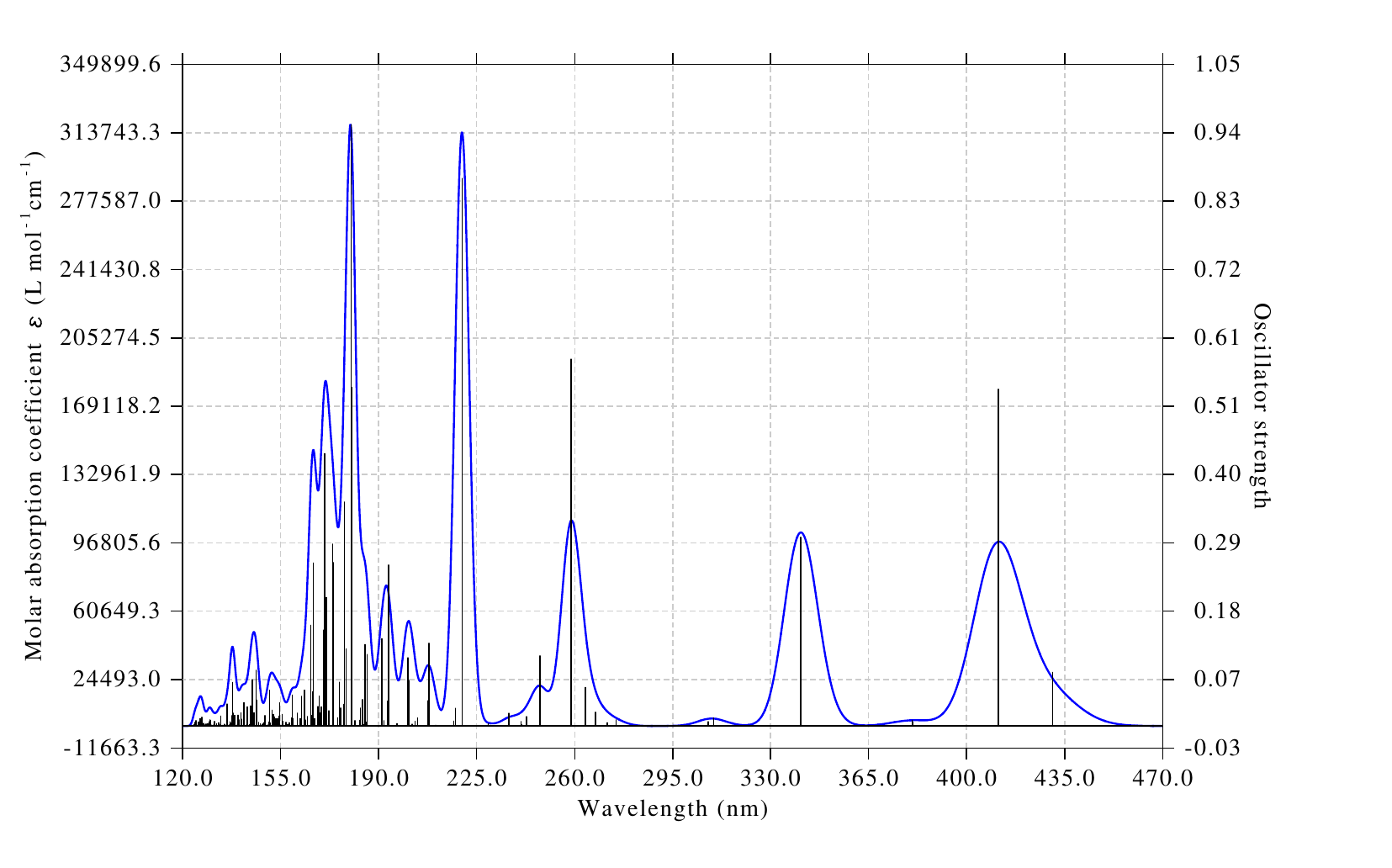}}\\
\subfloat[In toluene solvent: UV-Vis 
spectrum]{\includegraphics[width=0.4\textwidth]{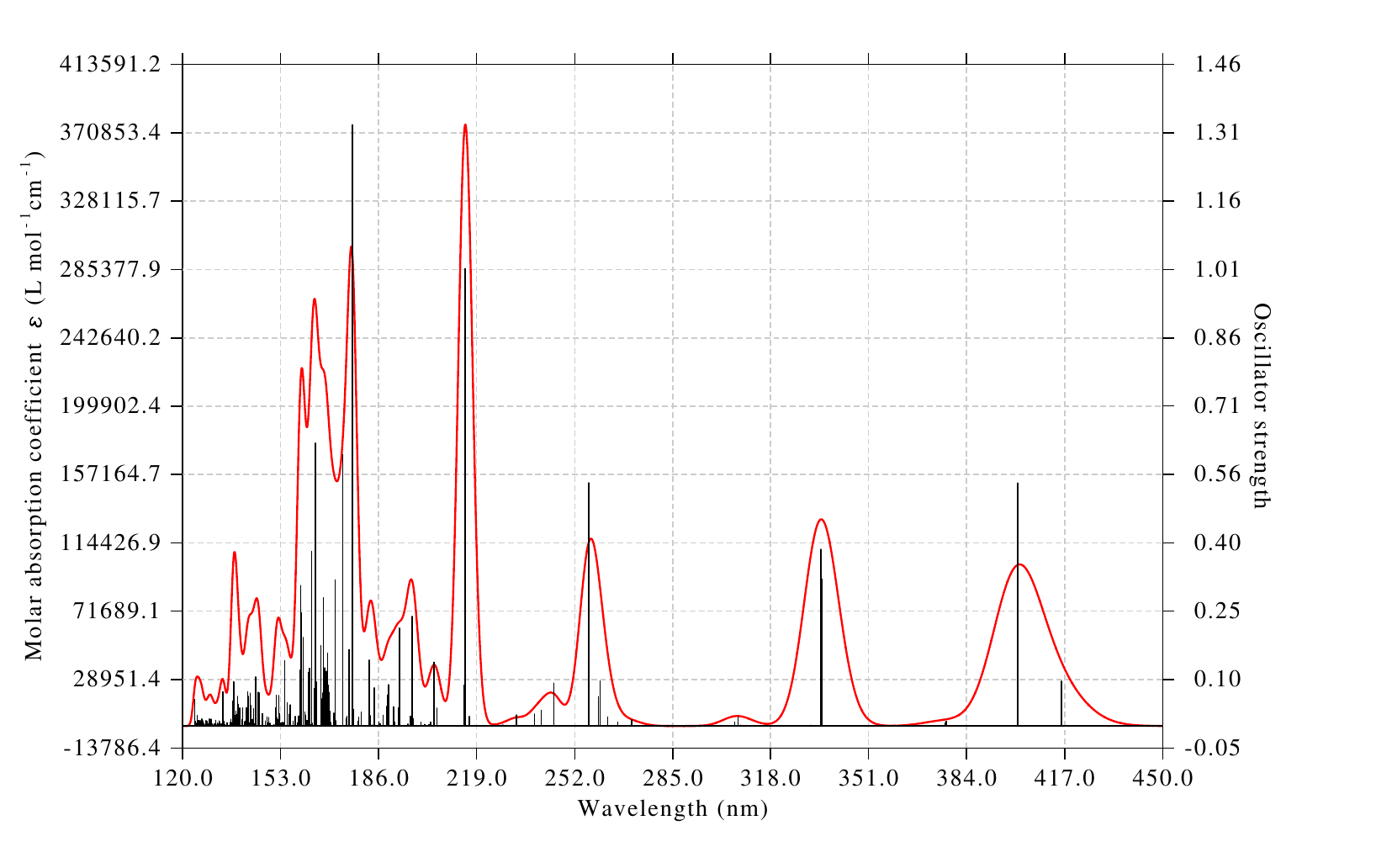}\includegraphics[width=0.4\textwidth]{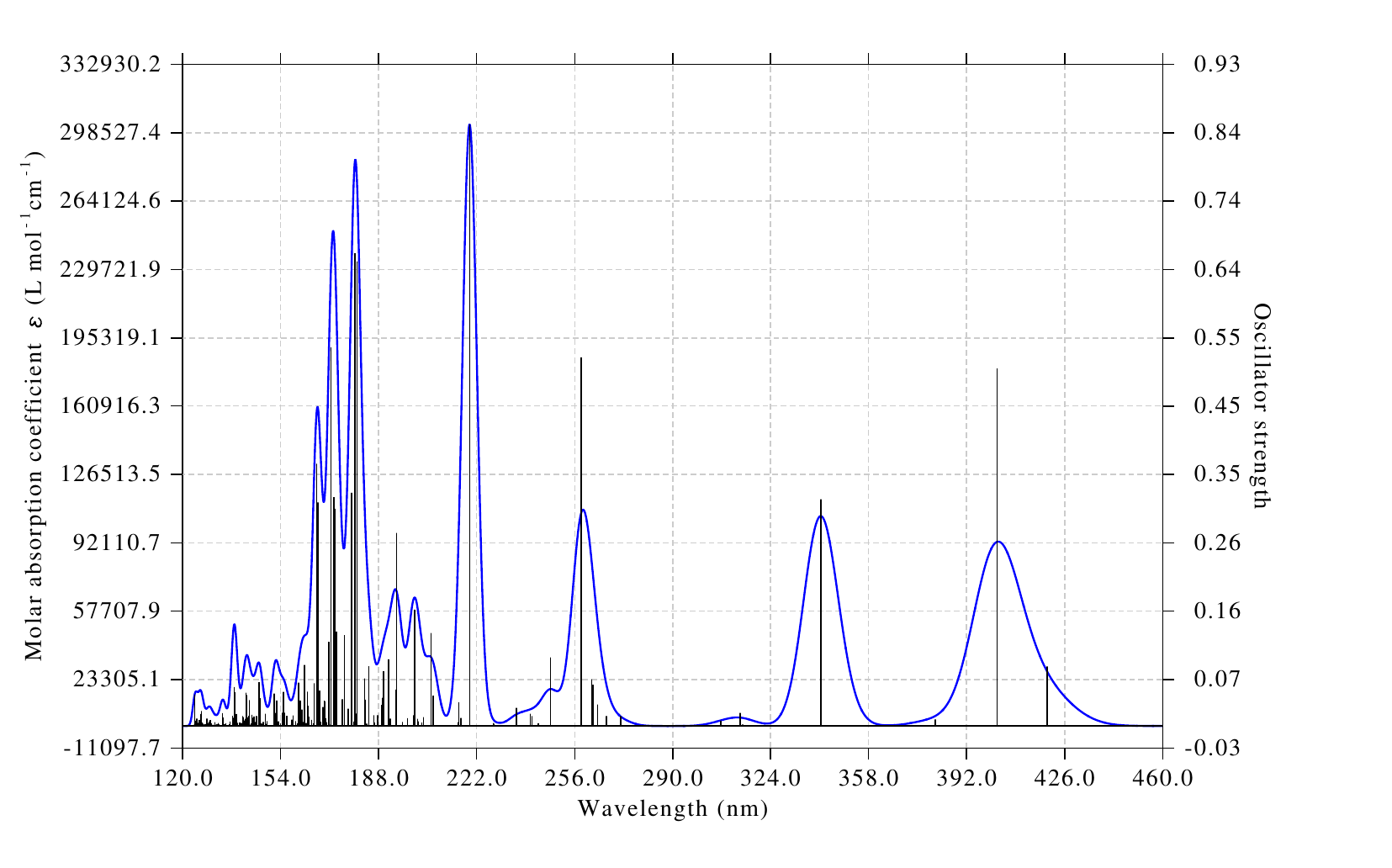}
}
\caption{UV-Vis absorption spectra of Px2BP: solvent and method dependence. Panels show the simulated UV-Vis absorption spectra in vacuum and toluene, 
calculated with \stda (top) and \stddft (bottom) methods. The absorption spectra of Px2BP exhibit a well-defined peak at around 420 nm, indicating a relatively 
strong absorption in the blue-violet region.}
\end{figure}

\begin{figure}[!htbp]
\centering
\leavevmode
\subfloat[In vacuum: ECD 
spectrum]{\includegraphics[width=0.4\textwidth]{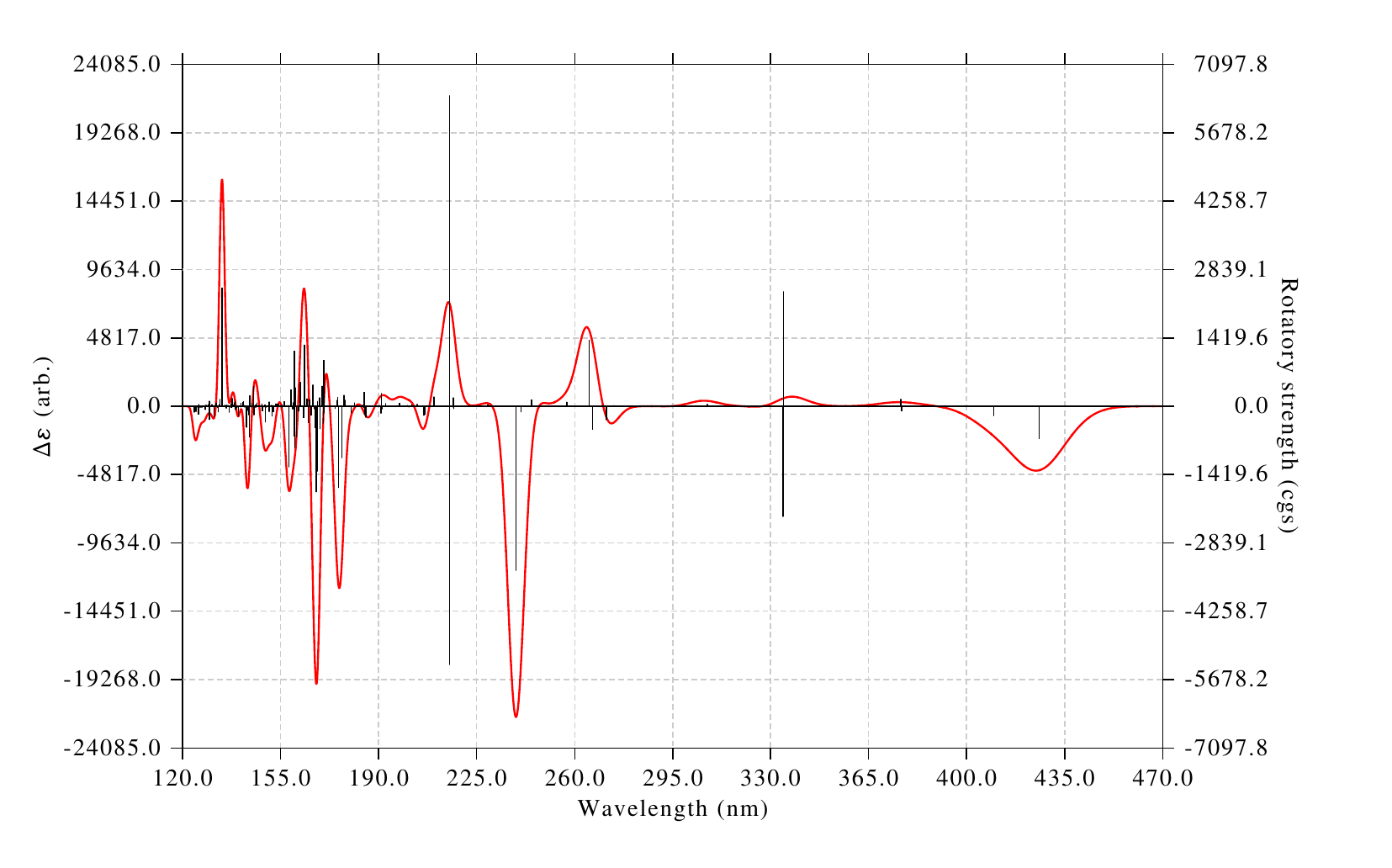}\includegraphics[width=0.4\textwidth]{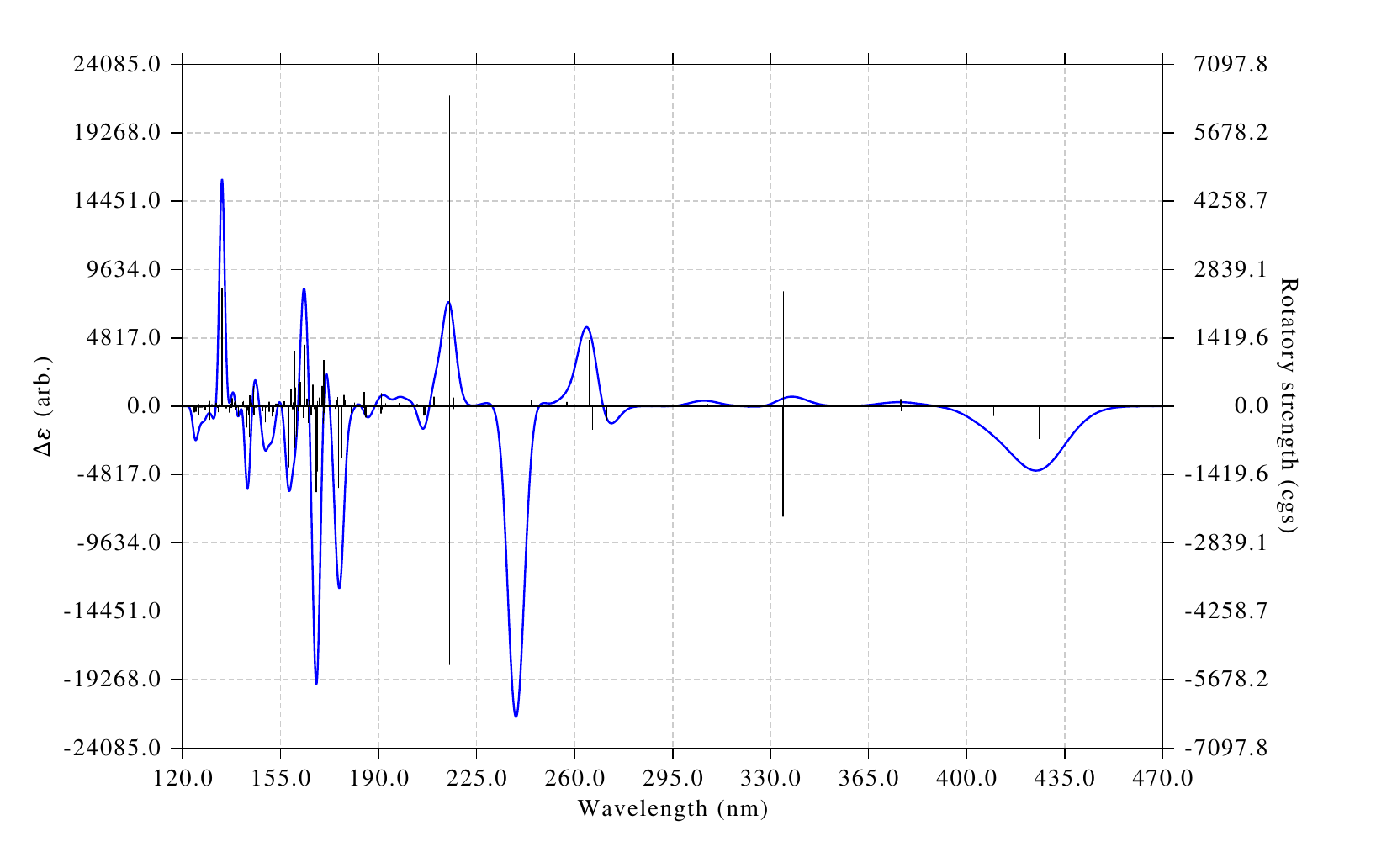}}\\
\subfloat[In toluene solvent: ECD 
spectrum]{\includegraphics[width=0.4\textwidth]{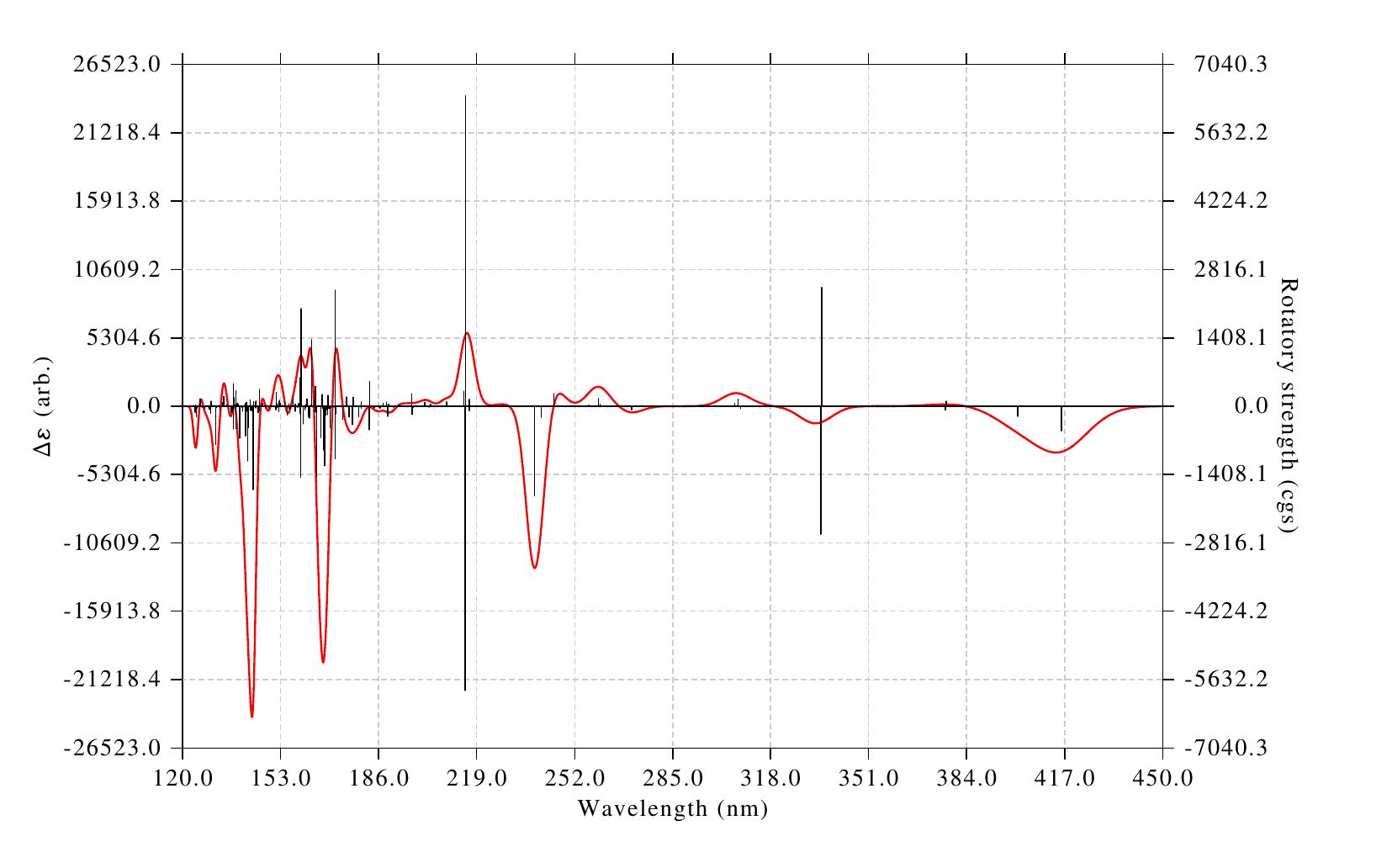}\includegraphics[width=0.4\textwidth]{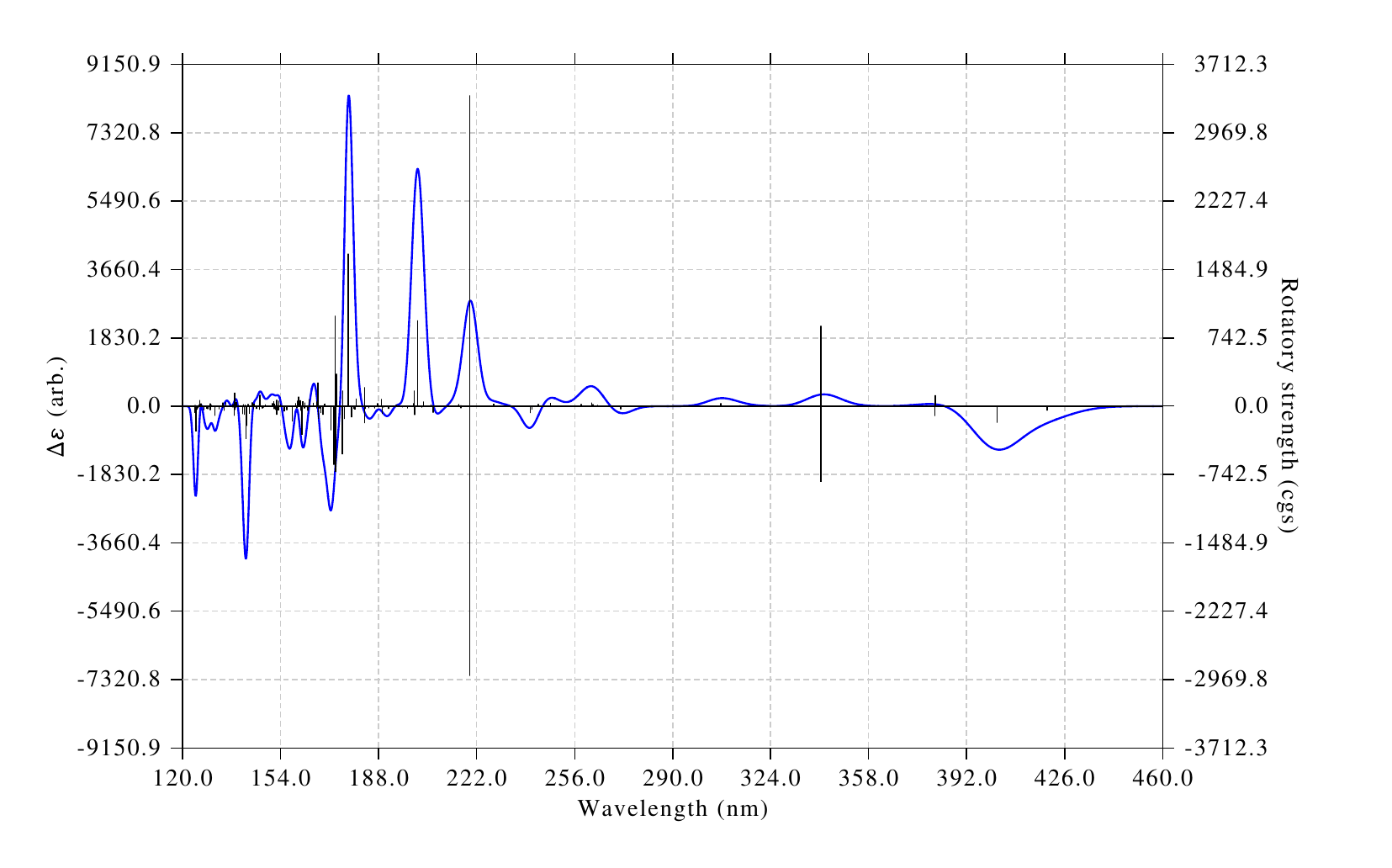}}
\caption{Electronic Circular Dichroism (ECD) spectra of Px2BP: solvent and method dependence. Panels show the simulated ECD spectra in vacuum and toluene, 
calculated with \stda (top) and \stddft (bottom) methods. The ECD spectra of Px2BP display a clear pattern of alternating positive and negative bands, 
suggesting the presence of chiral conformations.}
\end{figure}

\begin{table}[!htbp]
\centering
\caption{Colorimetric properties of Px2BP: wavelength maxima and CIE coordinates. This table summarizes the key colorimetric properties of Px2BP, including the 
wavelengths of maximum absorption and emission, CIE 1931 color space coordinates (X, Y, Z and x, y), and approximate sRGB color representation. The predicted 
emission color of Px2BP shifts from green to yellow-green upon solvation, indicating a potential for solvent-dependent color tuning.}
{\scriptsize
\begin{tabular}{m{1.2cm}m{1.8cm}@{\,}l*{3}{@{\,}c@{\,}}>{\columncolor{white}[0\tabcolsep][0pt]}c}
\toprule
&& Properties & {$\lambda_{max}(\unit{\nano\meter})$} & {$(X,Y,Z)$} & {$(x,y)$} & {(R,G,B)} \\
\midrule
\multirow{4}{=}{In vacuum} & \multirow{2}{=}{\emph{UV-vis} absorption} & \stda
&$425.9260$&$(95818.465009,4481.356426,470347.964126)$&$(0.1679117442,0.0078531040)$& \CcelCo{white}{37,0,255}\\
&& \stddft &$430.6768$&$(97190.744443,5383.702553,481807.676795)$ & $(0.1663136850,0.0092126407)$&\CcelCo{white}{35,0,255}\\
&\multirow{2}{=}{Fluorescence} & \stda
&$522.9441$&$(48648.024902,324507.000941,32903.858353)$&$(0.1198053455,0.7991624209)$&\CcelCo{black}{0,255,0}\\
&& \stddft &$527.8421$&$(58938.771173,299852.961542,20277.384442)$&$(0.1554829146,0.7910245071)$&\CcelCo{black}{0,255,0}\\
\midrule
\multirow{4}{=}{In toluene\\ solvent} & \multirow{2}{=}{\emph{UV-vis} absorption} & \stda
&$415.8868$&$(46737.618980,1647.272467,226053.988430)$&$(0.1703024695,0.0060023291)$&\CcelCo{white}{41,0,255}\\
&& \stddft &$419.9755$&$(54038.101184,2108.192111,262733.002632)$&$(0.1694625580,0.0066112543)$&\CcelCo{white}{40,0,255}\\
&\multirow{2}{=}{Fluorescence} & \stda
&$548.1411$&$(168034.392880,390818.813123,1303.913773)$&$(0.2999772509,0.6976949847)$&\CcelCo{black}{0,255,0}\\
&& \stddft &$552.5381$&$(167125.345717,336556.849028,2644.073115)$&$(0.3300744131,0.6647035131)$&\CcelCo{black}{17,255,0}\\
\bottomrule
\end{tabular}}
\label{tab:ColorPx2BP}
\end{table}


\begin{figure}[!htbp]
\centering
\leavevmode
\subfloat[In vacuum: UV-Vis 
spectrum]{\includegraphics[width=0.4\textwidth]{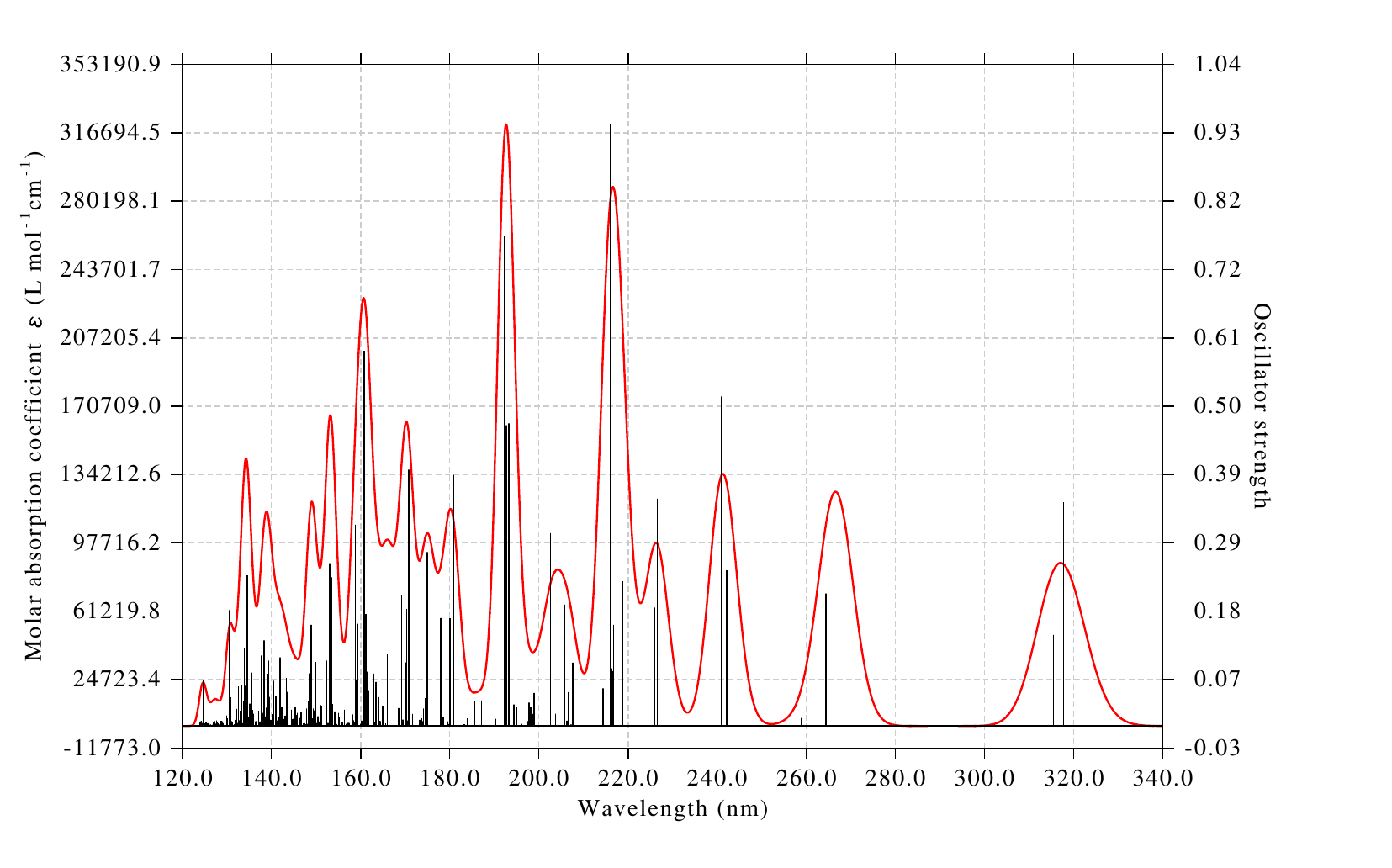}\includegraphics[width=0.4\textwidth]{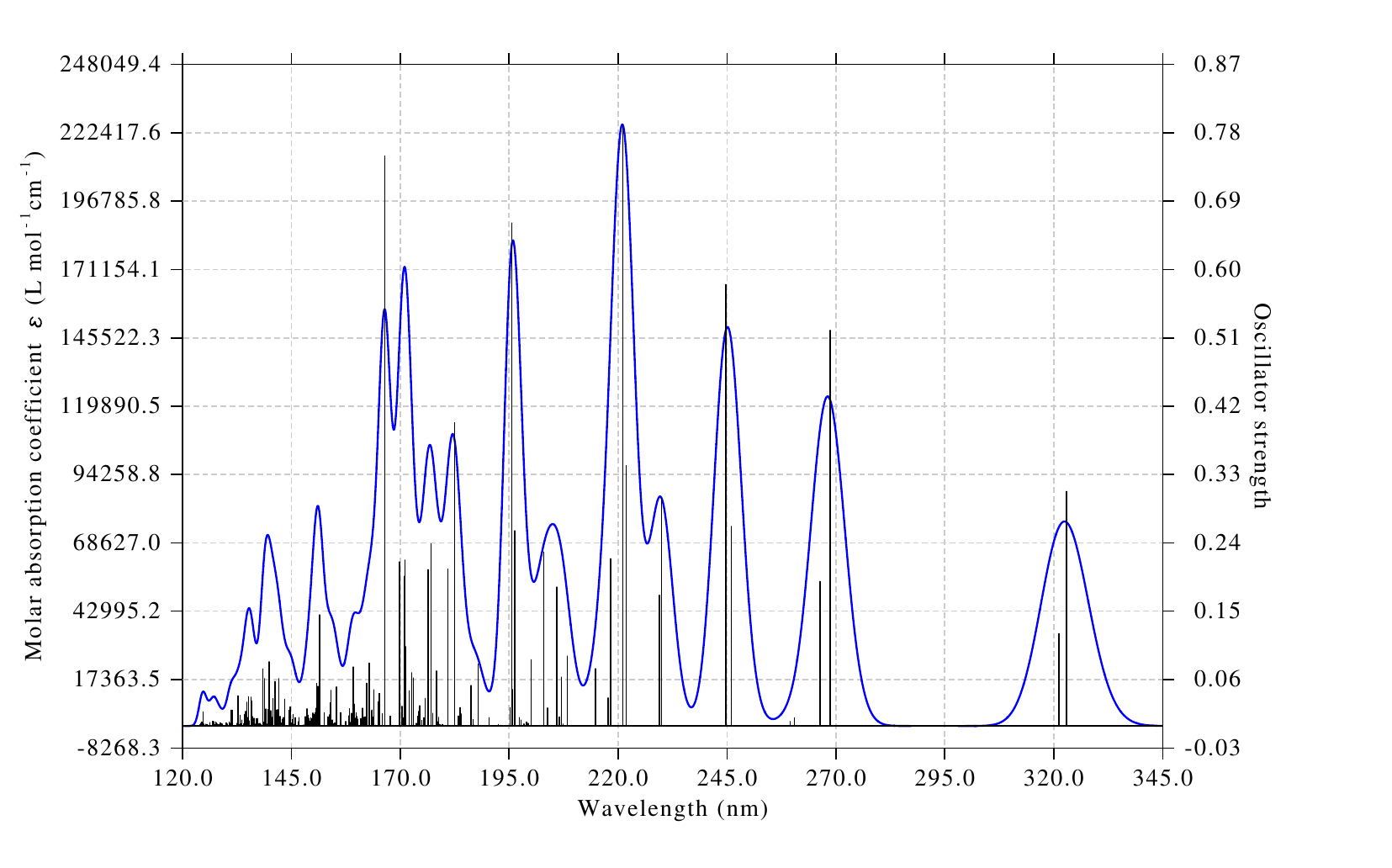}}\\
\subfloat[In toluene solvent: UV-Vis 
spectrum]{\includegraphics[width=0.4\textwidth]{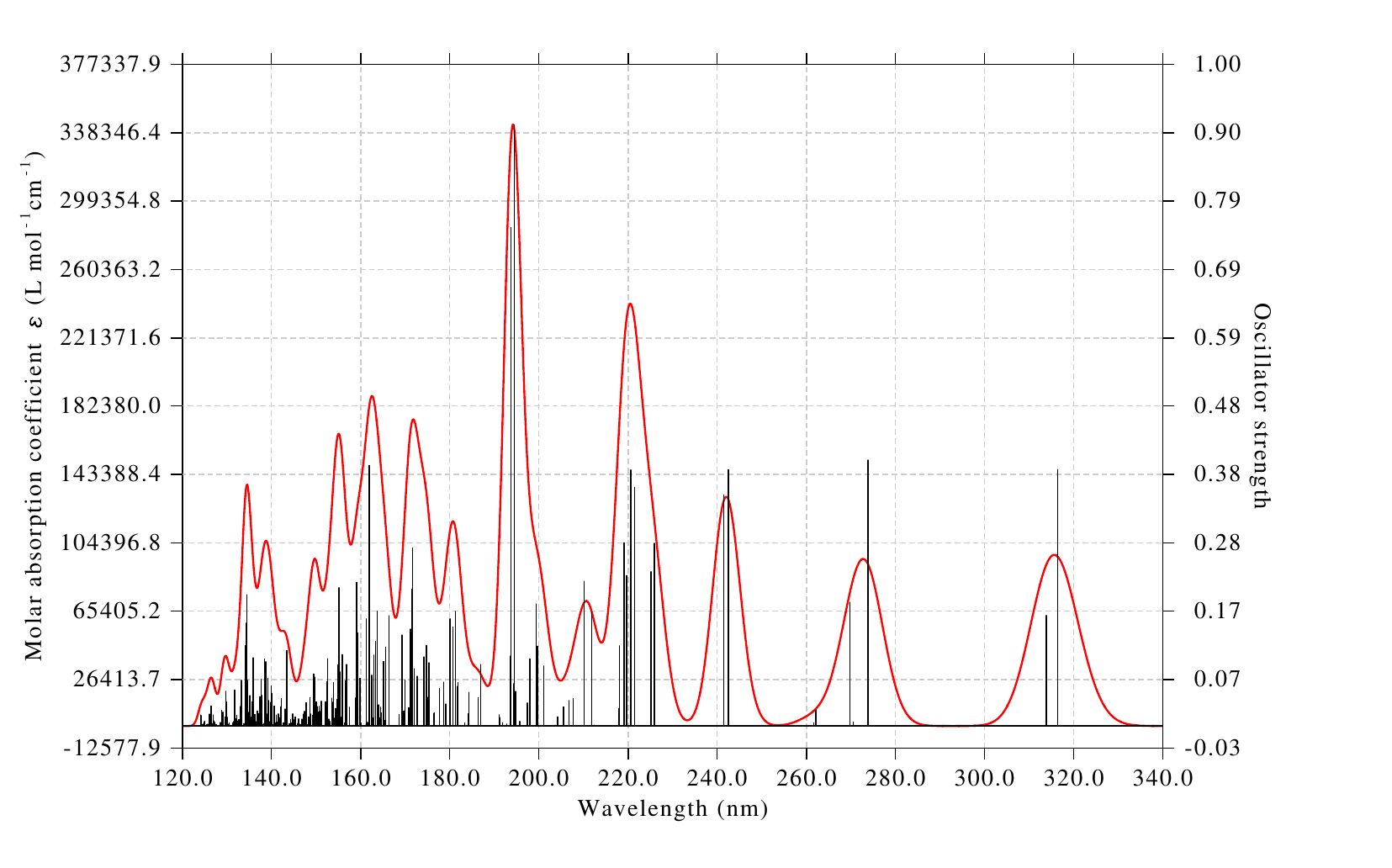}\includegraphics[width=0.4\textwidth]{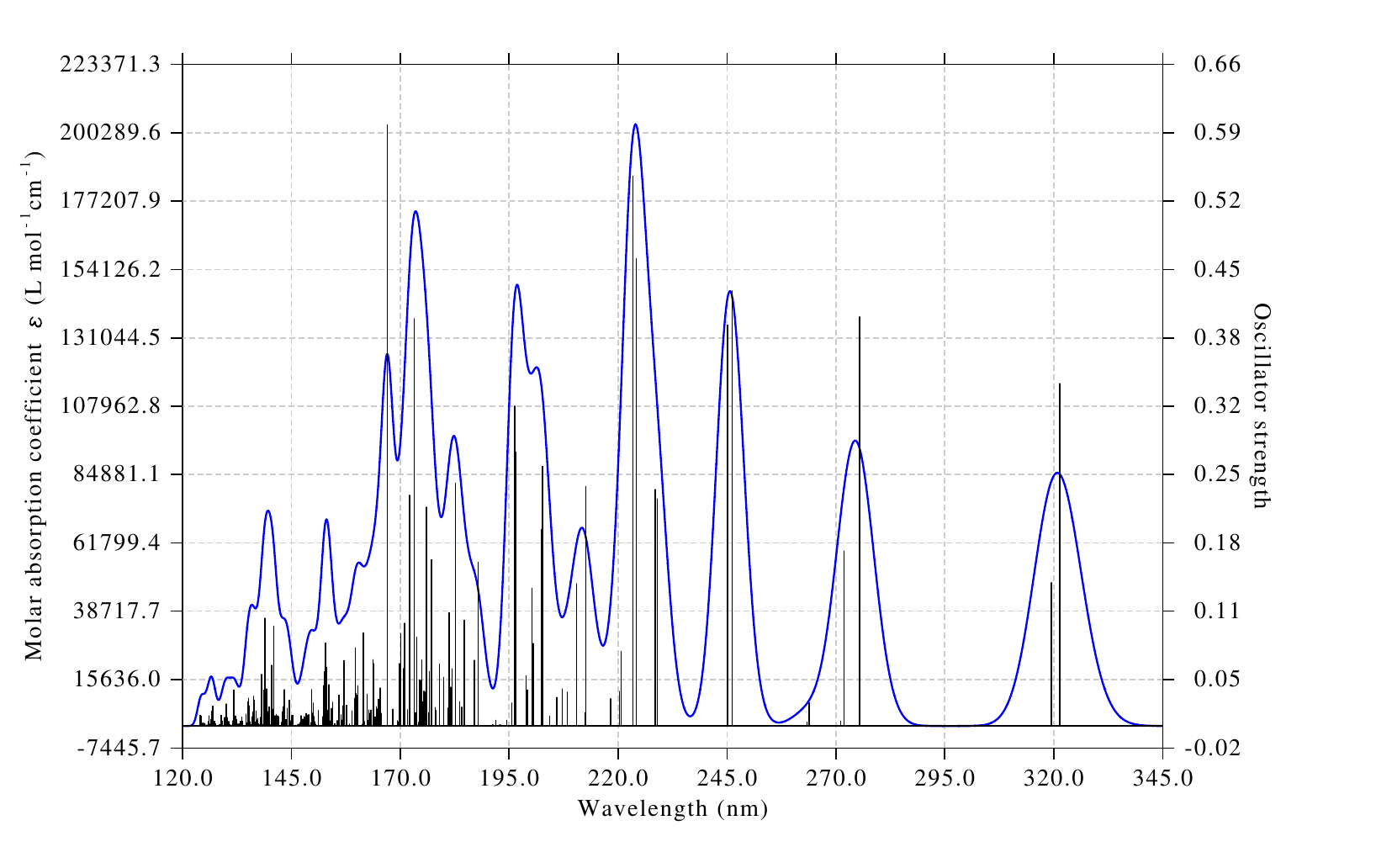}}
\caption{UV-Vis absorption spectra of CzS2: solvent and method dependence. Panels show the simulated UV-Vis absorption spectra in vacuum and toluene, calculated 
with \stda (top) and \stddft (bottom) methods. The CzS2 absorption spectra exhibit a shoulder peak in the high-energy region, potentially indicating the 
presence of multiple electronic transitions.}
\end{figure}

\begin{figure}[!htbp]
\centering
\leavevmode
\subfloat[In vacuum: ECD 
spectrum]{\includegraphics[width=0.4\textwidth]{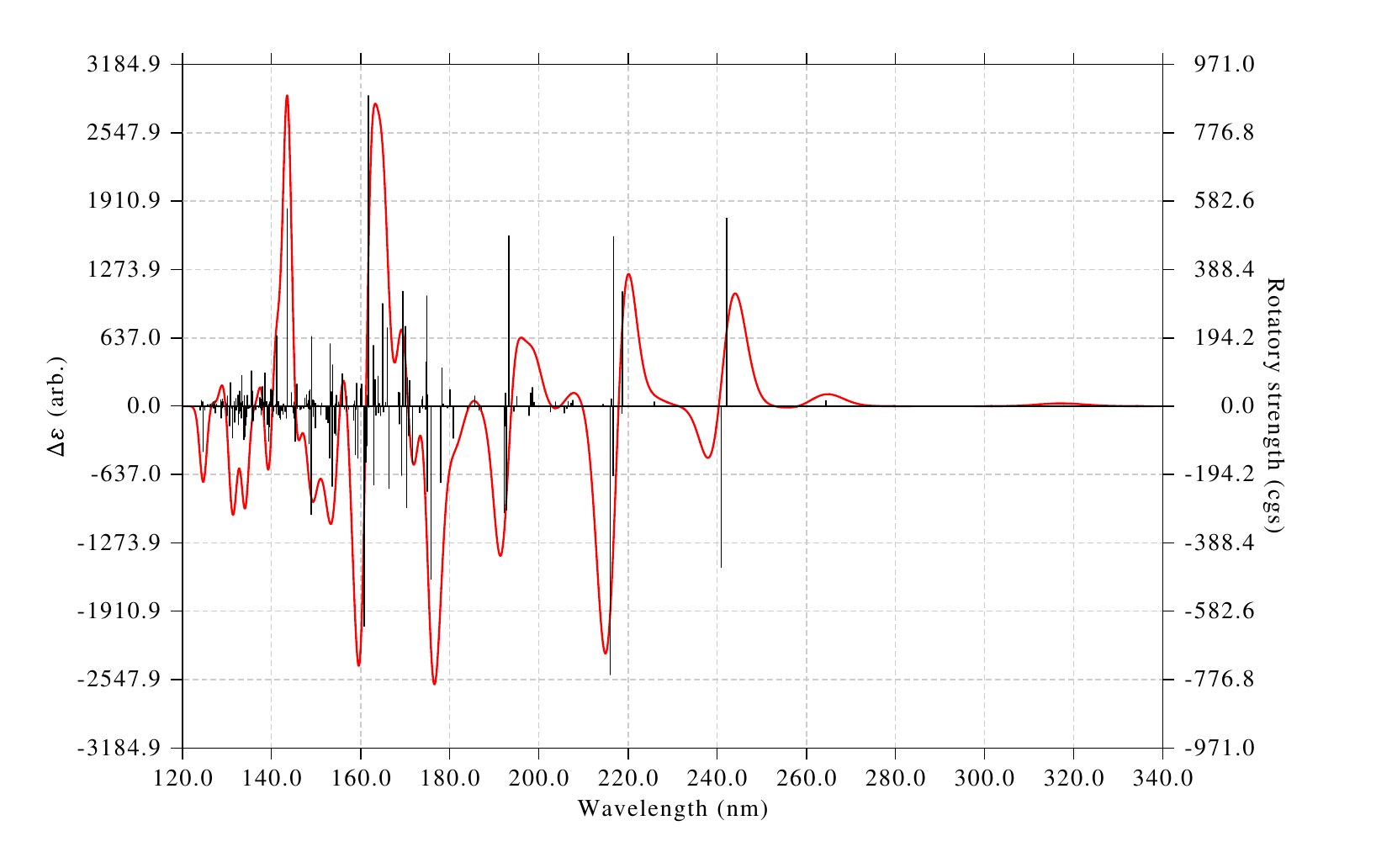}\includegraphics[width=0.4\textwidth]{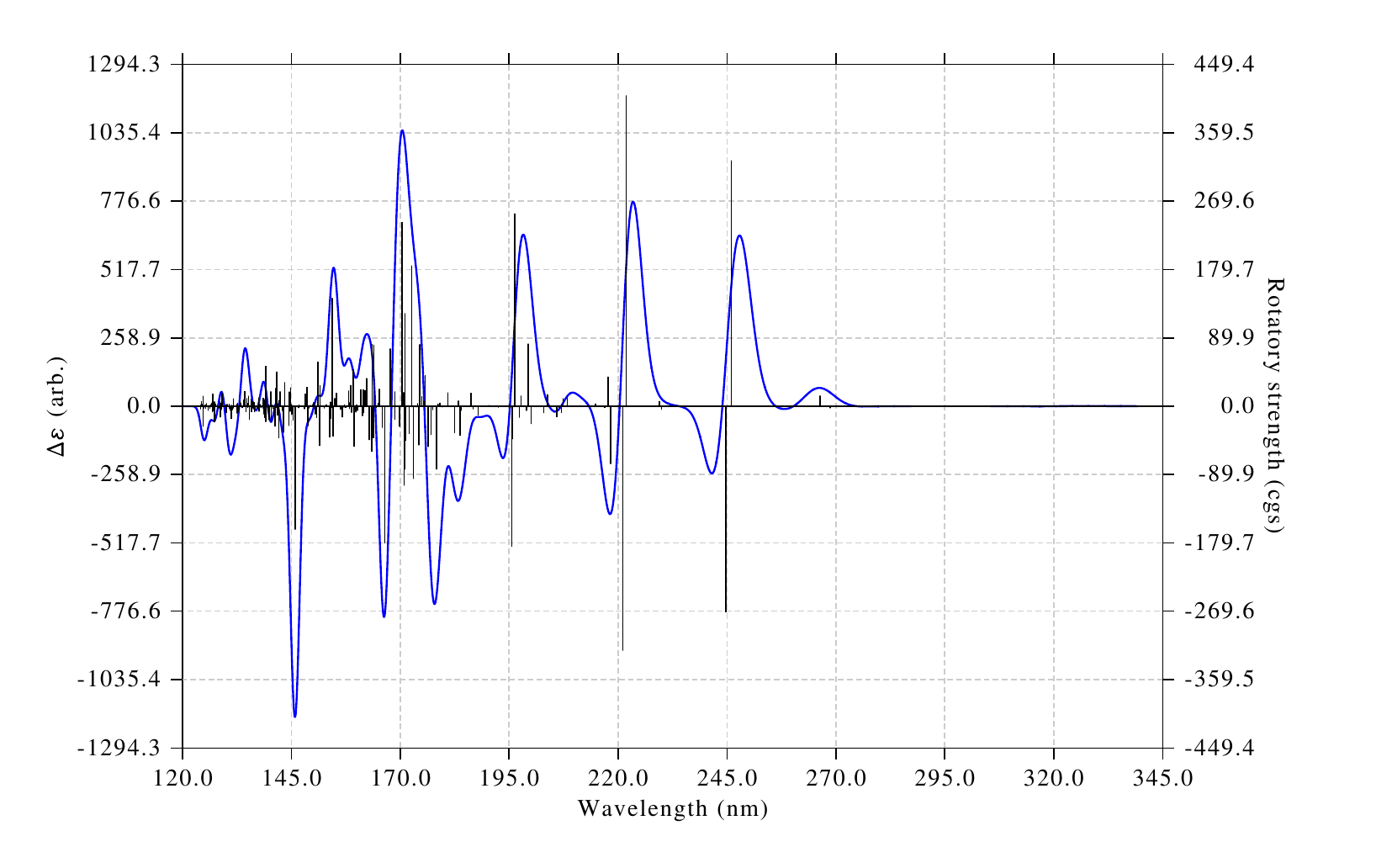}}\\
\subfloat[In toluene solvent: ECD 
spectrum]{\includegraphics[width=0.4\textwidth]{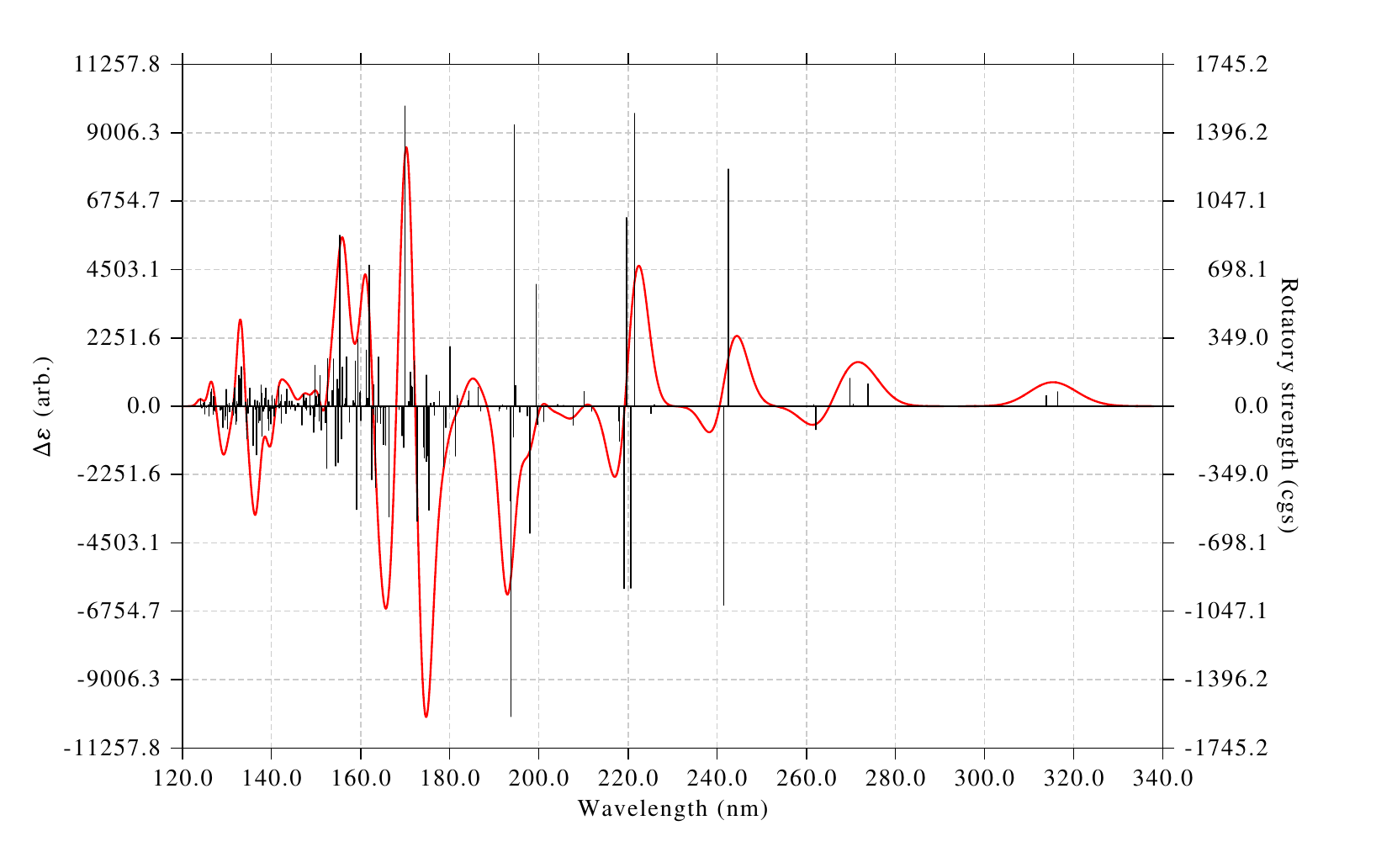}\includegraphics[width=0.4\textwidth]{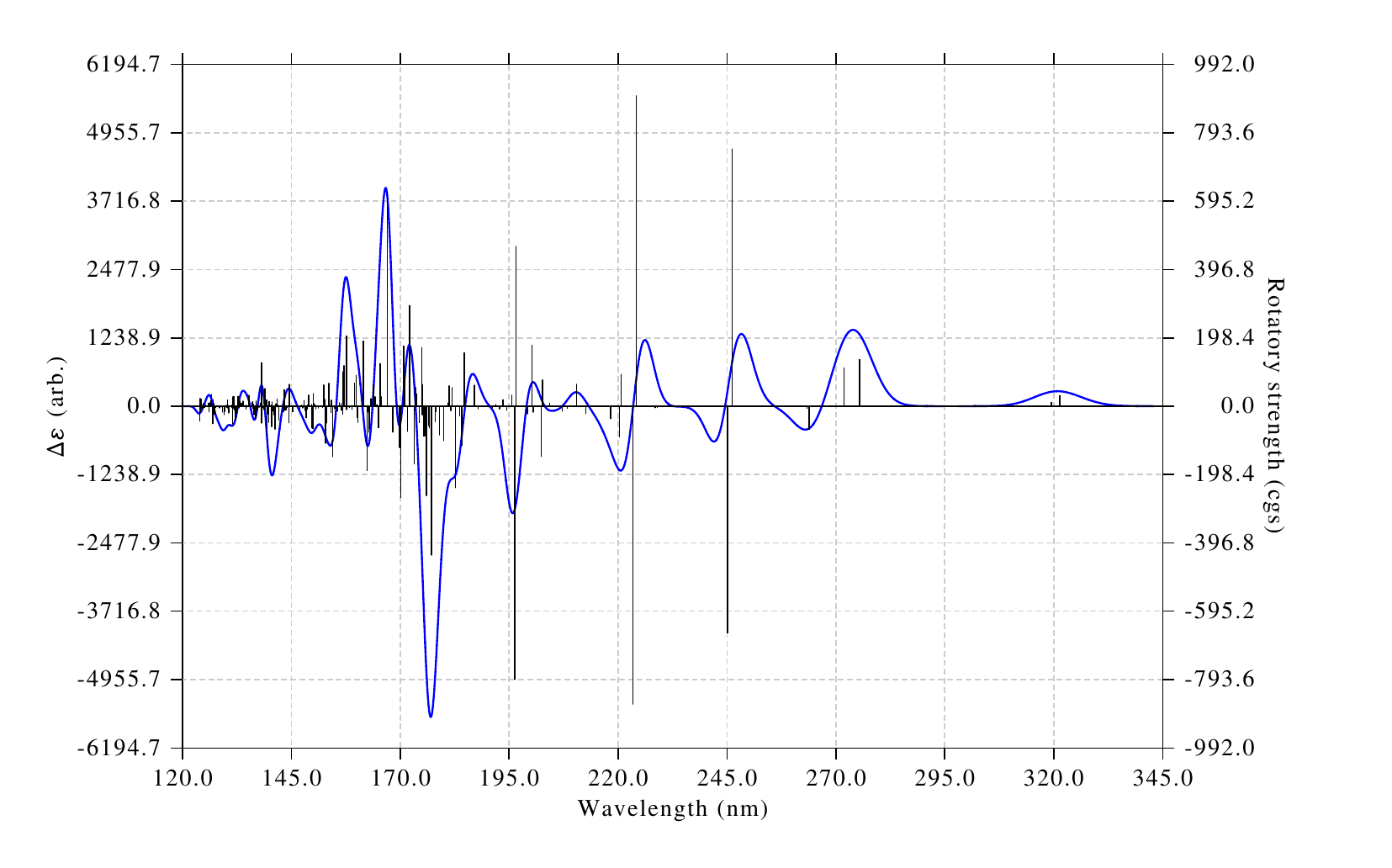}}
\caption{Electronic Circular Dichroism (ECD) spectra of CzS2: solvent and method dependence. Panels show the simulated ECD spectra in vacuum and toluene, 
calculated with \stda (top) and \stddft (bottom) methods. The ECD spectra of CzS2 display a clear pattern of alternating positive and negative bands, suggesting 
the presence of chiral conformations and strong electronic coupling between the carbazole units.}
\end{figure}

\begin{table}[!htbp]
\centering
\caption{Colorimetric properties of CzS2: wavelength maxima and CIE coordinates. This table summarizes the key colorimetric properties of CzS2, including the 
wavelengths of maximum absorption and emission, CIE 1931 color space coordinates (X, Y, Z and x, y), and approximate sRGB color representation. The predicted 
emission color is deep-blue. The low Y values suggest very little visible emission. The color stays roughly the same, and is more intense when a vacuum is 
used.}
{\scriptsize
\begin{tabular}{m{1.2cm}m{1.8cm}@{\,}l*{3}{@{\,}c@{\,}}>{\columncolor{white}[0\tabcolsep][0pt]}c}
\toprule
&& Properties & {$\lambda_{max}(\unit{\nano\meter})$} & {$(X,Y,Z)$} & {$(x,y)$} & {(R,G,B)} \\
\midrule
\multirow{4}{=}{In vacuum} & \multirow{2}{=}{\emph{UV-vis} absorption} & \stda
&$317.7118$&NA&NA&NA\\
&& \stddft &$322.9288$&NA&NA&NA\\
&\multirow{2}{=}{Fluorescence} & \stda
&$454.4447$&$(258349.493337,39781.120820,1414339.462559)$&$(0.15038636541,0.0232302575)$&\CcelCo{white}{13,0,255}\\
&& \stddft &$456.7887$&$(225964.230581,40319.400005,1262268.953105)$&$(0.1478288892, 0.0263775028)$&\CcelCo{white}{8,0,255}\\
\midrule
\multirow{4}{=}{In toluene\\ solvent} & \multirow{2}{=}{\emph{UV-vis} absorption} & \stda
&$316.4086$&NA&NA&NA\\
&& \stddft &$321.3936$&NA&NA&NA\\
&\multirow{2}{=}{Fluorescence} & \stda
&$435.5921$&$(289686.055708,16492.097125,1439309.486008)$&$(0.1659628228,0.0094484182)$&\CcelCo{white}{35,0,255}\\
&& \stddft &$451.2065$&$(266274.272154,33986.335253,1422808.541419)$&$(0.1545348730,0.0197243014)$&\CcelCo{white}{18,0,255}\\
\bottomrule
\end{tabular}}
\label{tab:ColorCzS2}
\end{table}


\begin{figure}[!htbp]
\centering
\leavevmode
\subfloat[In vacuum: UV-Vis 
spectrum]{\includegraphics[width=0.4\textwidth]{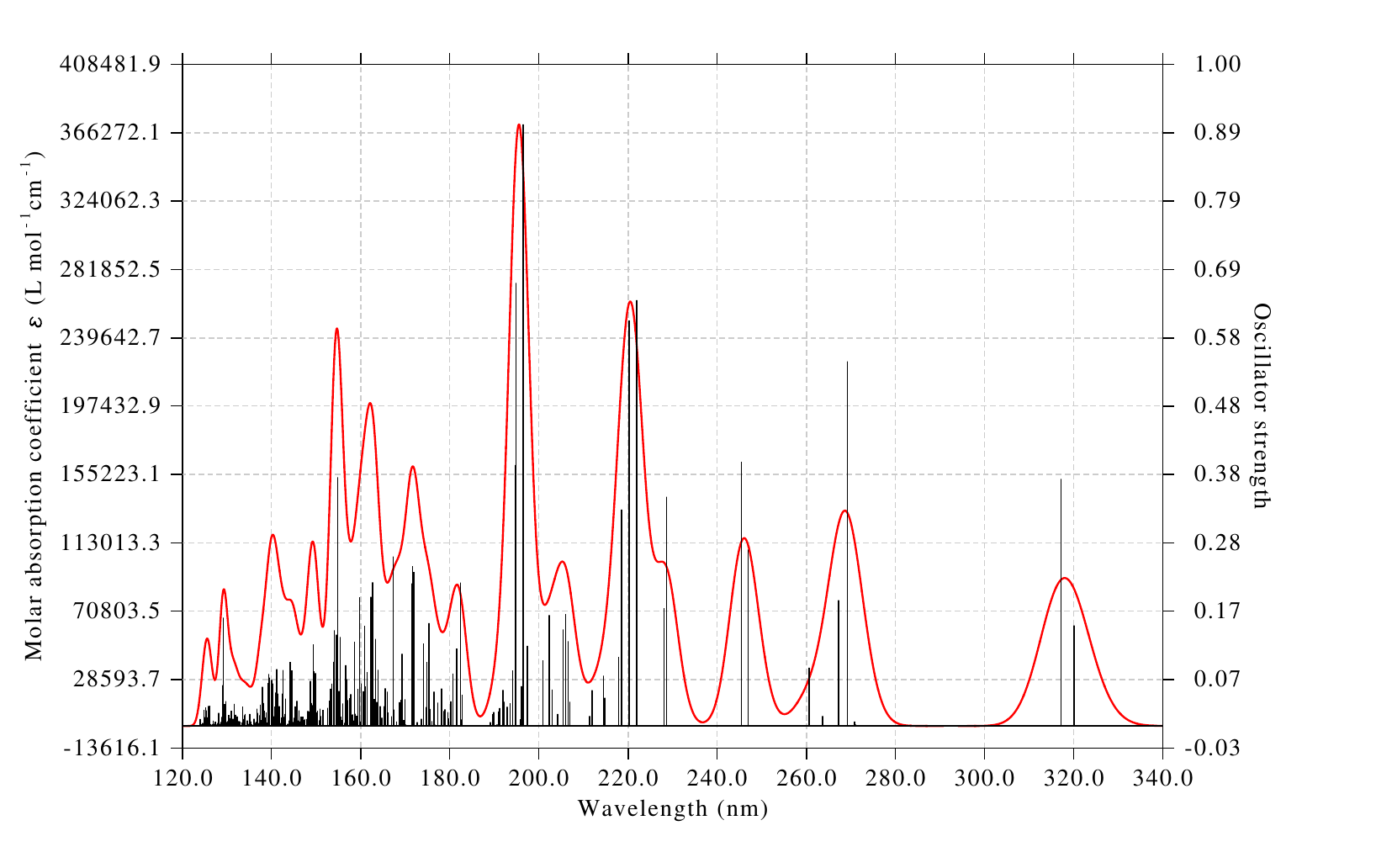}\includegraphics[width=0.4\textwidth]{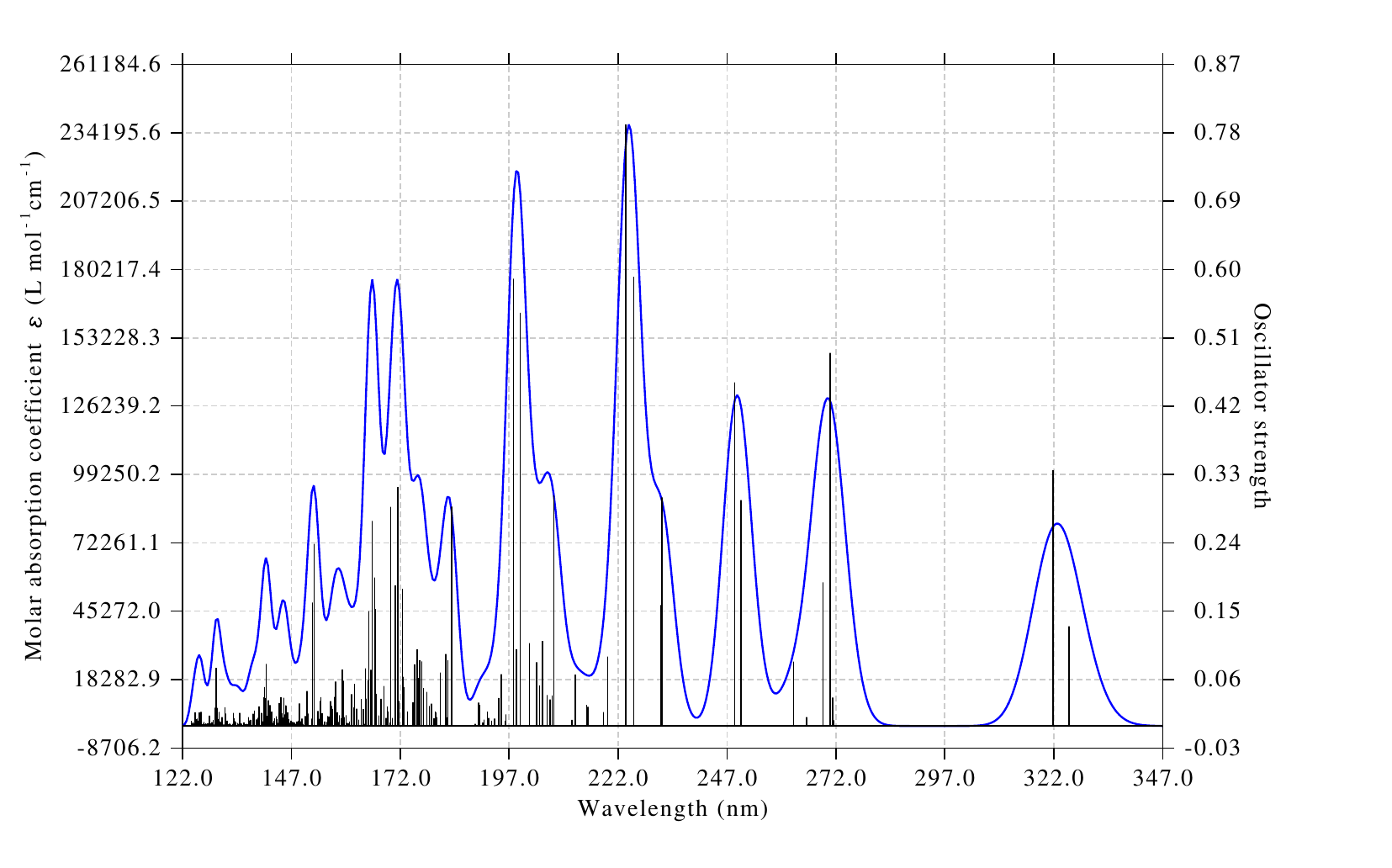}}\
\
\subfloat[In toluene solvent: UV-Vis 
spectrum]{\includegraphics[width=0.4\textwidth]{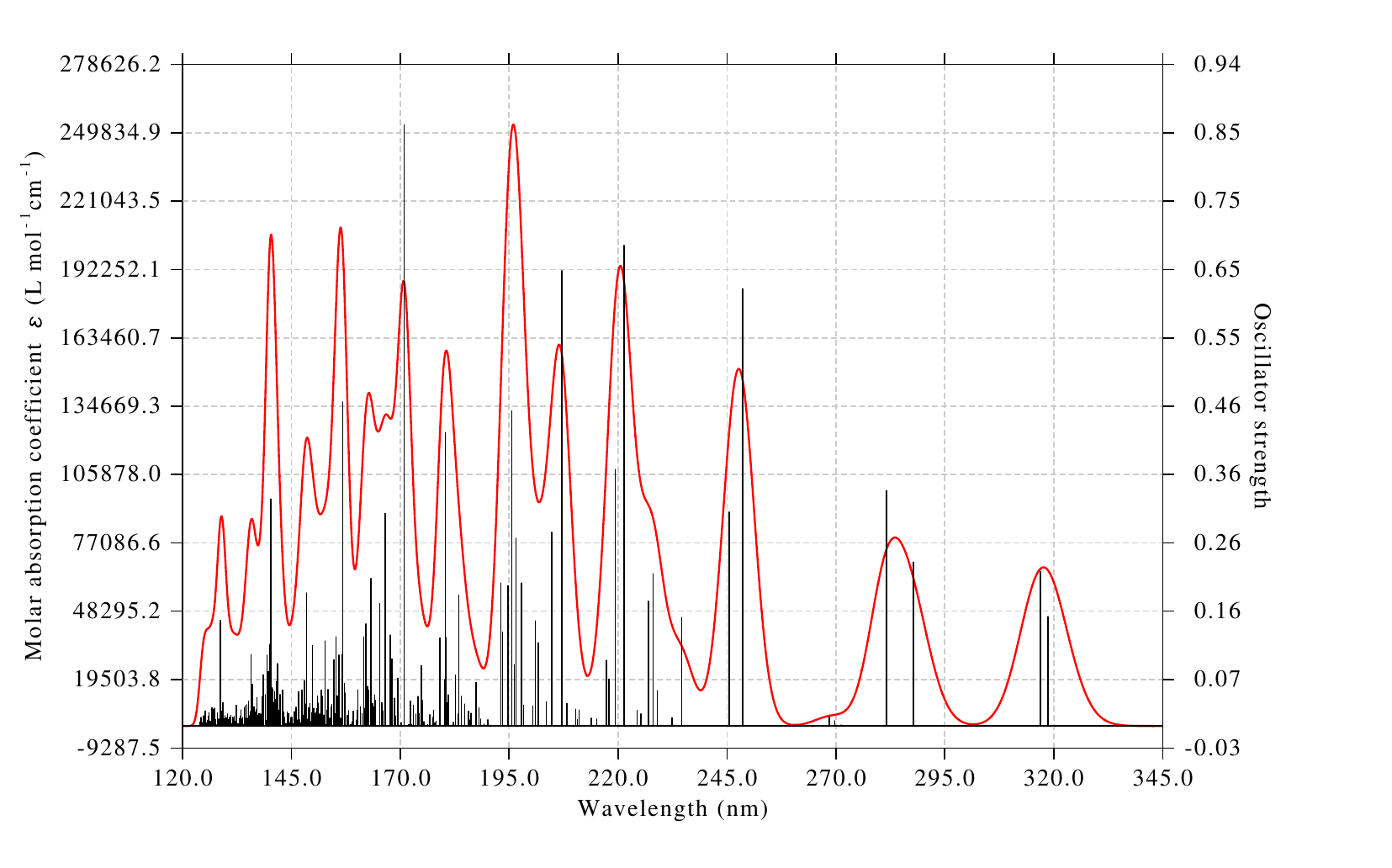}
\includegraphics[width=0.4\textwidth]{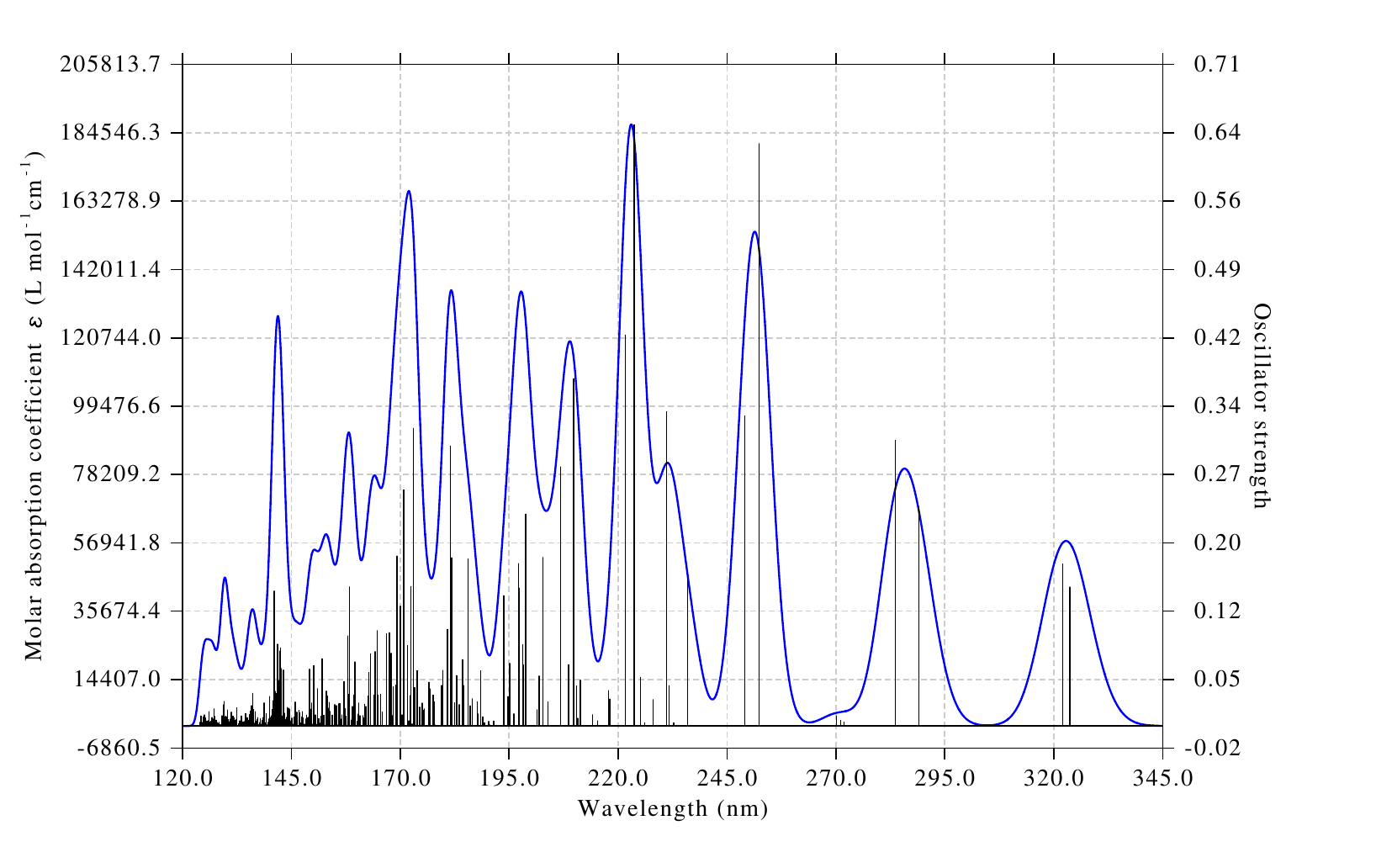}}
\caption{UV-Vis absorption spectra of 2TCz-DPS: solvent and method dependence. Panels show the simulated UV-Vis absorption spectra in vacuum and toluene, 
calculated with \stda (top) and \stddft (bottom) methods. A redshift in the main peak upon solvation is observed using both \stda and \stddft.}
\end{figure}

\begin{figure}[!htbp]
\centering
\leavevmode
\subfloat[In vacuum: ECD 
spectrum]{\includegraphics[width=0.4\textwidth]{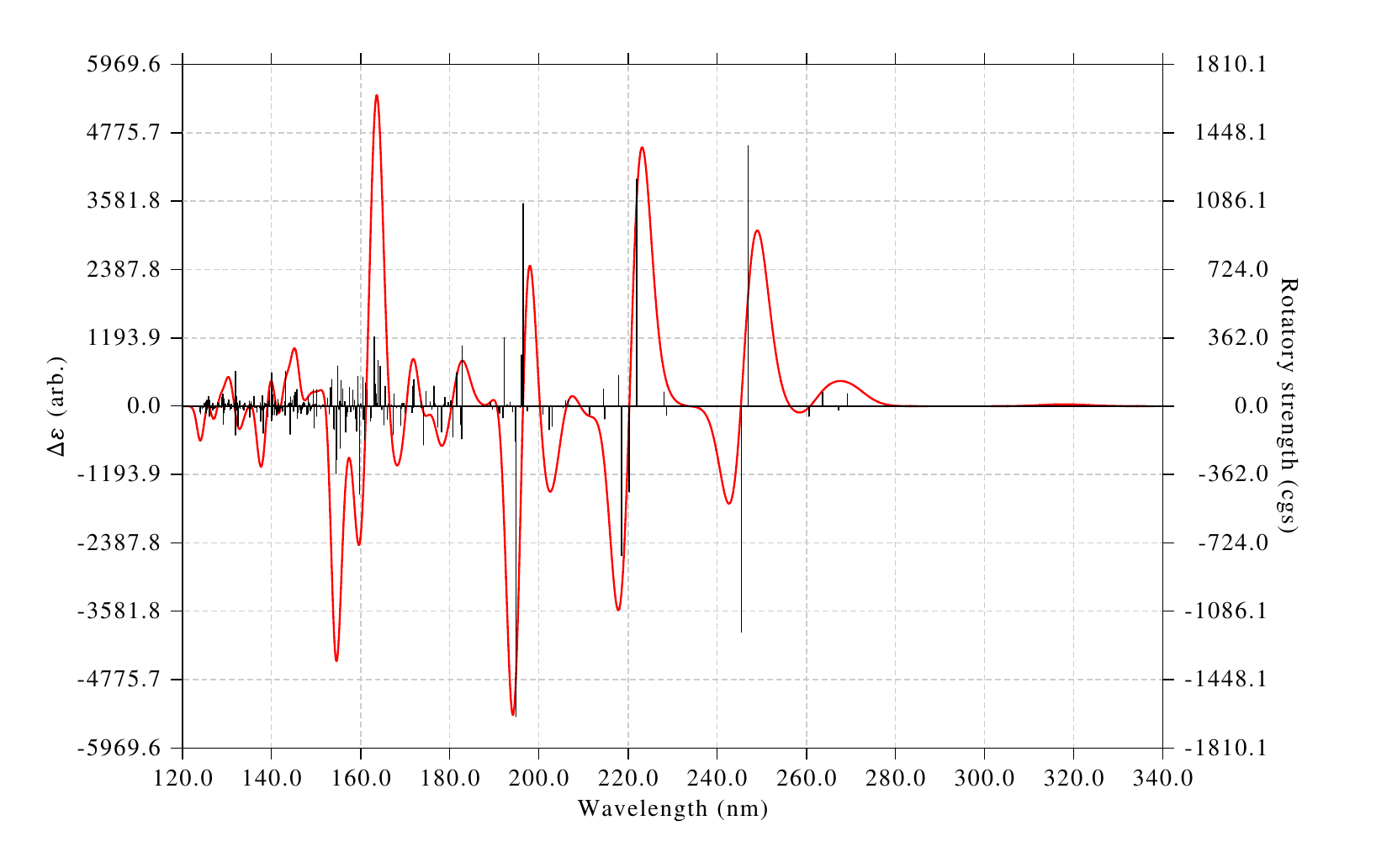}\includegraphics[width=0.4\textwidth]{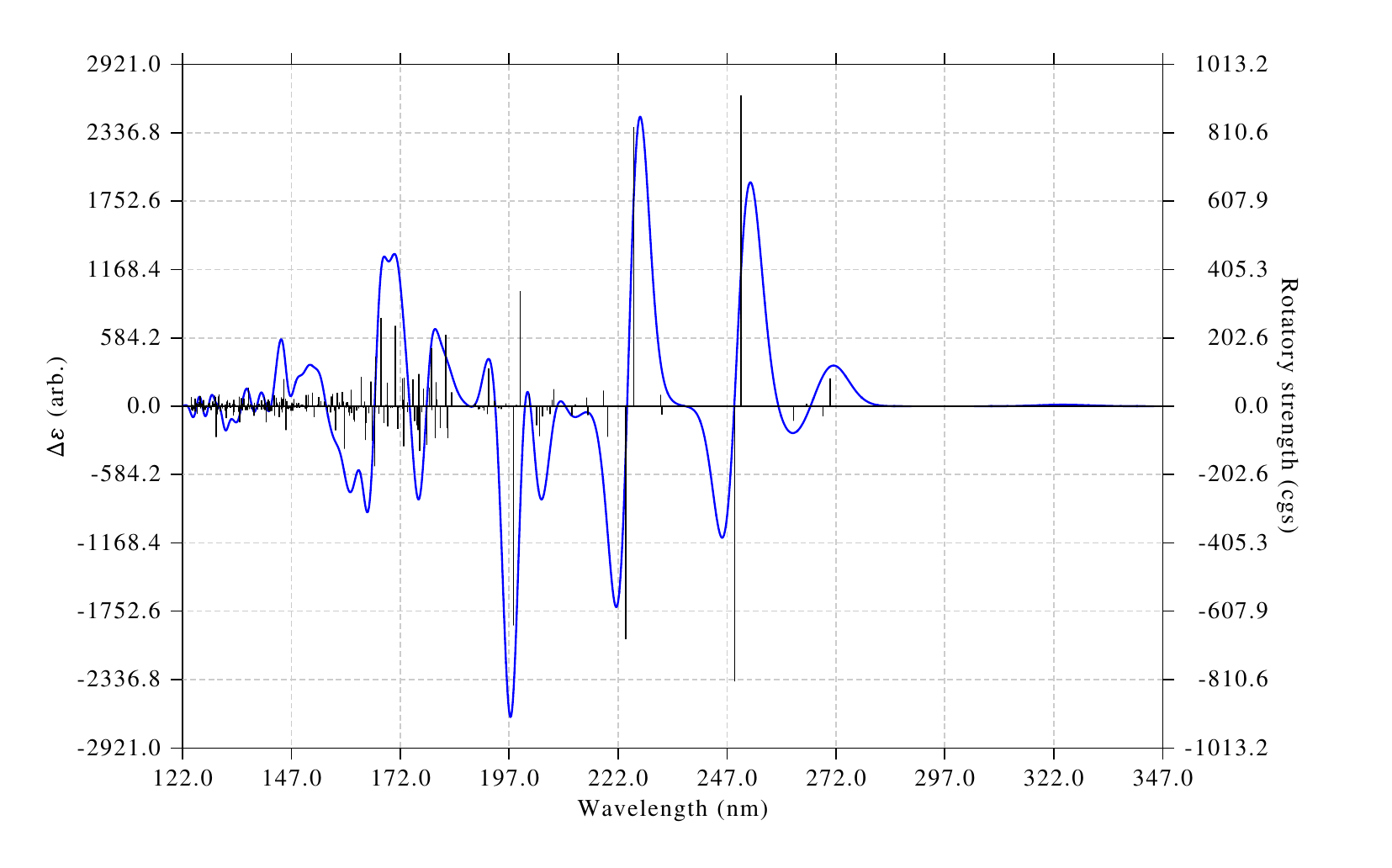}}\\
\subfloat[In toluene solvent: ECD 
spectrum]{\includegraphics[width=0.4\textwidth]{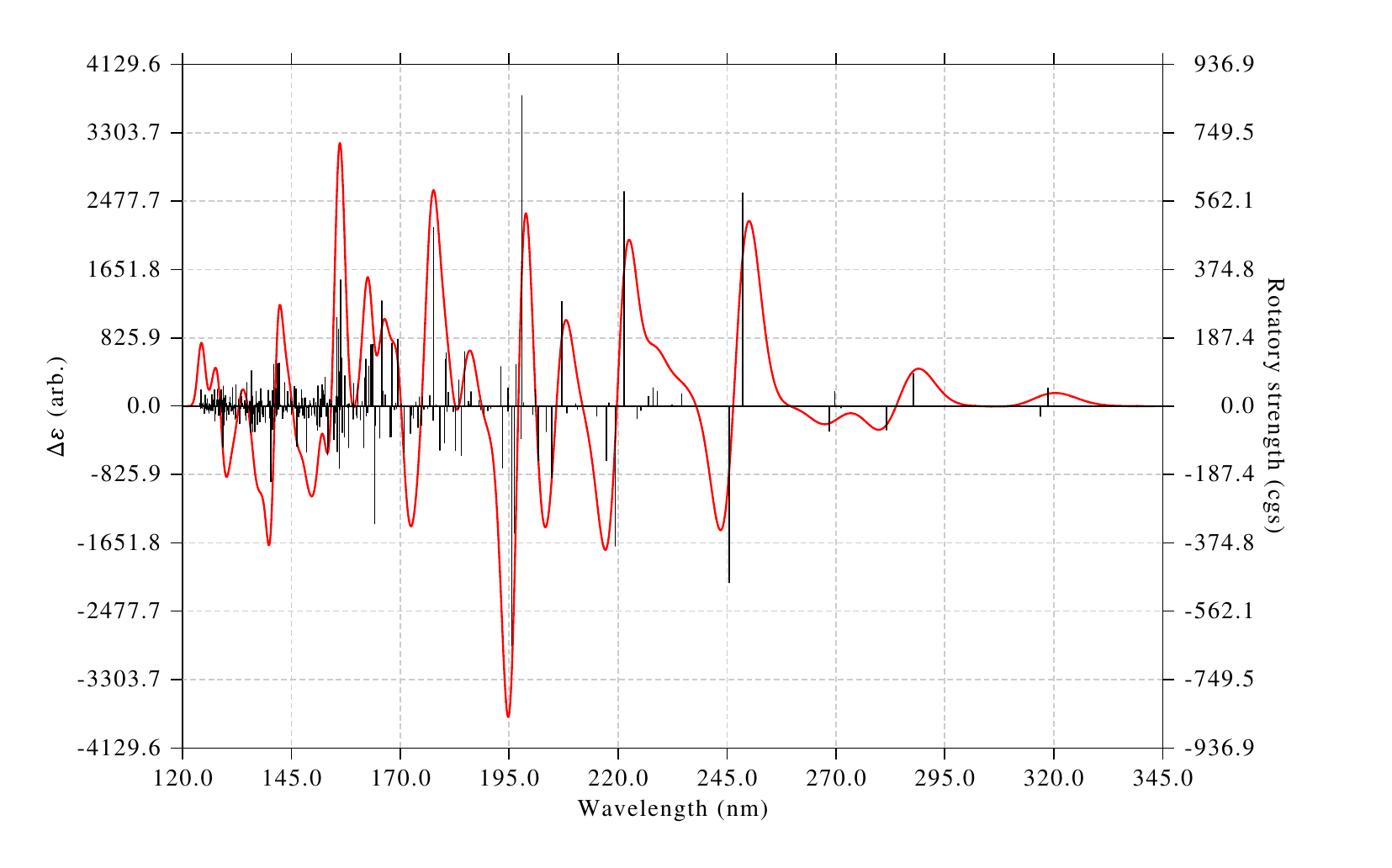}\includegraphics[width=0.4\textwidth]{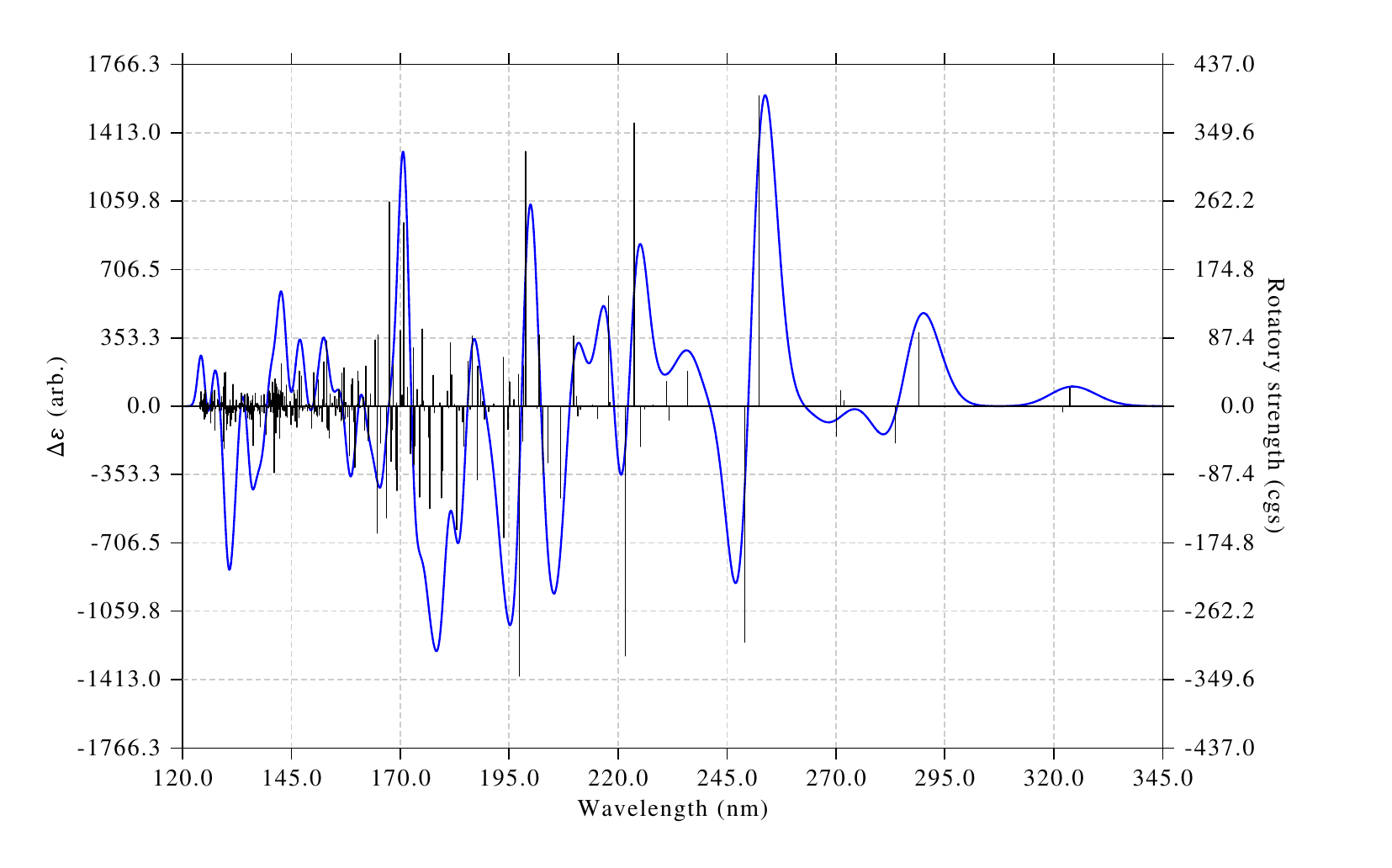}
}
\caption{Electronic Circular Dichroism (ECD) spectra of 2TCz-DPS: solvent and method dependence. Panels show the simulated ECD spectra in vacuum and toluene, 
calculated with \stda (top) and \stddft (bottom) methods. The ECD spectra indicate some changes to planarity occur due to solvation effects.}
\end{figure}

\begin{table}[!htbp]
\centering
\caption{Colorimetric properties of 2TCz-DPS: wavelength maxima and CIE coordinates. This table summarizes the key colorimetric properties of 2TCz-DPS, 
including the wavelengths of maximum absorption and emission, CIE 1931 color space coordinates (X, Y, Z and x, y), and approximate sRGB color representation. 
There does not appear to be a significant difference as a result of using toluene solvent. The low Y values suggest that 2TCz-DPS has a very minimal color 
output and weak light.}
{\scriptsize
\begin{tabular}{m{1.2cm}m{1.8cm}@{\,}l*{3}{@{\,}c@{\,}}>{\columncolor{white}[0\tabcolsep][0pt]}c}
\toprule
&& Properties & {$\lambda_{max}(\unit{\nano\meter})$} & {$(X,Y,Z)$} & {$(x,y)$} & {(R,G,B)} \\
\midrule
\multirow{4}{=}{In vacuum} & \multirow{2}{=}{\emph{UV-vis} absorption} & \stda
&$320.0697$&NA&NA&NA\\
&& \stddft &$325.5298$&NA&NA&NA\\
&\multirow{2}{=}{Fluorescence} & \stda
&$488.7999$&$(14840.653092,74056.693543,184126.398417)$&$(0.0543566388,0.2712463472)$&\CcelCo{white}{0,183,250}\\
&& \stddft &$492.6847$&$(8216.712948,78345.868615,133046.205790)$&$(0.0374152284,0.3567519750)$&\CcelCo{white}{0,255,240}\\
\midrule
\multirow{4}{=}{In toluene\\ solvent} & \multirow{2}{=}{\emph{UV-vis} absorption} & \stda
&$318.6898$&NA&NA&NA\\
&& \stddft &$323.6764$&NA&NA&NA\\
&\multirow{2}{=}{Fluorescence} & \stda
&$444.8775$&$(125617.885782,11316.842073,646441.372508)$&$(0.1603545037,0.0144462437)$&\CcelCo{white}{27,0,255}\\
&& \stddft &$492.3077$&$(9740.380621,87144.787118,153567.584509)$&$(0.0388910904,0.3479490097)$&\CcelCo{white}{0,255,241}\\
\bottomrule
\end{tabular}}
\label{tab:Color2TCz-DPS}
\end{table}


\begin{figure}[!htbp]
\centering
\leavevmode
\subfloat[In vacuum: UV-Vis 
spectrum]{\includegraphics[width=0.4\textwidth]{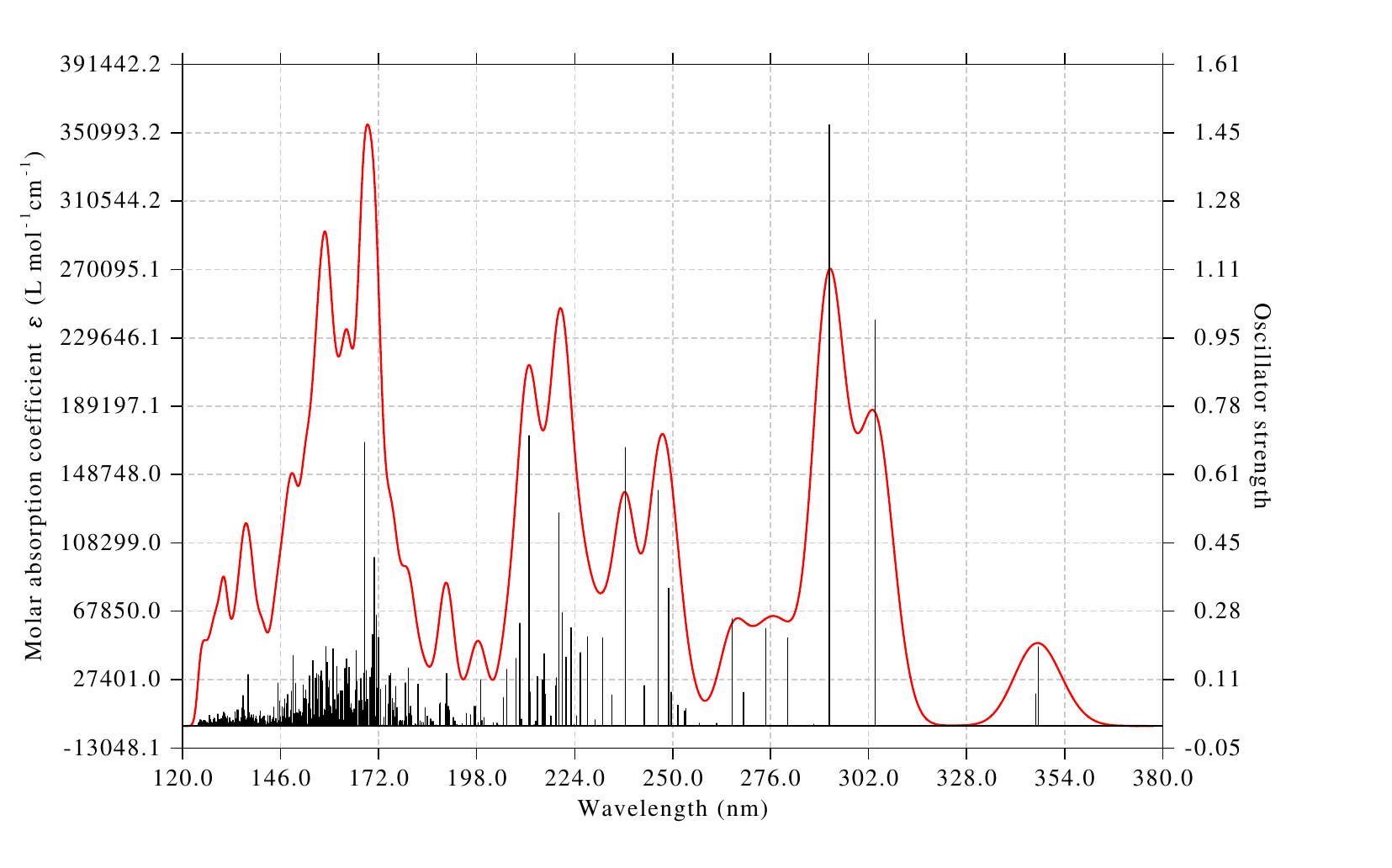}\includegraphics[width=0.4\textwidth]{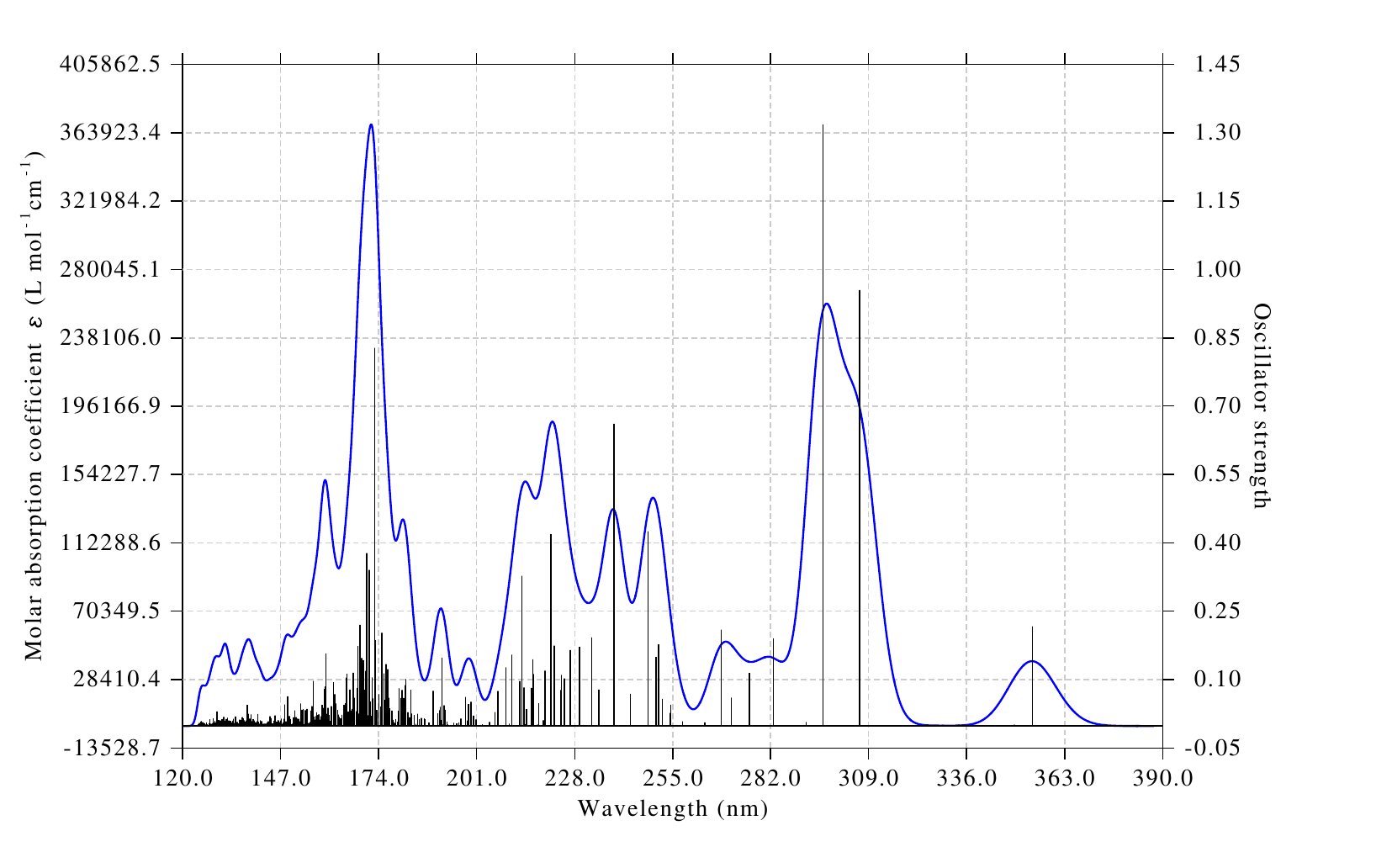}}\\
\subfloat[In toluene solvent: UV-Vis 
spectrum]{\includegraphics[width=0.4\textwidth]{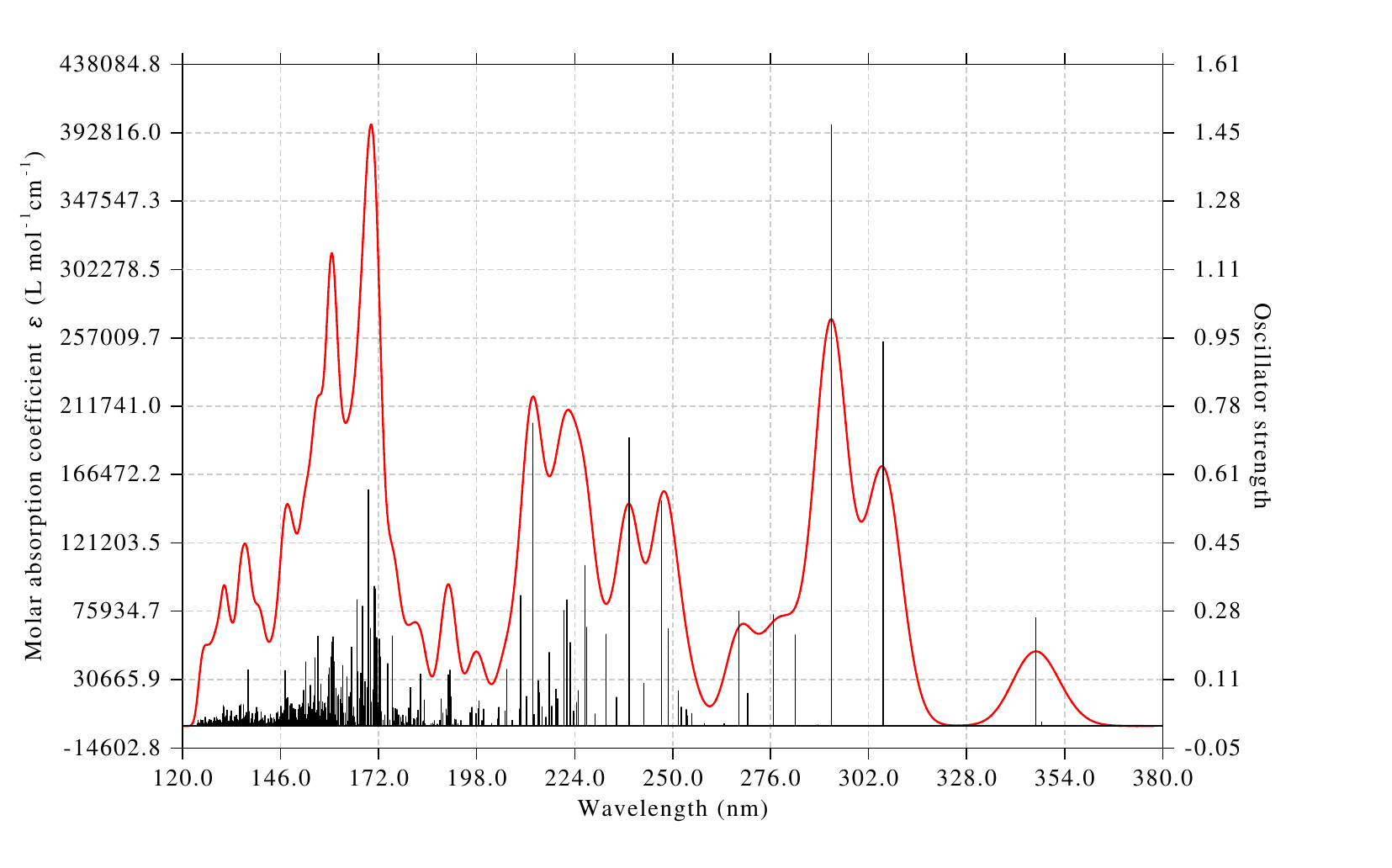}
\includegraphics[width=0.4\textwidth]{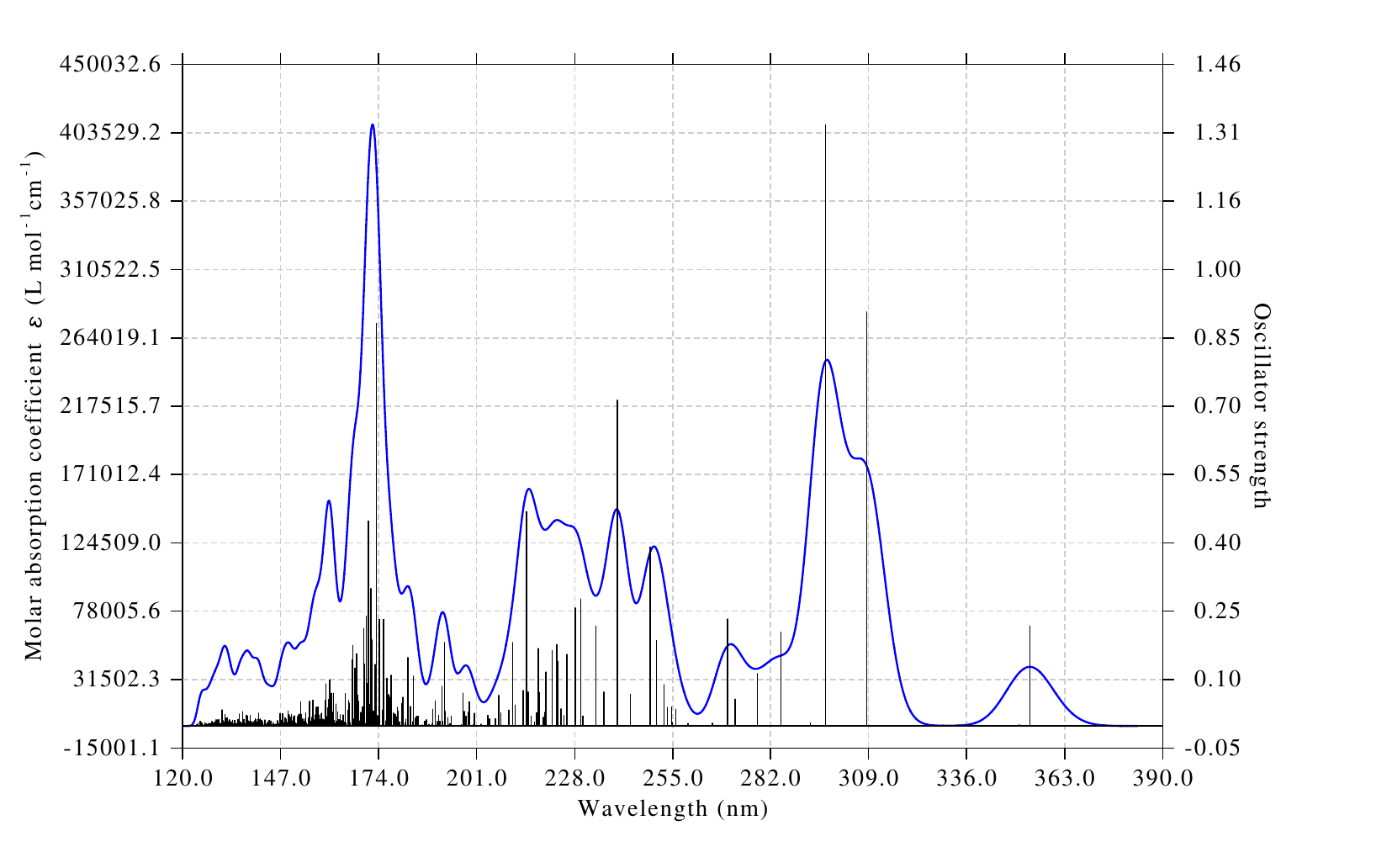}}
\caption{UV-Vis absorption spectra of TDBA-DI: solvent and method dependence. Panels show the simulated UV-Vis absorption spectra in vacuum and toluene, 
calculated with \stda (top) and \stddft (bottom) methods. Note that there are no real changes when a solvent (toluene) is used to perform the computations. This 
shows that TDBA-DI is very robust.}
\end{figure}

\begin{figure}[!htbp]
\centering
\leavevmode
\subfloat[In vacuum: ECD 
spectrum]{\includegraphics[width=0.4\textwidth]{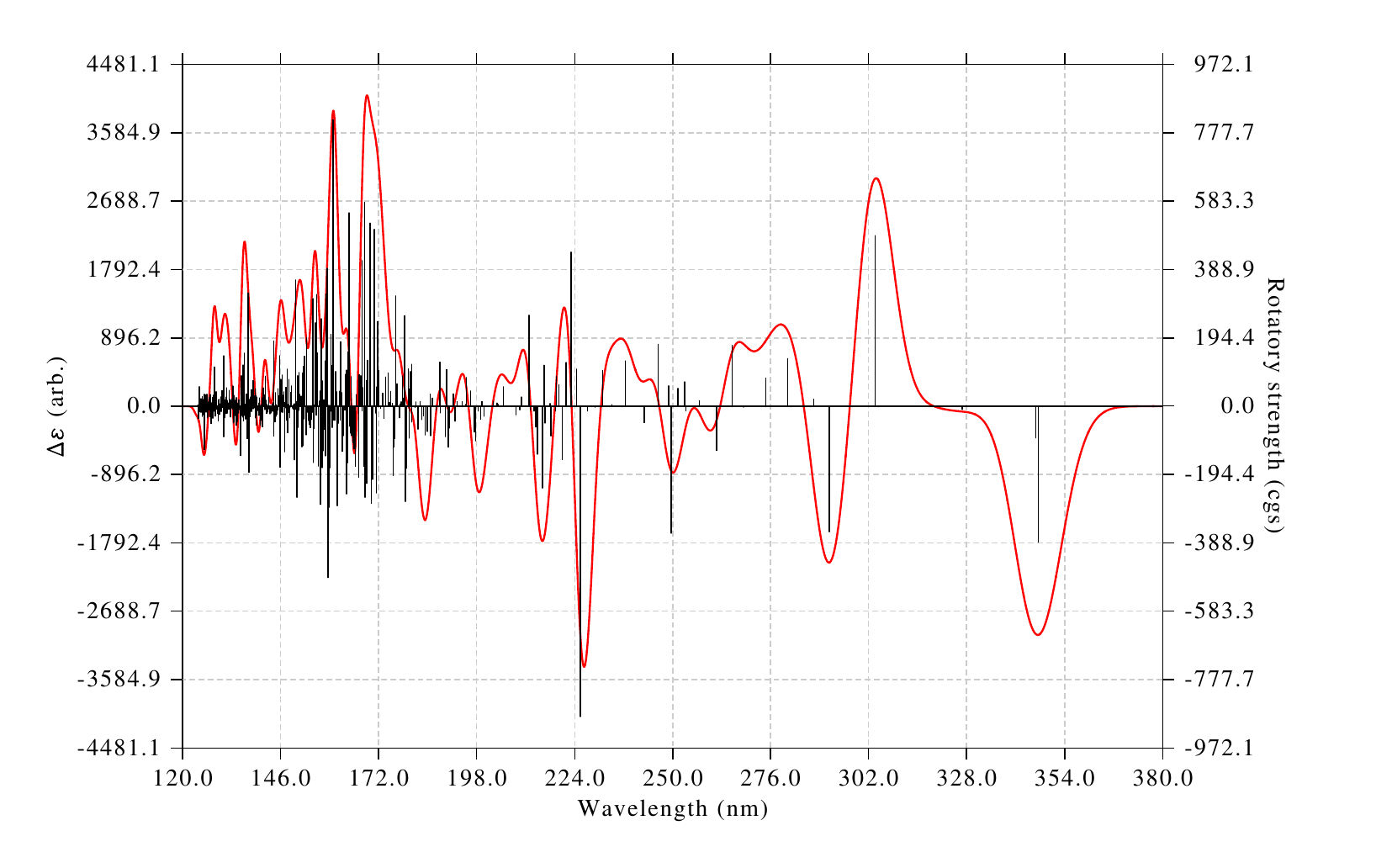}\includegraphics[width=0.4\textwidth]{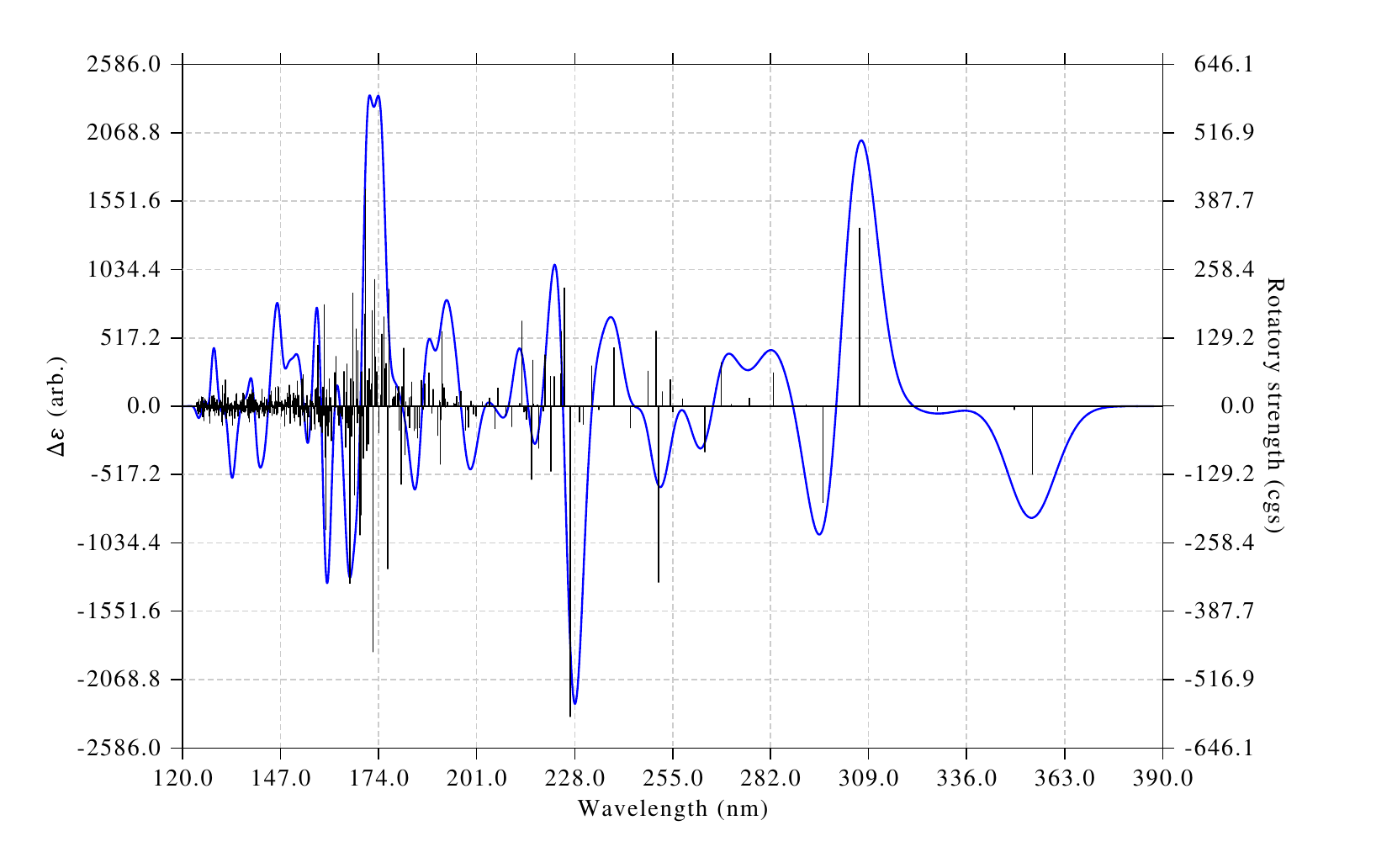}}\\
\subfloat[In toluene solvent: ECD 
spectrum]{\includegraphics[width=0.4\textwidth]{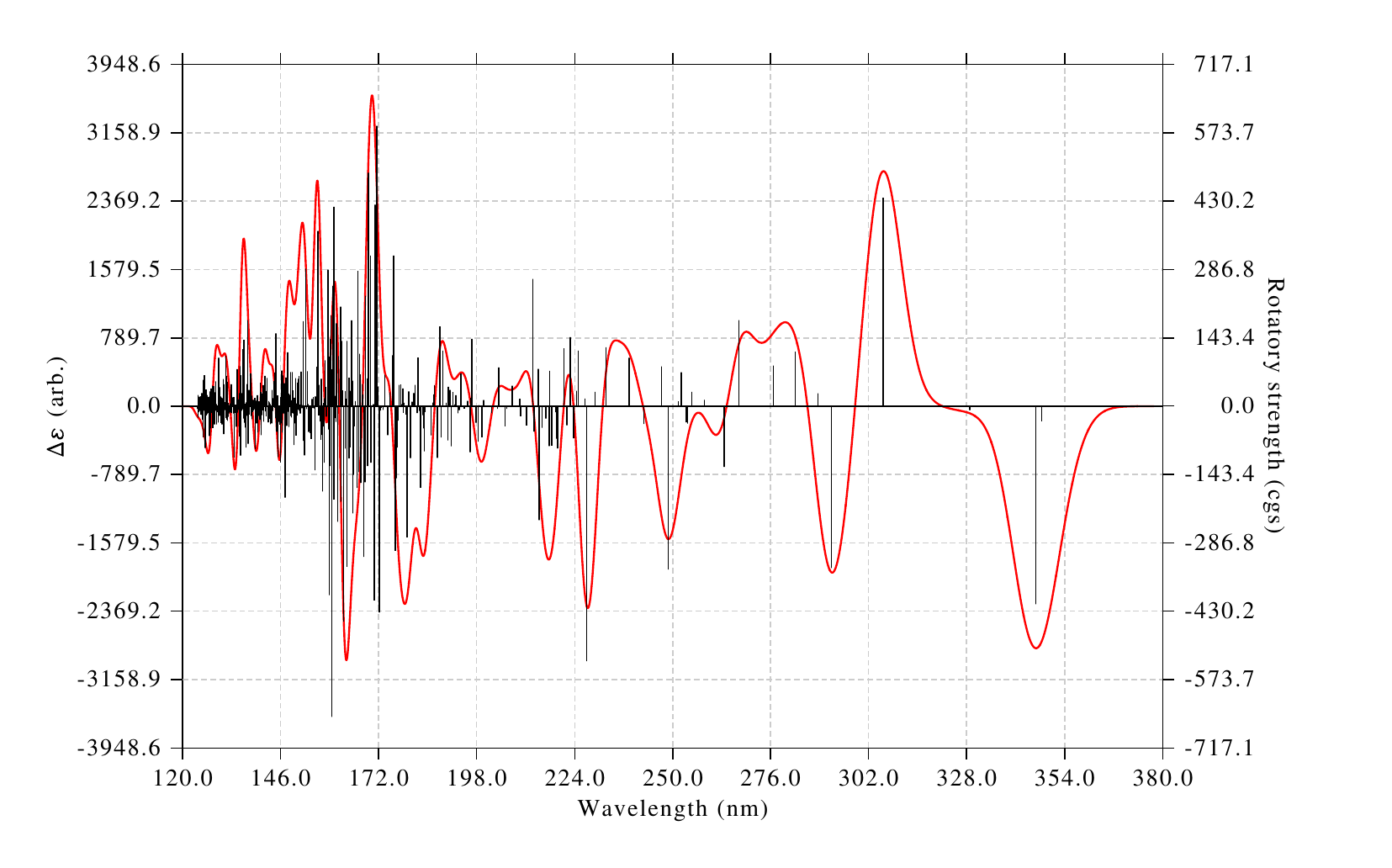}\includegraphics[width=0.4\textwidth]{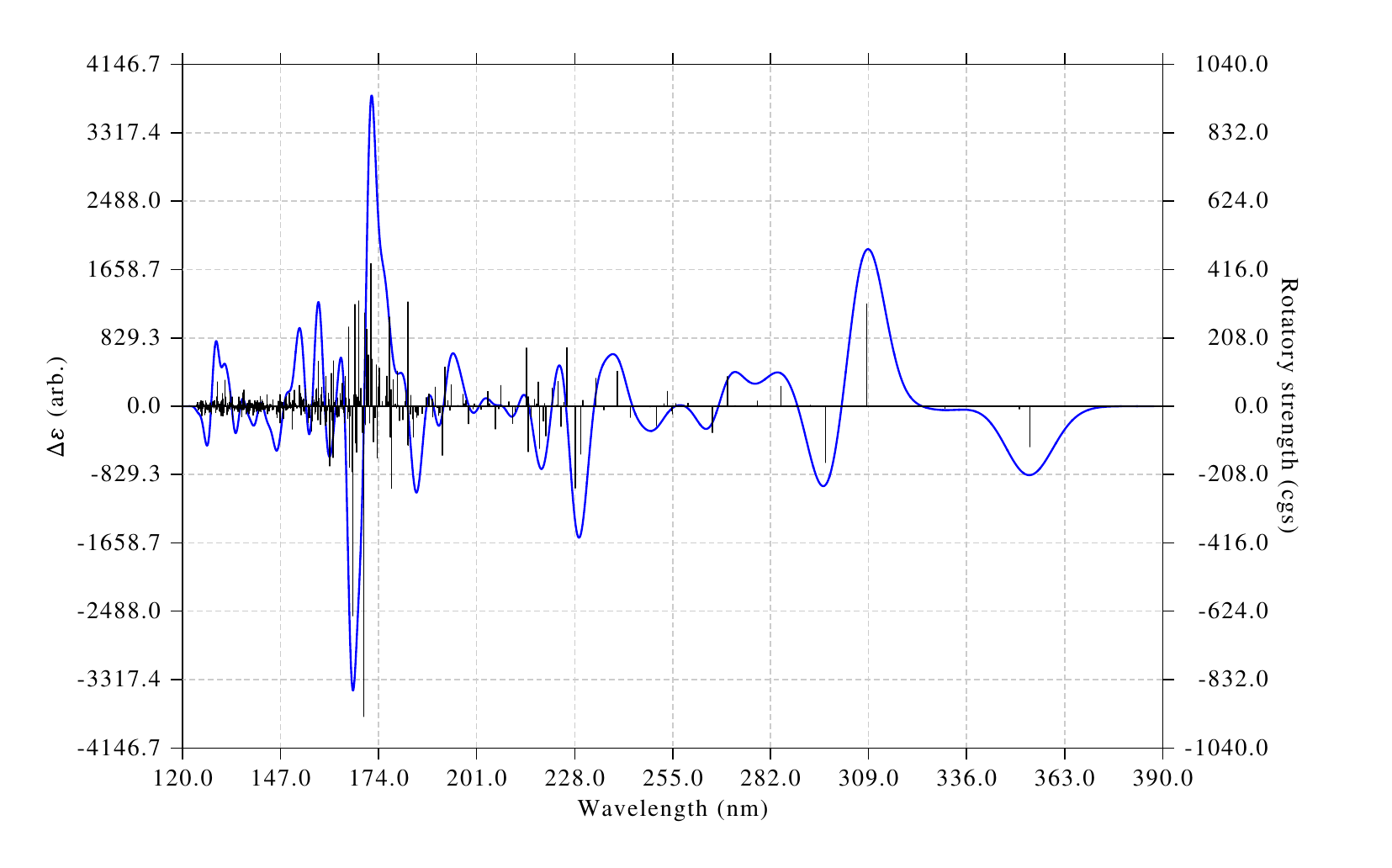}}
\caption{Electronic Circular Dichroism (ECD) spectra of TDBA-DI: solvent and method dependence. Panels show the simulated ECD spectra in vacuum and toluene, 
calculated with \stda (top) and \stddft (bottom) methods. The ECD spectra indicate that the optical activity of TDBA-DI is largely unaffect by the solvent 
environment.}
\end{figure}

\begin{table}[!htbp]
\centering
\caption{Colorimetric properties of TDBA-DI: wavelength maxima and CIE coordinates. This table summarizes the key colorimetric properties of TDBA-DI, including 
the wavelengths of maximum absorption and emission, CIE 1931 color space coordinates (X, Y, Z and x, y), and approximate sRGB color representation. The table 
indicates that TDBA-DI does not have intense visible light in either a vacuum or with toluene solvent. The data suggests that it has a relatively stable spectra 
under different conditions.}
{\scriptsize
\begin{tabular}{m{1.2cm}m{1.8cm}@{\,}l*{3}{@{\,}c@{\,}}>{\columncolor{white}[0\tabcolsep][0pt]}c}
\toprule
&& Properties & {$\lambda_{max}(\unit{\nano\meter})$} & {$(X,Y,Z)$} & {$(x,y)$} & {(R,G,B)}\\
\midrule
\multirow{4}{=}{In vacuum} & \multirow{2}{=}{\emph{UV-vis} absorption} & \stda
&$347.0120$&$(2.598548,0.078137,12.143628)$&$(0.1753368927,0.0052722678)$& \CcelCo{white}{47,0,255}\\
&& \stddft &$354.0573$&$(28.284089,0.848486,132.320941)$ & $(0.1751841007,0.0052552933)$&\CcelCo{white}{47,0,255}\\
&\multirow{2}{=}{Fluorescence} & \stda
&$407.9651$&$(24649.659973,704.505081,117790.587971)$&$(0.1722009326,0.0049216270)$&\CcelCo{white}{43,0,255}\\
&& \stddft &$415.0671$&$(62255.917001,1945.132272,299424.739677)$&$(0.1712087506,0.0053492693)$&\CcelCo{white}{42,0,255}\\
\midrule
\multirow{4}{=}{In toluene\\ solvent} & \multirow{2}{=}{\emph{UV-vis} absorption} & \stda
&$347.9288$&$(0.201153,0.006048,0.940124)$&$(0.1753234808,0.0052709907)$&\CcelCo{white}{47,0,255}\\
&& \stddft &$353.4456$&$(24.225603,0.727063,113.317577)$&$(0.1752047508,0.0052582725)$&\CcelCo{white}{47,0,255}\\
&\multirow{2}{=}{Fluorescence} & \stda
&$416.2007$&$(3197.424834,101.075516,15390.166478)$&$(0.1710889740,0.0054083856)$&\CcelCo{white}{42,0,255}\\
&& \stddft &$417.0407$&$(75059.780932,2429.076930,361728.702965)$&$(0.1708943076,0.0055304640)$&\CcelCo{white}{42,0,255}\\
\bottomrule
\end{tabular}}
\label{tab:ColorTDBA-DI}
\end{table}

\subsection{Approximation of the TADF efficiency}

To calculate the TADF  efficiency, we need to consider the radiative and non-radiative decay rates of the singlet and triplet states. The TADF efficiency 
($\eta_{\text{TADF}}$ derivated from \cite{Uoyama2012}) is typically calculated using the following formula:
\begin{equation}
\eta_{\text{TADF}} = \frac{k_{\text{rISC}} \cdot \Phi_{PF}}{k_{\text{rISC}} + k_{\text{ISC}} + k_{\text{nr}}},
\end{equation}
where,
\begin{itemize}
\item $k_{\text{rISC}}$ is the rate constant for reverse intersystem crossing (rISC) from the triplet state ($T_1$) to the singlet state ($S_1$);
\item $k_{\text{ISC}}$ is the rate constant for intersystem crossing (ISC) from the singlet state ($S_1$) to the triplet state ($T_1$);
\item $k_{\text{nr}}$ is the non-radiative decay rate;
\item $\Phi_{PF}$ is the prompt fluorescence quantum yield (efficiency of the singlet state emitting light).
\end{itemize}
The non-radiative decay rate ($k_{\text{nr}}$) can be calculated as:
     \begin{equation}
     k_{\text{nr}} = \frac{1-f_{12}}{\tau}.
     \end{equation}
Because we don't have spin-orbit coupling (SOC) data, we can estimate $k_{rISC}$ using the \emph{energy gap law} and the \emph{singlet-triplet energy gap} 
($\Delta E_{ST}$). A common approximation is:
\begin{equation}
k_{rISC} \approx A \cdot \exp\left(-\frac{\Delta E_{ST}}{k_B T}\right),
\end{equation}
where,
\begin{itemize}
 \item  $A$ is a pre-exponential factor (typically assumed to be $10^6 - 10^9 \, \text{s}^{-1}$ for organic molecules);
\item $\Delta E_{ST}$ is the singlet-triplet energy gap (in eV);
\item $k_B$ is the Boltzmann constant ($\qty{8.617e-5}{\electronvolt\per\kelvin}$);
\item $T$ is the temperature (in Kelvin, typically $\qty{300}{\kelvin}$ for room temperature).
\end{itemize}
For simplicity, we assume $A = \qty{1e7}{\per\second}$ and $\Phi_{PF} \approx 1$ $ (100\%)$.

We also assume that $k_{ISC}$ is similar to $k_{rISC}$ or slightly larger, depending on the molecule. For simplicity, we assume:
\begin{equation}
k_{ISC} \approx 10 \cdot k_{rISC}.
\end{equation}
This is a rough approximation, as $k_{ISC}$ is often faster than $k_{rISC}$ due to stronger spin-orbit coupling in the ISC process.

So \Cref{tab:EffiGas,tab:EffiTol} give us the approximated TADF efficiency values.

\begin{table}[!htbp]
 \centering
\caption{Calculated TADF emitters efficiencies in vacuum. Energies are given in \unit{\electronvolt} and radiative lifetimes in \unit{\nano\second}. Results are 
shown for both the Tamm-Dancoff Approximation (TDA) and Time Dependent Density Functional Theory (TD-DFT) within a simplified scheme (\stda and \stddft, 
respectively).}
\begin{adjustbox}{angle=90}
{\scriptsize
\begin{tabular}{l*{8}{@{ }S@{ }}}
\toprule
 Molecule & {DMAC-TRZ} & {DMAC-DPS} & {PSPCz} & {4CzIPN} & {Px2BP} & {CzS2} & 
{2TCz-DPS} & {TDBA-DI} \\
\midrule
$\Delta E_{ST}$ (\stda) & 0.097000 & 0.162500 & 0.120000 & 0.212000 & 0.338000 & 0.040500 & 0.077500 & 0.481000 \\
$f_{12}(S_0-S_1)$ (\stda) & 0.145800 & 0.020700 & 0.188900 & 0.152400 & 0.103300 & 0.351200 & 0.151400 & 0.194100 \\
$\tau$ (ns) (\stda) & 33.044831 & 148.121242 & 11.675089 & 24.590050 & 39.689274 & 8.815978 & 23.659157 & 12.855337 \\
$k_{nr}$ (\stda) & 25849731.395222 & 6611475.736268 & 69472703.817911 & 34469226.053696 & 22593005.642654 & 73593650.710875 & 35867719.374620 & 62689915.585706 
\\
$k_{rISC}$ (\stda) & 234678.695522 & 18626.163407 & 96403.129521 & 2745.117895 & 20.982474 & 2087520.539319 & 498952.101859 & 0.083093 \\
$k_{ISC}$ (\stda) & 2346786.955220 & 186261.634068 & 964031.295209 & 27451.178949 & 209.824738 & 20875205.393190 & 4989521.018594 & 0.830932 \\
$\eta_{TADF}$ (\stda) & 0.008254 & 0.002733 & 0.001367 & 0.000080 & 0.000001 & 0.021620 & 0.012065 & 0.000000 \\
$\Delta E_{ST}$ (\stddft) & 0.103000 & 0.162500 & 0.072000 & 0.191000 & 0.316000 & 0.026500 & 0.057500 & 0.429000 \\
$f_{12}(S_0-S_1)$ (\stddft) & 0.140000 & 0.019200 & 0.159100 & 0.137100 & 0.085500 & 0.309500 & 0.130600 & 0.219100 \\
$\tau$ (ns) (\stddft) & 36.015672 & 174.136364 & 14.011889 & 29.114825 & 52.684333 & 9.190989 & 22.719594 & 11.852942 \\
$k_{nr}$ (\stddft) & 23878493.455873 & 5632367.519303 & 60013320.674868 & 29637822.211727 & 17358101.366097 & 75127928.690406 & 38266528.452045 & 
65882376.071604 \\
$k_{rISC}$ (\stddft) & 186070.798829 & 18626.163407 & 617240.876844 & 6185.076174 & 49.140547 & 3587733.081291 & 1081542.693405 & 0.621049 \\
$k_{ISC}$ (\stddft) & 1860707.988290 & 186261.634068 & 6172408.768436 & 61850.761741 & 491.405465 & 35877330.812907 & 10815426.934046 & 6.210489 \\
$\eta_{TADF}$ (\stddft) & 0.007177 & 0.003191 & 0.009240 & 0.000208 & 0.000003 & 0.031308 & 0.021560 & 0.000000 \\
\bottomrule
\end{tabular}}
\end{adjustbox}
\label{tab:EffiGas}
\end{table}

\begin{table}[!htbp]
 \centering
\caption{Calculated TADF emitters efficiencies in toluene solvent. Energies are given in \unit{\electronvolt} and radiative lifetimes in \unit{\nano\second}. 
Results are shown for both the Tamm-Dancoff Approximation (TDA) and Time Dependent Density Functional Theory (TD-DFT) within a simplified scheme (\stda and 
\stddft, respectively).}
\begin{adjustbox}{angle=90}
{\scriptsize
\begin{tabular}{l*{8}{@{ }S@{ }}}
\toprule
 Molecule & {DMAC-TRZ} & {DMAC-DPS} & {PSPCz} & {4CzIPN} & {Px2BP} & {CzS2} & 
{2TCz-DPS} & {TDBA-DI} \\
\midrule
$\Delta E_{ST}$ (\stda) & 0.080000 & 0.242000 & 0.134000 & 0.217000 & 0.320000 & 0.232500 & 0.465000 & 0.458000 \\
$f_{12}(S_0-S_1)$ (\stda) & 0.065600 & 0.033500 & 0.212100 & 0.137500 & 0.100000 & 0.389000 & 0.156400 & 0.010600 \\
$\tau$ (ns) (\stda) & 73.042835 & 91.525663 & 10.713852 & 27.722247 & 41.770622 & 8.041629 & 23.268278 & 243.665284 \\
$k_{nr}$ (\stda) & 12792493.618432 & 10559879.778070 & 73540308.516424 & 31112196.221094 & 21546243.829253 & 75979632.232784 & 36255368.363384 & 4060488.156321 
\\
$k_{rISC}$ (\stda) & 452960.941838 & 860.168995 & 56092.108407 & 2262.376068 & 42.096212 & 1242.157224 & 0.154295 & 0.202278 \\
$k_{ISC}$ (\stda) & 4529609.418382 & 8601.689945 & 560921.084065 & 22623.760680 & 420.962122 & 12421.572238 & 1.542955 & 2.022776 \\
$\eta_{TADF}$ (\stda) & 0.025483 & 0.000081 & 0.000756 & 0.000073 & 0.000002 & 0.000016 & 0.000000 & 0.000000 \\
$\Delta E_{ST}$ (\stddft) & 0.108000 & 0.267000 & 0.134000 & 0.198000 & 0.302000 & 0.134000 & 0.196500 & 0.421000 \\
$f_{12}(S_0-S_1)$ (\stddft) & 0.063600 & 0.033100 & 0.178100 & 0.124800 & 0.083800 & 0.339900 & 0.149500 & 0.220800 \\
$\tau$ (ns) (\stddft) & 77.243232 & 99.113076 & 12.517078 & 32.496201 & 54.618954 & 8.979713 & 24.304962 & 11.809206 \\
$k_{nr}$ (\stddft) & 12122744.968175 & 9755524.104466 & 65662290.130933 & 26932378.701461 & 16774396.600655 & 73510144.072858 & 34992854.935097 & 
65982421.544256 \\
$k_{rISC}$ (\stddft) & 153349.378188 & 327.041428 & 56092.108407 & 4717.917787 & 84.455775 & 56092.108407 & 4999.760828 & 0.846291 \\
$k_{ISC}$ (\stddft) & 1533493.781879 & 3270.414276 & 560921.084065 & 47179.177871 & 844.557748 & 560921.084065 & 49997.608276 & 8.462910 \\
$\eta_{TADF}$ (\stddft) & 0.011105 & 0.000034 & 0.000846 & 0.000175 & 0.000005 & 0.000757 & 0.000143 & 0.000000 \\
\bottomrule
\end{tabular}}
\end{adjustbox}
\label{tab:EffiTol}
\end{table}

\subsection{Details on the computational resources}

Calculations were performed on a high-performance computing cluster consisting of nodes with dual Intel Xeon Gold 6130 CPUs (2.10 GHz), providing a total of 64 
physical cores and 128 threads. Each node was equipped with 92 GB of RAM.  OpenMP parallelization was employed for the \xtb, \crest, \stda, and \stddft 
calculations, utilizing 8 threads per job. The \tda calculations were performed with a single thread.

\end{document}